\documentclass[a4paper,12pt,oneside,openright]{report}

\usepackage{xcolor,sectsty}
\usepackage{pdfpages}
\usepackage[nodayofweek]{datetime}
\usepackage{amsmath}
\usepackage{amssymb}
\usepackage{natbib}
\usepackage{fancyhdr}
\usepackage{setspace}
\usepackage{footnote}
\usepackage{graphicx}
\usepackage{epstopdf} 
\usepackage{lscape}
\usepackage{rotating}
\usepackage{subfigure}
\usepackage{caption}
\usepackage{wrapfig,lipsum,booktabs}
\usepackage{afterpage}
\usepackage{pifont}
\usepackage{longtable}
\usepackage{latexsym}
\usepackage{amsfonts}
\usepackage{verbatim}
\usepackage[printonlyused]{acronym}
\usepackage{epigraph}
\usepackage{pifont}
\usepackage[lmargin=3.0cm,rmargin=2.3cm]{geometry}
\usepackage{color}
\usepackage{url}
\usepackage{float}
\usepackage{nicefrac}
\usepackage[section]{placeins}
\usepackage{multicol,multirow}
\usepackage[titletoc]{appendix}
\usepackage{pdflscape}

\usepackage{geometry}
\usepackage[colorinlistoftodos]{todonotes}
\usepackage[
	pdfauthor={H. F. Stevance},
	pdftitle={Specpol of CCSNe and WR stars}, 
	breaklinks=true, 
	colorlinks   = true, 
	urlcolor     = blue, 
	linkcolor    = black, 
	citecolor    = blue] 
	{hyperref}
\usepackage{afterpage}

\newcommand\blankpage{%
    \null
    \thispagestyle{empty}%
    \addtocounter{page}{-1}%
    \newpage}

\newcommand{\comments}[1]{}

\newcommand{\kms}{\ensuremath{\textrm{\,km s}^{-1}}}

\newcommand{\apj}{Astrophysical Journal}                                         
\newcommand{\apjs}{Astrophysical Journal Supplements}
\newcommand{\aap}{Astronomy and Astrophysics}
\newcommand{\aaps}{Astronomy and Astrophysics Supplements}
\newcommand{\mnras}{Monthly Notices of the Royal Astronomical Society}
\newcommand{\aj}{Astronomical Journal}
\newcommand{\apss}{Astrophysics and Space Science} 
\newcommand{\nat}{Nature} 
\newcommand{\pasp}{Publications of the Astronomical Society of the Pacific}
\newcommand{\araa}{Annual Review of Astronomy and Astrophysics}

\newcommand{\apjl}{Astrophysical Journal Letters}

\newcommand{\na}{New Astronomy}

\definecolor{sections}{RGB}{46,116,181} 

\chapterfont{\color{sections}}
\sectionfont{\color{sections}}
\subsectionfont{\color{sections}}
\subsubsectionfont{\color{sections}}

\def \cbat {CBAT}
\def \aj {AJ}
\def \mnras {MNRAS}
\def \pasp {PASP}
\def \apj {ApJ}
\def \apjs {ApJS}
\def \apjl {ApJL}
\def \aap {A\&A}
\def \nat {Nature}
\def \araa {ARAA}
\def \iaucirc {IAUC}
\def \aaps {A\&A Suppl.}
\def \apss {Ap\&SS}
\def \na {New Astronomy}
\def \pasa {PASA}

\newcommand{\msol} {M$_{\odot}$}
\newcommand{\lsol} {L$_{\odot}$}
\newcommand{\mza} {M$_{ZAMS}$}

\newcommand{\rsol} {R$_{\odot}$}
\newcommand{\about} {$\sim$}
\newcommand{\metal}{12 + log(O/H) }

\def\lesssim{\mathrel{\hbox{\rlap{\hbox{\lower4pt\hbox{$\sim$}}}\hbox{$<$}}}}
\def\gtrsim{\mathrel{\hbox{\rlap{\hbox{\lower4pt\hbox{$\sim$}}}\hbox{$>$}}}}

\newcommand{\ang} {$\mathrm{\AA}$}
\newcommand{\degree}{$^{\circ}$}
\newcommand{\deps}{$\Delta \epsilon\,$}

\newcommand{\halpha} {$\mathrm{H\alpha}$ }
\newcommand{\hbeta} {$\mathrm{H\beta}\,$}
\newcommand{\cmark}{\ding{51}}%
\newcommand{\xmark}{\ding{55}}%

\long\def\symbolfootnote[#1]#2{\begingroup%
\def\thefootnote{\fnsymbol{footnote}}\footnote[#1]{#2}\endgroup} 

\title{\textsc{Spectropolarimetry of Core Collapse Supernovae and their progenitors}}
\author{H. F. Stevance}

\begin{document}




%


\thispagestyle{empty}

\begin{center}
\fontsize{24.88}{57.6}
\vspace*{-1cm}

\textbf{Spectropolarimetry of stripped envelope core collapse supernovae and their progenitors}\\
\vspace*{2.5cm}
\LARGE
\text{H. F. Stevance}\\
\vspace{2cm}
\Large{Department of Physics \& Astronomy}\\
\Large{The University of Sheffield}\\
\vspace*{1cm}

\includegraphics[width=0.4\textwidth]{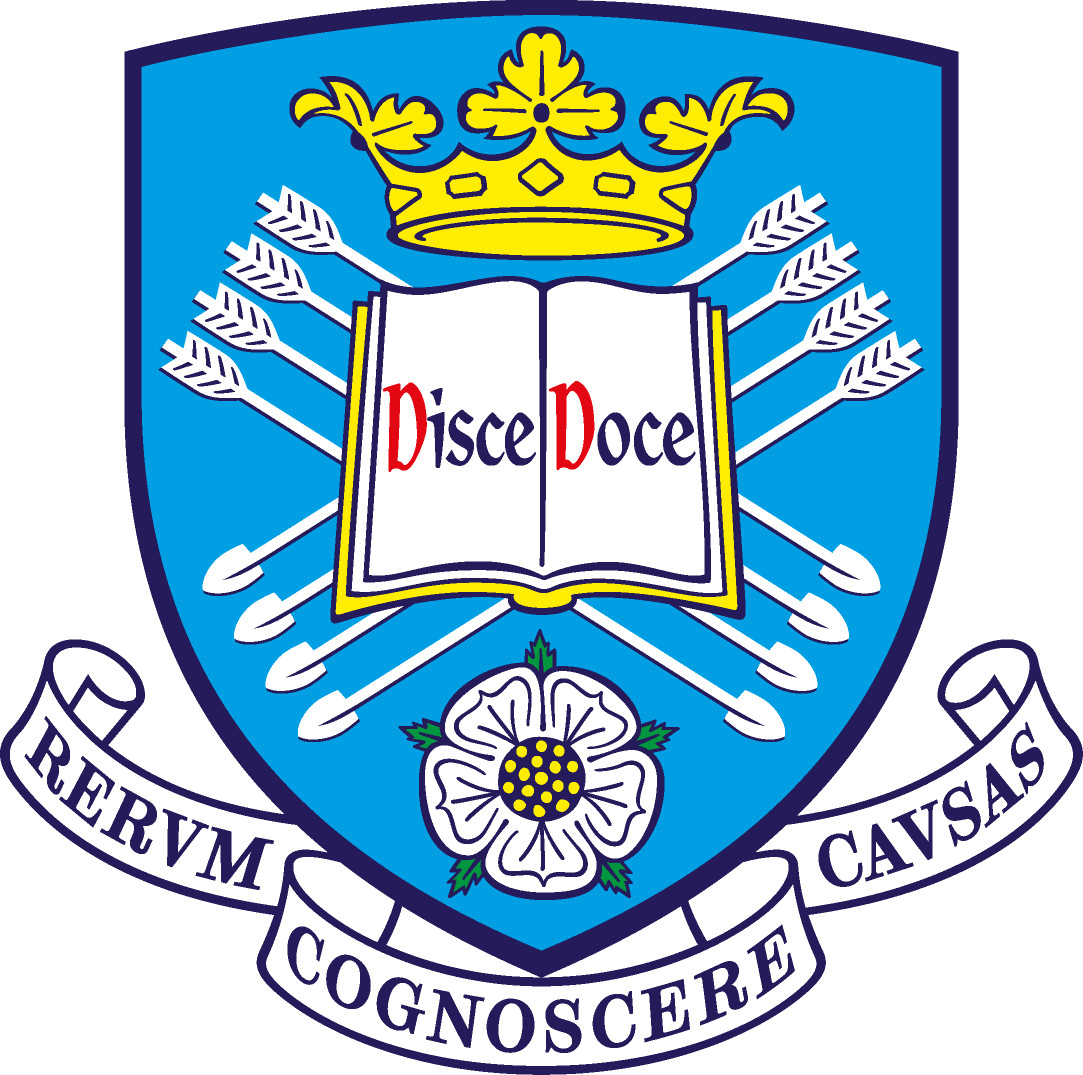}\\

\vspace*{1cm}

\large
{\it A dissertation submitted in candidature for the degree of}\\
{\it Doctor of Philosophy at the University of Sheffield}\\
\vspace*{1.5cm}
{April 2019}
\vfill
\end{center}
\pagenumbering{roman}

\afterpage{\blankpage}

\normalsize 

\onehalfspacing


\newgeometry{top=12cm, left=5cm, right=4cm}  
\vspace{4cm}
\begin{center}
{\Huge \noindent \textbf{\textcolor{sections}{``}}}\\
\vspace{0.6cm}
{\huge There is no problem so bad that}

{\vspace{0.5cm}\huge you can't make it worse.}\\

{\vspace{0.8cm}\Huge \noindent \textbf{\textcolor{sections}{"}}}\\

\end{center}

{\hspace{7cm}\large --- Chris Hadfield}

\restoregeometry 

\tableofcontents \clearpage
\listoffigures \clearpage
\listoftables \clearpage
\chapter*{Declaration}
\label{chpt:declaration} 
I declare that, unless otherwise stated, the work presented in this thesis is my own. 
No part of this thesis has been accepted or is currently being submitted for any other qualification at the University of Sheffield or elsewhere.

Much of the work presented here has already been published and can be found in \cite{stevance16}, \cite{stevance17}, \cite{stevance18} and \cite{stevance19}, see Appendix \ref{app:papers}. \clearpage

\chapter*{Acknowledgements}
I need to thank a great number of people for their support, this may take a while.

First and foremost, I need to thank the two people without whom this thesis could not have come to an end: Paul  and Clive. 
Under your guidance I learnt a lot more than science and have grown so much as a person. 
I am  more patient, understanding and calm in difficult situations than I ever could have been when I started three and a half years ago. 
This is because you lead by example and were great mentors I could look up to and rely on when I needed advice. 
There are no words for how grateful I am to have met you, and I want to thank you for your support and guidance over the past few years.

I am also grateful to the pastoral support of Richard, who was a reassuring and calming presence in the bleakest of times. Many thanks to Simon as well, I think I still owe you a Pinot Grigio; come collect.
Katie also needs a special mention. Her support ranged from proof-reading emails to double-checking my maths to being an amazing friend who was always kind and welcoming when I needed to talk. I wish I was more like you. 

I want to emphasise the support of all my collaborators, particularly Dietrich Baade, Craig Wheeler, James Bruten, Peter Hoflich, Lifan Wang, Aleksandar Cikota, Yi Yang, Jason Spyromilio, Ferdinando Patat and Richard Ignace. 

Many thanks to Pablo (The Computer Whisperer) for his saint-like patience. Thank you for fixing my messes, and for taking the time to explain even the most basic things to me.

Thank you Martin for teaching me how to read the Python manual when we were fresh PhD students. Also I wouldn't have a colour-scheme if it wasn't for you, and I will definitely be keeping in touch, if only to carry on our discussions on graphic design.    

I want to thank Liam for all the interesting conversations we had about statistics. I never did get to use MCMC in the end, but I learnt a lot.

I am also thankful to Gemma, for the great conversations about machine learning, for her help with Gaia data, and for the great time we had in Bariloche.  

I want to thank Stuart for something he said to me when I was a mere undergraduate; something that followed me throughout my PhD. I was telling him during my summer project that I felt like I had no idea what I was doing. He told me that everybody feels that way. Needless to say, as an undergraduate, I was not expecting this from a teacher who I thought was supposed to know everything. Whenever imposter syndrome starts setting in, I remember this.

I also need to thank: He is the reason I came to Sheffield as an undergraduate, he is the reason I came back, and he is also the reason Steven started working in Sheffield. 

Finally I want to thank my family. 
Thanks to Steven, my loving and amazing partner. You are my rock and you were always there for me in the worst of times. 

Et pour finir, evidemment, merci a Moman et a Jeremy\footnote{(Bravo pour ton Master... maintenant j'attends ta these!)}. Sortez le Champagne!\\

{\vspace{1cm} \hspace{6cm}Je vous aime tous fort -- I love you all }

\clearpage

\chapter*{Summary}
Over the past 30 years, spectropolarimetry has proven to be a great tool to probe the 3D geometry of core collapse supernovae (CCSNe).
The number of high-quality multi-epoch spectropolarimetric data sets for CCSNe remains quite low, however, due to the challenging nature of such observations.
In this work we present and analyse the data of two Type IIb SNe (SN 2008aq -- two epochs, SN 2011hs -- seven epochs) and one Type Ic-bl (SN 2014ad -- seven epochs).
The latter is the most complete spectropolarimetric data set for a Type Ic-bl to date.

SN 2011hs was found to have a geometry consistent with an off-axis energy source within an ellipsoidal envelope, and features similar to SN 2011dh.
In SN 2014ad we showed the presence of intermediate mass elements in outer parts of the ejecta as well as significant axi-symmetry, consistent with the presence of a jet. 
We also provided a re-analysis of the Type IIb SN 1993J, including a novel estimate of its interstellar polarisation. 
Diverse asymmetries were found in all SNe, both in the global geometry of the photosphere and in the distribution of the line forming regions.
They are discussed in the context of previous studies.

In order to simulate the geometry of the ejecta and the resulting observations, we re-created a pre-existing toy model, improving on previous work by devising a way to explore parameter space methodically. 
We found that such toy models are prone to degeneracies, and warn that they should be used with great caution, if at all. 

Spectropolarimetry can also be used to probe asymmetries caused by fast rotation in the potential progenitors of stripped envelope CCSNe.  
Although numerous Wolf-Rayet stars of type WN and WC have been observed with spectropolarimetry, no single WO stars have been studied to date. 
In our data of two Galactic WO stars, we found no line effect and could not sufficiently constrain the rotational velocity to exclude a collapsar scenario. 
Therefore, the absence of a line effect may not necessarily equate to a low rotational velocity. \cleardoublepage



\pagenumbering{arabic} \setcounter{page}{1}
\renewcommand{\chaptermark}[1]{\markboth{\bf{ #1}}{}} 
\renewcommand{\sectionmark}[1]{\markright{\thesection\emph{ #1}}{}} 

 \pagestyle{fancy}			 
 \fancyfoot{}				 
 \renewcommand{\chaptermark}[1]{	 
   \markright{{\bf\chaptername\ \thechapter.}\ #1}{}} %
 \renewcommand{\sectionmark}[1]{	 
   \markright{\thesection.\ #1}}	 %
 \fancyhead[RE,RO]{\bfseries\thepage}	 
\fancyhead[LE]{\bfseries\rightmark}	 
\fancyhead[LO]{\bfseries\rightmark}	 
 \renewcommand{\headrulewidth}{0.3pt}	 


%




\begin{onehalfspace}                                         


\chapter{Introduction}

\label{chpt:intro} 
\lhead{\emph{Introduction}} 

\section{Overview}
Core collapse supernovae (CCSNe) are the swansong of massive stars (M$_{ZAMS}\gtrsim$8 \msol), whose cores collapse once they have exhausted all their fuel and can no longer compete against gravity. 
Most of the energy of CCSNe (of order 10$^{53}$ erg) is released in the form of neutrinos; $\sim$ 1 percent of the total energy serves to accelerate the ejecta (kinetic energy), and only $\sim$ 0.01  percent is converted into electro-magnetic radiation  \citep{smartt09}. 
Still, they can be as bright as a billion suns and sometimes even outshine their host galaxy.

The explosive deaths of massive stars enrich the interstellar medium with the intermediate and heavy elements that were created in the progenitors or the supernovae themselves. 
Some of these elements (e.g. oxygen) are crucial to form planets and give rise to life as we know it.
The shock and energy released in the explosion causes positive and negative feedback with the nearby interstellar medium, playing a crucial role in star formation and in shaping galaxies (e.g.  \citealt{hensler11, scannapieco08}).

A variety of theoretical models (e.g. neutrino heating, jet-driven ; \citealt{janka12,woosley93}) have been developed, but understanding the explosion mechanism of SN is still work in progress. 
It has been known for nearly two decades that CCSNe are not spherically symmetric explosions, as evidenced by the high velocity kick of certain pulsars (e.g. \citealt{lyne94, cordes98, chatterjee05}), the asphericity of young SN remnants \citep{manchester87}, or the production of gamma ray bursts (GRBs) in certain broad-lined type Ic supernovae (Ic-bl SNe: e.g. \citealt{patat01, chornock10}).
Additionally, analytical models and simulations have shown that effects necessary for successful explosions such as magnetic fields, rotation and various instabilities, yield non-spherical explosions (e.g. \citealt{blondin03, kotake04, burrows06, takiwaki16}).
Hence, characterising the geometry of supernovae is crucial to testing and constraining explosion models.

In 1982, \citeauthor{ss82} were the first to propose polarimetry as a way to probe the geometry of the unresolved SNe envelopes, and, nearly 25 years later, spectropolarimetric data seemed to demonstrate that virtually all CCSNe were aspherical \citep{WW08}.
The work is ongoing to better understand the 3D geometry of SN ejecta through spectropolarimetry and this thesis focuses on stripped envelope SNe and their potential progenitors.

This chapter offers an introduction to the classification of SNe and their progenitors with an emphasis on stripped envelope CCSNe.
Additionally we provide some background on polarimetry, and how it allows us to probe the 3D geometry of SNe and asymmetry in Wolf-Rayet (WR) star winds.
 

\section{Supernova classification, explosion mechanisms and progenitors}
\subsection{Classification}
\label{introsec:classification}
The classification of SNe is an observational taxonomy that has been refined and extended over time as objects with different spectral and light-curve characteristics were observed.
Initially, SNe were divided into type I and type II according to whether their spectra was hydrogen poor or hydrogen rich (mainly Balmer lines), respectively \citep{minkowski41}. 

\begin{figure}
\centering
\includegraphics[width=14cm]{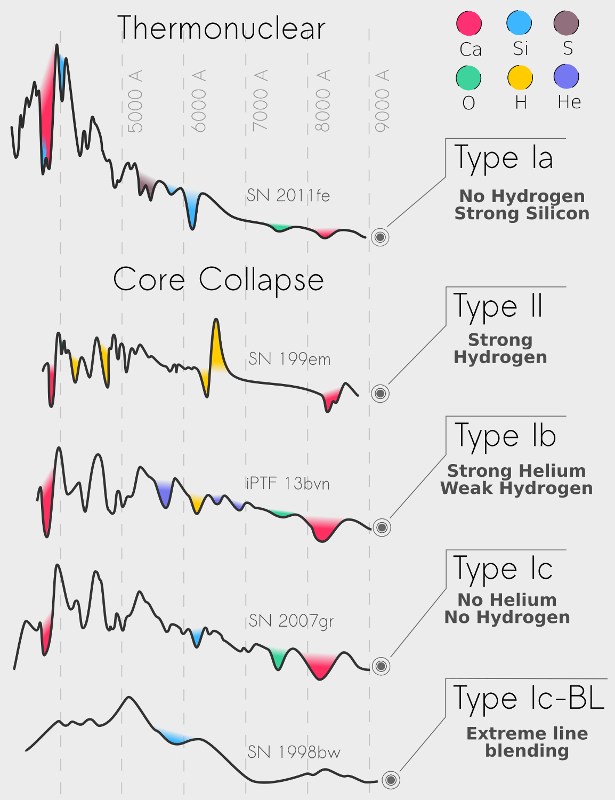}
\caption{Typical supernova spectra by types.}
\label{introfig:sn_pectra}
\end{figure}

Type I SNe are sub-divided into the Ia, Ib and Ic categories. 
The spectra of type Ia SNe are characterised by strong Si\,{\sc ii} $\lambda 6355$ and Ca\,{\sc ii} H\&K $\lambda\lambda 3934, 3968$.
Type Ib SNe, on the other hand, do not show strong silicon features, but exhibit significant helium ($\lambda 5876$, $\lambda 6678$, $\lambda 7065$) lines. 
Lastly, type Ic SNe show no helium, and silicon absorption features are much less prominent than in type Ia SNe \citep{filippenko97}. 
Additionally, some type Ic SNe exhibit very broad and blended spectral features, owing to very high ejecta speeds (e.g. SN~1997X, see \citealt{munari98}; SN~1997ef,  see \citealt{iwamoto00}; or SN~2002ap, see \citealt{mazzali02}), and are labelled broad-lined type Ic (Ic-bl).
Figure \ref{introfig:sn_pectra} gives examples of SN spectra by sub-types. 

Some of these Ic-bl SNe have also been associated with long Gamma-Ray Bursts (GRBs) and X-Ray Flashes.
SN~1998bw \citep{patat01} was the first convincing candidate for the association of a SN~to a subluminous long GRB (GRB980425). 
Since then, many other examples have been studied with various GRB energies -- e.g. SN~2003dh / GRB030329, a cosmologically normal GRB\citep{stanek03}; SN~2006aj / XRF060218, \citep{sollerman06}; SN~2010bh / GRB100316D, \citep{chornock10}.

\begin{figure}
\centering
\includegraphics[width=16cm]{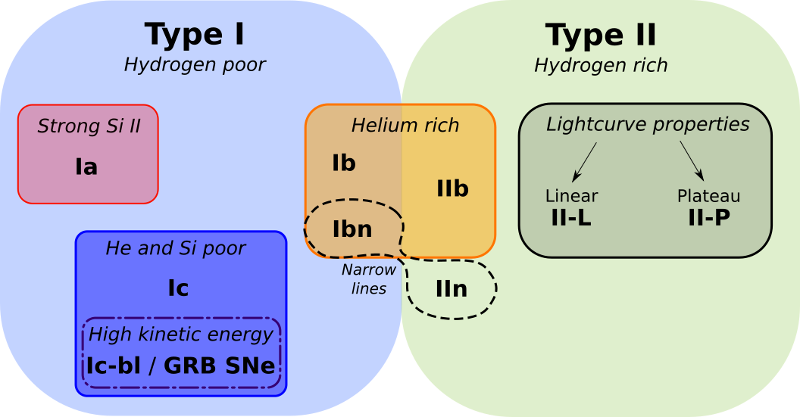}
\caption{Schematic representation of the supernova classification.}
\label{introfig:sn_classification}
\end{figure}

Hydrogen rich (type II) SNe are further divided into the type II-P, II-L, IIb and IIn. 
The type II-P and II-L are defined according to whether their lightcurve shows a plateau or decay linearly (in magnitudes), respectively \citep{filippenko97}.
In type IIb SNe, hydrogen features are present at early days, but diminish over time as helium lines gain in strength and the spectrum appears as a type Ib at late times \cite{filipenko87}. 
Finally, the spectra of type IIn SNe exhibit narrow (few hundred km\,s$^{-1}$) hydrogen emission lines, caused by interaction of the supernova shock with circumstellar medium (CSM -- \citealt{schlegel90}).
A summary of the SN classification is given in  Figure \ref{introfig:sn_classification}.

\subsection{Thermonuclear supernovae}
White dwarfs are the stripped, degenerate remnants of low and intermediate mass stars (Zero Age Main Sequence mass (M$_{ZAMS}$) $< $ 8 \msol). 
If they gain sufficient mass to reach the Chandrasekhar mass, M$_{ch} = 1.44$ \msol, the density and temperature in their interior reaches the point of carbon ignition which eventually leads to a thermonuclear explosion (and supernova).
Sub-Chandrasekhar mass white dwarfs can also undergo an explosion under the double-detonation scenario, whereby the helium layer surrounding the white dwarf reaches the point of ignition.
This detonation then triggers a second detonation in the interior of the star, causing the supernova \citep{woosley94}.
Thermonuclear SNe mainly produce iron-group elements, and are associated with type Ia SNe. 

Two main channels are preferred to explain how the progenitors of type Ia SNe gain sufficient mass to yield an explosion: the single and double degenerate channels.
In the single degenerate channel, mass is transferred to the white dwarf from a less evolved companion \citep{han04}, either through Roche-lobe overflow or stellar winds.
The double-degenerate channel involves the merging of two white dwarfs whose combined masses exceeds M$_{ch}$, resulting in an explosion \citep{webbink84}.

\newpage
\subsection{Core collapse supernovae}
\label{introsec:ccsne}

\subsubsection{Popular explosion mechanisms}
\label{introsec:explosion}
Most stars are supported against gravity by the outward pressure resulting from nuclear fusion. 
In massive stars (M$_{ZAMS}\gtrsim$10 \msol) fusion can continue until the core is composed of iron.
At this point, since the fusion of elements more massive than iron is not an exothermic process, the core is no longer supported against gravity.
As a result, the core collapses to a neutron star or black hole, and a supernova explosion ensues from the release of gravitational potential. 
Note that not all core collapse events result in an explosion; if the energy released is not sufficient to unbind the envelope of the star, it will collapse back onto the compact object. 
This is called a failed supernova, and is beyond the scope of this work.
The details of how energy is imparted to the outer envelope in order to result in a successful explosion are not fully understood. 
It was initially thought that the bounce of the forming neutron star launched a shock wave through the in-falling outer layers causing the explosion; however, this shock is insufficient to unbind the stellar material. 
One of the most popular ways to revive the shock is through tapping of the energy released in the form of neutrinos as the neutron star forms (neutrino heating; for a review see \citealt{janka12}).

Furthermore, an alternative model involving jets has been proposed in order to explain the existence of GRB-SNe: the collapsar model. 
In this scenario, the core of the massive star collapses to a black hole and accretion onto the black hole taps the rotational energy of the star via magnetic coupling, resulting in collimated jets  which power the explosion and yield the GRB (e.g. \citealt{woosley93,woosleymacfadyen99, callingham18}). 
Consequently this model requires the progenitor star to retain a high level of angular momentum for jet production (j $> 3\times10^{16} {\rm ergs/cm}^2$ -- \citealt{woosley93}).
Recent studies have shown that for central engines of short enough lifetime, the jets may fail to break out of the envelope of the progenitor, but would impart sufficient energy to drive a rapid expansion, hence yielding SNe Ic-bl without GRB counterparts (\citealt{bromberg11}, \citealt{lazzati12}). 

\subsubsection{Progenitors}
\begin{figure}
\centering
\includegraphics[width=12cm]{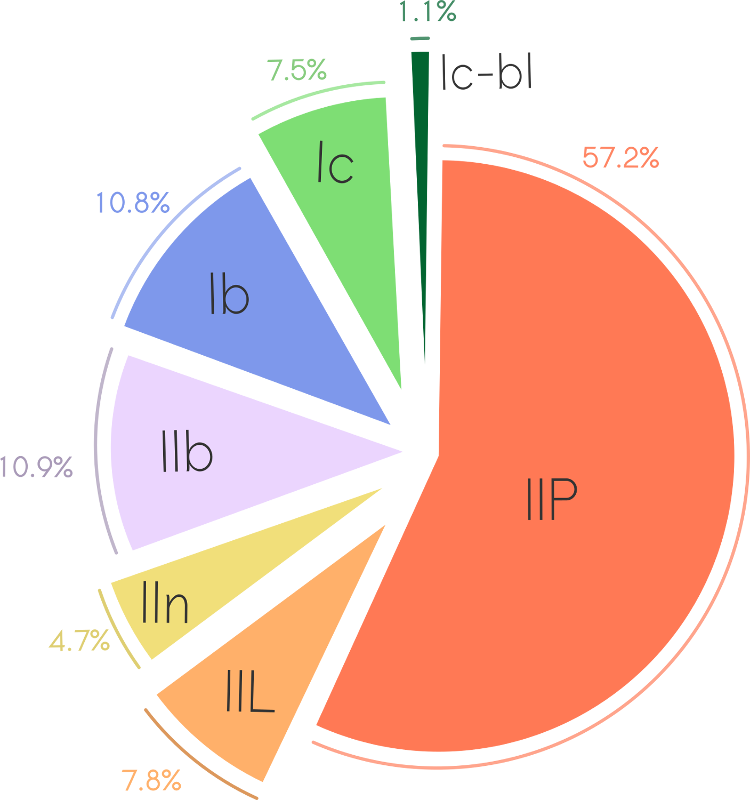}
\caption{Relative numbers of core collapse SNe as calculated from table 3 of \cite{shivvers17} and fig. 9 of \cite{li11}.}
\label{introfig:ccsn_rate}
\end{figure}

Amongst CCSNe, type II-P are the most prevalent, representing about 57 percent of the population by volume (see Figure \ref{introfig:ccsn_rate} -- \citealt{shivvers17}).
Their (single star) progenitors are red supergiants which would have started their lives with masses between 8 and \about30 \msol \citep{smartt09}.
Direct evidence as been found through the analysis of pre-images revealing  red supergiant stars at the location of the SN (e.g. \citealt{vandyk03, vandyk12}), and in some cases the late images taken after the SN event showed that the suspected progenitor had disappeared, like in the case of Type II-P SN 2003gd \citep{maund09}.
The presence of hydrogen in type IIn SNe points to progenitors of a similar mass range, although some type IIn SNe are known to have progenitors of higher mass (e.g. SN 2005gl; \citealt{galyam09}).
Type II-L SNe may also have slightly higher ZAMS progenitors masses than most type II-P, around 20\msol, although more data is required \citep{branch17}.

The type IIb, Ib, and Ic(-bl) SNe are collectively called stripped envelope SNe, owing to the nature of their progenitor stars which have been stripped of their outer H-rich envelopes to varying degrees.
Together they represent \about 30 percent of CCSNe.
Type IIb SNe progenitors have lost most of their hydrogen envelope, retaining less than 0.5 \msol\,of hydrogen \citep{smith11}, whereas type Ib SN progenitors have almost entirely lost their hydrogen envelope. 
Furthermore, the stars yielding type Ic SNe must have shed both their hydrogen and their helium layers. 

In a single star model, the best candidates for the progenitors of stripped-envelope SNe are WR stars.
The WR class is named after Wolf and Rayet who identified stars characterised by broad-emission lines \citep{wolf67}, which result from their fast (a few hundred to a few thousand km\,s$^{-1}$), dense (\about 10$^{-5}$\msol/yr) winds.
To reach the WR phase, an O star must have been heavily stripped during the luminous blue variable (LBV) or red super-giant (RSG) phase and in most cases have lost its hydrogen envelope.
This evolution can only be attained by the most massive stars (M$_{ZAMS}>$20-25\msol \, at solar metallicity -- \citealt{crowther07}).

There are three main sub-types of WR stars.
Nitrogen sequence WR stars (WN) are characterised by He\,{\sc i}-{\sc ii} and N\,{\sc iii}-{\sc v} emission lines; these stars are hydrogen poor but helium rich. 
They are anticipated to be the progenitors of type Ib or potentially type IIb SNe.
Carbon sequence WR stars (WC) show prominent C\,{\sc iii}-\,{\sc iv} emission lines, and oxygen sequence WR stars (WO) are defined by their strong O\,{\sc vi} $\lambda\lambda 3811-34$ emission lines. 
Both WC and WO are hydrogen and helium poor, and WO stars in particular are thought to be very close to core helium exhaustion (e.g. \citealt{langer12}, although see \citealt{tramper13}); they are also much rarer than WN and WC stars.
The degree of stripping of WC/WO stars make them candidates for the progenitors of type Ic SNe.
As of yet, there has been no confirmed detection of a WR star progenitor in pre-images of type Ib/c SNe; this could be due to the fact that WR stars are expected to become more compact as their temperature rises at the end of their lives. 
This results in progenitors with final masses greater than about 9 \msol\,  having fainter visual magnitudes ($M_v$ = $-1.5$ to $-2.5$) than most observed WR stars ($M_v \lesssim -4 $; \citealt{yoon12}).

In the case of Ic-bl SNe driven by jets in a scenario akin to that of a collapsar, the progenitor must retain sufficient angular momentum. 
This may be made easier if the progenitor has low metallicity (the winds are line driven, the lower metallicity implies less mass loss and therefore less angular momentum loss --  \citealt{vink01}).
This is supported observationally as Ic-bl SNe are found preferentially in low-metallicity environments, with GRB-SNe being found at lower metallicities than Ic-bl SNe without GRB counterpart (e.g. \citealt{modjaz08,levesquekewleyberger10,graham15}). 

The single star progenitor scenario, however, is difficult to reconcile with observations and stellar evolution.
At solar and sub-solar metallicity the mass loss rate of single WR stars could result in final stellar masses producing SN Ib/c with light-curves too broad and ejecta masses too high to fit observations \citep{yoon15}.
Additionally the WC/WN ratio is between 0.1 and 1.2 (at SMC and Milky Way metallicity, respectively -- \citealt{crowther07}), whereas the Ic/Ib ratio is \about 2 \citep{smartt09}.
\cite{crowther13} also found two Ib/c progenitors that lacked nebular emission, a host cluster or a nearby giant H\,{\sc ii} region, implying a lower mass progenitor. 
Furthermore, the presence of a binary companion to the progenitor of the type IIb SN 1993J was confirmed by \cite{maund93J}.  

\begin{figure}
\centering
\includegraphics[width=14cm]{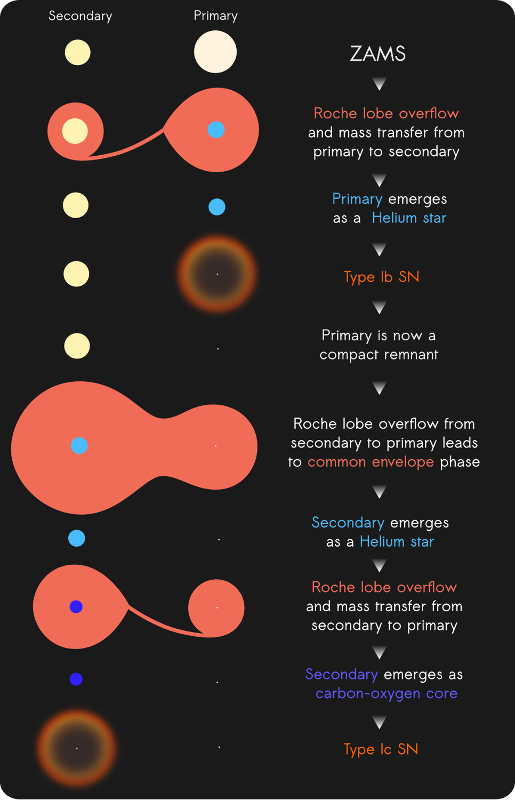}
\caption[Binary evolution to SNe]{Cartoon of a possible binary route to type Ib and type Ic SNe. Inspired by fig. 6 of \cite{yoon15}.}
\label{introfig:binary_route}
\end{figure}

The binary route has received a lot of attention in recent years, since binary interactions can strip donor stars effectively, and we know that most massive stars are in close binaries \citep{sana12}.
Mass transfer in a close binary via Roche Lobe overflow could strip the primary sufficiently to produce a type Ib (M$_{He} >$ 0.14\msol; \citealt{hachinger12}) or a type IIb.
Additionally, the progenitors that follow this evolutionary path could have ZAMS as low as 12.5\msol \, \citep{yoon15}.
It can be difficult to strip the primary star sufficiently to yield a type Ic SNe, but if the progenitor is the secondary star undergoing mass transfer to a compact companion, the mass loss rate could be high enough to strip the helium layers efficiently before the SN explosion \citep{yoon15}. 
An example of a binary route yielding type Ib/c SNe is given in Figure \ref{introfig:binary_route}).
\cite{eldridge08} pointed out that a combination of both single and binary systems shows better agreement with the rate of explosion of CCSNe.
In the past decade, increasing evidence has shown that the binary route is not only important, but dominates the evolution towards stripped-envelope SNe:
\cite{smith11} showed that stripped-envelope SN rates could not be solely  explained by single star evolution, and more recently \cite{prentice19} found that the average ejecta mass of stripped-envelope SN was 2.8$\pm1.5$\msol, which suggests that a majority of these SN arose from lower mass progenitors with \mza $<$25\msol.

\section{Polarisation}
The main focus of this thesis is the use of spectropolarimetry to gain insight into the geometry of CCSNe and WR star winds. 
It is therefore important to lay out the basics of polarimetry, and this section details the formalism that is used in this work, as well as the physical origin of polarisation in SNe and WR stars.
Additionally, the interpretation of spectropolarimetric data is briefly introduced, along with a short summary of the relevant literature.  

\subsection{Formalism}
\label{introsec:formalism}
\begin{figure}
\includegraphics[width=15cm]{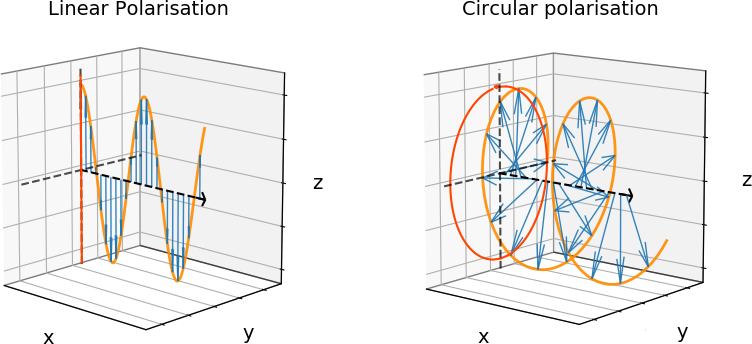}
\caption[Linear and Circular polarisation.]{Sketch of the oscillations of the electric field (blue arrows) in the cases of linear (left) and circular (right) polarisation. The waves are travelling in the +x direction. The dark orange trace shows the projection of the 3D electric field trace (light orange) onto the 2D y-z plane. }
\label{introfig:pol}
\end{figure}

Polarisation is a general property of transverse waves which characterises the orientation of oscillations. 
In the case of electro-magnetic radiation, it describes the orientation of the electric vector. 
There are two main types of polarisation: linear and circular. 
In the case of linear polarisation, the electric vector oscillates in one direction, and in the case of circular polarisation the direction of the electric vector rotates in the plane orthogonal to the direction of propagation (see Figure \ref{introfig:pol}).

In a homogeneous isotropic medium, the electric field vector $\vec{E}$ must verify
\begin{equation}\label{introeq:wave_equation}
\triangledown^2 \vec{E} - \frac{\varepsilon \mu}{c^2} \ddot{\vec{E}} = 0,
\end{equation}
where $\varepsilon$ is the dielectric permittivity, $\mu$ is the magnetic permeability and $c$ is the speed of light in a vacuum. 
The planewave solutions to Eq.\ref{introeq:wave_equation} are of the form
\begin{equation}
E_j =  A_j \; \text{e}^{-\text{i}(\omega t-\delta_j)},
\end{equation}
where j corresponds to the Cartesian components of the electric vector (x, y, or z), $A_j$ is the complex amplitude of each component and $\delta_j$ is the phase factor.
The complex amplitude has the form $A_j = a_j \text{e}^{\text{i}\vec{k}\vec{r}}$, where $a_j$ is the real amplitude, $\vec{k}$ is the wave vector and $\vec{r}$ is a given point in space. 
Therefore, if the light propagates in the +z$-$direction the components of the electric vector $\vec{E}$ are
\begin{equation}
E_x =  A_x\; \text{e}^{-\text{i}(\omega t-\delta_x)}, \; E_y =  A_y \; \text{e}^{-\text{i}(\omega t-\delta_y)}, \; E_z=0
\end{equation}

Detectors, however, are sensitive to the electromagnetic energy of visible light rather than its electric field. 
The polarisation tensor (also called coherency matrix), which characterises the energetic and vectorial properties of light is therefore considered.
For light propagating in the +z$-$direction, it takes the form
\begin{equation}
\textbf{C} \equiv \begin{pmatrix}
E_xE_x^* & E_xE_y^*\\
E_yE_x^* & E_yE_y^*
\end{pmatrix} = 
\begin{pmatrix}
a_x^2 & a_xa_y\text{e}^{\text{i}\delta}\\
a_xa_y\text{e}^{-\text{i}\delta} & a_y^2
\end{pmatrix},
\end{equation}
where $\delta \equiv \delta_x - \delta_y$ is the constant phase difference between the x and y components of the electric vector. 
$\textbf{C}_{01}$ and $\textbf{C}_{10}$ are complex quantities though, and are therefore not measurable.
It is however possible to measure real linear combinations of the coherency matrix components: The Stokes parameters
\begin{equation}\label{introeq:stokes_param}
\begin{split}
&I \equiv \kappa(\textbf{C}_{00}+\textbf{C}_{11}) = \kappa(a_x^2+a_y^2), 
\\
&Q \equiv \kappa(\textbf{C}_{00}-\textbf{C}_{11}) = \kappa(a_x^2-a_y^2),
\\
&U \equiv \kappa(\textbf{C}_{01}+\textbf{C}_{10}) = 2\kappa a_xa_y\cos \delta,
\\
&V \equiv i\kappa(\textbf{C}_{10}-\textbf{C}_{01}) = 2\kappa a_xa_y\sin \delta,
\end{split}
\end{equation}
where $\kappa$ is a dimensional constant that gives intensity units to the Stokes parameters, and cancels out when considering normalised Stokes parameters (see Section \ref{datredsec:linear_pol}).
Note that the equations presented here describe the idealised case of mono-chromatic waves.
In practice light is quasi-monochromatic, but the concepts described here translate to the quasi-monochromatic case.
Specifics on how the Stokes parameters are retrieved through observations are given in Section \ref{datredsec:specpol_datred}.

\begin{figure}
\centering
\includegraphics[width=8cm]{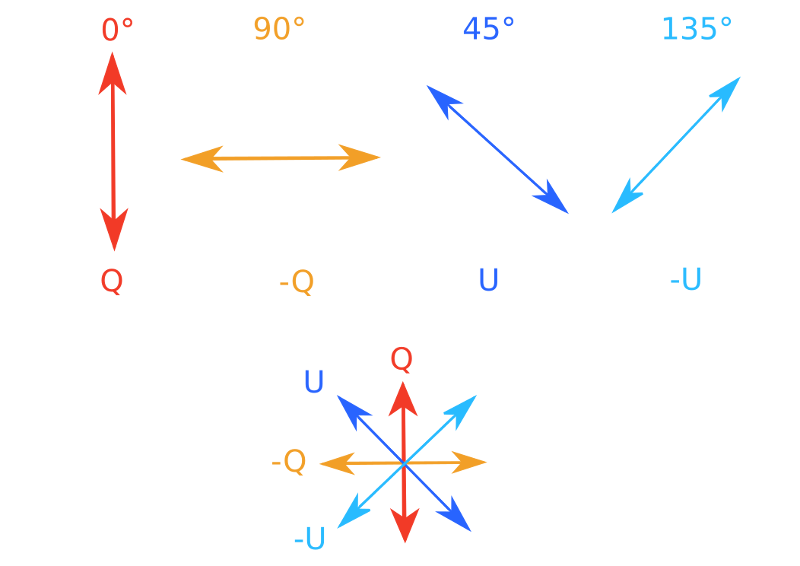}
\caption[Stokes parameters]{Schematic representation of the Stokes parameters, following the convention that the +Q direction is defined by the North-South orientation. }
\label{introfig:stokes_param}
\end{figure}

In the case of linear polarisation (most relevant in this thesis), $\delta$ will be zero and consequently $V$ will be zero too.
Therefore the Stokes parameters of interest when describing the direction of linear polarisation  are $Q$ and $U$, and $I$ is the intensity. 
By their nature, $Q$ and $U$ are quasi vectors, meaning that their 0$^{\circ}$ and 180$^{\circ}$ are identical. 
Furthermore, by convention we define $+Q$ as pointing in the North-South direction.
To facilitate visualisation, a sketch of the Stokes parameters is provided in Figure \ref{introfig:stokes_param}.

In practice, the normalised Stokes parameters are used. They are defined as $q$ = $Q$/$I$ and $u$ = $U$/$I$.
Lastly, when quantifying linear polarisation the degree of polarisation and polarisation angles are often used. 
They can be calculated from the Stokes parameters and are defined as:
\begin{equation}\label{introeq:pol}
p = \sqrt{q^2 + u^2},
\end{equation}
\begin{equation}\label{introeq:PA}
\theta = \frac{1}{2} \arctan \bigg( \frac{u}{q}\bigg),
\end{equation}
and their errors can be expressed as:
\begin{equation}\label{introeq:pol_err}
\Delta p = \frac{1}{p} \times \sqrt{(q \Delta q)^2 + (u \Delta u)^2},
\end{equation}
\begin{equation}\label{introeq:PA_err}
\Delta \theta = \frac{1}{2} \sqrt{\bigg[ \bigg( \frac{\Delta u}{u}\bigg)^2   + \bigg( \frac{\Delta q}{q}\bigg)^2   \bigg] \times \bigg( \frac{1}{1+(u/q)^2}  \bigg)^2}.
\end{equation}

\subsection{Polarisation in supernovae}
\label{introsec:pol_origin}
\begin{figure}
\centering
\includegraphics[width=6cm]{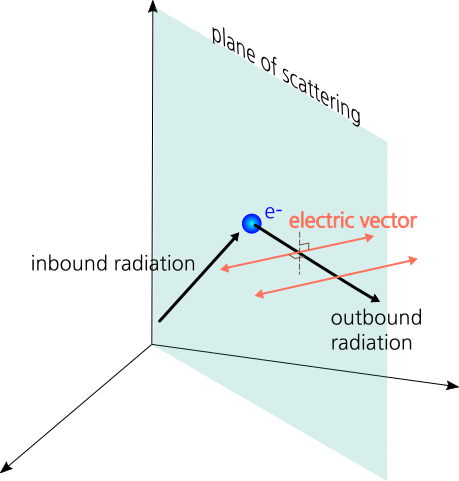}
\caption[Thomson scattering]{Sketch of a photon being scattered through Thomson (electron) scattering. The plane of scattering is defined as the plane containing the inbound and outbound radiation. The electric vector of the outgoing photon is perpendicular to the plane of scattering.}
\label{introfig:thomson}
\end{figure}

To be able to understand and interpret the observable spectropolarimetric properties of core collapse supernovae, it is crucial to understand how polarisation arises in SN ejecta. 
The opacity of the photosphere is dominated by electron (or Thomson) scattering, which is a wavelength independent linearly polarising process.
The direction of the resulting polarisation is orthogonal to the plane of scattering, which is defined as containing the trajectory of the incident and emitted photon (see Figure \ref{introfig:thomson} -- \citealt{chandra60}). 

Consequently, the plane of scattering will be orthogonal to the photosphere, and therefore the electric vector of the outgoing radiation will be tangential to the photosphere at the point of scattering. 
As mentioned in the previous section, the orientation of linear polarisation is characterised by the Stokes parameters $Q$ and $U$, and since supernovae are unresolved at early days the observed polarisation will be a sum of all the polarisation components.

Additionally, it is very important to note that after a few hours to a few days after explosion (which will be the case for all the observations discussed in this thesis), the ejecta of SNe can be assumed to be in homologous expansion ($r = v \times t$).

\begin{figure}
\centering
\includegraphics[width=10cm]{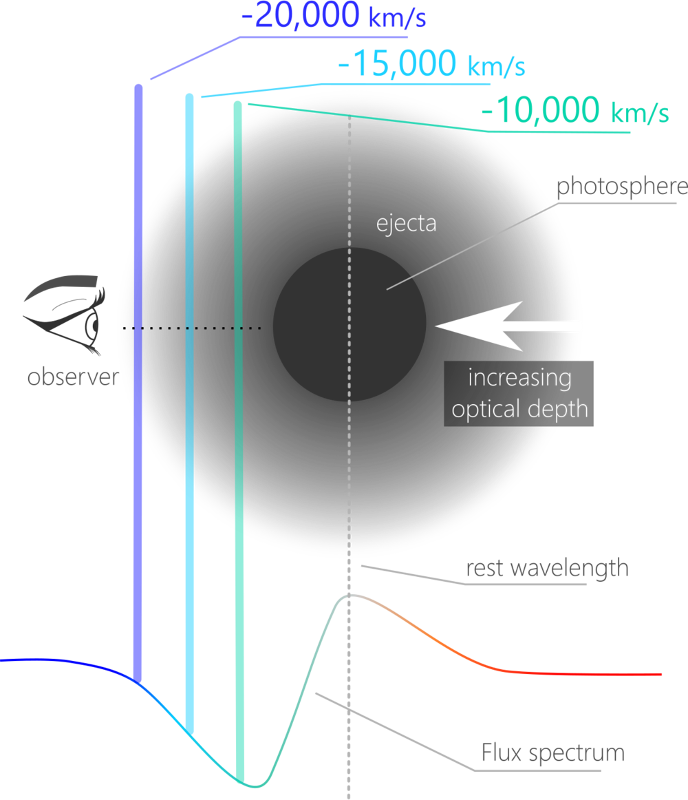}
\caption[Velocity Slices]{Sketch illustrating the concept of velocity slices (represented by the vertical blue lines).}
\label{introfig:vel_slice}
\end{figure}

\begin{landscape}
\begin{figure}
\centering
\includegraphics[width=23cm]{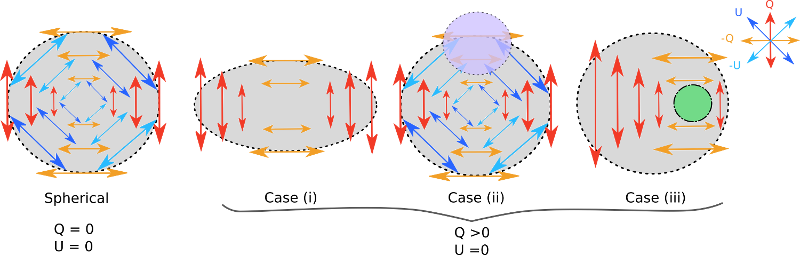}
\caption[Stokes parameters in case (i), (ii) and  (iii).]{Schematic representation of the Stokes parameters arising from scattering out of a spherical and aspherical configuration of the photosphere. The convention described in Section \ref{introsec:formalism} and Figure \ref{introfig:stokes_param} is adopted. We note that for case (i) and (iii), the  $U$ and $-U$ Stokes parameters are not shown for clarity and to emphasise the difference between the $Q$ and $-Q$ components. The purple circle represents a region of absorption whereas the green circle denotes the presence of an additional source of energy, for example a nickel plume. }
\label{introfig:spherical_phot}
\end{figure}
\end{landscape}

Therefore, the 3D geometry of the envelope does not change over time, and variations in $p$, $q$, $u$ are the result of the photosphere receding in a hydro-dynamically inert structure.
Furthermore, the combined effect of homologous expansion and Doppler shift across P Cygni profiles means that different wavelengths across a line feature will probe different “velocity slices”, which are flat projections on the sky. 
Each velocity slice therefore samples a range of optical depths, as is illustrated in Figure \ref{introfig:vel_slice}.
Consequently, a velocity slice at high (low) radial velocity will sample lower (greater) optical depths on average.
Additionally, the degree of polarisation is dependent on optical depth, as exemplified by the Monte Carlo simulations of \cite{hoflich91}.
Note that in the rest of this Thesis, depth an velocity will be used interchangeably, although it is not a one to one relation. 
In this picture, the integrated polarisation of a spherical unresolved ejecta is 0, whereas aspherical geometries result in incomplete cancellation of the Stokes parameters and a net polarisation signal.
Polarisation probes deviations from sphericity in both the structure of the photosphere and the direction and focus of the radiation reaching the photosphere. 
This leads to three classical base cases:
\begin{enumerate}
\item[(i)] Aspherical electron distributions, such as ellipsoidal photosphere, resulting in continuum polarisation \citep{1957lssp.book.....V, hoflich91}.
\item[(ii)] Partial obscuration of the underlying Thomson-scattering photosphere leading to line polarisation (e.g. \citealt{kasen03}).
\item[(iii)] Asymmetric energy input, e.g. heating by off-centre radioactive decay \citep{hoflich95}.
\end{enumerate}
We illustrate these cases as well as the spherical configuration in Figure \ref{introfig:spherical_phot}.
Continuum polarisation will arise from case (i) and (iii), but not from case (ii), as it is dependent on the distribution of the line forming region of a specific element, and will therefore only affect the polarisation in the wavelengths associated with that element. 

These 3 base cases offer a framework within which to interpret the polarisation. 
Naturally we could imagine several or all these configurations arising simultaneously (e.g. an oblate photosphere with an off-centre energy source).

\subsubsection{$q$-$u$ plots and Loops}
\begin{figure}
\centering
\includegraphics[width=15cm]{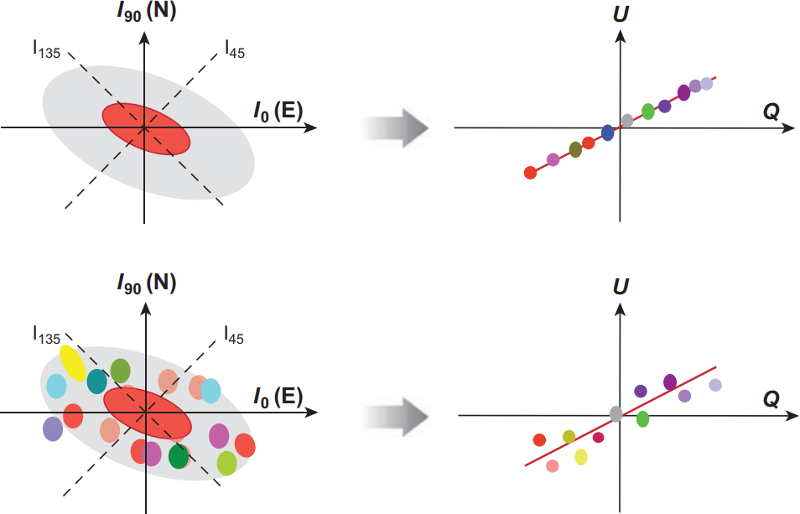}
\caption[Dominant Axis sketch]{\textbf{Top left}: Sky projection of a smooth axi-symmetric ellipsoidal ejecta. The blobs of different colours represent different wavelengths (velocity slices). \textbf{Top right}: Resulting $Q-U$ plot; the dominant axis is indicated by the red line and the coloured blobs represent the wavelength dependent polarisation. \textbf{Bottom left}: Sky projection of axi-symmetric photosphere obscured by clumps of material with high optical depth, resulting in a break of axi-symmetry. The resulting data shows scatter on the $Q-U$ plane around a dominant axis. \textbf{Reference}: This figure is taken from \cite{WW08} -- their figure 1. }
\label{introfig:qu_plot}
\end{figure}

One useful way to show spectropolarimetric data is on a $q$-$u$ diagram. 
For an axi-symmetric ejecta geometry, the observed polarisation data will fall along a line (or dominant axis) on the $q$-$u$ plane (see top panel of Figure \ref{introfig:qu_plot}).
That is because the polarisation angle (P.A.) will be constant across all wavelengths, but since the degree of polarisation is a function of optical depth, which is itself wavelength dependent, the data  will span a range of $p$ values across the observed spectral range. 
When axial symmetry is broken, however, departures from the dominant axis will be observed (see bottom panel of Figure \ref{introfig:qu_plot}).

\begin{figure}
\centering
\includegraphics[width=15cm]{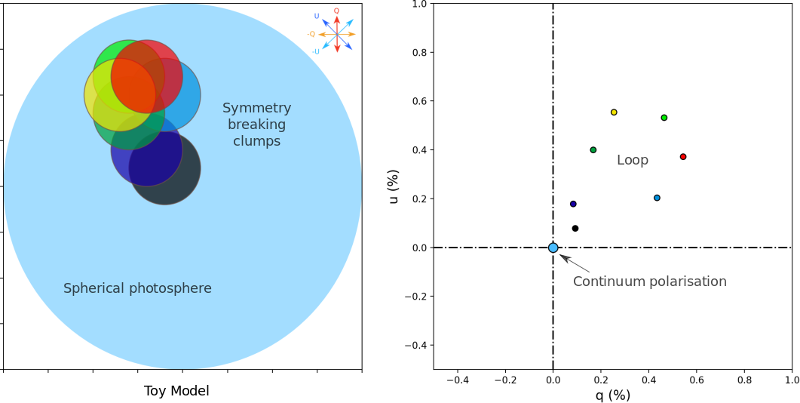}
\caption[Loop toy model]{Toy Model of a loop on the $q-u$ plot arising from asymmetrically distributed clumps obscuring a spherical photosphere (case (ii) -- idealised scenario). The different colours represent different wavelengths. \textbf{Left}: Picture of the toy ejecta. \textbf{Right}: Resulting data on the $q-u$ plot.}
\label{introfig:loop_model}
\end{figure}

The effects of axi-symmetry breaking geometries can result in (relatively) smooth changes in P.A. across the wavelength ranges associated with absorption line regions. 
This can be understood as being the result of different wavelengths probing different velocity slices of the ejecta geometry due to the Doppler shift.
Let's first consider our case (ii) described earlier and imagine a scenario in which the distribution of a line forming region varies with depth.
Even in the case of a spherical photosphere, the ejecta as a whole is not axi-symmetric (see left panel Figure \ref{introfig:loop_model}), and the polarisation will change as we move through each velocity slice (i.e.wavelength), resulting in a loop on the $q-u$ plane. 

\begin{figure}
\centering
\includegraphics[width=15cm]{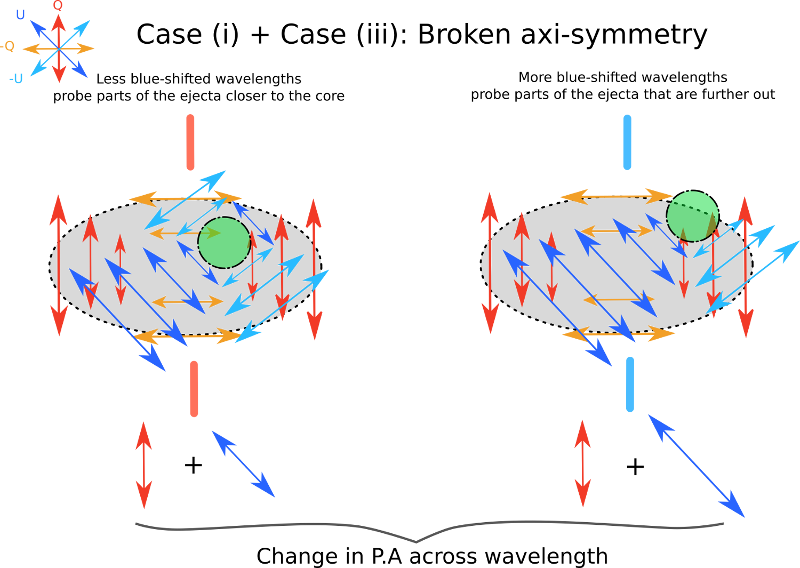}
\caption[Loop case (i) + case (iii)]{Illustration of the same ejecta probed at two different wavelengths (redder wavelength on the left side and bluer wavelength on the right side). Note that not all the Stokes vectors are shown to preserve clarity: We only show the $Q$ and $-Q$ components of the oblate photosphere, and the $U$ and $-U$ components of the additional light from the off-centre energy source. Note that the $U$ and $-U$ components of the oblate photosphere cancel completely, but the $Q$ and $-Q$ components from the off-centre energy source contribution do not. However, this simplified picture already provides some intuition as to why the P.A. changes with wavelength, since it illustrates the variation in the mixing of the Stokes parameters from the oblate photosphere and off-centre energy source.}
\label{introfig:off_centre_loop}
\end{figure}

But small scale asymmetries related to specific line forming regions are not the only way in which loops can form.
A mixture of the cases described in the previous section can be sufficient to provide a departure from axi-symmetry. 
For example an off-centre energy source in an oblate photosphere -- case (i) + case (iii) -- can result in a change in P.A with wavelength (as is illustrated by Figure \ref{introfig:off_centre_loop}), which will manifest as a loop on the $q-u$ plot.

\subsubsection{Brief summary of CCSN spectropolarimetry}
One of the first spectropolarimetric studies of a CCSN unquestionably showing intrinsic signal, was that of Type II-pec SN 1987A, which exploded in the Large Magellanic Cloud on 1987 February 24.23 \citep{kunkel87}. 
One of the remarkable features of the spectropolarimetry of SN 1987A as reported by \cite{jeffery91b} was the jump in polarisation to 1.1$\pm0.08$ percent \about 3 months after explosion, in stark contrast with the low polarisation observed at early times (\about 0.32$\pm0.3$ percent on February 27).
This sudden increase in polarisation is interpreted as a manifestation of the photosphere receding through the outer hydrogen envelope, revealing an inner more aspherical core.

A similar evolution was seen a decade and a half later in the type II-P SN 2004dj, which showed no polarisation in the middle and at the end of the plateau phase but exhibited a jump to $p$\about 0.6  percent as the light-curve came off the plateau around 91 days post-explosion. 
Again, this was interpreted as an inner aspherical core being revealed as the photosphere receded past the hydrogen envelope \citep{leonard06}.

Polarisation changes when the photosphere transitions from the hydrogen envelope to deeper layers can also be seen in Type IIb SNe.
This was observed for the first time in SN 1993J whose degree of polarisation rose from \about 0.3  percent a few days after discovery to \about 1  percent at +20 days \citep{tran97}.
In the case of SN 2001ig, the continuum polarisation increased from \about 0.2 percent 13 days after explosion to \about 0.5 percent by 31 days, when helium lines started dominating the spectrum.
This is also accompanied by a rotation of the P.A. between those epochs, and \cite{maund01ig} interpreted the polarisation of SN 2001ig as being the result of an outer envelope decoupled from the inner ejecta.

The tendency for Type IIb observations to resemble that of Type II is, however, not universal, and they show interesting diversity. 
SN 2008ax, for example, exhibited very high line polarisation (up to \about 3.4 percent in $\mathrm{H\alpha}$) about 2 weeks before maximum.
In the case of the Type IIb SN 2011dh, the continuum polarisation remained about constant (p = 0.47$\pm 0.02$ percent) when the photosphere entered the helium layer, although a major rotation of the polarisation vector occurred from $\theta = 23.5 \pm 1.5$\degree at 9 days after explosion to $\theta=2.2\pm1.4$\degree at +14 days. 
Subsequently the degree of polarisation dropped drastically (p = 0.18$\pm 0.04$ percent) by day 30 when helium dominated the spectrum \citep{mauerhan15}.
Additionally, the correlation between the P.A. of hydrogen and helium with that of the continuum suggested a common geometry.
The authors made several suggestions as to what phenomenon could explain these observations, with their preferred alternative being the presence of fast-rising plumes of $^{56}$Ni causing clumpy excitation in the outer layers of the ejecta.
This would fall under our case (iii) described above.  

Spectropolarimetric observations of Type Ib/c SNe show diversity in the degree of polarisation observed, with the case of SN 1997X \citep{wang01} exhibiting $p$ at least 4 percent, whilst other examples such as SN 2005bf or SN 2008D showed much lower continuum levels: SN 2005bf revealed $p$\about 0.8 percent at $-$6 days with respect to the second B-band maximum and $p$\about 0.5 percent 14 days later \citep{tanaka09}, while SN 2008D showed even lower continuum polarisation with $p<0.4$ percent both at V-band maximum and 15 days later \citep{maund08D}.

Additionally, careful study of the spectropolarimetric signatures in the region of strong spectral lines (in particular calcium H\&K and infrared triplet, He I features, hydrogen features and oxygen features when present) allow a picture of the ejecta to be constructed to better understand the mechanism that resulted in the observations.
This allowed \cite{maund05bf} and \cite{tanaka09} to deduce the potential presence of jets in SN 2005bf, while \cite{maund08D} presented two possible scenarios to explain the data of SN 2008D: a jet-like flow stalled inside the the CO core or a jet that broke out of the envelope.
Their analysis favoured the former picture, and a similar interpretation was proposed by \cite{wang03} to explain the spectropolarimetric behaviour of the Ic-bl SN 2002ap, which also exhibited high velocity and symmetry breaking calcium features. 

This  section is not meant to be an exhaustive summary of the spectropolarimetric observations of CCSNe.
It simply illustrates the power of spectropolarimetry, and the fact that each spectropolarimetric data set brings us closer to understanding the diversity and similarities in CCSN ejecta geometries, and can help constrain their explosion models.

Due to the challenging nature of the observations, the number of high-quality multi-epoch spectropolarimetric data sets of CCSNe remains quite low. 
In Table \ref{introtab:sn_specpol} we give a summary of such data sets available in the literature at the time this project began. 
We exclude the Type Ic SN 1997X, since only the measured continuum polarisation is presented in \cite{wang01}. 
Additionally, spectropolarimetric data were obtained at two epochs for Type Ib/c (XRF) SN 2006aj, the second epoch only provided a detection limit of \about 2 percent, which is not a strong constraint \citep{maund06aj, mazzali07}.

\begin{table}
\centering
\caption[CCSNe spectropolarimetric data sets]{\label{introtab:sn_specpol} Core Collapse SNe with available multi-epoch spectropolarimetric data sets.}
\begin{tabular}{l l c l}
\hline
Name & Type & Number of epochs & Reference \\
\hline
1987A & II-pec &9& \cite{cropper88} \\
1993J & IIb &9& \cite{trammell93, tran97} \\
1998bw & Ic-bl (GRB) &2& \cite{patat01} \\
1998em & II-P &5&  \cite{leonard01} \\
2001ig & IIb &3& \cite{maund01ig} \\
2002ap & Ic-bl &4& \cite{kawabata02, wang03} \\
2004dj & II-P &9& \cite{leonard06} \\
2008D & Ib/c (XRF) &2& \cite{maund08D} \\
2008ax & IIb &3& \cite{chornock11} \\
2011dh & IIb &7& \cite{mauerhan15} \\
2009ip & IIn &6& \cite{reilly17} \\
iPTF13bvn & Ib &6& \cite{reilly16} \\ 

\hline
\end{tabular}
\end{table}


\subsection{Polarisation in Wolf-Rayet Stars}
\label{introsec:wrpol}

Spectropolarimetry can also be used to investigate the progenitors of stripped envelope SNe.
Indeed, the hot (highly ionised) wind of Wolf-Rayet stars has a high electron scattering opacity, resulting in polarisation. 
In the case of a spherical wind and unresolved star, just like in the case of a spherical SN photosphere (see previous Section), the polarisation components are created isotropically and will therefore cancel out completely, yielding no net polarisation.

\begin{figure}
\centering
\includegraphics[width=10cm]{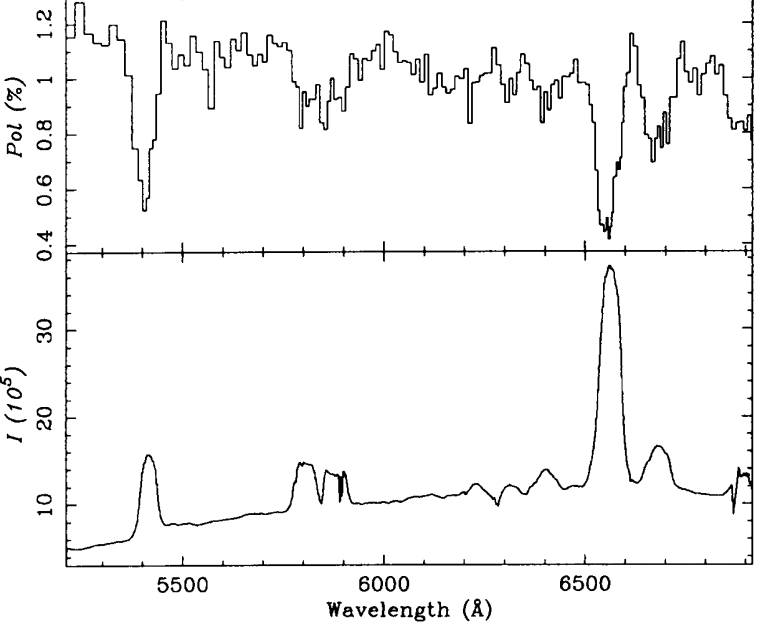}
\caption[Line effect]{Polarisation of the WN WR 134 showing the line effect in He\,{\sc ii} $\lambda$ 5412\r{A}, 6560\r{A}, as presented in Figure 1e of \cite{harries98}}
\label{introfig:line_effect}
\end{figure}

If the wind is aspherical due to disk like structures (e.g. \citealt{stlouis87}) or wind compressed zones (e.g. \citealt{ignace96}), continuum polarisation will result from incomplete cancellation of the Stokes parameters. 
Additionally, the flux of strong emission lines (emitted further out in the wind and therefore undergoing less electron scattering) will be less polarised than the continuum flux, therefore diluting the polarisation and causing a line-effect (e.g. \citealt{schulte92,harries98, vink17}).
This can be seen as peaks or troughs in the polarisation (uncorrected for ISP, see Section \ref{introsec:intro_ISP}) at the wavelengths associated with strong emission lines. 
An example of the line effect is presented in Figure \ref{introfig:line_effect}.

Over the past two decades, continued studies of Galactic and Magellanic Cloud WR stars (see \citealt{harries98} for a summary of Galactic WR star spectropolarimetry and \citealt{vink17} for a summary of Small and Large Magellanic Cloud WR star spectropolarimetry) has revealed that a significant number of WR stars exhibit a line effect (\about 20 percent of 39 stars observed in the Milky Way  and \about 10  percent in the Magellanic Clouds -- with 39 and 12 WR stars observed for the Large and Small Magellanic Clouds, respectively).
This has mostly been attributed to the effects of fast rotation on the 3D geometry of the WR winds, and a possible link with long gamma-ray bursts (which require progenitors with fast rotating cores -- see Section \ref{introsec:ccsne}) has been suggested (e.g. \citealt{grafener12}).

\subsection{Interstellar Polarisation}
\label{introsec:intro_ISP}
Dichroic absorption of incoming radiation by dust grains aligned along the magnetic field of the galaxy causes interstellar polarisation (ISP) which must be removed in order to study the polarisation vector intrinsic to the object of interest. 

\cite{serkowski73} derived an empirical relationship between the ISP and wavelength by observing Galactic stars in the optical range. 
The resulting Serkowski law is of the form:
\begin{equation}
\label{introeq:serk}
p(\lambda)\,  = p_{\text{max}}\exp\bigg[-\text{K}\text{ln}^2\bigg(\frac{\lambda_{\text{max}}}{\lambda}\bigg)\bigg]\, {\rm  percent},
\end{equation}
where $p(\lambda)$ is the polarisation at a given wavelength $\lambda$ (in microns), $p_{\text{max}}$ is the maximum polarisation, $\lambda_{\text{max}}$ is the wavelength at maximum polarisation in microns, and K is a constant. \cite{serkowski75} determined a value of K=1.15; further studies by \cite{whittet92} found a wavelength dependent value, where K = 0.01 + 1.66$\lambda_{\rm max}$. 
For best results when applying a Serkowski-law, it is worth performing a 3-parameter fit with K as a free parameter along with $p_{\rm max}$ and $\lambda_{\rm max}$ \citep{martin99}, particularly when performing Serkowski law fits to extra-galactic targets whose host galaxies may have dust properties different from the Milky Way (e.g. \citealt{patat15}).  

Furthermore, \cite{serkowski75} found that $\lambda_{\rm max}$ was correlated with the ratio of total to selective extinction R (=$A_{V}/E_{B-V}$) through the relation R\about$5.5\times\lambda_{\rm max}$($\mu$m). 
Consequently for a standard value of R\about3.1 \citep{cardelli89}, $\lambda_{\rm max}$ \about 5600\r{A}.

Additionally, an empirical relation between the upper limit on the degree of interstellar polarisation and the reddening is given in \cite{serkowski75}:
\begin{equation}
\label{eq:serk_plim}
p_{\rm ISP} < 9 \times E_{B-V} \,{\rm  percent}.
\end{equation}
Similarly to Eq. \ref{introeq:serk}, this relationship was obtained from observing Galactic targets, and may not apply to other galaxies whose dust properties may differ significantly.

\section{This thesis}
This thesis focuses on the spectropolarimetry of stripped envelope SNe, particularly type IIb and Ic-bl SNe.
These two types are of particular interests, as type IIb SNe are a transitional type between the type II and type Ib SNe, whereas Ic-bl are sometimes (but not always) associated with GRBs (see Section \ref{introsec:classification}). 
Ultimately, a better observational understanding of their ejecta geometry would help us constrain theoretical explosion models. 
Unfortunately, due to the time consuming nature of spectropolarimetry observations, high quality data sets is scarce.
This work adds to a growing library of spectropolarimetric studies of SNe by presenting one of the most extensive data set for a type IIb SNe as well as the best (to date) spectropolarimetric data set for a Ic-bl SN.
Another aim is to explore and further develop pre-existing spectropolarimetric data analysis techniques.
For example, a new ISP removal technique is introduced, and  a method was devised to examine the degeneracies and applications of pre-exisiting toy models.

In the next Chapter, a description will be given of the hardware used for observation and the data reduction techniques.
In Chapter \ref{chpt:08aq}, an updated analysis of the data of the type IIb SN 2008aq is provided (the original data reduction and results can be found in \citealt{stevance16}).
The data analysis and interpretation for the type IIb SN 2011hs are presented in Chapter \ref{chpt:11hs} as they were in \cite{stevance19}. 
In Chapter \ref{chpt:tm}, we present a toy model very similar to that of \cite{maund05hk, reilly16} and discuss the results we obtain for SN 2011hs, the limitations of this approach as well as the need to fully explore parameter space when employing such models. 
A new analysis of the spectropolarimetric data of SN 1993J and comparison to previous studies is given in Chapter \ref{chpt:93J}. 
Then in Chapter \ref{chpt:14ad}, we present our results for the data of SN 2014ad, which is the best (to date) spectropolarimetric data set obtained for a type Ic-bl SN. 
This Chapter is an update on our results published in \cite{stevance17}.
Finally, in Chapter \ref{chpt:wos} we veer away from SNe and look at two Galactic WO stars, WR93b and WR102. 
WR stars of this type are potential progenitors of Ic-bl SNe if they can conserve enough angular momentum (see Section \ref{introsec:ccsne}). 
We use spectropolarimetry to search for rapid rotation in our targets and place limits on their rotational velocities to compare to explosion models.
This work was published in \cite{stevance18}.
Concluding remarks are presented in Chapter \ref{chpt:conclusions}.

\chapter{Hardware and Data Reduction}
\label{chpt:datred} 

\lhead{\emph{Observations and Data Reduction}} 
\section{Hardware}
\subsection{The VLT}
All of our spectroscopic and spectropolarimetric observations were taken at the Very Large Telescope (VLT).
It is situated at the Cerro Paranal observatory of the European Southern Observatory (ESO), at a latitude of 24.6\degree \, South. 
The VLT is composed of four units (UTs), each containing alt-azimuth Ritchey-Chretien telescopes with 8.2 m primary mirrors of diameter.
All observations were taken in service mode.

\subsection{FORS}
\label{datredsec:FORS}
The FOcal Reduced and low dispersion Spectrograph (FORS; \citealt{appenzeller98}) is installed on UT1 of the VLT. 
Two versions were built: FORS1 and FORS2.
FORS1 was dismounted to make room for Xshooter on UT2 in April 2009 and only FORS2 remains on UT1, mounted at Cassegrain. 
The instrument works in the range 3300$-$11000\r{A}.  
The current default detectors are two 2048$\times$4096 MIT CCDs which are optimised in the red, providing low levels of fringing; for a summary of the specifications of the FORS1 E2V and FORS2 MIT detectors, see Table \ref{datredtab:detectors}.   
The spectropolarimetric mode of FORS (PMOS) allows for measurements of linear and  circular polarisation through the use of a half or quarter wave retarder plate and a Wollaston prism.
The instrument allows the degree of polarisation to be determined with a relative error of $<3\times10^{-4}$ and the position angle to $\sim0.2$\degree.

\begin{table}[!h]
\centering
\caption[CCSNe spectropolarimetric data sets]{\label{datredtab:detectors} Specifications of the detectors used.}
\begin{tabular}{l l c c p{30mm} p{30mm}}
\hline
\hline
Instrument & Detector & Size & Pixel scale & Comments & Targets \\
\hline
FORS1 & E2V & 2k$\times$4k & 0.25"/pixel & bad fringing $>$ 650 nm  & SN2008aq, WR93b, WR102 \\
FORS2 & MIT & 2k$\times$4k & 0.25"/pixel & & SN2011hs, SN2014ad \\
\hline
\hline
\end{tabular}
\end{table}

The retarder plates can rotate to adjust the orientation of the optical axis.
We used 4 half-wave retarder plate angles for linear polarisation (0\degree, 22.5\degree, 45\degree and 67.5\degree) and 2 quarter-wave retarder plate angles (-45\degree and 45\degree) for circular polarisation.
The choice of angles is motivated in Section \ref{datredsec:linear_pol} and \ref{datredsec:circ_pol}.
The Wollaston prism splits the light into two rays with polarisation directions parallel and orthogonal to the optical axis of the retarder plate: the ordinary ray and the extra-ordinary ray, respectively.

Unlike normal spectroscopic observations where the slit is oriented along the paralactic angle, our spectropolarimetric observations required the slit be kept at a fixed orientation ( 0\degree -- pointing North) to avoid inducing rotation in the polarisation angle.

\subsection{Xshooter}
Xshooter is a medium resolution spectrograph mounted on Unit 2 at the Cassegrain Focus.
It has 3 arms with optics optimised for their respective wavelength range: UVB (3000-5595\r{A}), VIS (5595-10240\r{A}) and NIR (10240-24800\r{A}). 
The pixel scales for the three arms are 0.164"/pixel, 0.154"/pixel and 0.245"/pixel, respectively. 
We only needed observations from the blue and visible arms, which were reduced by Dr. Marvin Rose. 
Late time Xshooter observation of SN 2014ad are presented in Chapter \ref{chpt:14ad} and were used to calculate the metallicity at the location of SN 2014ad.


\section{Spectroscopic data reduction}
\subsection{Bias subtraction}
During the analogue to digital conversion an offset is added to pixel values in order to prevent negative digital counts, and it is necessary to correct for this instrumental bias when performing data reduction. In order to do so, bias frames (zero-second exposure) are taken. 
They are averaged into a Master Bias, which is then subtracted from all images.
Multiple bias frames are combined in order to  avoid introducing statistical noise in the science images.  

\subsection{Flat-fielding}
The pixel response across a CCD is not homogeneous.
This is mainly due to variations in quantum efficiencies between individual pixels.
In order to correct for this, flat-fields are taken with the same optical train as the science images.
In our case this means that the retarder plate, Wollaston prism and order sorting filter (when used) should be included.
Several flats were averaged together and normalised to create a master flat from which we extracted traces in the regions of the CCD where we knew our spectra were located.
The master flat needs to be normalised because the traces extracted from it will be divided from the science images to correct for sensitivity variations; if the master flat had a high number of counts, the procedure would reduce the number of counts in our science frames, which is to be avoided. 

\subsection{Aperture extraction and background removal}
\begin{figure}
\centering
\includegraphics[width=16cm]{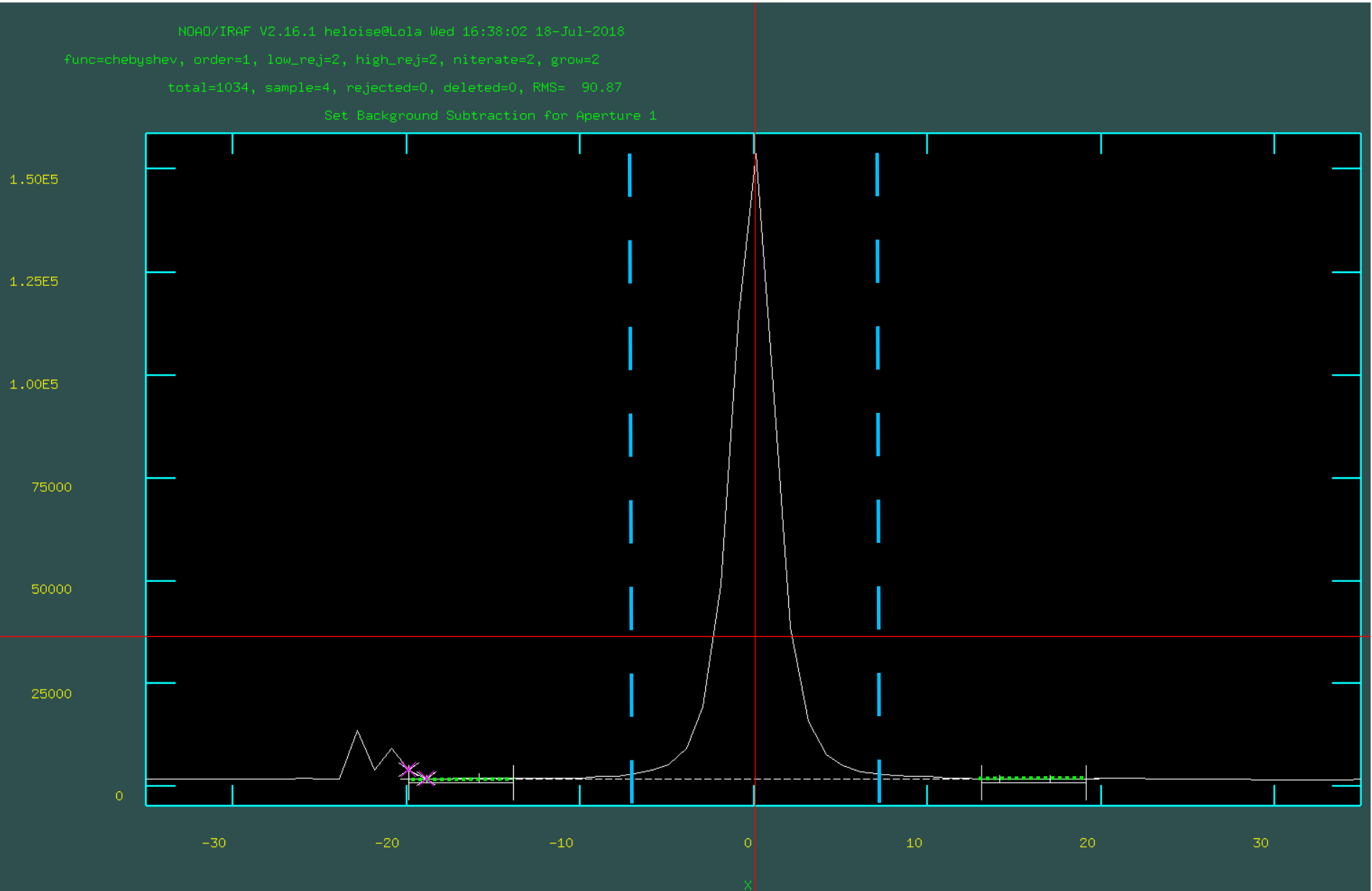}
\caption[2D spectrum cut in IRAF]{Screenshot of a spacial cut of a 2D spectrum in IRAF (solid white line). The dashed blue lines show the aperture that would be chosen for this profile (\about 7 pixels either side of the centre in this case), the green lines show where the sky signal is sampled, the magenta crosses show points that were manually removed from the sky samples as they belonged to a spurious feature. The resulting background fit is showed as the dashed white line. The big red cross is the IRAF pointer and was positioned so the vertical line corresponded to the centre of the spacial profile.}
\label{datredfig:aperture}
\end{figure}

After performing bias subtraction and flat removal, the spectra were extracted.
{\sc IRAF}\footnote{IRAF is distributed by the National Optical Astronomy Observatory, which is operated by the Association of Universities for Research in Astronomy (AURA) under a cooperative agreement with the National Science Foundation.} can be used to place apertures around the spectra to fit their spatial profiles, as well as fit the background by sampling the signal in nearby regions.
This last step was crucial to isolate the supernova light from that of the sky and the host galaxy.  
The counts within the apertures traced were then summed and the background removed to obtain the flux spectra.
The size of the apertures was chosen to include the wings of the spatial profile but not extend so far as to unnecessarily increase the noise levels. 
An example of aperture and background window is shown in Figure \ref{datredfig:aperture}.

\subsection{Wavelength Calibration}
Arc-lamp frames were obtained with optical trains matching those of the science images in order to wavelength calibrate. 
The  emission lines from the HeHgArCd lamps were identified and a low order polynomial was used to fit a dispersion function to the pixel and wavelength coordinates.
The corresponding transformation to wavelength space was then applied to the science images.  
For our typical set up using a slit width of 1'' and the 300V grism we achieved a spectral resolution of \about 12\r{A} -- as determined using arc lamp calibration frames.

\subsection{Flux Calibration}
Flux calibration is not necessary in order to retrieve the Stokes parameters as the procedure involves flux differences (see Sections \ref{datredsec:linear_pol} and \ref{datredsec:circ_pol}).
It is however important to obtain accurate flux spectra and it was therefore performed on the unpolarised total flux spectra.
The spectrophotometric standards were observed with the polarimetric optics in place, and the extracted spectra were compared to standard data files in order to estimate the sensitivity function and apply to right transformation from counts to flux units. 
When the standard data files were not found in the {\sc IRAF} database they were created from the data files found on the ESO website or from white dwarf models fit in the VO Sed Analyser. 


\section{Spectropolarimetric data reduction}
\label{datredsec:specpol_datred}
In order to extract the Stokes parameters the light has to be processed through a retarder plate with an adjustable direction of the optical axis ($\phi$), and a Wollaston prism which splits the light into the ordinary and the extra-ordinary ray (denoted with the subscripts o and e, respectively). 
The flux spectra of the ordinary ($f_{o}$) and extra-ordinary ray ($f_{e}$) can be extracted in the same way as we would a standard spectrum.
The following section provides a derivation of the formulae used to extract the Stokes parameters from combinations of the polarised flux spectra.

\subsection{Mueller Matrices}
Mueller calculus is used to describe the transformation of Stokes vectors for unpolarised or partially polarised light \citep{iniesta03}. 
This method uses $4\times4$ matrices called Mueller matrices; each polarising optical component has an associated Mueller matrix.
The changes in the polarisation of the beam as it travels through the optical train are described by combining the Mueller matrices of the individual components. 
Generally speaking, if there are n elements in an optical train in the order 1, 2, .., n-1, n,  then the Mueller matrix of the system is :
\begin{equation}
\mathbf{M} = \mathbf{M_n} \; \mathbf{M_{n-1}} \; ... \; \mathbf{M_2} \; \mathbf{M_1}
\end{equation}
So if the original Stokes parameters are $\mathbf{S} = (I, Q, U, V)^T$, the Mueller matrix of a retarder plate with optical axis $\phi$ and retardance $\delta$ is $\mathbf{R(\phi,\delta)}$, and the Wollaston prism Mueller matrix is $\mathbf{W_o}$ and $\mathbf{W_e}$, for the ordinary and the extra-ordinary ray respectively, then the Stokes parameters of the light reaching the detector are $\mathbf{S'_{o,e}}$ such that:

\begin{equation}
\mathbf{S'_{o,e}} =  \mathbf{W_{o,e}} \; \mathbf{R(\phi, \delta)} \; \mathbf{S}
\end{equation}
Additionally, as given by \cite{iniesta03}, the general Mueller matrix of the linear retarder plate is:
\begin{equation}\label{datredeq:retarder}
\mathbf{R(\phi, \delta) = }
\begin{pmatrix}
1 & 0 & 0 & 0\\
0 & c_2^2+s_2^2\cos\delta & c_2s_2(1-\cos\delta) & -s_2 \sin(\delta)\\
0 & c_2s_2(1-\cos\delta) & s_2^2+c_2^2\cos\delta & c_2 \sin(\delta)\\
0 & s_2 \sin(\delta)& -c_2 \sin(\delta) & \cos(\delta) \\
\end{pmatrix}
\end{equation}
Where $c_2 = \cos2\phi$ and $s_2 = \sin2\phi$. And the Mueller matrices of the Wollaston prism for the ordinary and extra-ordinary rays are:
\begin{equation}
\mathbf{W_o} = \frac{1}{2}
\begin{pmatrix}
1 & 1 & 0 & 0\\
1 & 1 & 0 & 0\\
0 & 0 & 0 & 0\\
0 & 0 & 0 & 0\\
\end{pmatrix}
\end{equation}
\begin{equation}
\mathbf{W_e} = \frac{1}{2}
\begin{pmatrix}
1 & -1 & 0 & 0\\
-1 & 1 & 0 & 0\\
0 & 0 & 0 & 0\\
0 & 0 & 0 & 0\\
\end{pmatrix}
\end{equation}

In the following sections we make use of this knowledge to derive the combinations of polarised fluxes required to retrieve the Stokes parameters.

\subsection{Linear polarisation}
\label{datredsec:linear_pol}
When measuring linear polarisation we only consider Stokes I, Q and U and a half-wave retarder plate with retardance $\delta=\pi$ is used. Consequently, the Mueller matrix of the retarder plate (Equation \ref{datredeq:retarder}) becomes:
\begin{equation}
\mathbf{R(\phi, \delta=\pi) = }
\begin{pmatrix}
1 & 0 & 0 \\
0 & \cos4\phi & \sin4\phi \\
0 & \sin4\phi & -\cos4\phi
\end{pmatrix}
\end{equation} 
Hence, for a beam with initial Stokes parameters $\mathbf{S} = (I, Q, U)^T$ and final Stokes parameters $\mathbf{S'} = (I', Q', U')^T$, the ordinary ray has Stokes vectors:
\begin{equation}
\mathbf{S'_o} = 
\begin{pmatrix}
I'_o \\
Q'_o \\
U'_o \\
\end{pmatrix}
=\frac{1}{2}
\begin{pmatrix}
1 & 1 & 0\\
1 & 1 & 0\\
0 & 0 & 0\\
\end{pmatrix}
\begin{pmatrix}
1 & 0 & 0 \\
0 & \cos4\phi & \sin4\phi\\
0 & \sin4\phi & -\cos4\phi
\end{pmatrix}
\begin{pmatrix}
I \\
Q\\
U\\
\end{pmatrix}
\end{equation}
Yielding:
\begin{equation}
I'_o = \frac{1}{2} \bigg[ I + Q\cos4\phi + U\sin 4\phi \bigg]
\end{equation}
Similarly, for the extra-ordinary ray we find:
\begin{equation}
I'_e = \frac{1}{2} \bigg[ I - Q\cos4\phi- U\sin 4\phi \bigg]
\end{equation}
Where $I'_o$ and $I'_e$ are the ordinary and extra-ordinary fluxes measured on the detector $f_{o}$ and $f_{e}$, respectively.

Now, the normalised flux difference at a given half-wave plate optical axis angle $\phi_i$ can be defined as in \cite{patat06}:
\begin{equation}\label{datredeq:norm_flux}
F_i \equiv \frac{f_{o,i}-f_{e,i}}{f_{o,i}+f_{e,i}} 
\end{equation}
Then:
\begin{equation}
F_i = \frac{I + Q\cos4\phi_i + U\sin 4\phi_i - I + Q \cos4\phi_i + U\sin 4\phi_i}{I + Q\cos4\phi_i + U\sin 4\phi_i +I - Q\cos4\phi_i - U\sin 4\phi_i} =\frac{Q\cos4\phi_i + U\sin 4\phi_i}{I}
\end{equation}
And if we use the normalised Stokes parameters as defined in Section \ref{introsec:formalism}, we obtain:
\begin{equation}\label{datredeq:flux_diff}
 F_i = q \cos 4\phi_i + u \sin 4\phi_i
\end{equation}

From Equation \ref{datredeq:flux_diff}, it is easily seen that for $\phi_i = \{0^{\circ}, 22.5^{\circ}, 45^{\circ}, 67.5^{\circ}\}, F_i=\{q, u, -q, -u\}$. 
And consequently we can combine $F_0$ with $F_2$, and $F_1$ with $F_3$ to cancel out the instrumental signature and obtain the normalised Stokes parameters:
\begin{equation}\label{datredeq:q}
q = \frac{1}{2} F_0 - \frac{1}{2} F_2
\end{equation}
\begin{equation}\label{datredeq:u}
u = \frac{1}{2} F_1 - \frac{1}{2} F_3. 
\end{equation}
Once $q$ and $u$ are known, the degree of polarisation $p$ and the polarisation angle $\theta$ can be calculated using Equations \ref{introeq:pol} and \ref{introeq:PA}.

The use of 4 retarder plate angles may seem redundant as only observing with $\phi_i = \{0^{\circ}, 22.5^{\circ}\}$ would provide $F_i=\{q, u\}$.
This redundancy is called beam-switching. The second set of angle $\phi_i = \{45^{\circ}, 67.5^{\circ}\}$ effectively switches the ordinary and extra-ordinary rays on the detector. 
Then combining $F_0$ with $F_2$ and $F_1$ with $F_3$ allows us to cancel spurious effects like the polarisation induced by post-analyser optics (e.g. filters, grisms, camera lenses).
Additionally, the removal of flats taken with the polarimetry optics in place can induce significant polarisation, and these effects are also accounted for when using the beam-switching method \citep{patat06}.

Furthermore, the measured $q$ and $u$ also need to be corrected for wave plate chromatism, which is the variation in the polarisation zero-angle with wavelength. 
In our case this is done by interpolating the zero-angles provided on the ESO web site\footnote{http://www.eso.org/sci/facilities/paranal/instruments/fors/inst/pola.html} to our set of wavelengths, subtracting the zero-angles from our initial $\theta$, and then re-calculating $q$ and $u$:
\begin{equation}
q_{\mathrm{corr}} = p  \cos\big(2 \times (\theta - \theta_{\mathrm{zero\ angle}})\big) \; ; \; u_{corr} = p  \sin\big(2 \times (\theta - \theta_{\mathrm{zero\ angle}})\big)
\end{equation}
The degree of polarisation and the polarisation angle can then be calculated again using the new values of the normalised Stokes parameters using Equation \ref{introeq:pol} and \ref{introeq:PA}, and their errors can be derived using Equation \ref{introeq:pol_err} and \ref{introeq:PA_err}.

\subsubsection{Debiasing the degree of polarisation}

Lastly, since the degree of polarisation $p$ is found by adding $q$ and $u$ in quadrature, low values of $p$ and high levels of noise will results in a bias towards values greater than the true degree of polarisation. 
One way to correct for this effect is using the method described in \cite{wang97} where
\begin{equation}
p_{\mathrm{corr}} = p - \frac{\sigma_p ^2}{p} \times h(p - \sigma_p)
\end{equation}

Where $\sigma_p$ is the uncertainty in $p$, and $h(p - \sigma_p)$ is the Heaviside function, defined as
\[
h(p - \sigma_p) = 
\begin{cases}
1 \; & \text{if} \;  p - \sigma_p > 0.\\
0 \; & \text{if} \;  p - \sigma_p < 0. \\
\end{cases}
\]


\subsection{Circular polarisation}
\label{datredsec:circ_pol}
In the case of Circular polarisation, a quarter wave plate with retardance $\delta = \pi/2$ is used instead of a half-wave plate, and therefore Equation \ref{datredeq:retarder} becomes
\begin{equation}
\mathbf{R(\phi, \delta=\pi /2) = }
\begin{pmatrix}
1 & 0 & 0 & 0\\
0 & c_2^2 & c_2s_2 & -s_2\\
0 & c_2s_2 & s_2^2 & c_2\\
0 & s_2 & -c_2 & 0 \\
\end{pmatrix}.
\end{equation}
Considering a beam with initial Stokes parameters $\mathbf{S} = (I, Q, U, V)^T$ and final Stokes parameters $\mathbf{S'} = (I', Q', U', V')^T$, the ordinary ray will have polarisation characterised by
\begin{equation}
\mathbf{S'_o}=
\begin{pmatrix}
I'_o\\ 
Q'_o\\ 
U'_o\\ 
V'_o\\
\end{pmatrix} = \frac{1}{2}
\begin{pmatrix}
1 & 1 & 0 & 0\\
1 & 1 & 0 & 0\\
0 & 0 & 0 & 0\\
0 & 0 & 0 & 0\\
\end{pmatrix}
\begin{pmatrix}
1 & 0 & 0 & 0\\
0 & c_2^2 & c_2s_2 & -s_2\\
0 & c_2s_2 & s_2^2 & c_2\\
0 & s_2 & -c_2 & 0 \\
\end{pmatrix}
\begin{pmatrix}
I\\ 
Q\\ 
U\\ 
V\\
\end{pmatrix},
\end{equation}
yielding:
\begin{equation}
I'_o = \frac{1}{2} \bigg[ I + Q\cos^22\phi + U\cos2\phi \sin 2\phi - V\sin2\phi\bigg]
\end{equation}
Similarly, for the extra-ordinary ray:
\begin{equation}
I'_e = \frac{1}{2} \bigg[ I - Q\cos^22\phi - U\cos2\phi \sin 2\phi + V\sin2\phi\bigg]
\end{equation}

So for  $\phi_i$ = \{45\degree, -45\degree\}, the normalised fluxes (calculated as in Equation \ref{datredeq:norm_flux}) are:
\begin{equation}
F_0 = v  \; \text{and} \; F_1 = -v 
\end{equation}
Where $v$ is the normalised V Stokes parameters. 
We therefore obtain:
\begin{equation}
v = \frac{1}{2}[ F_0 - F_1].
\end{equation}

\subsection{Instrumental signature correction}
\label{datred:deps}
As previously mentioned, instrumental effects are accounted for and cancelled out by beam-switching. 
For a given observation though, the measured normalised flux difference will be the sum of the ideal normalised flux difference and the instrumental signature correction, as defined by \cite{maund08}:
\begin{equation}
F^m_i = F_i + \epsilon
\end{equation}
The instrumental signature corrections for Stokes $q$ and $u$ are then defined as:
\begin{equation}\label{epsilonQU}
\epsilon_Q = \frac{1}{2}(F^m_0 + F^m_2)\;{\rm and }\;\epsilon_U = \frac{1}{2}(F^m_1 + F^m_3)
\end{equation}

It was also found by \cite{maund08} that $\epsilon_Q - \epsilon_U \approx 0$ and that the difference in instrumental signatures $\Delta \epsilon \equiv \epsilon_Q - \epsilon_U$ is dependent on the level of signal-to-noise ratio.

This can be used as quality control after data reduction to identify data points that do not represent true signal. 
We note that deviations in \deps can arise if $p\ge20$ percent, but that never occurred in our data sets. 

\subsection{FUSS}
\label{datredsec:FUSS}
In practice, the polarised spectra were reduced and extracted using IRAF. 
They were then combined according to the procedure described in Sections \ref{datredsec:linear_pol} and \ref{datredsec:circ_pol} using my Python package {\sc FUSS} (i.e. my package For Use with Supernova Spectropolarimetry -- note that despite the name one \emph{can} use it on the data of objects other than supernovae).
The errors on $p$ and $\theta$ in {\sc FUSS} are propagated according to Equations \ref{introeq:pol_err} and \ref{introeq:PA_err}. 
The code  is available on GitHub\footnote{https://github.com/HeloiseS/FUSS} under an open source license. 

One of the features of FUSS is the ability to easily create $q-u$ plots and to automatically fit the dominant axis. 

Usually fits of linear models are performed through a least-squares minimisation,  but this technique assumes no error in the independent variable. 
In the case of data represented on the $q-u$ plane however, both $q$ and $u$ are subject to uncertainties which need to be accounted for.
We therefore choose an Orthogonal Distance Regression (ODR) method, implemented using the Scipy ODR package\footnote{https://docs.scipy.org/doc/scipy/reference/odr.html\#r12d0b3321264-1},
and performed the regression on a model of the form:
\begin{equation}\label{datredeq:line}
u = \beta_0 + \beta_1 q
\end{equation}

The underlying assumption is that the true distribution of the data is defined by Eq. \ref{datredeq:line}, which corresponds to the expected behaviour of spectropolarimetric resulting from a perfectly smooth ellipsoidal ejecta (see the top panel of Figure \ref{introfig:qu_plot}). 
In practice a number of inhomogeneities are present in supernova ejecta, and this fit is only considered an approximation of the global behaviour of the photosphere and of the ejecta as a whole. 
Additionally it should be noted that strong polarisation features caused by deviation from axial symmetry can skew the ODR fit, and caution is advised when interpreting fits of the dominant axis.

\chapter{The Type IIb SN 2008aq}

\label{chpt:08aq} 

\lhead{\emph{SN 2008aq}} 

\section{Introduction}
\label{08aqsec:intro}
Type IIb SNe, despite representing only a small fraction of CCSNe ($\sim$ 11 percent), are an essential transitional class whose members evolve from type II into type Ib/c SNe. 
Their progenitors are stripped of nearly all of their hydrogen envelope, retaining less than 0.5 M$_{\odot}$ \citep{smith11}. As such, type IIb SNe are sensitive probes of mass loss processes, particularly binary interactions (e.g. \citealt{maund93J}, \citealt{fox14}).
Some type IIb SNe exhibit early polarisation levels similar to that of type IIP/L SNe -- e.g. SN 2001ig \citep{maund01ig} -- but more typical cases such as SN 1993J, SN 1996cb, SN 2008ax, or SN 2011dh show continuum polarisation around $p \sim$ 0.5  percent to $p \sim$ 1  percent \citep{trammell93,wang01,chornock11,silverman09,mauerhan15}.

\begin{figure}
\centering
\includegraphics[width=6cm]{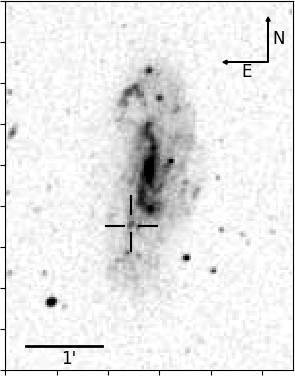}
\caption{Location of SN 2008aq in MCG-02-33-20. This is a Digitized Sky Survey image, retrieved via Aladin.}
\label{08aqfig:host}
\end{figure}

An analysis of the spectropolarimetric data of SN 2008aq was published in \cite{stevance16}, using data that had been pre-reduced.
Unfortunately, a number of issues were identified in the original reduction after publication. 
This chapter presents a new reduction for this data, as well as an updated analysis based on this reduction. 
Further details on the discrepancies with the original reductions that motivated a re-analysis are given in Appendix \ref{app:08aq}.
The details of the observation and of the new data reduction are given in Section \ref{08aqsec:obs}, followed by the presentation of our spectroscopic data in Section \ref{08aqsec:sepctroscopy}.
The analysis of the spectropolarimetric data is then given in Section \ref{08aqsec:specpol}, and our results are discussed and summarised in Section \ref{08aqsec:disc}.

\section{Observations and Data Reduction}
\label{08aqsec:obs}

SN 2008aq was discovered by \cite{08aq}, on February 27.44 2008, in the galaxy MCG-02-33-20 (see Figure \ref{08aqfig:host}), which has a recessional velocity \footnote{Found on https://ned.ipac.caltech.edu} of 2407 km\,s$^{-1}$, and the SN was subsequently classified by \cite{modjaz14} as a type IIb. 
Based on a visual comparison of the lightcurve of SN 2008aq published by \cite{bianco14} to that of similar type IIb SNe (SN 1993J and SN 2008ax), we estimate that SN 2008aq was discovered approximately 8 days before V-band maximum. 
The date of the explosion was taken to be 20 days prior to V-band maximum, as done by \cite{kumar13}, following the study of type IIb and type Ib/c SNe lightcurves by \cite{richardson06}. 
Consequently, we conclude that SN 2008aq exploded on February 16 2008.

Spectropolarimetric observations of SN 2008aq were acquired with FORS1 (see Section \ref{datredsec:FORS}) at two epochs: 4.3 and 15.2 March 2008, a few days before and about a week after V-band maximum, respectively. 
A summary of our observations is given in Table \ref{08aqtab:obs} (ESO ID program 080.D-0107; PI: D. Baade).
Based on our estimate of the explosion date, the observations correspond to $\sim$16 and 27 days post explosion. 
Both sets of observations used the $300V$ grism, providing a spectral resolution of 12.5\r{A} at 6000\r{A} (as determined from arc lamp calibration frames). 
These observations did not use an order separation filter, such that the observations covered a wavelength of 3400-9300\r{A} at the expense, however, of possible second order contamination at redder wavelengths\footnote{Dr. Jason Spyromilio kindly checked the effect of the contamination of higher orders. It was found that in standard stars observed with and without order sorting filters, the second order effects at 7000\r{A} (the cut-on wavelength of the grism efficiency) and long-wards are well below the noise level in our data.}.

The data were reduced following the description of Chapter \ref{chpt:datred}.  
The Stokes parameters were calculated following the  routines of \citet{patat06}, with the data rebinned to 15\r{A} to improve levels of signal-to-noise.  

Flux spectra of SN 2008aq were calibrated against observations of flux standard stars acquired with the polarimetry optics in place. 

\begin{table}
\centering
\caption{\label{08aqtab:obs} VLT FORS1 Observations of SN~2008aq. The epochs are given with respect to the explosion date.}
\begin{tabular}{lcccc}
\hline\hline
Object    &  Date   & Exposure   & Epoch  & Airmass \\
               &(UT)     & (s)              &  (d)       & (Avg.)   \\
\hline
SN 2008aq  & 2008 Mar 04.3 & $4\times 900$   & +16 & 1.119\\
EG274        & 2008 Mar 04.4   & $10$     & +16 & 1.047\\
\\
SN 2008aq & 2008 Mar 15.2 & $2 \times 4 \times 900$  & +27 & 1.073\\
GD108        & 2008 Mar 15.2   & 60 & +27 & 1.067\\
\hline\hline               
\end{tabular}
\end{table}


\section{Optical Spectroscopy}
\label{08aqsec:sepctroscopy}

\begin{figure}
\centering
\includegraphics[width=15cm]{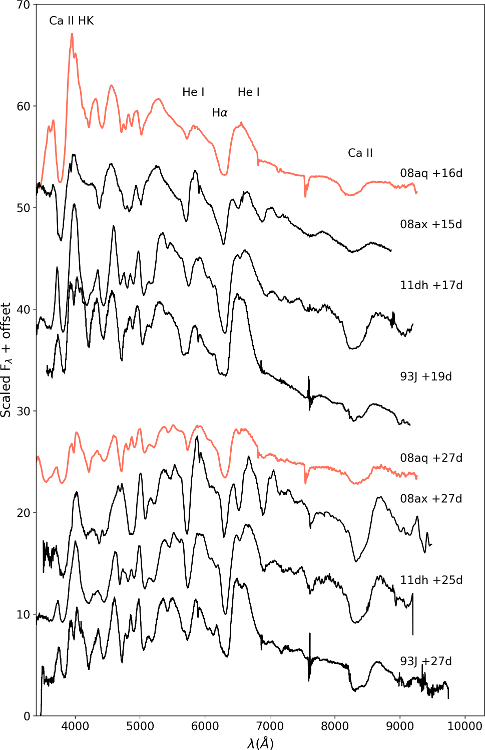}
\caption[SN 2008aq flux comparison and line IDs.]{Flux spectra of SN 2008aq (shown in red) at 16 and 27 days post-explosion. Also shown are spectra of SN 1993J, SN 2008ax, and SN 2011dh (in black). All of the spectra are corrected for the recessional velocity of their respective host galaxies.}
\label{08aqfig:comp}
\end{figure}

The flux spectra of SN 2008aq on 4 and 15 March 2008 are plotted in Figure \ref{08aqfig:comp}, along with the flux spectra of SN 1993J, SN 2008ax and SN 2011dh (all type IIb SNe; obtained from WISeREP\footnote{http://wiserep.weizmann.ac.il/}) at similar epochs, for comparison.

At 16 days post explosion, the spectrum of SN 2008aq  is dominated by broad P Cygni profiles of Ca II H\&K and $\mathrm{H\alpha}$; their absorption minima correspond to velocities of $-$13,700 km\,s$^{-1}$ and $-$12,000 km\,s$^{-1}$, respectively. 
The $\mathrm{H\alpha}$ emission appears to be flat topped, due to a weak blue-shifted He\,{\sc i}\,$\lambda$6678 absorption feature, which is common to the other type IIb SNe.  
This feature evolves into a 'notch' at 6530 \r{A} by the second epoch. The velocity at the absorption minimum of $\mathrm{H\alpha}$ was found to be  $-$11,700 km\,s$^{-1}$ at +27 days. 
An absorption feature due to He\,{\sc i}\,$\lambda$5876 was observed with a velocity of $-$7,800 km\,s$^{-1}$ and $-$7,350 km\,s$^{-1}$ in the first and second epochs, respectively. 
The strength of this feature increases by the second epoch, and a He\,{\sc i}\,$\lambda$7065 feature emerges, with a velocity of $-$7,000 km\,s$^{-1}$. 
Additionally, two narrow absorption lines can be seen superposed onto the emission component of He\,{\sc i}\,$\lambda$5876, due to Na\,{\sc i}\,D originating in the Milky Way and at the recessional velocity of the host galaxy. 
The absorption observed around 8200 \r{A} is attributed to the Ca II Infra-Red (IR) triplet, with a velocity of $-$12,000 km\,s$^{-1}$ at 16 days post-explosion. 
The Ca II IR triplet P Cygni profile is peculiar in that it exhibits a relatively deep absorption paired with a weak emission component. 
\cite{branch02} noted that this behaviour is characteristic of a "detached" line forming region at velocities significantly greater than that of the photosphere. By the second epoch, 10 days later, the velocity of the Ca II IR triplet has decreased dramatically to a value of $-$9,000 km\,s$^{-1}$ and the P Cygni profile shows a prominent emission component.
In the second epoch, the velocity of the Ca II IR triplet and Ca H \& K features has decreased more noticeably than that of other elements. 

The most remarkable difference between SN 2008aq and the comparison SNe (see Figure \ref{08aqfig:comp}) is the relative weakness of the He\,{\sc i}\,features. 
In the first epoch (+16 days) the He\,{\sc i}\,$\lambda$5876 line is much shallower in SN 2008aq than in SN 2008ax, SN 1993J or SN 2011dh.
 Also the presence of a He\,{\sc i}\,$\lambda$6678 feature can be inferred from the flat-topped profile of the $\mathrm{H\alpha}$ emission component, whereas in SN 2008ax, SN 2011dh and SN 1993J, this feature is clearly visible in the form of a "notch." 
 In the second epoch (27 days post-explosion), the He\,{\sc i}\,lines of SN 2008aq have significantly increased in strength, but are still much weaker than in the spectra of the other type IIb SNe, and $\mathrm{H\alpha}$ remains the dominant feature.
 Additionally, the Ca II IR triplet feature at the second epoch is much shallower in the spectrum of SN 2008aq than in the spectra of 1993J, 2011dh and 2008ax, whereas at the first epoch the strength of that feature is similar for SN 2008aq, SN 2008ax and SN 2011dh. 


\section{Spectropolarimetry}
\label{08aqsec:specpol}

\begin{figure}
\centering
\includegraphics[width=15cm]{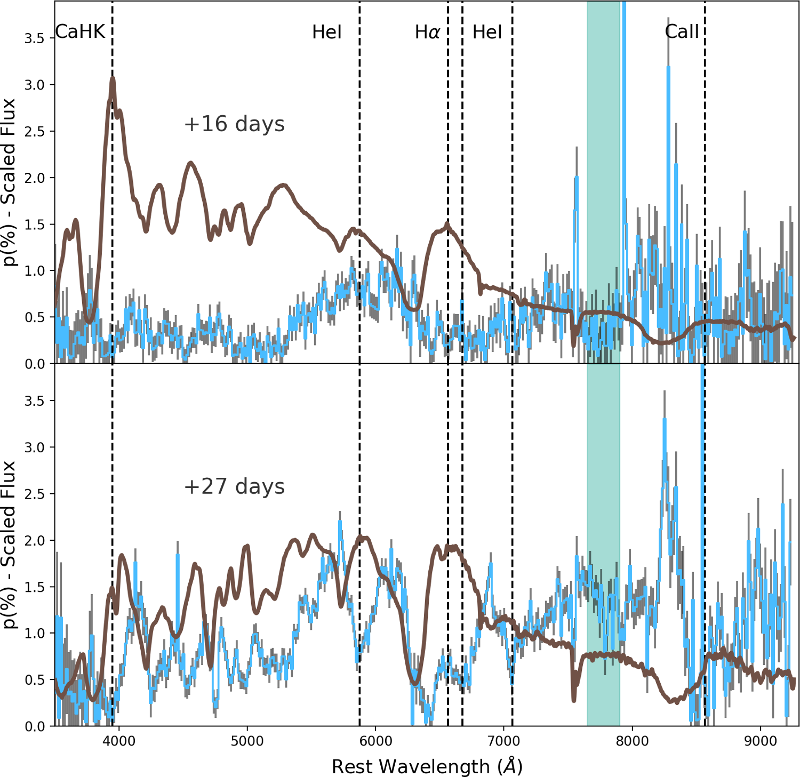}
\caption[SN 2008aq flux and polarisation (ISP not removed)]{Scaled Flux spectra (brown) and polarisation spectra (blue) of SN 2008aq at 16 days and 27 days post-explosion. The polarisation spectra were not corrected for interstellar polarisation. The regions highlighted in green were used to estimate the continuum polarisation in Sections \ref{08aqsec:ep1_pol} and \ref{08aqsec:ep2_pol}.}
\label{08aqfig:pol}
\end{figure}

The polarisation and flux spectra for SN 2008aq are shown in Figure \ref{08aqfig:pol}. 
Even before ISP correction, the polarisation of SN 2008aq exhibits strong line features, particularly at +27 days. 

\subsection{Interstellar Polarisation}
\label{08aqsec:isp}
As described in Section \ref{introsec:intro_ISP} under the assumption of a \citeauthor{serkowski75} type ISP the maximum degree of polarisation can be constrained by Equation \ref{eq:serk_plim}, where in the case of an extra Galactic source the colour excess $E(B-V)$ used in the formula is the sum of the colour excesses of the Milky and of the Host galaxy. 
An empirical relationship relating the equivalent width of the sodium lines and the reddening was derived by \cite{poznanski12}. 
From the Na\,{\sc i}\,D of the Milky Way component, we estimate that the reddening associated with dust in our Galaxy is $E(B-V)_{\rm MW}$ = 0.045 mag, which is in agreement with the estimates of foreground reddening of $E(B-V)_{\rm MW}$ = 0.04 mag by \cite{schlafly11}\footnote{https://ned.ipac.caltech.edu/}. 
The reddening associated with the host galaxy was estimated from the corresponding Na\,{\sc i}\,D line, yielding $E(B-V)_{\rm host}$ = 0.027 mag. 
Consequently, the total reddening, is $E(B-V)_{\rm total}$ = 0.072 mag, which is similar to the value found and used by \cite{stritzinger09}. 
The upper limit on the ISP associated with our data is therefore 0.65 percent.

To further characterise the ISP we make the assumption that the emission component of $\mathrm{H\alpha}$ is intrinsically unpolarised at early times \citep{tran97}.
It then follows that the average polarisation over that feature must be representative of the ISP. 
Because the profile of He\,{\sc i}\, $\lambda 6678$  is less apparent in the spectrum of SN 2008aq at +16 days, we favour the  $\mathrm{H\alpha}$ emission component of the first epoch for our ISP estimate.
The values of the Stokes parameters in the range 6700-6900 \r{A} we find $q_{\rm ISP}$ = 0.29 ($\pm$ 0.07)  percent and $u_{\rm ISP}$ = 0.13 ($\pm$ 0.07)  percent, corresponding to $p_{\rm ISP}$ = 0.32  ($\pm$ 0.07)  percent. 
This is well within the upper limit derived from the Serkowski relation.

\begin{figure}
\centering
\includegraphics[width=15cm]{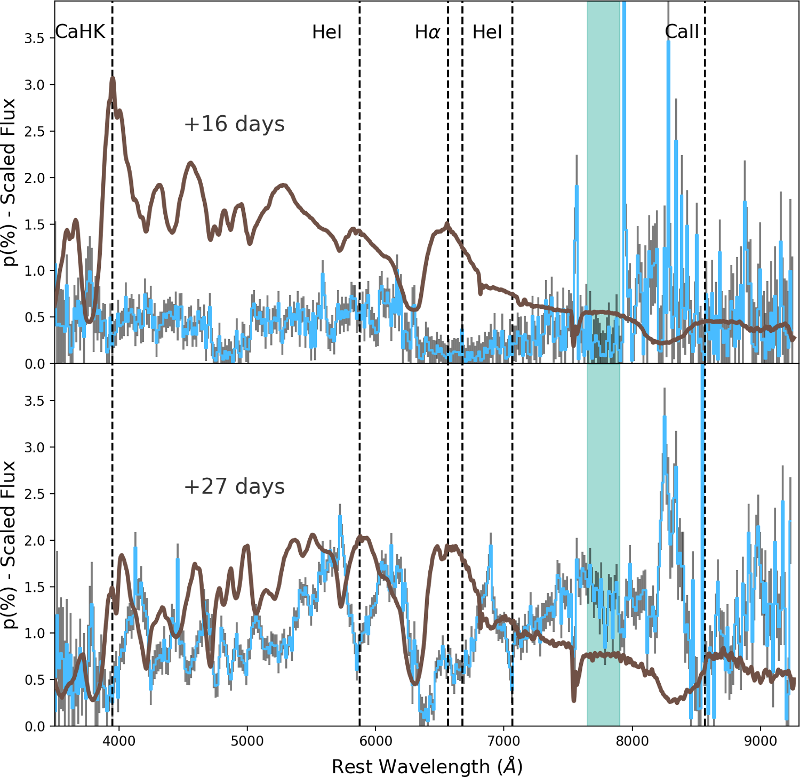}
\caption[SN 2008aq flux and polarisation (ISP removed)]{Scaled Flux spectra (brown) and polarisation spectra (blue) of SN 2008aq at 16 days and 27 days post-explosion. The polarisation spectra were corrected for interstellar polarisation. The regions highlighted in green were used to estimate the continuum polarisation in Sections \ref{08aqsec:ep1_pol} and \ref{08aqsec:ep2_pol}.}
\label{08aqfig:polnosip}
\end{figure}

Using these estimates to correct for the ISP, we find a polarisation for SN 2008aq showed in Figure \ref{08aqfig:polnosip}.
Before proceeding to the analysis, a number of caveats must be discussed.

Firstly, the profile of the degree of polarisation of SN 2008aq after ISP correction is somewhat peculiar, particularly in the blue.
Indeed, we expect line blanketing in the blue and near-UV to lead to a strong dilution of the intrinsic polarisation, and so after ISP removal we would expect $p$ to decrease in this region of the spectrum, rather than increase (as is clear when comparing the top panel of Figures \ref{08aqfig:pol} and \ref{08aqfig:polnosip}).

Additionally, the assumption of complete depolarisation of strong emission lines is not necessarily accurate \citep{tanaka09}.
In particular, as mentioned in Section \ref{08aqsec:sepctroscopy}, the flattened profile of the $\mathrm{H\alpha}$ emission component at +16 days indicates the presence of an unresolved He\,{\sc i}\, $\lambda6678$ absorption feature.
Since absorption features are typically associated associated with excess polarisation, this is an indication that our assumption of complete depolarisation in the emission region of $\mathrm{H\alpha}$ may not be valid.

Unfortunately the lack of observation at earlier phases (where the $\mathrm{H\alpha}$ emission component may be less contaminated by He\,{\sc i}\, $\lambda6678$) means that it is not possible at this time to check the veracity of our ISP estimate any further. 

Consequently our derived ISP values should be considered with great caution and the following analysis is performed both on the corrected and uncorrected data sets. 
\newpage
\subsection{Line and Continuum polarisation}

\subsubsection{+16 days}
\label{08aqsec:ep1_pol}
At the first epoch, the uncorrected polarisation data are dominated by a large feature starting at 5330\r{A} with an initial polarisation of $p$\about 0.5 percent, rising to a 1.24$\pm0.15$ percent peak at 6168\r{A}, and decreasing back down to the \about 0.1 percent level at 6430\r{A}. 
This feature can be interpreted as a blend of $\mathrm{H\alpha}$ and He\,{\sc i}\,$\lambda 5876$ with the main peak being correlated to the $\mathrm{H\alpha}$ absorption component, as has been seen previously in type IIb SNe (e.g. \cite{mauerhan15}). 

In the ISP corrected data this broad feature is not seen, and the polarisation is roughly constant at \about 0.5 percent.
A narrow peak is seen at 5587\r{A} with $p=0.96\pm0.15$ percent, which could be associated with a high velocity ($-14,500$\kms) component of He\,{\sc i}\, $\lambda 5876$ although it is unclear why the polarisation would settle back down to 0.5 percent and remain constant across the main absorption and emission components of the P Cygni profile. 
The $\mathrm{H\alpha}$ peak is found in the same location as in the ISP uncorrected data (6178\r{A}), and with a polarisation $p=1.01\pm0.15$ which is within uncertainties of the uncorrected value.

Another feature visible at +16 days after explosion, albeit more subtle, is the polarisation peak at 3772\r{A} associated with the absorption minimum of the large Ca H\&K feature.
The peak reaches a polarisation of $p=0.79\pm0.37$  percent in the uncorrected data and $p=0.99\pm0.38$ in the ISP corrected data. 

In the red, the high levels of noise beyond \about 7900 \r{A} caused by fringing make the interpretation of the polarisation associated with the absorption feature of the Ca IR triplet difficult. 
In particular it should be noted that the isolated peaks at 7630, 8005, 8350, and 8410 \r{A} are outliers and not representative of a true signal (see Section \ref{app08aqsec:deps}). 
There may be a hint of polarisation excess correlated with the absorption component of the Ca IR triplet, but values of peak polarisation cannot be accurately derived. 

The continuum polarisation of SN 2008aq was evaluated by producing the average of the polarisation measured in a wavelength range identified as devoid of strong lines. 
We chose the spectral region extending from 7650 to 7900\r{A} (highlighted in green in Figures \ref{08aqfig:pol} and \ref{08aqfig:polnosip}), avoiding strong telluric lines and the onset of the Ca II IR region. 
In this first epoch, the continuum polarisation is estimated at $p = 0.22 \pm 0.12$ percent in the ISP corrected data and $p=0.34 \pm 0.14$ percent for the uncorrected data. 
Within uncertainties, these two values are the same, and correspond to an axis ratio of \about 0.90 (\citealt{hoflich91} -- their figure 4).

\subsubsection{+27 days}
\label{08aqsec:ep2_pol}
The evolution of the polarisation by 27 days after explosion is drastic, and shows features that are less blended and have stronger amplitudes, both in the ISP corrected and ISP uncorrected data.
The He\,{\sc i}\, $\lambda 5876$ and $\mathrm{H\alpha}$ are now separate.
The helium peak is found at 5722 \r{A} and $p=2.21\pm0.11$ percent ($p=2.26\pm0.13$ percent) for the ISP uncorrected (corrected) data.
Hydrogen on the other hand shows a relatively flat peak between 6040 and 6210\r{A}; averaging the polarisation in this wavelength range we find $p=1.62 \pm0.10$ percent and $p=1.66\pm0.11$ percent in the ISP uncorrected and corrected data, respectively. 

Additionally, a clear He\,{\sc i}\, $\lambda 7065$ feature is now found at 6882\r{A} with a peak polarisation of $p=1.60\pm0.12$ percent ($p=1.71\pm0.14$ percent) in the ISP uncorrected (corrected) data.

As opposed to the first epoch, the calcium infrared triplet now clearly dominates over the noise, and has become the strongest polarisation feature. 
The infrared calcium peak is found at 8251\r{A} with $p=3.3 \pm 0.3$ percent in both the ISP corrected and uncorrected data sets. 
The Ca H\&K feature on the other hand, has not significantly strengthened, with a peak now reaching $p=0.97\pm0.45$ percent ($p=1.31\pm0.45$ percent) at 3787\r{A} in the ISP uncorrected (corrected) data set.

Lastly, the continuum polarisation is calculated in the same way as for epoch 1, and is found to be $p = 1.36 \pm 0.10$ percent / $p = 1.33 \pm 0.140$ percent for the ISP corrected and uncorrected data, respectively. 
This increase in polarisation indicates a significant decrease in axis ratio, which is now \about 0.77 (\citealt{hoflich91} -- their figure 4).

\subsection{$q-u$ plots}
\label{08aqsec:quline}

\begin{figure}
\centering
\includegraphics[width=15cm]{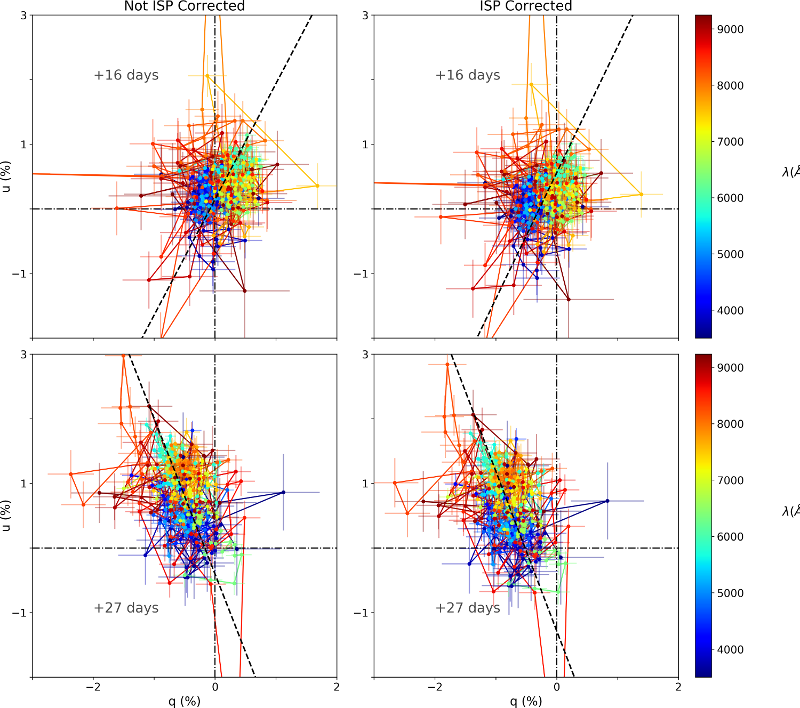}
\caption[SN 2008aq $q-u$ plot -- whole wavelength range]{Spectropolarimetric data of SN 2008aq presented on $q-u$ plots. The top panels show the data at +16 days post-explosion while the data at +27 days are shown on the bottom panels. The left hand side panels show the ISP uncorrected data while the right hand side panels show the ISP corrected data. The colour bar represents wavelength, and the dashed line shows the dominant axis of the data as determined with Orthogonal Distance Regression.}
\label{08aqfig:quwhole}
\end{figure}

Firstly we present in Figure \ref{08aqfig:quwhole} the $q-u$ plots for the spectropolarimetric data of SN 2008aq across the full wavelength range. 
The dominant axis is fitted using an ODR method as described in Section \ref{datredsec:FUSS}.

At +16 days the data shows a subtle elongation.
This is expected for low global asymmetry \citep{WW08} and is therefore consistent with the low axis ratio determined in Section \ref{08aqsec:ep1_pol}.
The angle of the dominant axis (calculated anti-clockwise from the $u=0$ line) is $61\pm2$\degree in the uncorrected data and $63\pm2$\degree in the ISP corrected data. 
At +27 days, the elongation is more pronounced and the angle of the dominant axis is found to be $113\pm1$\degree ($112\pm1$\degree) for the ISP uncorrected (corrected) data.
A global rotation of 52\degree (49\degree) is therefore observed between the ISP uncorrected (corrected) data at +16 days and +27 days, which corresponds to a polarisation angle (P.A) rotation of 26\degree(24.5\degree), from \about 30.5\degree (31\degree) at +16 days to 56.5\degree (56\degree) at +27 days.

\begin{figure}
\centering
\includegraphics[height=20cm]{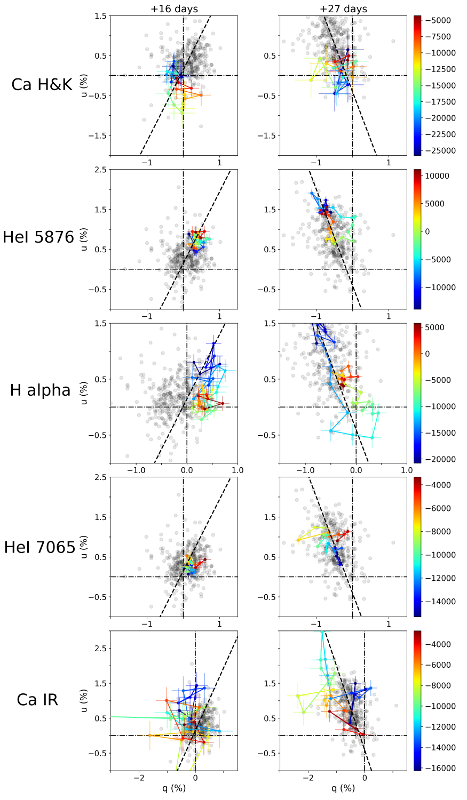}
\caption[SN 2008aq $q-u$ plot -- lines -- ISP not removed]{Uncorrected spectropolarimetric data associated with the strong spectral lines of SN 2008aq. The colour bars represent the velocity space corresponding to each line.
The dominant axis is plotted as the dashed line and the light grey points show the whole data for reference. }
\label{08aqfig:quline_uncorr}
\end{figure}
\begin{figure}
\centering
\includegraphics[height=20cm]{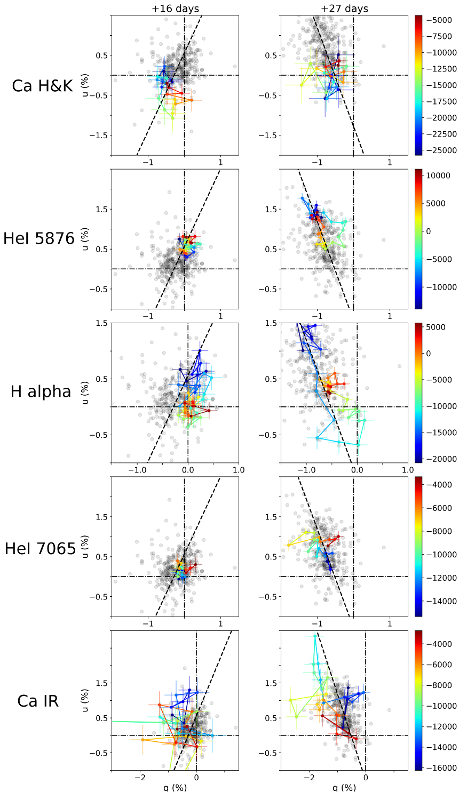}
\caption[SN 2008aq $q-u$ plot -- lines -- ISP removed]{ISP corrected spectropolarimetric data associated with the strong spectral lines of SN 2008aq. The colour bars represent the velocity space corresponding to each line.
The dominant axis is plotted as the dashed line and the light grey points show the whole data for reference. }
\label{08aqfig:quline}
\end{figure}

The spectropolarimetric data of SN 2008aq in the region of Ca H\&K $\lambda 3948$, He\,{\sc i}\, $\lambda 5876$, $\mathrm{H\alpha}$, He\,{\sc i}\, $\lambda 7065$ and Ca\,{\sc ii} $\lambda 8567$ are shown in Figure \ref{08aqfig:quline_uncorr} and \ref{08aqfig:quline} for the ISP uncorrected and corrected case, respectively.

At the first epoch we find that only the Ca H\&K and the $\mathrm{H\alpha}$ polarisation show significant departure from the rest of the data. 
Their orientation off-axis with respect to the dominant axis indicates the presence of small scale asymmetries that can be caused by clumping. 
At +27 days post-explosion all of the plotted lines show features on the $q-u$ plot. 
The Ca H\&K and He\,{\sc i}\, $\lambda 7065$ features are both very narrow and therefore quite subtle, but their peak polarisation correspond to significant departures from the dominant axis, 
suggesting asymmetrically distributed line forming regions. 

The He\,{\sc i}\, $\lambda 5876$ loop at +27 days is very well defined and is the consequence of a change in the sky projected location of the line forming region with velocity (i.e. depth). 
A similar interpretation is consistent with the $\mathrm{H\alpha}$ polarisation feature, although the chirality is reversed compared to the helium feature (clockwise in He\,{\sc i}\, and anti-clockwise in $\mathrm{H\alpha}$). 
Additionally the loop is more extended suggesting greater geometry changes with depth. 

Finally the Ca\,{\sc ii} IR feature shows an extended and complex polarisation feature that roughly follows the dominant axis. 
The complexity of the loop may reflect the complexity of the geometry of the line forming region in 3D, although this interpretation should be considered with caution given the size of the error bars.

\section{Discussion \& Conclusion}
\label{08aqsec:disc}
 
Optical spectra and spectropolarimetric data of SN 2008aq were presented at 2 epochs: 16 days and 27 days post-explosion. 
The intrinsic polarisation calculated for SN 2008aq at +16 days ($p_{\rm cont}$= 0.22 $\pm$ 0.12  percent ) is much lower than found in other type IIb SNe at earlier and similar epochs: e.g. SN 2008ax at +9 days ($p_{\rm cont}$= 0.64 $\pm$ 0.02  percent) and SN 2011dh at +14 days ($p_{\rm cont} \sim$ 0.5  percent -- \citep{chornock11,mauerhan15}). 

By the second epoch, the continuum polarisation of SN 2008aq had reached a value of $p_{\rm cont}$=1.33 ($\pm$ 0.33)  percent, which is similar to the continuum polarisation calculated for SN 1993J at +29 days ($p_{\rm cont} \sim$ 1  percent) and that of SN 2001ig at 31 days (\about 1 percent -- \citealt{tran97, maund01ig}). 

A characteristic of type IIb SNe is the increase in strength of their He features with time, as they transition to type Ib SNe. 
This behaviour is observed in SN 2008aq (see Section \ref{08aqsec:sepctroscopy}), however, comparison with other type IIb SNe at similar epochs reveals that the He features in the spectra of SN 2008aq are significantly weaker (see Figure \ref{08aqfig:comp}). 
Additionally, we find that the pseudo equivalent width of the He\,{\sc i}\,$\lambda$5876 absorption component at both epochs was roughly 4 to 5 times smaller in SN 2008aq than in other type IIb SNe \citep{liu16}. 
The scarcity of helium indicates that the receding photosphere of SN 2008aq reached the helium layer at a later date than in SN2008ax, SN 2011dh, SN 1993J, and other previously studied type IIb SNe; this may suggest that the progenitor of SN 2008aq was less stripped than other type IIb SNe.

Similarly to SN 2001ig \citep{maund01ig}, SN 2008aq showed a drastic dominant axis rotation between +16 and +27 days (see Figure \ref{08aqfig:quwhole}) suggesting a change in the axis of symmetry deeper into the envelope. 
Additionally, as we have seen in Section \ref{08aqsec:quline}, departures from axial symmetry are seen at both epochs, as indicated by the presence of loops in the $q-u$ plots of CaH\&K and $\mathrm{H\alpha}$ at +16 days, and in the $q-u$ plots of calcium, $\mathrm{H\alpha}$, He\,{\sc i}\, $\lambda5876$ and He\,{\sc i}\, $\lambda7065$ at +27 days after explosion. 
Therefore, both the overall geometry and smaller scale geometry of the envelope of SN 2008aq vary with depth.

As mentioned in Section \ref{08aqsec:isp}, our determination of the ISP is dependent on the assumption that the $\mathrm{H\alpha}$ P Cygni profile at early times is completely  depolarised (e.g. \citealt{trammell93}, \citealt{tran97}).
Given the suspected contamination of the $\mathrm{H\alpha}$ emission by He\,{\sc i}\, $\lambda6678$, this assumption is to be taken with great caution. 
On the whole, the interpretation of the data is mostly unchanged whether we consider the ISP removed or uncorrected data as the continuum polarisation values and the $q-u$ plot features are all consistent.
The only major change between the ISP corrected and uncorrected data is the profile of the polarisation spectrum at +16 days (see the top panels of Figure \ref{08aqfig:pol} and \ref{08aqfig:polnosip}). 
The shape of the ISP corrected data being a lot more difficult to interpret, it either suggests the need for a novel interpretation, or simply be a reflection of the inaccurate ISP removal. 
Given the current context, the latter explanation seems  more likely.
Additionally, since this work studies the variations of the polarisation across the lines, the absolute value of the derived polarisation is of interest but not critical to our conclusions. 

In the next chapter we will focus on another type IIb SN: SN 2011hs.

\chapter{The 3D shape of Type IIb SN 2011hs}

\label{chpt:11hs} 

\lhead{\emph{SN 2011hs}} 

\section{Introduction}
Type IIb SNe are better represented than other types in the spectropolarimetric literature, and the previous chapter on SN 2008aq added a new object to our sample. 
However, the number of high-quality multi-epoch data sets remains limited (5 SNe, including SN 2008aq -- see Table \ref{introtab:sn_specpol}).
Increasing our sample is ultimately necessary to compare the statistical distribution of the observed polarisation to available models (e.g. \citealt{hoflich91, kasen03, dessart11, tanaka17}).
In this chapter we analyse the data for another type IIb SN: SN 2011hs.
We present 7 epochs of spectropolarimetry, from $-3$ days to +40 days with respect to V-band maximum. 
This, along with SN 2011dh, is one of the best data sets ever obtained for a type IIb SN, in terms of time coverage, cadence and signal-to-noise. 
In Section \ref{11hssec:obs} we give the details of the observations. 
In Section \ref{11hssec:isp} a thorough analysis of the interstellar polarisation is provided,  and we present the intrinsic polarisation of SN 2011hs in Section \ref{11hssec:pol}.
The results are discussed and compared to the literature in Section \ref{11hssec:disc}, and we summarise the results in Section \ref{11hssec:conclusions}.

This work has been published in the Monthly Notices of the Royal Astronomical Society \citep{stevance19}.

\section{Observations and data reduction}
\label{11hssec:obs}
\begin{table}
\centering
\caption[VLT observations of SN 2011hs]{\label{11hstab:obs} VLT Observations of SN~2011hs.  The epochs are given relative to the estimated V-band maximum.  \\$^a$Flux Standard. }
\begin{tabular}{c c c c c}
\hline
Object & Date & Exp. Time & Epoch & Airmass \\
 & (UT) & (s) & (days) & \\
\hline

SN~2011hs & 2011 Nov. 19 & 8 $\times$ 850 & $-3$ & $1.05-1.23$ \\
EGGR 141$^a$  & 2011 Nov. 19 & 2$\times$60 & $-$ & 1.26\\
 \\ 
SN~2011hs & 2011 Nov. 24 & 8 $\times$ 900 & +2 &  $1.1-1.36$ \\
EGGR 150$^a$& 2011 Nov. 24 & 2 $\times$ 191 & $-$ & 1.31 \\
 \\
SN~2011hs & 2011 Dec. 02 & 8 $\times$ 900 & +10 & $1.16-1.56$ \\
LTT1788$^a$ & 2011 Dec. 02 & 2 $\times$ 60 & $-$ & 1.04 \\
 \\
SN~2011hs & 2011 Dec. 10 & 8 $\times$ 900 & +18 & $1.3-2.0$ \\
LTT1788$^a$  & 2011 Dec. 10  & 2 $\times$ 60  & $-$ & 1.8 \\
 \\
SN~2011hs & 2011 Dec. 16 & 4 $\times$ 855 & +24 & $1.4-2.3$ \\
LTT1788$^a$ & 2011 Dec. 16 & 2 $\times$ 120 & $-$ & 1.14 \\
  \\
SN~2011hs & 2011 Dec. 23  & 4 $\times$ 1100 & +31 & $1.4-1.77$ \\
LTT1788$^a$ & 2011 Dec. 23 & 2 $\times$ 60  & $-$ & 1.034 \\
 \\
SN~2011hs & 2012 Jan. 01  & 4 $\times$ 1100  & +40 & $1.6-2.1$ \\
LTT1788$^a$  & 2011 Dec. 16 & 2 $\times$ 120 & $-$ & 1.14 \\
\hline
\end{tabular}
\end{table}

SN 2011hs was discovered by Stuart Parker on 12.476 November 2011 and classified by \cite{2011hs}.
It is located at R.A. = 22$^{\text{h}}$57$^{\text{m}}$11$^{\text{s}}$ and $\delta$ = -43\degree23'04" in the galaxy IC 5267 (face-on sA0) with redshift z = 0.005711, corresponding to a recessional velocity of 1714 \kms \citep{koribalski04}.
A series of spectropolarimetric observations of SN 2011hs were taken at the VLT with FORS2 (ESO ID program 088.D0761; PI: J. Maund). 
Linear spectropolarimetry was obtained for 4 half-wave retarder angles  (0\degree, 22.5\degree, 45\degree, 67.5\degree) at seven epochs between 19 November 2011 and 01 January 2012. 
A summary of observations is given in Table \ref{11hstab:obs}.
All observations were performed with the 300V grism, providing a spectral resolution of 12 \r{A}. 
The GG435 order sorting filter was used to avoid contamination of our data by higher spectral orders.
Our data are thus limited to the wavelength range 4450\r{A} -- 9330\r{A}.
The data were reduced as described in Chapter \ref{chpt:datred}.
The data acquired on 19 November, 24 November and 02 December 2011 were binned to 15 \r{A}, the data captured on 10 December 2011 were binned to 30\r{A} and the data obtained for subsequent dates were binned to 45 \r{A} to increase the signal-to-noise ratio. 
The spectra were flux calibrated using the standard stars reported in Table \ref{11hstab:obs} in order to remove the instrumental response.
It should be noted, however, that due to unknown wavelength dependent slit losses, absolute flux calibration was not possible.

\subsection{Observed flux and spectropolarimetry}
\subsubsection{Flux spectrum and photospheric velocity}
\begin{figure}
\centering
	\includegraphics[width=11cm]{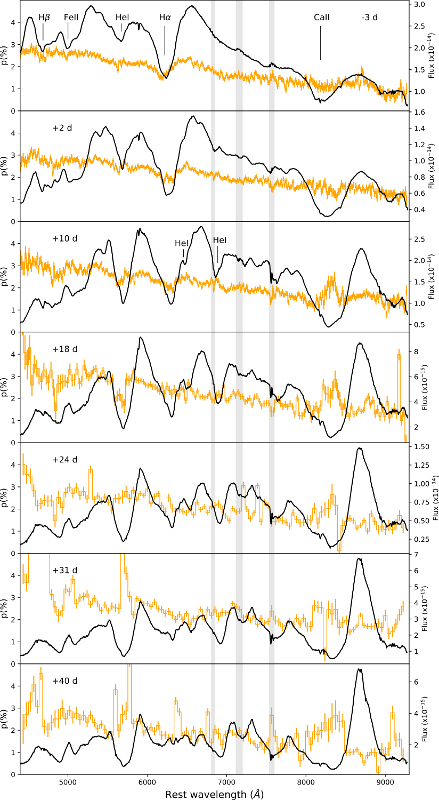}
    \caption[Polarisation of SN 2011hs before ISP removal]{\label{11hsfig:flu_n_pol0}Flux spectrum (black)  and degree of polarisation (not corrected for ISP - orange) of SN 2011hs  ranging from $-$3 days to +40 days. The grey shaded areas represent the regions affected by telluric lines. The data at $-$3, +2 and +10 days were binned to 15\r{A}, the data at +18 days were binned to 30\r{A} and the following epochs were binned to 45\r{A}. Note that the strong peaks of polarisation near 4600\r{A} and 5700\r{A} at +31 and +40 days are not considered to be real (see Section \ref{11hssec:obs_pol}). }
\end{figure}

In Figure \ref{11hsfig:flu_n_pol0}, we present the observed spectropolarimetry and flux spectrum at seven new epochs for SN 2011hs.

As noted by \cite{bufano14}, early epochs of spectroscopy are dominated by $\mathrm{H\alpha}$ ($-3$ days), but as the supernova evolves through maximum the feature decreases in strength and  a blue shoulder appears (+2 days).
We also see that from a week after maximum, the spectrum is dominated by He\,{\sc i} features. 
An exhaustive spectroscopic analysis is beyond the scope of this chapter, and more details on the spectroscopy of SN 2011hs and comparison to other SNe can be found in \cite{bufano14}.
Nonetheless, for the needs of this work we estimate the photospheric velocities at the time of observations (see Section \ref{11hssec:pol}).
The velocities of the Fe\,{\sc ii} lines in the blue part of the spectrum can be used as a proxy to probe photospheric velocity (e.g. \citealt{bufano14}), so we perform fits of Fe\,{\sc ii} $\lambda5169$ at all epochs in order to find the absorption minima. 
The resulting velocities are given in Table \ref{11hstab:vel_table}, and are consistent with the estimates of \cite{bufano14} from fits of the same line, see their figure 9.
Note that errors are not provided because the formal uncertainties on these values resulting from the Gaussian fits would be misleadingly small. 
The main source of error here would be from the blending of multiple lines, which is not taken into account.
Consequently, the values provided in Table \ref{11hstab:vel_table} should be used with caution.

\begin{table}
\centering
\begin{tabular}{r c }
\hline 
Epoch  & Fe\,{\sc ii}  \\
\hline 
$-$3 days & $-$9,860 \kms \\ 
+2 days & $-$8,760 \kms \\ 
+10 days & $-$6,320 \kms \\ 
+18 days & $-$5,570 \kms \\ 
+24 days & $-$5,390 \kms  \\  
+31 days & $-$4,810 \kms \\ 
+40 days & $-$4,470 \kms\\ 
\hline
\end{tabular}
\caption{\label{11hstab:vel_table}Line velocities at the absorption minimum for Fe\,{\sc ii} $\lambda 5169$ at all epochs,  which is used as a proxy for photospheric velocity.}
\end{table}

\subsubsection{Observed polarisation}
\label{11hssec:obs_pol}

As seen in Figure \ref{11hsfig:flu_n_pol0}, we detect significant polarisation in the direction of SN 2011hs, showing an underlying slope with a higher degree of polarisation at the blue end of the spectrum (\about3 percent) than at  the red end (\about1 percent).
This behaviour is not expected for the continuum polarisation of SNe since Thomson scattering is wavelength independent.
Additionally, some excess polarisation is associated with the absorption component of Ca\,{\sc ii} from 10 days after V-band maximum, and polarisation troughs are seen corresponding to $\mathrm{H\alpha}$ at the first 3 epochs, as well as with He $\lambda5876$ at +10 and +18 days.
The presence of troughs is inconsistent with the expected correlation of polarisation peaks with absorption components (e.g. \citealt{WW08}).
These discrepancies seem to indicate that interstellar polarisation contributes very significantly to the continuum polarisation observed (see Section \ref{11hssec:isp} for a more complete discussion). 

Lastly, it should be noted that the strong peaks near 4600\r{A} and 5700\r{A} at +31 and +40 days are coincident with significant variations from zero in our instrumental signature correction $\Delta \epsilon$ (see \citealt{maund08}), reducing our confidence in our data at these wavelengths. We therefore do not consider these peaks to be real.


\section{Interstellar polarisation}
\label{11hssec:isp}

\subsection{Galactic ISP}
\label{11hssec:gal_isp}
In order to estimate the contribution of the Galactic ISP we searched for Milky Way field stars in the vicinity of SN 2011hs. 
If we assume that they are intrinsically unpolarized, any polarisation measured for these stars is due to Galactic ISP. In the \cite{Heiles} catalogue, we found two stars within 2 degrees of SN 2011hs: HD218227 and HD215544 with polarisation $p = 0.012 $ and $p = 0.116$ percent, respectively. 
Using the parallaxes found in the Gaia data release DR2 \citep{gaia_parallaxes} we calculate distances of 36.1$\pm0.7$ pc and 501$\pm10$ pc for HD218227 and HD215544, respectively.
HD218227 being such a nearby object explains its low polarisation degree compared to HD215544, and the latter better samples the Galactic dust column and the ISP in the direction of SN 2011hs.

It is possible to put an upper limit on the Galactic ISP following Eq. \ref{eq:serk_plim}. 
The Galactic reddening in the direction of SN 2011hs is E(B-V)$=0.011 \pm 0.002$ \citep{schlafly11}, yielding  $p_{\text{ISP}} < 0.099 \pm 0.018$ percent. 
The polarisation of HD215544 is at the upper end of this limit (within uncertainties). 
This is much lower than the observed polarisation and the Galactic ISP is therefore a small contribution to the total observed ISP. 

\subsection{ISP determination from late time data}
\label{11hssec:ispep7_fit}
\begin{figure}
\centering
	\includegraphics[width=13cm]{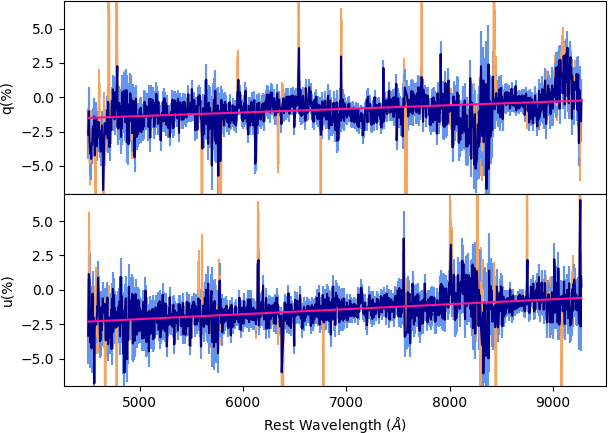}
    \caption[isgma clipped stokes parameters of SN 2011hs]{\label{11hsfig:ep7_isp}Stokes parameters $q$ and $u$ at +40 days (dark blue; errors in light blue) 3$\sigma$-clipped according to $\Delta \epsilon$; the discarded points are shown in orange. The bin size for the data is 3.3\r{A}. The fits to the $\sigma$-clipped data are shown in magenta.}
\end{figure}

In order to estimate the ISP from our observations, some assumptions must be made.
Any such estimate is inevitably model dependent. In the context of electron scattering by the SN ejecta, at late times the polarisation intrinsic to the SN will tend towards zero since the electron density decreases as the ejecta expand, and the observed polarisation therefore tends to the ISP. 
By assuming complete depolarisation at our last epoch (+40 days) and fitting a straight line to the normalised Stokes parameters $q$ and $u$, the wavelength dependent $q_{\text{ISP}}$ and $u_{\text{ISP}}$ can then be found (see Figure \ref{11hsfig:ep7_isp}). 
Best results were obtained when using the original data at +40 days (not rebinned to 45\r{A}), where spurious points were removed according to their corresponding $\Delta \epsilon$.
Data points whose $\Delta \epsilon$ showed a 3$\sigma$ or greater deviation from zero were not included in our fits.

The resulting ISP-wavelength relationships are: 
\begin{equation}\label{11hseq:qisp}
q_{\text{ISP}}\, {\rm (percent)} = 2.63 (\pm 0.19) \times 10^{-4} \times \lambda - 2.68 (\pm 0.14),
\end{equation}

\begin{equation}\label{11hseq:uisp}
u_{\text{ISP}} \, {\rm (percent)} = 3.61 (\pm 0.17) \times 10^{-4} \times \lambda - 3.94 (\pm 0.12),
\end{equation}
where $\lambda$ is in \r{A}, the gradients have units of \r{A}$^{-1}$ and the intercepts are unit-less. 
The reduced chi-squared ($\chi^2_{\nu}$) values on the $q$ and $u$ fits were 1.39 and 0.93, respectively. 
The fits are therefore satisfactory and the derived wavelength dependent relationships for $q$ and $u$ were used to retrieve the intrinsic polarisation of SN 2011hs presented and analysed in Section \ref{11hssec:pol}. 

\subsection{Host galaxy of SN 2011hs and Serkowski law}
\label{11hssec:serk}
From the upper limits on the Galactic ISP derived in Section \ref{11hssec:gal_isp} we can conclude that most of the ISP in the data of SN 2011hs originated from IC 5267.
Under the assumption that the size and composition of interstellar dust grains are similar in the host galaxy of SN 2011hs as to those in the Milky Way, the Serkowski limit $p_{\text{ISP}} < 9 \times E(B-V) $ can be applied to derive a value for the maximum $p_{\text{ISP}}$. 
In order to do so, a value for the reddening of IC 5267 must be calculated. 
To this end, we measured and averaged the equivalent widths of the Na\,{\sc i} D line in the data of epochs 1 to 3 (at later epochs the blend of sodium with He\,{\sc i} made our equivalent width measurements unreliable). 
We then used the relationship between equivalent width and reddening described by \cite{poznanski12}, and calculated $E(B-V)_{\text{host}} = 0.11\,(\pm 0.02)$. 
According to Eq. \ref{eq:serk_plim}, this yields $p_{\text{ISP}} < 1.09 \,(\pm 0.18)$. 
At 5000 \r{A}, however, we find $p_{\text{ISP}} = 2.50 (\pm0.15)$ percent, as calculated from $q_{\text{ISP}}$ and $u_{\text{ISP}}$ derived in Section \ref{11hssec:ispep7_fit}. 
Therefore, the assumption that the dust of IC 5267 or in the vicinity of SN 2011hs is similar to that of the Milky Way may not be correct.
It is however worth noting that the Na\,{\sc i} D lines are caused by discrete gas clouds and do not probe the more diffuse ISM, therefore only providing a lower limit on the dust extinction. Consequently, the \cite{poznanski12} estimate may not be completely adequate. 

Another approach is to fit our data at +40 days with the Serkowski law (see Eq. \ref{introeq:serk}).
We attempted to find new values of the parameters $p_{\text{max}}$ and $\lambda_{\text{max}}$ by minimising the $\chi^2$ to better fit our data.
Two forms of the constant K were considered: the original value derived by \cite{serkowski75}  $K_S=1.15$, and the $\lambda_{\text{max}}$ dependent form of  \cite{whittet92}  $K_W=0.01 + 1.66\lambda_{\text{max}}$.

The normalised Stokes parameters $q$ and $u$ can be expressed as functions of the wavelength dependent degree of polarisation $p(\lambda)$ and the polarisation angle $\theta$ as
\begin{equation}\label{11hseq:qu}
q = p(\lambda)\cos(2\theta)\; \text{and} \; u = p(\lambda)\sin(2\theta),
\end{equation}
where $p(\lambda)$ is defined by Eq. \ref{introeq:serk} and here $\theta = 120$\degree\, (as from Eqs. \ref{11hseq:qisp} and \ref{11hseq:uisp} the polarisation angle $\theta$ is found to be 120\degree$\pm2$\degree\, across our wavelength range). 
Note that we did not attempt to remove the Galactic component, since it is $<0.1$ percent and therefore contributes very little to the total ISP.

\begin{figure*}
	\includegraphics[width=17cm]{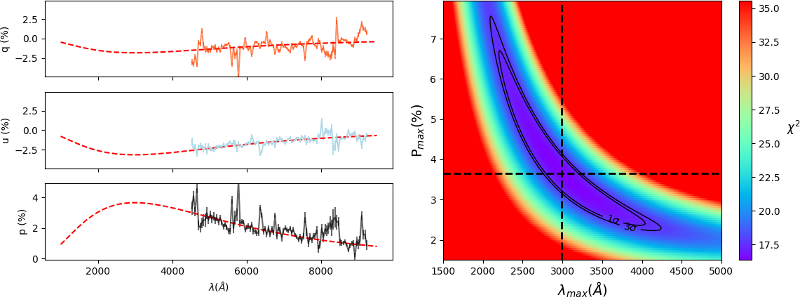}
    \caption[Serkowski law fit with Ks]{\label{11hsfig:fit_serk}\textbf{Left Panel:} The red dashed lines represent the Serkowski law fits ($K=1.15$ ) to the normalised Stokes parameters q and u at +40 days, and the corresponding fit to $p$. \textbf{Right Panel:} Reduced $\chi^2$ in parameter space. The 1$\sigma$  and 3$\sigma$ contours are plotted as thick black lines, the dashed lines indicate the values of p$_{\text{max}}$ and $\lambda_{\text{max}}$ for which $\chi^2_{\nu}$ is minimised. The colorscale shows the evolution of $\chi^2_{\nu}$ in parameter space. }
\end{figure*}

\begin{figure*}
	\includegraphics[width=17cm]{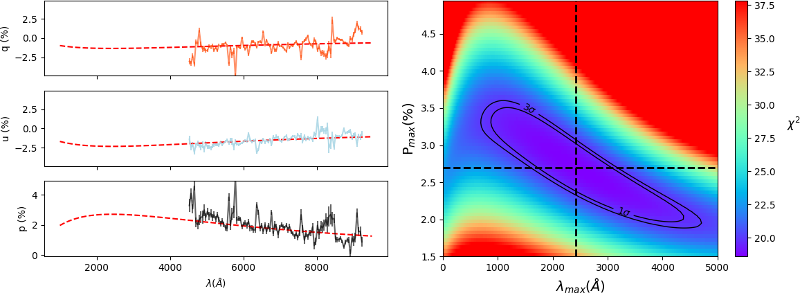}
    \caption[Serkowski law fit with Kw]{\label{11hsfig:fit_whittet}\textbf{Left Panel:} The red dashed lines represent  Serkowski law fits ($K=0.01 + 1.66\lambda_{\text{max}}$) to the normalised Stokes parameters q and u at +40 days, and the corresponding fit to $p$. \textbf{Right Panel:} Reduced $\chi^2$ in parameter space. The 1$\sigma$  and 3$\sigma$ contours are plotted as thick black lines, the dashed lines indicate the values of p$_{\text{max}}$ and $\lambda_{\text{max}}$ for which $\chi^2_{\nu}$ is minimised. The colorscale shows the evolution of $\chi^2_{\nu}$ in parameter space.  }
\end{figure*}

The fits to $q$, $u$ and $p$ performed using $K_S$, as well as  values of $\chi^2_{\nu}$ in parameter space are shown in Figure \ref{11hsfig:fit_serk}. 
The values of the reduced chi-squared (16.4 at best) indicate a relatively poor fit, and the size of the 1 and 3$\sigma$ contours in Figure \ref{11hsfig:fit_serk} reflect the uncertainty on the best parameter values. 
Nevertheless, it is possible using the 3$\sigma$ contours to put limits on $p_{\text{max}}$ and $\lambda_{\text{max}}$.
We find that 2090\r{A} $<\lambda_{\text{max}}<$ 4245\r{A} and 2.24 percent $<$ $p_{\text{max}}<$ 7.55 percent, with 99.7 percent confidence. 

We also fitted a Serkowski law with $K_{W}$ \citep{whittet92}, as shown in Figure \ref{11hsfig:fit_whittet}, along with the plot of $\chi^2_{\nu}$ in parameter space. 
Once again, the reduced chi-squared shows that the fit is relatively poor and the 1$\sigma$ contour covers a large portion of parameter space. 
We can however put limits on the values of $p_{\text{max}}$ and $\lambda_{\text{max}}$ from the 3$\sigma$ contour: 690\r{A} $<\lambda_{\text{max}}<$ 4700\r{A} and 1.88 percent $<$ $p_{\text{max}}<$ 3.66 percent with 99.7 per cent confidence. 

\subsection{Applicability of Serkowski's law}
\label{11hssec:disc_serk}

The $\chi^2_{\nu}$ values across parameter space showed very large 1$-\sigma$ contours, encompassing many $p_{\text{max}} - \lambda_{\text{max}}$ pairs.
This is mostly due to the fact that the  $\lambda_{\text{max}}$ peak is not in the range covered by our observations, hence making it difficult to better constrain the fit. 
The limits placed on the values of $\lambda_{\text{max}}$ extended well beyond the optical range into the far UV.
This is a concern as Serkowski's law was empirically defined from optical data \citep{serkowski75}, and it has been found that the law is not necessarily applicable in the UV (e.g. \citealt{anderson96, martin99, patat15}).
Furthermore \cite{martin99} explain that the Serkowski equation (see Eq. \ref{introeq:serk}) is not flexible enough to provide adequate fits all the way from UV to IR for a given $K$, and that, often, performing a three parameter fit to find p$_{\text{max}}$, $\lambda_{\text{max}}$ and $K$ gives better results. \cite{patat15} applied this method, and for the case of SN 2006X found $K=1.47\pm0.05$, which is very different from $K_{\text{W}}$ and $K_{\text{S}}$ which we used for our fits.
Therefore, fitting data which peak in the UV with optical Serkowski laws as is done here may not be adequate.

The very different shapes of the $\chi^2_{\nu}$ contours in parameter space for the two values of $K$ used in this work (see Figures \ref{11hsfig:fit_serk} and \ref{11hsfig:fit_whittet}) shows the importance of the choice of $K$. 
Given the nature of our data, in particular the fact that $\lambda_{\text{max}}$ is visibly outside of the range we cover, a three parameter fit cannot help constrain our estimate of the parameters.

\begin{figure}
\centering
	\includegraphics[width=8cm]{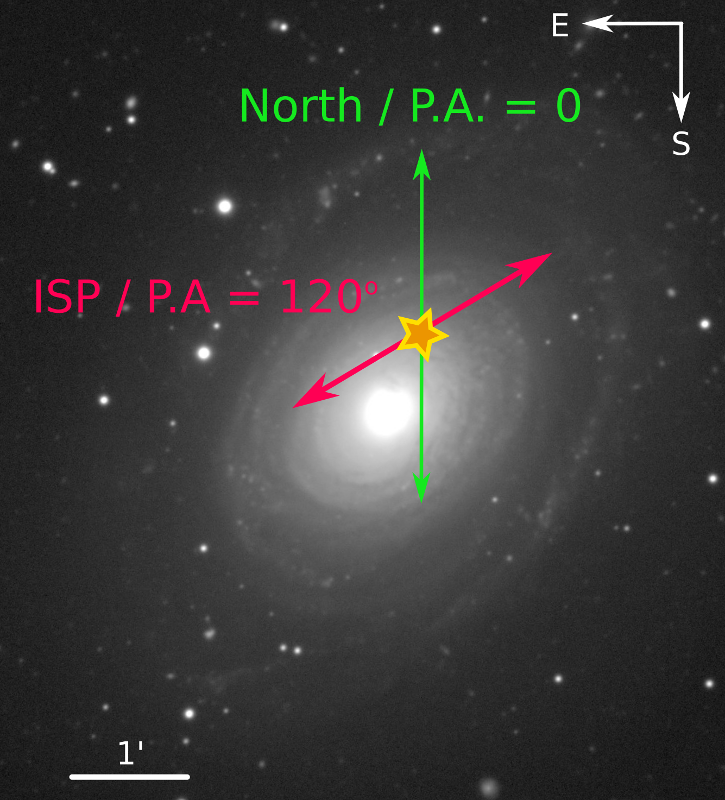}
    \caption[IC 5267 and P.A.]{\label{11hsfig:ic5267} Orientation of the ISP polarisation angle at the location of SN 2011hs (marked by the star) in IC 5267. The image of the host galaxy is a B, V, R, I composite from the Carnegie-Irvine Galaxy Survey (CGS -- \citealt{ho11}).}
\end{figure}

On the whole, the observed polarisation curves are very different from what is expected from Galactic type ISP. 
Furthermore, it is anticipated that the ISP P.A. should be parallel to the local spiral arm of the host galaxy at the location of the SN, due to the alignment of the dust with the magnetic field of the host. 
However, the ISP angle of \about 120\degree\, is found not to be parallel to the local spiral arm (see Figure \ref{11hsfig:ic5267}).
This could suggest that the dust in IC 5267 or in the local neighbourhood of SN 2011hs may be different from that in the Milky Way.

An alternative explanation to the observed ISP profile and P.A. is the presence of circumstellar dust. 
Polarisation curves increasing towards short wavelengths and with $\lambda_{\text{max}} < 4000$ \r{A} have been see in type Ia SNe (e.g. 1986G, SN 2006X, SN 2014J -- \citealt{patat15}). 
\cite{hoang17} modelled the observed polarisation curves and extinction and found this behaviour to be the result of an enhancement in small silicate grains in the dust. 
They suggest this could be due to cloud-cloud collisions resulting from SN radiation pressure.
Alternatively \cite{hoang18} show that large silicate grains can be destroyed by large radiation fields such as those around massive stars and SNe.
Additionally, \cite{cikota17}, remarked that the polarisation curves of type Ia SNe are similar to those of proto-planetary nebulae, whose polarisation curves are produced by scattering in the circumstellar medium. 
Consequently, the peculiar ISP found in the observations of SN 2011hs may hint at the presence of circumstellar dust around SN 2011hs.

\section{Intrinsic polarisation}
\label{11hssec:pol}
\begin{figure*}
	\includegraphics[width=17cm]{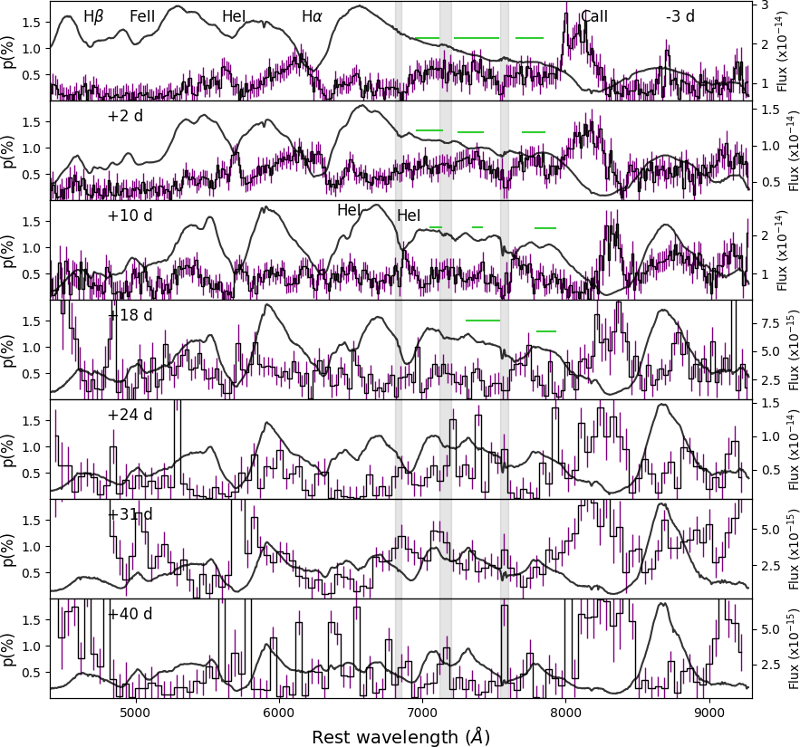}
    \caption[Polarisation of SN 2011hs corrected for ISP]{\label{11hsfig:pol} Degree of polarisation (corrected for ISP) of SN 2011hs  ranging from $-$3 days to +40 days (purple). The flux spectrum is shown in black, the light grey shaded areas highlight the regions of telluric lines, and the green horizontal lines show the wavelength ranges used to calculate the continuum polarisation. The data at $-$3, +2 and +10 days were binned to 15\r{A}, the data at +18 days was binned to 30\r{A} and the following epochs were binned to 45\r{A}. Note that the strong peaks of polarisation near 4600\r{A} and 5700\r{A} at +31 and +40 days are not considered to be real (see Section \ref{11hssec:obs_pol}).}
\end{figure*}

The degree of polarisation of SN 2011hs after ISP correction (see Section \ref{11hssec:isp}) is shown alongside the flux spectra in Figure \ref{11hsfig:pol}.

\begin{table*}
\caption{\label{11hstab:pol_table} Summary of the polarisation and polarisation angle (P.A.) of the continuum and strong lines of SN 2011hs by epoch.}
\begin{tabular}{ r  c l  }
\hline 
\hline
Epoch &  $p$ in percent  (wavelength range) & P.A. \\
\hline
\hline
\multicolumn{3}{c}{Continuum} \\
\hline

$-$3 days  & 0.55 $\pm$ 0.12 (6955-7115 / 7225-7530 / 7655-7840) & 170\degree$\pm$7\degree \\  
+2 days  & 0.75 $\pm$ 0.11 (6960-7140 / 7250-7425 / 7700 -7850) & 158\degree$\pm$4\degree\\  
+10 days & 0.48 $\pm$ 0.09 (7055-7130 / 7350-7415 / 7790-7930) & 141\degree$\pm$7\degree  \\  
+18 days & 0.29 $\pm$ 0.18 (7310-7535 / 7800-7925) & 120\degree$\pm$9\degree \\ 

\hline
\multicolumn{3}{c}{He\,{\sc i}$\lambda 5876$} \\
\hline

$-$3 days& 0.59 $\pm$ 0.05 (5590-5670) & 173\degree$\pm$8\degree \\  
+2 days& 0.89 $\pm$ 0.03 (5660-5715) & 155\degree$\pm$4\degree \\  
+10 days&  0.57$\pm$ 0.05 (5685-5771) & 141\degree$\pm$10\degree  \\  
+18 days&  1.08 $\pm$ 0.36 (5712) & 54\degree$\pm$3\degree \\ 

\hline
\multicolumn{3}{c}{$\mathrm{H\alpha}$} \\
\hline

$-$3 days&  0.75 $\pm$ 0.08 (6070-6200) & 5\degree$\pm$6\degree \\  
+2 days& 0.89 $\pm$ 0.10 (6085-6160) / 0.87 $\pm$ 0.08 (6225-6295) & 167\degree$\pm$1\degree / 168\degree$\pm$3.4\degree \\  
+10 days& 0.62 $\pm$ 0.02 (6150-6200) / 0.59 $\pm$ 0.01 (6285-6305)  & 155\degree$\pm$3\degree / 163\degree$\pm$1\degree \\  

\hline
\multicolumn{3}{c}{Ca\,{\sc ii}} \\
\hline

$-$3 days& 1.17$\pm$ 0.20 (8030-8180) & 177$^{\circ}\pm$4$^{\circ}$ \\  
+2 days& 1.25$\pm$ 0.15 (8055-8250) & 166$^{\circ}\pm$4$^{\circ}$ \\  
+10 days& 1.20$\pm$ 0.25 (8275-8375) & 134$^{\circ}\pm$7$^{\circ}$ \\  
+18 days& 1.35$\pm$ 0.30 (8205-8415) & 134$^{\circ}\pm$13$^{\circ}$ \\ 
+24 days& 1.30$\pm$ 0.26 (8100-8400) &  75$^{\circ}\pm$23$^{\circ}$ \\ 
+31 days& 1.79$\pm$ 0.22 (8100-8300) & 41$^{\circ}\pm$18$^{\circ}$ \\ 
+40 days& 1.64$\pm$ 0.16 (8100-8300) &  107$^{\circ}\pm$19$^{\circ}$ \\ 
\hline
\hline
\end{tabular}
\end{table*}

We derived the degree of polarisation and P.A. for the prominent polarisation features and the continuum where possible.
Each value is associated with a particular wavelength or wavelength range. 
In the former case, we simply recorded the polarisation at the corresponding wavelength bin, whereas when a range is provided we averaged the values within that range and used the standard deviation as the error. 
In order to estimate the continuum polarisation, we average the polarisation values in spectral regions devoid of strong lines and telluric lines.
We could only find suitable regions of the spectra in epochs 1 to 4, and the wavelength ranges considered are indicated by the green lines in Figure \ref{11hsfig:pol}.
The precise ranges used, and the continuum polarisation and P.A. derived can be found in Table \ref{11hstab:pol_table}.

The continuum polarisation is found to be constant within errors at epochs 1 and 2, with $p = 0.55\pm0.12$ percent and $p = 0.75\pm0.11$ percent, respectively.
At those dates the P.A. of the continuum is also constant within error (170\degree $\pm$ 7\degree and 158\degree $\pm$ 4\degree at $-$2 and +3 days, respectively).
The degree of polarisation then decreases to $p = 0.48\pm0.09$ percent by +10 days and $p = 0.29\pm0.18$ percent by +18 days, and a significant rotation of the P.A. is also observed (141\degree $\pm$ 7\degree and 120\degree $\pm$ 9\degree at +10 and +18 days, respectively).

Prominent polarisation features of He\,$\lambda5876$ are seen in Figure \ref{11hsfig:pol} from $-$3 days to +18 days with degree of polarisation as high as $p=1.08\pm0.36$ percent at the latter epoch.
The P.A. of helium follows that of the continuum at the first 3 epochs (see Table \ref{11hstab:pol_table}), but a sudden \about90\degree rotation is seen from +10 days to +18 days (from 141\degree to 54\degree, respectively).

The $\mathrm{H\alpha}$ polarisation at $-3$ days and +2 days exhibits a very broad feature  extending from \about 5700\r{A} to \about 6300\r{A}.
The broad hydrogen profile rises to $p=0.75\pm0.08$ percent  in the first epoch (see Figure \ref{11hsfig:pol} and Table \ref{11hstab:pol_table}).
By +2 days  two distinct peaks can be seen centred around 6125\r{A} and 6260\r{A}, which share the same degree of polarisation within errors ($p=0.89\pm0.10$ percent and  $p=0.87\pm0.08$ percent, respectively). 
These two peaks become better defined at +10 days, as the underlying broad profile fades, although their amplitude is seen to decrease by \about 0.3 percent with respect to epoch 2. 
In subsequent epochs, no significant hydrogen feature is detected above the noise. 
Over time  $\mathrm{H\alpha}$ exhibits a decrease in P.A., as seen in helium and the continuum.

Calcium shows by far the most prominent polarisation features at all epochs, and is the only element to continue exhibiting polarisation above noise levels up to our last epoch at +40 days (see Figure \ref{11hsfig:pol} and Table \ref{11hstab:pol_table}). 
The amplitude of the feature remains roughly constant from $-$3 days to +24 days, between 1 and 1.5 percent, although the P.A. is seen to slowly decrease over time.
In the last two epochs, however, the polarisation degree is well above the 1.5 percent level, and a change in the evolution of the P.A. is also seen, as between +31 days and +40 days it increases by \about60\degree.

\subsection{Polar Plots}

\begin{figure*}
	\includegraphics[width=16cm]{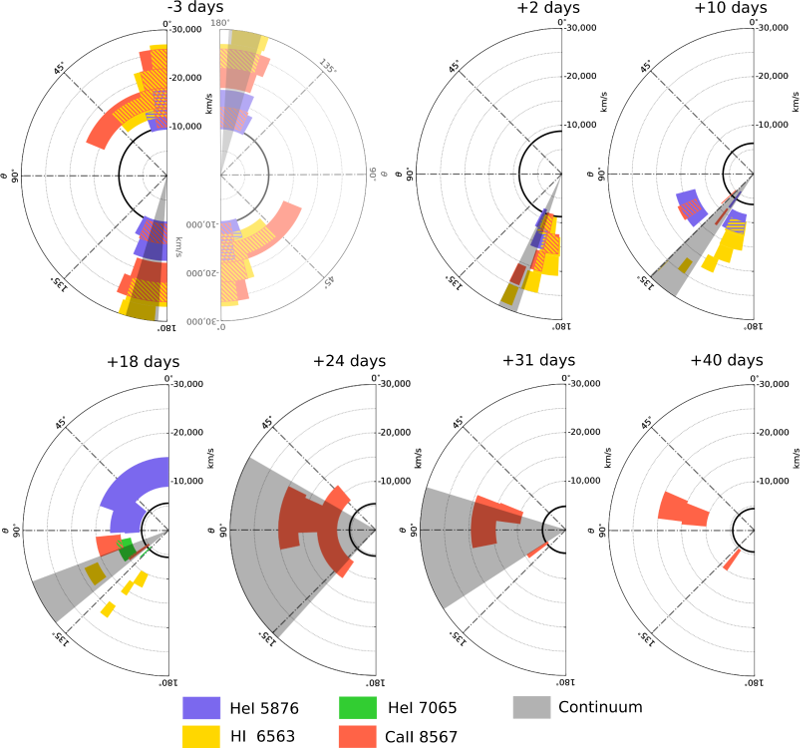}
    \caption[Polar Plots of SN 2011hs]{\label{11hsfig:polplot}  Polar plots of SN 2011hs from $-$3 days to +40 days with respect to V-band maximum light. The polarisation angles of He\,{\sc i}, $\mathrm{H\alpha}$ and Ca\,{\sc ii} are shown as a function of radial velocity. The velocity of Fe\,{\sc ii}$\lambda 5169$ (proxy for photospheric velocity) is shown in black. The hatched areas represent the overlap between elements. The continuum polarisation angle is shown in grey. It should be noted that the apparent symmetry seen at $-$3 days is a consequence of the nature of the Stokes parameters being quasi-vectors. The upper and lower arcs are joined at the 0\degree$-$180\degree boundary, as shown by the faded addition of the rotated diagram of the $-$3 days plot. }
\end{figure*}

To help visualise the behaviour of the P.A. in velocity space, we create polar plots, which are diagrams showing the relative positions of the P.A. associated with strong polarisation features.
They were first proposed by \cite{maund08D}, and use the radial velocity position and the angle on the plane of the sky as coordinates. 
In Figure \ref{11hsfig:polplot} we show the polar plots at all epochs for helium, hydrogen and calcium.
Note that the data were binned for better visualisation; the centres of the bins are the average P.A. calculated within the range.
The angular extent of the bin on the polar plot represents the error, so that a larger angular extent shows greater uncertainty on the P.A. for a particular line, not a greater physical extent. 

\begin{landscape}
\begin{figure*}
	\includegraphics[width=20cm]{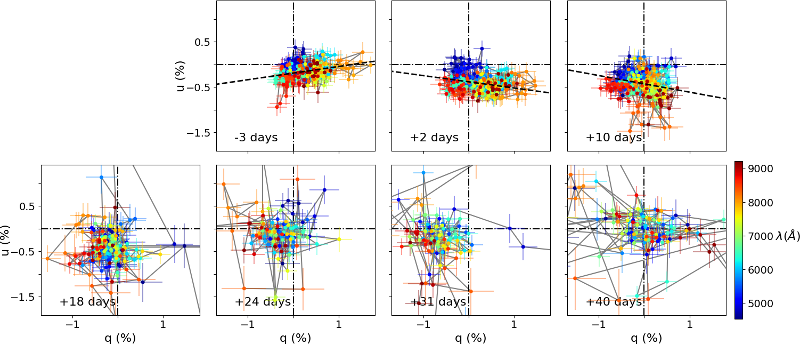}
    \caption[$q-u$ plots of SN 2011hs -- whole data]{\label{11hsfig:qu_all} Stokes $q-u$ planes of the ISP corrected data of SN 2011hs from $-$3 days to 40 days after V-band maximum. The data were binned to 15 \r{A} and the colour scale represents wavelength. The dominant axes of epoch 1 to 3 are shown as the dashed black line, and were calculated using ODR. }
\end{figure*}
\end{landscape}

The locations of the continuum shown on Figure \ref{11hsfig:polplot} for epochs 1 through 4 correspond to the polarisation angles calculated for the continuum values given in Table \ref{11hstab:pol_table}.
Unfortunately the continuum polarisation could not be formally calculated at the last 3 epochs, since we could not isolate spectral regions devoid of strong lines. 
We attempted to estimate the polarisation angle of the continuum by identifying its possible locus in $q-u$ space (see Section \ref{11hssec:loops} and Figure \ref{11hsfig:qu_all}) by using a $2$-$\sigma$ clipping method to eliminate outliers and the contribution of strong lines. 
This method was successfully used for the data at +24 and +31 days, yielding angles of 99\degree$\pm$39\degree and 98\degree$\pm$25\degree, respectively. 
This is consistent with an absence of rotation from +18 days to +24 and +31 days.
At the last epoch, the average values of $q$ and $u$ and their standard deviations are consistent with null continuum polarisation.

The change in P.A. of the continuum described in the previous section is visualised as a progressive clockwise rotation that takes place at epochs 2 to 4.
Additionally, the P.A. of helium, hydrogen and calcium clearly follow the clockwise rotation of the continuum over time, apart from He\,$\lambda5876$ at +18 days.
In Section \ref{11hssec:pol} we highlighted the 90\degree rotation of this feature from +10 days to +18 days.
This behaviour is clearly seen in Figure \ref{11hsfig:polplot}, as at +18 days the He\,$\lambda5876$ bins now fall into the upper quadrant, when the helium data were located in the bottom quadrant at +2 and +10 days.
It is also very distinct from the behaviour of the other line features and the continuum. 

\subsection{$q-u$ plane}
\label{11hssec:loops}

\begin{figure*}
	\includegraphics[width=15cm]{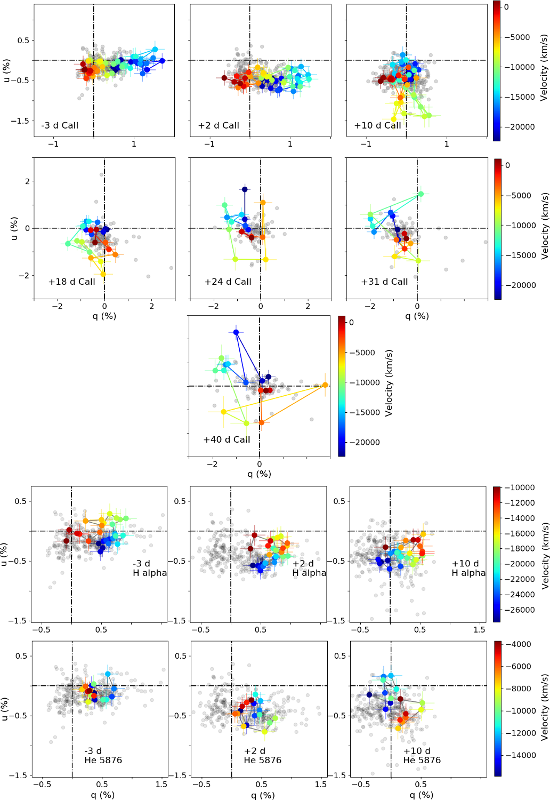}
    \caption[$q-u$ plots of SN 2011hs -- loops]{\label{11hsfig:loops} Stokes $q-u$ plots of the ISP corrected data of He\,$\lambda5876$ and $\mathrm{H\alpha}$ and Ca\,{\sc ii}.  The colour scales represent velocity in km\,s$^{-1}$ and vary between species for better visualisation. The grey points show the rest of the data across the whole wavelength range covered by our observations. }
\end{figure*}

In Figures \ref{11hsfig:qu_all} and \ref{11hsfig:loops} we show the ISP corrected polarisation data of SN 2011hs plotted on the Stokes $q-u$ plane.
This is a very useful tool to try to understand the geometries of the ejecta \citep{wang01}. 
Bi-axial geometries will result in a linear structure of the polarisation data on the $q-u$ plane.
More complex geometries (such as the addition of a third axis of symmetry) will cause departures from such an alignment, sometimes in the form of a smooth rotation of the P.A. across the wavelength range of spectral lines, i.e. loops \citep{WW08}.
Potential physical causes for these geometries and interpretations of the $q-u$ plots will be discussed in Section \ref{11hssec:pol_origin}.

The spectropolarimetric data of SN 2011hs at $-3$ days, +2 and +10 days show elongated ellipses on the  $q-u$ plane, and the data were fitted with a dominant axis by performing ODR (see Section \ref{datredsec:FUSS}) on the whole data range. 
The lines of best fit at $-3$, +2 and + 10 days were found to have inclinations on the $q-u$ plane of $8.3\pm1.4$\degree, $172.3 \pm1.4$\degree and $169.6 \pm2.2$\degree, respectively.  
Therefore a clockwise rotation is observed in the orientation of the dominant axis as well as in the continuum P.A. (see Figure \ref{11hsfig:polplot} and Table \ref{11hstab:pol_table}).

The data at epochs 4 to 7 (+18 to +40 days) have a lower signal-to-noise ratio and show no significant alignment along a dominant axis.
The data at +18, +24 and +31 days present no significant elongation, but their slight offset from the origin of the  $q-u$ plane is due to a residual level of intrinsic polarisation. 
At +40 days no elongation is seen and the data are centred  around the origin. which is expected since it was used to quantify the ISP.

In order to look for departures from bi-axial geometry it is worth isolating the spectral regions corresponding to strong lines and seek loops. 
In Figure \ref{11hsfig:loops}, we show the $q-u$ plots of Ca\,{\sc ii}, $\mathrm{H\alpha}$ and He\,$\lambda5876$.
The calcium data at the first two epochs follow a clear linear configuration.
A striking evolution is then seen between +2 days and +10 days, with the appearance of a loop; loops are seen at all subsequent epochs. 
This evolution indicates that at early days the calcium polarisation probes a bi-axial geometry and a third axis comes into play at epoch 3, resulting in the loops seen at and after +10 days.
Hydrogen, on the other hand, shows the clearest loop at $-$3 days, where the $\mathrm{H\alpha}$ line is strongest. 
This indicates a break of the bi-axial geometry by hydrogen that is not shared by the calcium.
Lastly, as the helium lines suddenly strengthen in the flux spectrum by +2 days, a loop also becomes visible in the wavelengths associated with He\,$\lambda5876$.
This loop becomes more prominent by +10 days.

The observables presented are discussed and compared to those of other IIb SNe in the following section.

\section{Discussion}
\label{11hssec:disc}

\subsection{Potential interpretations for the polarisation}
\label{11hssec:pol_origin}

In Section \ref{introsec:pol_origin} we summarised three classical base cases for the possible physical causes of the observed polarisation.
We repeat these below since we will use them extensively in this section to try to understand the data of SN 2011hs.
Note that visual aids are provided in Figure \ref{introfig:spherical_phot}.
\begin{enumerate}
\item[(i)] Aspherical electron distributions, such as an ellipsoidal photosphere, resulting in continuum polarisation \citep{1957lssp.book.....V, hoflich91}.
\item[(ii)] Partial obscuration of the underlying Thomson-scattering photosphere leading to line polarisation (e.g. \citealt{kasen03}).
\item[(iii)] Asymmetric energy input, e.g. heating by off-centre radioactive decay \citep{chugai92, hoflich95}.
\end{enumerate}

These three base cases provide a framework within which to understand changes in the continuum P.A. over time, as well as loops on the $q$-$u$ plane at specific epochs.

Changes in the continuum P.A. over time arise if the geometry of the ejecta is more complex than a simple bi-axial configuration. 
In the case of an ellipsoidal photosphere, for example, the overall evolution of the density gradient over time causes a change from a prolate to an oblate configuration, as the photosphere traverses layers with less steep density structures or a recombination front \citep{hoflich99}. 
This can be seen as a mixture of two case (i) scenarios (oblate + prolate), and this departure from the simple bi-axial geometry will result in a rotation in continuum P.A. across epochs. 
Another possible configuration causing such a rotation is a mixture of case (i) and case (iii), whereby the photosphere has a bi-axial geometry but an off-axis energy source also comes into play, resulting in a tri-axial geometry.

Rotation of the P.A. across spectral lines (loops) at a particular epoch can also be understood as being the result of a tri-axial component. 
In the mixture of case (i) and (iii) described above (an off-axis energy source in a bi-axial photosphere) the change of P.A. across wavelength arises as a result of the frequency dependence of the thermalisation depth (below which information about the geometry is lost due to multiple scattering -- \citealt{hoflich95}). 
Alternatively, loops will also form if the distribution of the line forming region blocking the photosphere is not aligned with the symmetry axis created by an underlying bi-axial photosphere -- case (i) + case (ii) -- or by an off-centre energy source -- case (ii) + case (iii).

In Section \ref{11hssec:pol} the continuum polarisation was seen to be significant and constant within errors at the first two epochs ($p = 0.55\pm0.12$ and $p = 0.75\pm0.11$ percent at $-$3 and +2 days, respectively), but then decreased down to \about 0.3 percent by +18 days (see Table \ref{11hstab:pol_table}).
Simultaneously, the $q-u$ plots showed significant elongation in the first 2 epochs, which then becomes less prominent by +10 days, finally leading to an absence of clear dominant axis by +18 days (see Figure \ref{11hsfig:qu_all}).
This suggests that the ejecta initially showed significant bi-axial geometry and overall appeared more spherical at later dates. 
In the context of an oblate ellipsoid, the early time ejecta differed from sphericity by \about10 percent \citep{hoflich91}.

Now focusing on P.A., we saw in Table \ref{11hstab:pol_table} and Figure \ref{11hsfig:polplot} that the continuum P.A. showed a gradual clockwise rotation (until +18 days, after which no precise estimate could be made). 
The average P.A. of the polarisation features associated with helium, hydrogen and calcium followed the behaviour of the continuum at the first 4 epochs (apart from He\,{\sc i} $\lambda5876$ at +18 days, which we discuss below).
Additionally, a clockwise rotation of the dominant axis is also seen in the $q-u$ planes between epochs 1 and 3.
The homogeneous behaviour of the continuum and line polarisation suggests that they are in some way coupled.
In a scenario where line polarisation is solely due to partial blocking of the photosphere by a non-isotropic distribution of the line forming regions -- our case (ii) described above -- it would be very unlikely to see the P.A. of multiple lines follow that of the continuum. 
Therefore this could suggest that a major contributor to the observed line polarisation is the global geometry.

In order to see a rotation in polarisation angle over time, a tri-axial geometry is required. 
A possible configuration could be an off-axis energy source within initially ellipsoidal ejecta. 
In the framework presented above, this would be a mixture of a case (iii) and a case (i).
As the photosphere recedes through the ejecta, the off-axis energy source would be revealed and add a third axis to the original bi-axial geometry, causing a change in P.A. over time.
Our estimates of the photospheric velocities (see Table \ref{11hstab:vel_table}) showed that at the first two epochs the photosphere was found at $-$9,860 km\,s$^{-1}$ and $-$8,760 km\,s$^{-1}$, respectively.
At epoch 3, when the rotation of the P.A. becomes significant, the photosphere had considerably receded down to $-$6,320 km\,s$^{-1}$, which could have started to reveal a deeper off-centre energy source. 

As previously mentioned, the emergence of an off-axis energy source at +10 days, breaking the initial bi-axial symmetry of the ejecta, would result in the loops on the $q-u$ plane.
Indeed, we saw that the calcium feature formed clear lines on the $q-u$ plots at $-$3 and +2 days, and then exhibited a clear loop at +10 days (see Figure \ref{11hsfig:loops}).
This, however, was not seen in hydrogen or helium, which showed loops at earlier dates. 
As described in Section \ref{11hssec:loops}, $\mathrm{H\alpha}$ has the clearest loop at $-$3 days, where hydrogen dominates the spectrum. 
The clarity of this loop diminished over the next 2 epochs, whereas the He\,{\sc i} $\lambda5876$ loop (non-existent at $-$3 days), became more distinct as helium started dominating the spectrum.
This seems to suggest that in addition to the global geometry effects dominating the P.A. behaviour of the line polarisation features, there could be line specific effects.
Therefore, some anisotropies in the distribution of the line forming regions of hydrogen and helium -- case (ii) -- may also be present. 

We want to emphasise that the possible solutions detailed in this section are non-unique. 
Additionally, they do not explain all of the observational characteristics described in Section \ref{11hssec:pol}. 
The drastic rotation of the P.A. of He\,{\sc i} $\lambda5876$ at +18 days, which differs from the rotation of the continuum and other elements, remains unexplained.
The cause for the increase in polarisation and rotation of the P.A. by \about60\degree \, in the calcium data at later times is also unclear.
Careful modelling is required, but is beyond the scope of this study.

\subsection{Comparison to previous studies}

\begin{figure}
	\includegraphics[width=15cm]{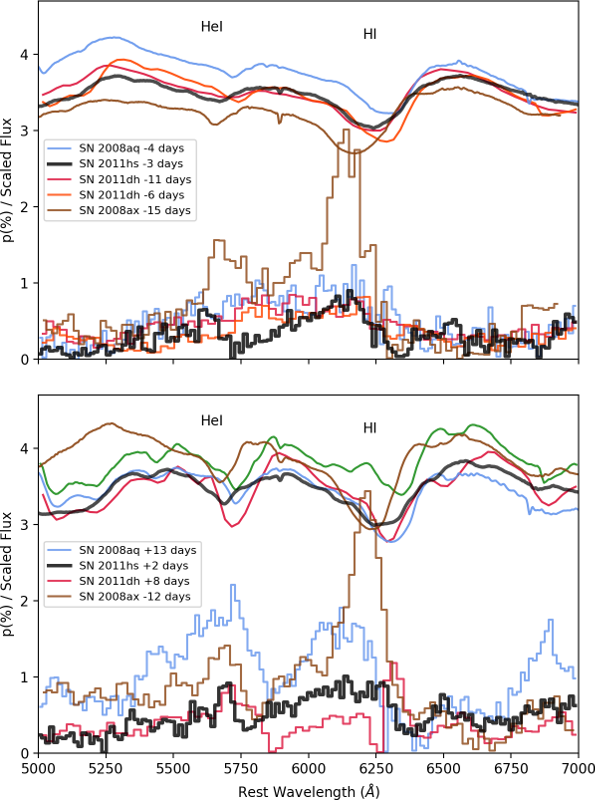}
    \caption[Comparison of helium and hydrogen polarisation in IIb SNe]{\label{11hsfig:broad_hhe} $\mathrm{H\alpha}$ and He\,{\sc i} $\lambda5876$ polarisation features in phases where hydrogen dominates (top panel) and after helium has strengthened (bottom panel). We show SN 2011hs, SN 2008ax \citep{chornock11}, SN 2011dh \citep{mauerhan15} and SN 2008aq (Chapter \ref{chpt:08aq}). The polarisation data are shown as step plots and the flux spectra are plotted for comparison as smooth curves. Note that the phases are quoted with respect to V-band maximum instead of explosion date.}
\end{figure}

As previously mentioned, type IIb SNe are relatively well represented in the spectropolarimetry literature and a variety of spectropolarimetric evolutions have been observed for different objects.
SN 2001ig only showed 0.2 percent continuum polarisation at early times, which then rose to \about 1 percent after maximum light \citep{maund01ig}. 
Other cases, exhibited strong continuum polarisation early on, with \cite{chornock11} recording 0.64 percent continuum polarisation in SN 2008ax 9 days after explosion, or 12 days before V-band maximum. 
In Chapter \ref{chpt:08aq} we also found some continuum polarisation (\about 0.2 -- 0.3 percent) in SN 2008aq 16 days after explosion, or 4 days before V-band maximum. 
Subsequently, the continuum polarisation of SN 2008aq rose above 1.3 percent a week after maximum, therefore following a similar pattern to SN 2001ig. 

The increase in polarisation as the photosphere probes the deeper ejecta is not universal.
Indeed SN 2011dh \citep{mauerhan15} showed significant continuum polarisation (0.45 percent) at 9 and 14 days after explosion (i.e. 11 and 6 days before V-band maximum) and decreased to $p<$0.2 percent by 30 days after explosion (i.e. a week after V-band maximum). 
A similar behaviour is observed in SN 2011hs, where  significant polarisation $p = 0.55\pm0.12$ and $p = 0.75\pm0.11$ percent was seen at $-3$ and +2 days with respect to V-band maximum, and then decreased by \about 0.25 percent by +10 days, down to \about 0.3 percent at +18 days.

Another common feature of the SN 2011hs and SN 2011dh data is the coupling between the P.A. of the $\mathrm{H\alpha}$ and He\,{\sc i} polarisation and that of the continuum.
\cite{mauerhan15} suggested this was best explained by clumpy excitation by plumes of nickel rising from the core.
This could also be the case in SN 2011hs, and would fit in with the scenario of an off-axis energy source described in Section \ref{11hssec:pol_origin}, where $^{56}$Ni plumes rising from the core could break the global bi-axial geometry of the outer ejecta. 

The presence of an asymmetric nickel distribution have numerous explanations.
In the case of the jet powered bipolar explosion of SN 1987A, asymmetric nickel distribution was suggested to cause polarisation \citep{chugai92}.
Alternatively,  3D simulations by \cite{wongwathanarat13} showed asymmetric distributions of nickel in neutron star kick scenarios, where the iron group elements are ejected in the direction opposite that  of the kick imparted on the remnant. 
These simulations, however, used progenitors of mass 15-20 \msol \,at explosion, which is not consistent with the mass limits derived by \cite{bufano14} for SN 2011hs (\mza = 12-15\msol).
This relatively low mass for the progenitor of SN 2011hs could imply the presence of a binary companion. 
Indeed, SN progenitors with \mza $<20$ \msol\, require binary interaction to lose enough mass to result in stripped envelope SNe \citep{dessart11}.
It was proposed by \cite{hoflich95} for the case of SN 1993J that an off-centre energy source could be the result of binary interaction, where the asymmetry results from the inner region of the ejecta still being accelerated by the gravitational potential of the binary whereas the faster outer regions are driven far from the orbit of the system very early.
Furthermore, the binary scenario could also explain the early asphericity through binary interaction, similarly to the case of SN 2001ig \citep{maund01ig}.
Note, however, that this is a non-unique solution to our data. 

Certainly one of the most commonly shared characteristics of type IIb SNe is the presence of line polarisation associated with He\,{\sc i} $\lambda5876$ and $\mathrm{H\alpha}$. 
In Figure \ref{11hsfig:broad_hhe} we place SN 2011hs in the context of other SNe at similar spectral stages (where hydrogen dominates and where helium starts becoming prominent). 
We would like to emphasise that the spectropolarimetric behaviour of type IIb SNe is non-uniform. 
Some SNe (SN 2008ax, SN 2011dh) show a strong blend of the helium and hydrogen polarisation features before helium features start strengthening as the photosphere recedes through the ejecta. 
At later dates, once helium lines become stronger, the $\mathrm{H\alpha}$ feature separates from the He\,{\sc i} $\lambda5876$ and two distinct peaks are visible. 
In other cases, e.g. SN 2008ax and SN 2011hs, two distinct peaks of $\mathrm{H\alpha}$ and  He\,{\sc i} $\lambda5876$ can be seen even before helium spectral features start deepening. 
It is interesting that SN 2011hs and SN 2011dh behave differently in this respect, despite the other commonalities in their spectropolarimetric characteristics. 
These disparities should be reproduced by future models of the spectropolarimetry of type IIb SNe.

The presence of loops associated with the polarisation of strong lines has also been observed repeatedly in type IIb SNe.
One of the most extreme cases was $\mathrm{H\alpha}$ in SN 2008ax (\citealt{chornock11} see their figures 13 and 14). 
In SN 2011hs, we saw loops of $\mathrm{H\alpha}$ and He\,{\sc i} $\lambda5876$ at early times. 
Hydrogen exhibited the most prominent loop at $-3$ days when hydrogen was strongest in the spectrum, whereas the helium loop only arose at +2 days and grew at +10 days as the helium lines strengthened. 
In SN 2011dh the best defined hydrogen loops are also found before helium starts showing prominent features in the flux spectrum (see figure 4 of \citealt{mauerhan15}). 
Contrary to SN 2011hs, however, SN 2011dh shows significant loops in He\,{\sc i} $\lambda5876$ even at the earliest times.
Another major difference between these two objects is in their calcium loops. 
As we saw in Section \ref{11hssec:loops}, the calcium data in SN 2011hs has a linear structure on the $q-u$ plane around maximum light, and only starts exhibiting a loop 10 days after V-band maximum. 
In contrast, SN 2011dh shows its strongest calcium loop at the earliest epoch (9 days after explosion or 11 days before V-band maximum).

These differences in the loop behaviour of type IIb SNe, and the precise origin of the loops of $\mathrm{H\alpha}$ and He\,{\sc i}$\lambda5876$ in SN 2011hs at the first two epochs, are difficult to understand without modelling. 
Toy models have been used in the past (e.g. \citealt{maund05hk}, \citealt{reilly16}), to try to reproduce line polarisation and constrain the ejecta geometry.
In an attempt to understand the early hydrogen and helium loops (and potentially extend this to other SNe), we created a model based on the same assumptions, but used a more methodical approach to explore parameter space (see Chapter \ref{chpt:tm}). 
The result of this work, however, was to demonstrate the great number of degeneracies that arise in such models, even when considering a small number of free parameters, as well as the fact that it can
result in good fits to the data even in cases were the original assumptions are invalid. 
This highlights the fact that simplified models must be considered very carefully as they may yield misleading results.

Fully hydrodynamic simulations with radiative transfer that can simultaneously reproduce the flux spectrum and the corresponding polarisation features will be necessary to better understand the varied geometry of type IIb and other core-collapse SNe. 
Additionally, higher cadence observations, especially around and soon after maximum light, could help better understand the variations of the ejecta geometry with depth.


\section{Conclusions}
\label{11hssec:conclusions}
We presented seven epochs of spectropolarimetry for the type IIb SN 2011hs from $-$3 days to +40 days with respect to V-band maximum. 
The observed polarimetry data showed very high levels of polarisation increasing towards blue wavelengths (up to \about 3 percent).
We quantified the ISP and identified that most of the observed polarisation was caused by the interstellar component.
Fits of the Serkowski law \citep{serkowski75} allowed us to constrain $\lambda_{\rm max}$ to wavelengths $<$ 4245\r{A} or $<$4700\r{A}, depending on the value of K used, either from \cite{serkowski75} or \cite{whittet92}, respectively.
Such levels of ISP have never been observed in a type IIb SN before. 
Similar behaviours of the interstellar component, with low values of $\lambda_{\rm max}$ have been seen in some type Ia SNe \citep{patat15}.
This may suggest enhanced levels of small silicate grains, either resulting from cloud-cloud collisions caused by SN radiation pressure, or due to the destruction of large grains by the radiation field \citep{hoang17,hoang18}. 
Consequently, the behaviour and level of the ISP in the spectropolarimetric data of SN 2011hs seem to indicate the presence of dust in the vicinity of the SN, which could reflect the mass loss history of the progenitor. 

The intrinsic polarisation of SN 2011hs was retrieved after removal of the ISP component. 
Significant continuum polarisation was observed at the first two epochs, with $p=0.55\pm0.12$ percent and $p=0.75\pm0.11$ percent, respectively, corresponding to \about10 percent departure from spherical geometry, in the context of an oblate spheroid \citep{hoflich91}. 
The continuum polarisation then decreased by \about 0.25 percent by +10 days and declined further by +18 days.
A strong correlation was found between the behaviour of the P.A. of hydrogen, helium, calcium and that of the continuum, indicating they share a common geometry.
The progressive rotation of the continuum P.A. after epoch 2 can be interpreted as the presence of an off-centre energy source being revealed.
This is supported by the calcium data on the $q-u$ plane, where a dichotomy exists between the first two epochs at which the data form a line (indicating bi-axial geometry), and the following epochs at which loops are observed (evidence for a departure from bi-axial geometry). 

On the other hand, $\mathrm{H\alpha}$ shows the clearest loop at the first epoch when hydrogen is strongest in the spectrum.
It then diminished as He\,{\sc i}$\lambda5876$ starts to show a loop by +3 days, which strengthens by +10 days, where helium is starting to dominate the spectrum.
These characteristics show that the lines of hydrogen and helium probe tri-axial geometries at early times when calcium does not, and therefore line specific geometries must also be a contributor to their polarisation. 
A possibility would be the presence of anisotropies in the distribution of their line forming regions.

Compared to previously studied type IIb SNe, SN 2011hs is most similar to SN 2011dh \citep{mauerhan15}, where a decrease in continuum polarisation over time and a correlation between the P.A. of hydrogen, helium and the continuum was observed. 
They also favoured an off-centre source of energy to explain their observations, namely in the form of plumes of nickel.
This could also be consistent with SN 2011hs, but is not unique and is probably an insufficient solution to our observations.

Lastly, there are a number of features whose origins remain unknown.
At +18 days the P.A. of He\,{\sc i}$\lambda5876$ undergoes a drastic rotation that is not coupled with that of the continuum or other elements.
Additionally, the calcium data at later dates see an increase in polarisation and a rotation in P.A. by +60\degree.

On the whole, SN 2011hs brings to the sample of type IIb SN spectropolarimetry data a number of features that have been previously seen as well as new disparities that need to be explained. 
It is clear from examples such as SN 1993J, SN 2001ig, SN 2008aq, SN 2008ax, SN 2011dh and SN 2011hs that there is great variety in the observed spectropolarimetric properties of type IIb SNe \citep{tran97, hoflich95, maund01ig, stevance16, chornock11, mauerhan15}.
More observations, especially with a higher cadence around and after maximum, and detailed hydrostatic models with radiative transfer of the current and future spectropolarimetric data, are required for us to better understand the similarities and the diversity in the geometries of type IIb SNe. 

As mentioned earlier, we will dedicate the next Chapter to reproducing a toy model previously used in the literature, and provide a more methodical approach to determining solutions and exploring parameter space. 
We will use SN 2011hs as a test case for this toy model. 

\chapter{A toy model to simulate supernova polarisation}

\label{chpt:tm} 

\lhead{\emph{Toy model}} 
\section{Introduction}
One of the possible ways in which line polarisation and loops can form in SNe is through non-isotropic distributions of line forming regions -- see case (ii) in Section \ref{introsec:pol_origin}.
In an attempt to better understand the ejecta geometry of the studied supernovae, we created a Monte Carlo toy model that could in principle be used to reproduce observations and therefore help us pin down possible ejecta geometries.
Similar models have been created in the past (e.g. \citealt{wang07, maund05hk, reilly16}), and this model is most similar to that of \cite{reilly16}, with minor differences in the physics and a major step towards the systematic exploration and determination of the best parameters. 

Firstly, the model itself and the assumptions made are described in Section \ref{tmsec:model}, the way we quantify goodness of fit is laid out in Section \ref{tmsec:goodness}, then the sampling method used to determine the best parameters is detailed in Section \ref{tmsec:param_space}.
The implementation and testing are discussed in Section \ref{tmsec:code}, the model is trialed using our SN 2011hs data in Section \ref{tmsec:11hs} and finally limitations are discussed in Section \ref{tmsec:limits}.

\section{The model}
\label{tmsec:model}

The first step in the creation of the model is to simulate the radiation coming from the photosphere (surface of last scattering) of the supernova.
The photosphere is defined in two dimensions by an ellipse with semi-major axis $a$ and semi-minor axis $b$. 
Photon packets are generated by randomly assigning them a position within the ellipse (hence "Monte Carlo").
The projected radius in 2D at a given $x, y$ position\footnote{In Cartesian coordinates with the origin placed at the centre of the photosphere.} is then: 
\begin{equation}\label{tmeq:r}
r = \sqrt{x^2 + y^2}, 
\end{equation}
and the radius of the ellipsoidal photosphere at a given $x$ coordinate is:
\begin{equation}\label{tmeq:R}
R = \sqrt{x^2 + b^2 \times \frac{a^2-x^2}{a^2}}.
\end{equation}

\begin{figure}
\centering
\includegraphics[height=8cm]{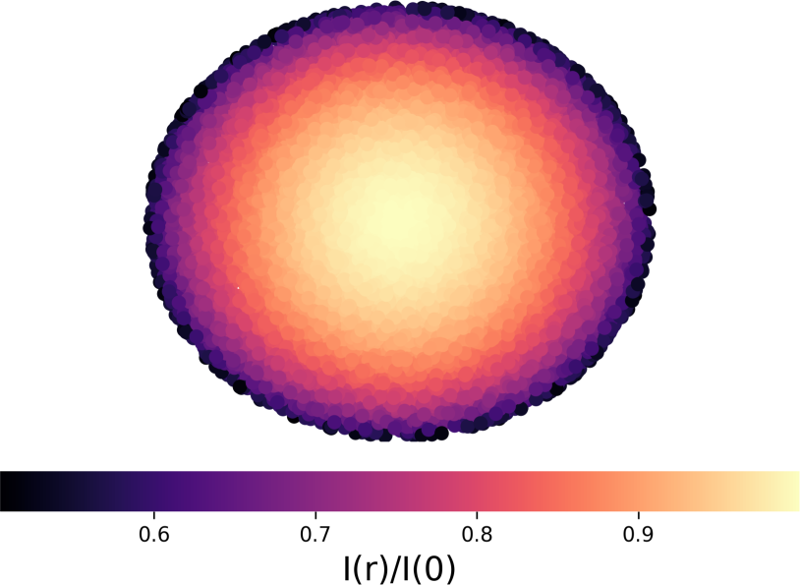}
\caption[Toy model limb darkening ]{Intensity given to individual photon packets in a photosphere modelled with $10^5$ photons, an axis ratio $E=0.86$ and PTPL = 30\%.}
\label{tmfig:proba_limb_dark}
\end{figure}

Following \cite{maund05hk}, it is then possible to map a linear spherical limb darkening model to calculate the intensity $I(r)$ (ranging from 0 to 1) of a packet of photons emitted at a distance $r$ from the origin:
\begin{equation}\label{tmeq:limbdark}
\frac{I(r)}{I(0)} = 1-k \times \bigg(1-\sqrt{1-\frac{r^2}{R^2}} \bigg),
\end{equation}
where k = 0.5 \citep{maund05hk}, and $I(0)$ is the intensity at the centre, which is set to 1. 
This intensity profile is illustrated in Figure \ref{tmfig:proba_limb_dark}.

\begin{figure}
\centering
\includegraphics[width=15cm]{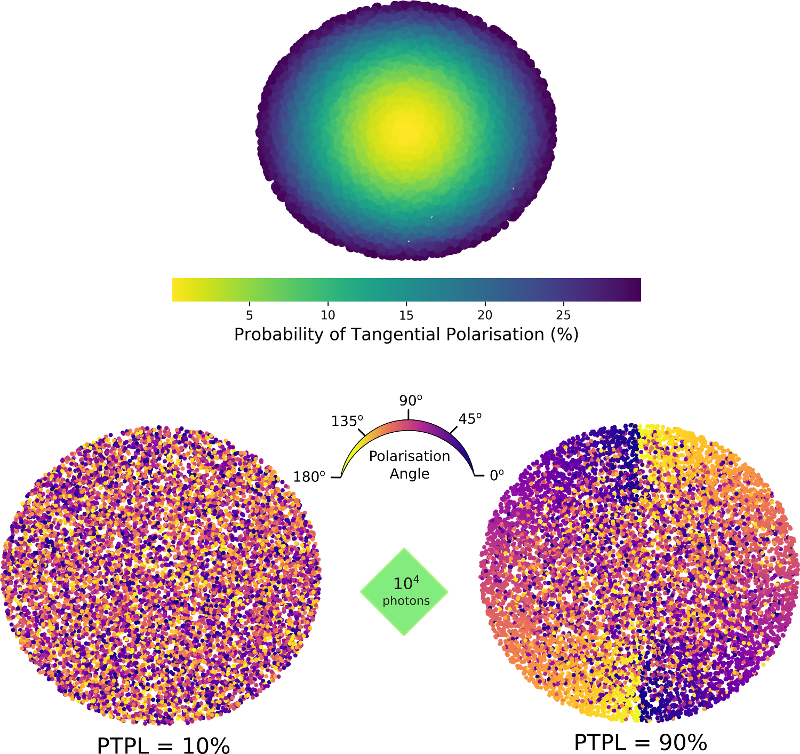}
\caption[Polarisation angles in model]{\textbf{Top:} Probability of individual photon packets to be given a polarisation tangential to the photosphere. Modelled with $10^5$ photons, an axis ratio $E=0.86$ and PTPL = 30\%; \textbf{Bottom:} Polarisation angles of 10,000 photon packets in circular photospheres with PTPL 10 percent (left) and 90 percent (right). }
\label{tmfig:test_pa}
\end{figure}

Additionally, each photon packet is assigned a polarisation (ranging from 0\degree to 180\degree) that is either randomly selected from a uniform distribution or set to be tangential to the photosphere at the position of the packet. 
The probability for a photon packet to be tangentially polarised ($P_{\top}$) will increase from 0 at the centre of the photosphere to a maximum value at the limb, according to the following quadratic probability function:
\begin{equation}\label{tmeq:proba}
P_{\top}  = {\rm PTPL} \times \bigg(\frac{r}{R}\bigg)^2
\end{equation}
where PTPL is the Probability of Tangential Polarisation at the Limb, which is fine tuned for a given axis ratio $E=a/b$ to produce a net degree of polarisation (before absorption) that is consistent with the results of \cite{hoflich91}, see Section \ref{tmsec:finetuning}.
A map of the probability for a packet of photon to be polarised tangentially to the photosphere as a function of radius is given in the top panel of Figure \ref{tmfig:test_pa}, for an arbitrary PTPL \about 30\% and axis ratio $E=0.86$. 
Additionally, we show in Figure \ref{tmfig:test_pa} the polarisation of the photon packets within a circular photosphere for a PTPL of 10 percent and 90 percent. 
This clearly demonstrates that in practice the packets from the centre of the photosphere are indeed randomly polarised and that a high PTPL imparts strong order to the polarisation of the photons at the limb. 

%
%

Line forming regions  were assumed to have very large optical depth $\tau \gg 1$ such that all the photons behind these regions would be absorbed.
Previous models by \cite{reilly16} include a number of geometries for the absorbing regions: circles, ellipses, triangular lobes, edge-on disk-like CSM. 
However, the solutions for these models were found through manual trial and error and therefore did not fully explore parameter space.  
This is understandable as the inclusion of complex shapes increases the number of dimensions of parameter space (compared to e.g. a circle). 
For example, a singular circular absorbing region is defined by 3 parameters in 2 dimensions: the coordinates for the centre ($x,\,y$) and a radius ($r_{\rm abs}$). 
An ellipse on the other hand is defined by a centre ($x,\,y$), a semi-major axis $a$, a semi-minor axis $b$ and an orientation $\theta$, bringing the dimensionality of parameter space to 5. 

One of the goals of this investigation, however, is to perform a systematic parameter search to identify potential degeneracies.
Consequently, a good starting point to this exploration is to focus on the most simple configuration: a single, circular absorption region. 

In summary, each model is defined by 5 parameters, 2  of which are set by the observations and the literature ($E$ and PTPL), and the other 3 are free parameters that define the absorption region  $\mathbf{x}_{\rm free} = (x, \, y, \, r_{\rm abs})$.

\section{Quantifying the goodness of a model}
\label{tmsec:goodness}
After creating a model, it is important to quantify how well it reproduces the observed data. 
There are 3 quantities that are taken into account when comparing a model to observations: the net Stokes parameters $q$ and $u$ that result from the model, and the ratio of the flux after and before absorption ($F_{\rm line}/F_{\rm cont}$).

\begin{figure}
\centering
\includegraphics[width=15cm]{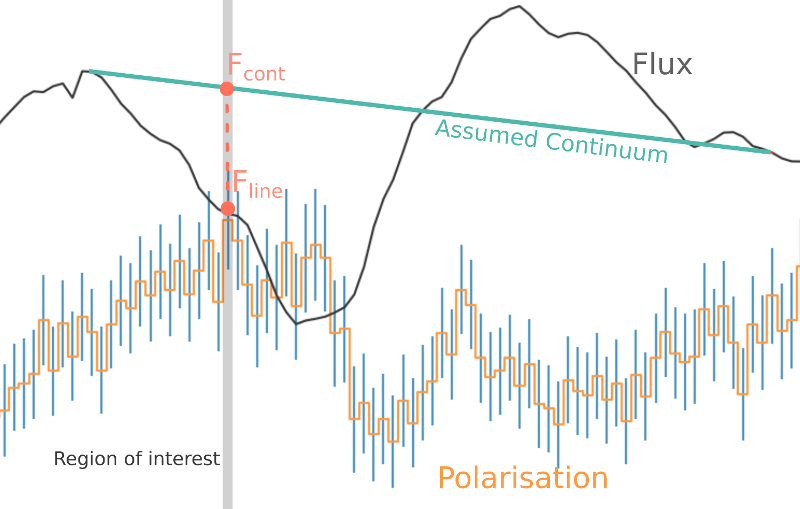}
\caption[Toy model continuum estimate]{Illustration of how the continuum flux of the spectral region being modelled is estimated. The data shown are those of the $\mathrm{H\alpha}$ line for SN 2011hs at +3 days, see Chapter \ref{chpt:11hs}.}
\label{tmfig:find_cont}
\end{figure}

The net Stokes parameters of a model are calculated by averaging the Stokes parameters of all the photon packets, weighted by intensity. 
As for $F_{\rm line}/F_{\rm cont}$, it is trivial to compute in the model by dividing the flux after absorption by the flux before absorption, but it can be difficult to accurately estimate it from the data. 
As a true continuum cannot be derived for the spectroscopic data, it is approximated as a straight line traced between 2 points on either side of the P Cygni profile of interest. 
This process is done manually and visually; an illustration of continuum determination is given in Figure \ref{tmfig:find_cont}.

Once the Stokes parameters and flux ratio of the model ($q_{\rm model}, \, u_{\rm model}$ and $(F_{\rm line}/F_{\rm cont})_{\rm model}$) and of the observed data ($q_{\rm obs}, \, u_{\rm obs}$ and $(F_{\rm line}/F_{\rm cont})_{\rm obs}$) have been calculated, the goodness of the model can be quantified. 
This is done by simply computing the residuals between the observed quantities and the modelled quantities.
The models were considered ``good" when the results fell within 1$\sigma$ of the observed data\footnote{Although see exceptions in Section \ref{tmsec:11hs}} .

\begin{figure}
\centering
\includegraphics[width=15cm]{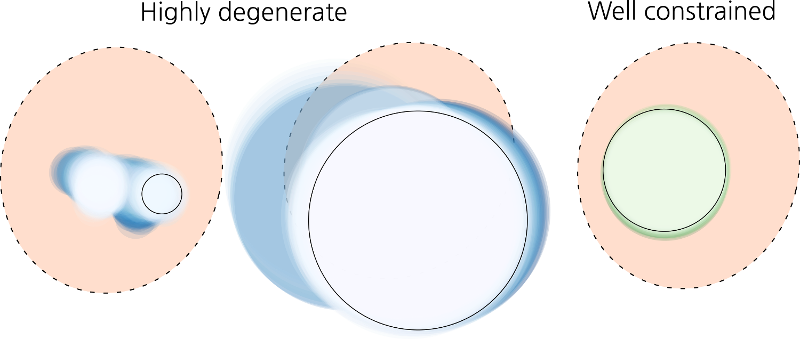}
\caption[Toy model degeneracy example]{Example of highly degenerate and well constrained solutions to the model. The photosphere is represented by the orange ellipse, the 1$\sigma$ solutions are illustrated by the coloured circles (blue or green), and the intensity of the colour represents the goodness of fit, with lighter hues indicating better solutions. The best solution for each model is highlighted by a black contour. From left to right the models correspond to the data for $\mathrm{H\alpha}$ at 6048\r{A} for SN 20011hs at $-$3 days, Ca II at 8240\r{A} for SN 2011hs at +10 days and $\mathrm{H\alpha}$ at 6152\r{A} for SN 2011hs at $-$3 days (see Chapter \ref{tmsec:11hs} for more details).}
\label{tmfig:example_models}
\end{figure}

Because the parameters in $\mathbf{x}_{\rm free}$ are correlated, providing error bars on individual parameters would not be very informative and potentially misleading.
To help visualise the results, the models found within 1$\sigma$ of the observations are plotted with a colour map used to indicate the better solutions (the lighter the colour the less deviation from observed values).
This is particularly effective at illustrating the level of degeneracy for a specific observation/model, as can be seen in Figure \ref{tmfig:example_models}

\section{Exploring parameter space}
\label{tmsec:param_space}
As a first approach, the models were run with \textit{one (and only one)} circular line forming region, which is defined by the location of its centre and its radius.
This is the geometry that is defined by the smallest number of free parameters, which is why we chose it to start our exploration of parameters space.
There are only three free parameters $\mathbf{x}_{\rm free} = (x, \, y, \, r_{\rm abs})$ -- see Section \ref{tmsec:model}.

In order to find the best solutions for $\mathbf{x}_{\rm free}$, parameter space needs to be explored effectively.
Firstly, it is possible to restrict the parameters that need to be investigated by making use of the fact that Stokes parameters are quasi vectors (i.e. the 0\degree and 180\degree direction are the same). 
This means that for each $\mathbf{x}_{\rm free}$, there is an equivalent set of parameters $\mathbf{x}_{\rm free}'$ (rotated by 180\degree) that returns the exact same solutions.
As a result, all possible values of $y$ and half of the possible values of $x$ were sampled.

We then perform a grid search and create look-up tables to summarise the output ($q_{\rm model}, \, u_{\rm model}$ and $(F_{\rm line}/F_{\rm cont})_{\rm model}$) for each $\mathbf{x}_{\rm free}$.
These tables are specific to a particular axis ratio, but can otherwise be used no matter which element or velocity slice is being investigated. 
This avoids duplicating simulations.

In practice it was computationally time consuming to create look-up tables that explore parameter space finely enough to consistently provide 1$\sigma$ solutions. 
To address this problem, look-up tables were created for the whole of parameter space with a moderately fine grid.
Region (or regions) of parameter space that yielded good solutions for a particular set of target values ($q_{\rm obs}, \, u_{\rm obs}$ and $(F_{\rm line}/F_{\rm cont})_{\rm obs}$) were then identified, and a finer grid was produced for this particular region.

\section{Implementation}
\label{tmsec:code}

None of the previous codes used for the toy models of \cite{reilly16} or \cite{maund05bf} were available, and I therefore had to re-create the model described in Section \ref{tmsec:model}, in addition to developing the parameter search code. 
This was done entirely in Python. 

The photosphere is implemented as an object that contains information about each individual photon packets, mainly: the Cartesian coordinates of the photon, the Stokes parameters $q$ and $u$ of each photon, and their intensity.
Before adding absorption regions and trying to model some data, simple tests were performed to ensure that the integrated polarisation of the generated photosphere was consistent with expectations.

\subsection{Testing}
\begin{figure}
\centering
\includegraphics[height=20cm]{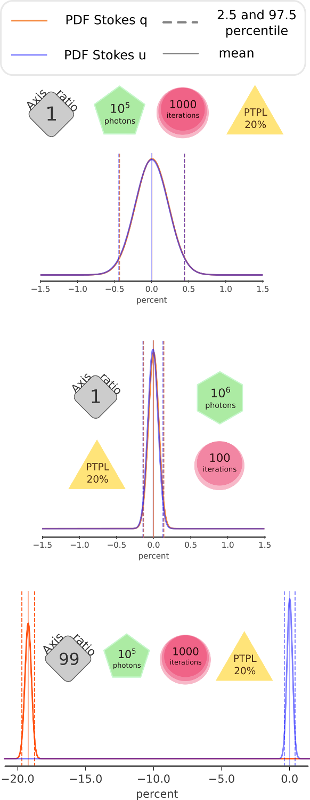}
\caption[PDF of the polarisation in tests]{Probability density functions of the Stokes parameters for 1000 circular photospheres with $10^5$ photon packets (top panel); 100 circular photospheres with $10^6$ packets (middle panel); 1000 photospheres with 99 percent asymmetry and $10^5$ photon packets (bottom panel).  }
\label{tmfig:pdf_stokes}
\end{figure}

One of the first steps in testing this model was to ensure that the integrated polarisation of a circular photosphere was null. 
In order to check this I created a large number of photospheres (1000), with axis ratio of 1,  $10^5$ photon packets each, and a PTPL of 20 percent.
Their respective $q$ and $u$ were calculated and used to infer the probability density function (PDF) of the Stokes parameters. 
The resulting PDF for $q$ and $u$ is plotted on the top panel of Figure \ref{tmfig:pdf_stokes}. 
On average we find the Stokes parameters to be $q=0.007 \pm 0.22$ and $u=0.002 \pm0.23$, where the uncertainties here are the standard deviations of the individual Stokes parameters recorded for the 1000 photospheres. 

Although these results show that when modelling a circular photosphere null polarisation is indeed retrieved, the level of uncertainty is too high for most practical uses, since the polarisation levels are usually a few tenth of a percent. 
Increasing the number of iterations does not affect these uncertainties, as they are related to the number of photon packets being created for each photosphere. 
We can check this by creating 100 photospheres with $10^6$ photon packets.
The resulting PDF of the Stokes parameters is shown in the middle panel of Figure \ref{tmfig:pdf_stokes}. 

In this case the average Stokes parameters are $q = 0.005\pm 0.070$ percent and $u = -0.004 \pm 0.069$ percent.
The errors are reduced by a factor of 3.14 for $q$ and 3.33 for $u$. 
This is consistent with the expectation from Poisson statistics that the noise level should decrease by a factor of $\sqrt{N}$ given an increase of the number of photon packets by a factor of $N$.

After showing that the circular photosphere case was be well-behaved, it was important to test the polarisation resulting from the introduction of asymmetry. 
In order to do this, I chose to simulate a very extreme configuration with an axis ratio of 99. 
In the case of this very elongated photosphere, most of the photon packets are near the limb, and therefore the polarisation should tend to the chosen PTPL. 
Additionally, the polarisation excess will be found to be tangential to the symmetry axis, such that if the photosphere is elongated in the 0\degree direction, we should expect the polarisation to be found in the $-q$ direction (and null in $u$).

For these tests 1000 photospheres with $10^5$ photon packets and an elongation in the 0\degree direction were simulated. 
As before, a PTPL of 20 percent was used. 

The average Stokes parameters are found to be $q = -19.22 \pm 0.24$ and $u = 0.009 \pm 0.20$ percent and their PDF is shown in the bottom panel of Figure \ref{tmfig:pdf_stokes}. 
As expected, the polarisation tends to the PTPL (20 percent) in the $-q$ direction, and Stokes $u$ is consistent with zero polarisation. 

Consequently, we are confident that the model imparts polarisation as expected. 

\subsection{Fine tuning the PTPL}
\label{tmsec:finetuning}

As mentioned in Section \ref{tmsec:model} the PTPL is fine tuned to a given axis ratio such that the resulting polarisation before considering absorption regions is consistent with the results of \cite{hoflich91}, see their figure 4. 
This step is necessary for consistency, as we derive the axis ratios from the cross-matching of our observations to the results of \cite{hoflich91}. 

\begin{figure}
\centering
\includegraphics[width=10cm]{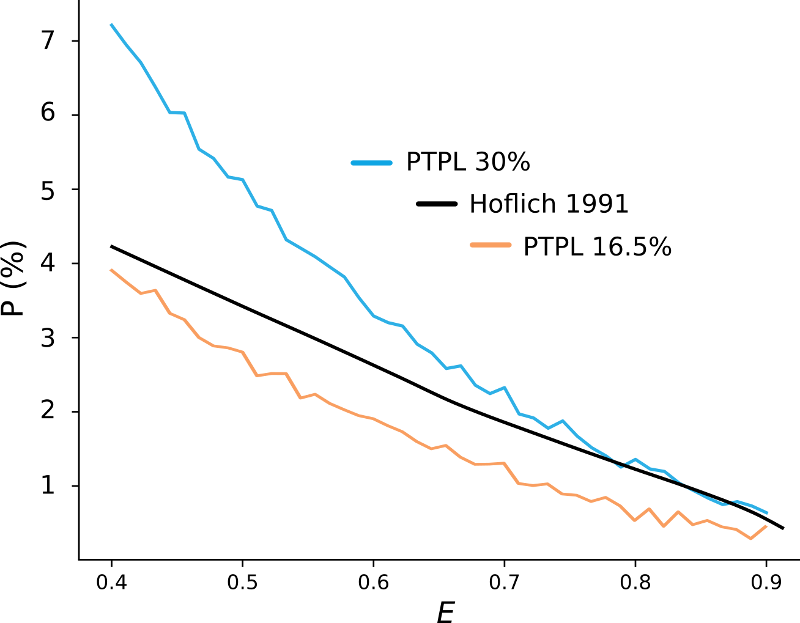}
\caption[Comparison with Hoflich 1991]{Comparison of the relation between polarisation and axis ratio $E$ for our model  using a PTPL of 16.5 percent (orange) 30 percent (blue) and that of \cite{hoflich91}. The black line was reproduced from their Figure 4. Note that the minor peaks and troughs in our polarisation are caused by noise. }
\label{tmfig:comp}
\end{figure}

The relation between $E$ and $p$ in this model and in that of \cite{hoflich91} differ (see Figure \ref{tmfig:comp}), since here we consider no physics and simply impart polarisation according to a power-law (see Eq. \ref{tmeq:proba}).
As a result, a single value of PTPL is not sufficient to reproduce the expected photospheric polarisation for all axis ratios and fine tuning to specific values of E is necessary.

In practice this was done by simulating 100 photospheres with $10^6$ photon packets for a range of PTPL of with increments of 0.5 percent.  
For each PTPL the average degree of polarisation $p$ and associated errors were calculated and compared to the continuum polarisation measured from the data.

\subsection{Implementing absorption regions and parameter space exploration}
After fine tuning the PTPL to ensure that an appropriate level of continuum polarisation is being generated, absorption regions can be added to model line polarisation.
The absorption region, since it is assumed to have very high optical depth such that all underlying photon packets are removed from the simulation, can be implemented with a simple mask. 

As previously mentioned, one of our main objectives is to systematically explore parameter space to find all good solutions and understand potential degeneracies. 
Our grid search approach for 3 free parameters required thousands of simulated systems (photosphere + absorption region), but due to the nature of Monte Carlo methods, photospheres simulated with the same parameters may yield slightly different results.
Additionally, re-generating a million photon packets and their associated polarisation is computationally intensive. 

Therefore, we favoured an approach where the photosphere were not re-created for every simulation. 
Instead, a single instance of the photosphere object was created and its state was saved to a file (``Pickled").
A range of absorption regions with parameters $\mathbf{x}_{\rm free}$ could then be modelled by calling in the object, masking the appropriate areas, and calculating the integrated polarisation.

%

\section{Application to SN 2011hs}
\label{tmsec:11hs}
We applied the toy model to SN 2011hs for the polarisation features of He\,{\sc i} $\lambda5876$, $\mathrm{H\alpha}$ and Ca\,{\sc ii} at $-$3, and +2 days, where the assumption of a bi-axial ellipsoid for the geometry of the photosphere seems robust (see Section \ref{11hssec:pol}). 
We also applied our models to epoch 3 (+10 days), where this assumption is not expected to be valid, for reasons that are discussed in the next section. 
For these epochs the asymmetry was \about 10 percent, and the fine tuned PTPL for an axis ratio of 0.87 was found to be 30 percent. 

The number of photon packets used in each simulation was determined by the precision required given the uncertainty on the observations that were being modelled. 
From Section \ref{tmsec:code} we know that for $10^6$ photons the errors on $q$ and $u$ are \about 0.07 percent.
Better precision can be achieved with more photons (e.g. \citealt{reilly16} used $10^7$  photons), but the run time scales linearly with photon number. 
Additionally, due to Poisson statistics, an increase in run-time by a factor of 10 would only result in an increase in SNR by a factor of \about$\sqrt{10}$.
Given the errors on the observational data investigated, models with $10^6$ photons optimised for shorter run times whilst providing sufficient precision.

The toy model solves for one wavelength bin at a time, but in an attempt to visualise the 3D structure of the line forming regions we selected 2 to 3 wavelength bins per line to model. 
This is an alternative between modelling only one bin per line feature (resulting in no depth information), and running models for all bins, which is unnecessarily laborious as successive bins will have very similar solutions given the size of the errors on our data.
Additionally, if a polarisation feature showed two components, we made sure to select bins sampling them both. 
Within those guidelines, the choice of bins is arbitrary, and other combinations of bins could have been chosen to obtain similar results. 
A summary of the target values for the bins chosen is given in Table \ref{tab:tm}.

\begin{figure}
	\includegraphics[width=15cm]{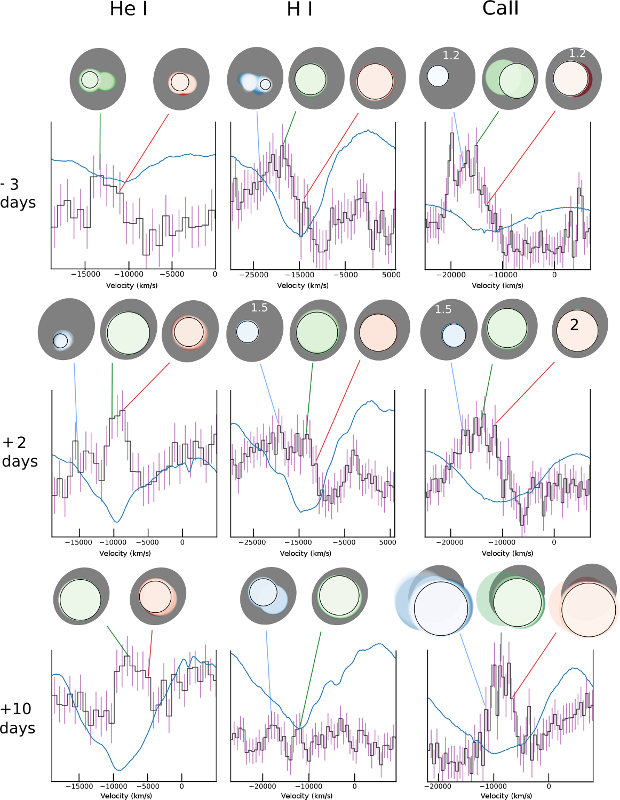}
    \caption[Toy model SN 2011hs solutions]{\label{fig:tm} Best solutions of our toy model for He\,{\sc i}$\lambda 5876$, $\mathrm{H\alpha}$ and Ca\,{\sc ii}$\lambda 8567$ at $-$3, +2 and +10 days. The target values and bins chosen are summarises in Table \ref{tab:tm}. The colour pallet represents the goodness of fit, where the lighter colours indicate the most likely solutions. The grey ellipse shows the photosphere, onto which solutions yielding good fits to the observed polarisation are superposed. The black circles indicate the absorption regions giving the best fit. In most cases the solutions within 1 $\sigma$ are shown, although some results are presented to 1.2, 1.5 and even 2 $\sigma$, as indicated by the numbers on the visual representation of the modelled ejecta. The coloured lines link the modelled bins to their visual representation. The flux spectrum (blue), and the degree of polarisation is shown (purple) are shown for comparison with the models.}
\end{figure}

In Figure \ref{fig:tm}, we show the models that could reproduce the observed polarisation data.
In most cases the solutions within 1 $\sigma$ are shown, although some results are presented to 1.2, 1.5 and even 2 $\sigma$ (as indicated by the numbers on the visual representation of the models shown in Figure  \ref{fig:tm}). 
These cases arose when parameters could not be found down to the 1-$\sigma$ level, and were the result of solutions being confined to very small regions of parameter space. 
Finer grids were explored, but due to time constrains we did not always reach the 1-$\sigma$ level, and halted simulations when sufficient precision was obtained. 
As can be seen in Figure \ref{fig:tm}, these cases yielded the most consistent solutions, with best constrained free parameters. 
This suggests that no solution exist within 1-$\sigma$. 
In these particular cases we find good agreement with the observed parameters, and the fitting issues arise in the line to continuum flux ratio. 
This is most likely due to the inaccuracy of the continuum flux estimate and the small formal errors on  $(F_{\rm line}/F_{\rm cont})_{\rm obs}$.

This type of visualisation, if accurate, can be complementary with Polar Plots (see Section \ref{11hssec:pol}), as they provide a view of the distribution of line forming regions as projected on the sky whilst polar plots show the P.A. of polarisation line features with depth.

We want to emphasise here that a model contains only one circular absorption region, and the multiplicity observed in some cases in Figure \ref{fig:tm} is just a result of the uncertainty on the possible location of the absorption region and shows the degeneracy of the solutions. 

\begin{landscape}
\begin{table}
\scriptsize
\caption{\label{tab:tm} Measured values to be reproduced by the models shown in Figure \ref{fig:tm}.}
\begin{tabular}{ r   c c c c c c c c c}

\hline
  &  \multicolumn{3}{c}{He I 5876} &\multicolumn{3}{c}{H$\alpha$} &\multicolumn{3}{c}{CaII}\\
  \hline
  \\
 \multicolumn{10}{c}{$-3$ days }\\
 \\
 & 5615\r{A} & 5660\r{A} & - & 6048\r{A} & 6152\r{A} & 6257\r{A} & 8046\r{A} & 8136\r{A} & 8195\r{A}\\
$q$ & 0.69 $\pm 0.18$ & 0.51 $\pm 0.17$ & - & 0.62 $\pm 0.18$ & 0.91 $\pm 0.19$ & 0.51 $\pm 0.19$ & 1.22 $\pm 0.22$ & 1.53 $\pm 0.24$ & 0.79 $\pm 0.24$ \\
$u$ & -0.051 $\pm 0.16$ & -0.24 $\pm 0.16$ & - & -0.21 $\pm 0.16$ & 0.21 $\pm 0.17$ & 0.09 $\pm 0.17$ & -0.09 $\pm 0.20$ & 0.27 $\pm 0.22$ & -0.14 $\pm 0.22$ \\
$F_{\rm line}/F_{\rm cont}$  & 0.91 $\pm 0.02$ & 0.886 $\pm 0.020$ & - & 0.948 $\pm 0.022$ & 0.690 $\pm 0.018$ & 0.592 $\pm 0.016$ & 0.81 $\pm 0.20$ & 0.604 $\pm 0.046$ & 0.604 $\pm 0.007$ \\
 \\
 \multicolumn{10}{c}{$+2$ days}\\
 \\
 & 5571\r{A} & 5675\r{A} & 5705\r{A} & 6137\r{A} & 6271\r{A} & 6316\r{A} & 8061\r{A} & 8165\r{A} & 8240\r{A}\\
$q$ & 0.25 $\pm 0.16$ & 0.64 $\pm 0.18$ & 0.55 $\pm 0.18$ &  0.96 $\pm 0.18$ & 0.95 $\pm 0.20$ & 0.66 $\pm 0.19$ & 1.17 $\pm 0.21$ & 1.25 $\pm 0.23$ & 1.25 $\pm 0.27$ \\
$u$ & -0.66 $\pm 0.15$ & -0.62 $\pm 0.16$ & -0.77 $\pm 0.16$ & -0.42 $\pm 0.16$ & -0.20 $\pm 0.18$ &  -0.17 $\pm 0.17$ & -0.63 $\pm 0.19$ & -0.56 $\pm 0.21$ & -0.66 $\pm 0.26$ \\
$F_{\rm line}/F_{\rm cont}$ & 0.94 $\pm 0.03$ & 0.72 $\pm 0.02$ & 0.81 $\pm 0.03$ & 0.523 $\pm 0.017$ & 0.556 $\pm 0.018$ & 0.78 $\pm 0.03$ &  0.495 $\pm 0.017$ & 0.443 $\pm 0.015$ \\
 \\
  \multicolumn{10}{c}{$+10$ days}\\
  \\
 & - & 5720\r{A} & 5779\r{A} & 6167\r{A} & 6301\r{A} & - & 8240\r{A} & 8315\r{A} & 8374\r{A}\\  
$q$ & - &  0.12 $\pm 0.20$ & 0.25 $\pm 0.19$ & 0.55 $\pm 0.19$ & 0.53 $\pm 0.21$ &  - & 0.20 $\pm 0.3$ &  0.45 $\pm 0.29$ & -0.136 $\pm 0.29$  \\
$u$ & - & -0.71 $\pm 0.19$ & -0.47 $\pm 0.17$ & -0.41 $\pm 0.17$ & -0.38 $\pm 0.19$ & - & -0.92 $\pm 0.3$ & -1.40 $\pm 0.28$ & -1.046 $\pm 0.27$  \\
$F_{\rm line}/F_{\rm cont}$ & - & 0.441 $\pm 0.013$ & 0.630 $\pm 0.018$ & 0.73$\pm 0.02$ &  0.450 $\pm 0.014$ & - & 0.36 $\pm 0.01$ & 0.324 $\pm 0.011$ & 0.404 $\pm 0.014$   \\
\hline
\end{tabular}
\end{table}
\end{landscape}

\section{Degeneracies and Limitations}
\label{tmsec:limits}
As can be seen in Figure \ref{fig:tm}, a number of cases show very high levels of degeneracy (e.g. calcium at +10 days, helium and hydrogen at $-$3 days).
This is a concern because our choice of only simulating one circular absorbing region resulted in a low number of free parameters (three), which is the simplest test case.
Using more complex geometries as done in \cite{reilly16}, or multiple absorbing regions, would result in a greater number of free parameters and a greater number of degeneracies. 
Consequently, the veracity of similar toy models of polarisation with more complex geometries is a cause for concern, and such results should be taken with great caution if parameter space has not been fully explored. 
It should also be noted that the method employed here to explore parameter space is not scalable to high numbers of parameters. 
The run-time of the current implementation, and the need for human input to constrain parameter space before fine solutions can be achieved, does not lend itself to a wider application.

Additionally, it is important to note that good fits were also found at +10 days, whereas in Chapter \ref{chpt:11hs} we saw a departure from axial symmetry in the photosphere at that epoch, therefore invalidating some of the assumptions in our model.
This highlights the importance of considering the effect of the continuum on the polarisation observed in the lines before attempting to interpret it in the context of our case (ii), since good fits may result from the application of such models despite potential issues with the original assumptions.  

On the whole, if the degeneracies of these models and the effects of the continuum polarisation are not fully understood, the results of such simulations could lend themselves to the over-interpretation, or an erroneous interpretation, of the origin of the polarisation  associated with strong spectral lines. 
More complex physical models are beyond the scope of this study, but are needed for a more accurate interpretation of the spectropolarimetric data of SN 2011hs and other SNe.

\chapter{A new analysis of SN 1993J}

\label{chpt:93J} 

\lhead{\emph{SN 1993J}} 

\section{Introduction}

SN 1993J was one of the first examples of the transitional Type IIb SNe, whose spectra resemble those of type II SNe at early days and later transition to helium dominated spectra like those of type Ib SNe \citep{swartz93, filippenko93}.
In addition to its peculiar spectral behaviour (at the time), SN 1993J also exhibited a double-peaked light curve. 
This phenomenon was explained as being the result of a late-type supergiant progenitor which had lost most of its hydrogen envelope through binary interaction \citep{podsiadlowski93, nomoto93}.
A massive binary companion was later discovered by \cite{maund93J}.

In addition to the photometric and spectroscopic studies of SN 1993J, spectropolarimetry analysis was also performed to detect asymmetries in this SN.  
Observational spectropolarimetric analyses of SN 1993J were published by \cite{trammell93} and \cite{tran97}. 
However, there has been debate over the level of interstellar polarisation (ISP) present in the data.
Both studies used the assumption that the emission component of the $\mathrm{H\alpha}$ line was completely depolarised. 
But this has since been shown not to be necessarily true (e.g. \citealt{tanaka09}). 

Additionally, the method used in \cite{trammell93} assumed that the ISP in NGC 3031 could be approximated by a Serkowski law with $p_{\rm max}$ equal to the degree of polarisation derived for the $\mathrm{H\alpha}$ emission component and  $\lambda_{\rm max}$ = 5500\r{A}. 
This means that a Galactic type ISP (with a Serkowski function and a Galactic value of $\lambda_{\rm max}$) was used, and this is not necessarily a good assumption, as seen in Chapter \ref{chpt:11hs}.

The method used by \cite{tran97} was not dependent on a valid Serkowski ISP, but required an estimate of the continuum, which the authors remarked is difficult to do and is a large cause of uncertainty.
We found continuum estimates to also be problematic in Chapters \ref{chpt:11hs} and \ref{chpt:tm}.

Overall, previous ISP estimates have relied on assumptions that we now know can be erroneous. 
A new investigation of the ISP of SN 1993J is therefore required, and is the aim of this work.
Using data obtained from private correspondence with H. Tran (see Section \ref{93Jobs}), we produced our own estimate of the ISP of SN 1993J, using a similar method to that developed for SN 2011hs (see Chapter \ref{chpt:11hs}), and compared our results to expectations and literature values (Section \ref{93Jisp}).
The new ISP was then removed and a new intrinsic polarisation for SN 1993J is presented and analysed in Section \ref{93Jpol}.
We compare our results to the literature and discuss in Section \ref{93Jdisc} how they affect previous conclusions.
A summary of our findings is given  in Section \ref{93Jconc}.

\section{Observations and data reduction}
\label{93Jobs}

\begin{table}
\centering
\caption[SN 1993J data]{\label{93Jtab:obs}Data obtained from private correspondence with H. Tran. Previously published in \cite{tran97}, although the coverage for some data sets differs from the original publication.\\$^*$Resolution \\$^{\dagger}$ Bin size} 
\begin{tabular}{c c c c c}

\hline
Date & Phase  & Telescope & Coverage (\r{A}) & Pixel Size (\r{A})\\
\hline
April 20 & +26 days & 3.0m Lick & 3900 - 5335 + 5865-7260 & 1.78* \\
April 26 & +30 days & 4.0m KPNO & 4000 - 7250 & 8$^{\dagger}$ \\
April 30 & +34 days & 3.0m Lick & 4600 - 7394 & 2.36*  \\
May 11 & +45 days & 3.0m Lick & 4600 - 7410 & 2.36* \\
May 14 & +48 days & 2.3m Steward & 4100 - 7440 & 4$^{\dagger}$ \\
\hline

\end{tabular}
\end{table}

We obtained from H. Tran (through J. R. Maund) the reduced spectropolarimetric and spectroscopic data of SN 1993J at 5 epochs (summarised in Table \ref{93Jtab:obs}). 
These data were first presented in \cite{tran97}, and we refer the reader to their section 2 for the details of the observations and data reduction.
It should be noted, however, that for this work the degree of polarisation $p$ (percent) and its error were re-calculated (when possible) from the errors on $q$ and $u$, and $p$ was debiased using a step function as done by \cite{wang97}.

\section{A new ISP}
\label{93Jisp}
\begin{figure}
	\includegraphics[width=\columnwidth]{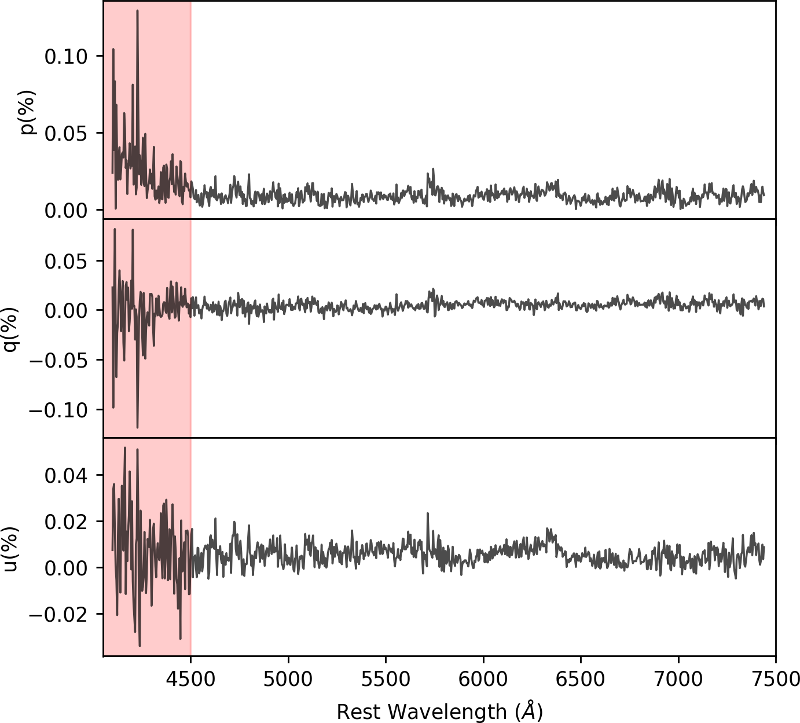}
    \caption[SN 1993J on May 14]{\label{fig:may14}Degree of polarization $p$ and Stokes $q$ and $u$ of SN 1993J on 14 May 1993 uncorrected for ISP. We highlight the spectral region below $<$4500\r{A} which were not considered when fitting the ISP. Note that significant increase in noise in the Stokes parameters data results in an upward slope in $p$.}
\end{figure}

Removing the interstellar component of the polarisation is crucial to understanding the intrinsic signal from the SN. 
In order to derive the interstellar polarisation, assumptions must be made. 
As previously mentioned, we now know that the $\mathrm{H\alpha}$ emission component is not necessarily completely depolarised. 
Instead we make the assumption that at our latest epoch (14 May 1993 / 48 days after explosion), the polarisation observed is dominated by the ISP rather than intrinsic SN polarisation. 
This is true if the electron density has decreased sufficiently such that very little light is being polarised in the ejecta by electron scattering. 
Similarly to Chapter \ref{chpt:11hs}, we want to fit the late time data with a straight line to estimate a wavelength dependent ISP. 
As in Chapter \ref{chpt:11hs}, we use a sigma clipping method since the data on 14 May 1993 have a low SNR.
Additionally, the data below 4500\r{A} at this epoch show a notable increase in noise (see Figure \ref{fig:may14}).
Since the spectral features of interest are located above 4500\r{A}, and given that 2 of our data sets have no data below 4600\ang\, (see Table \ref{93Jtab:obs}), we choose to fit only the data with wavelength $>$ 4500\r{A}.

A full discussion of the caveats of this method is given in Section \ref{93Jsec:disc_isp},  as well as a comparison to historical derivations of the ISP for SN 1993J and alternative ISP determination schemes.

\begin{figure}
\center
	\includegraphics[width=14cm]{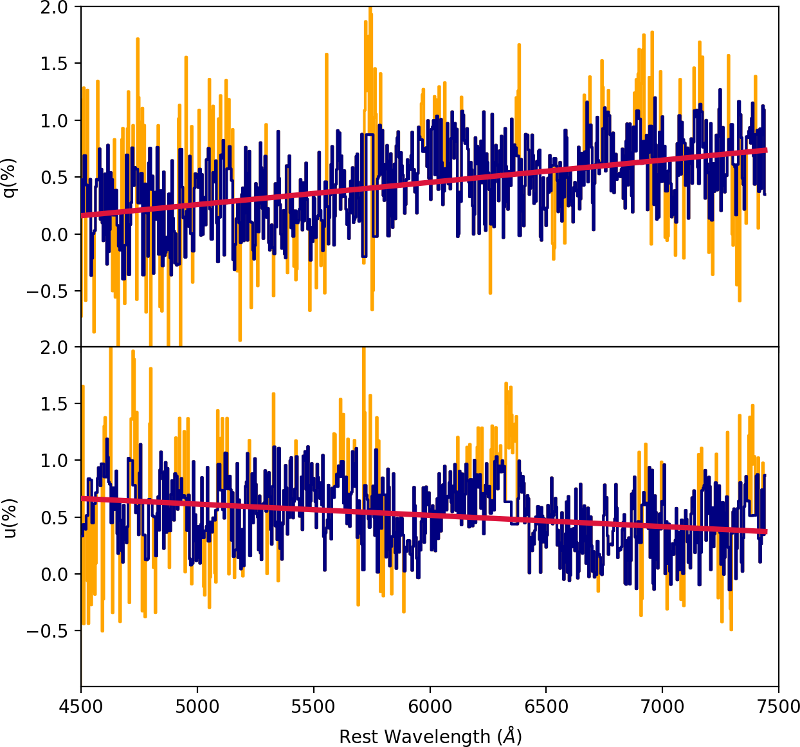}
    \caption[SN 1993J ISP fit]{\label{93Jfig:isp}$\sigma$-clipped Stokes parameters $q$ and $u$ at +48 days (dark blue) w.r.t explosion; the discarded points are shown in orange. The fits to the $\sigma$-clipped data are shown in magenta.}
\end{figure}

The fits to the $q$ and $u$ data are showed in Figure \ref{93Jfig:isp}. 
We find the following ISP-wavelength relationships: 
\begin{equation}\label{93Jeq:qisp}
q_{\text{ISP}}\, {\rm (percent)} = 1.96 (\pm 0.14) \times 10^{-4} \times \lambda - 0.72 (\pm 0.08)
\end{equation} 
\begin{equation}\label{93Jeq:uisp}
u_{\text{ISP}}\, {\rm (percent)} = -0.99 (\pm 0.13) \times 10^{-4} \times \lambda + 1.11 (\pm 0.08)
\end{equation}
where $\lambda$ is in \r{A}, the gradients have units of \r{A}$^{-1}$ and the intercepts are unit-less. 

\begin{figure}
\center
	\includegraphics[width=13cm]{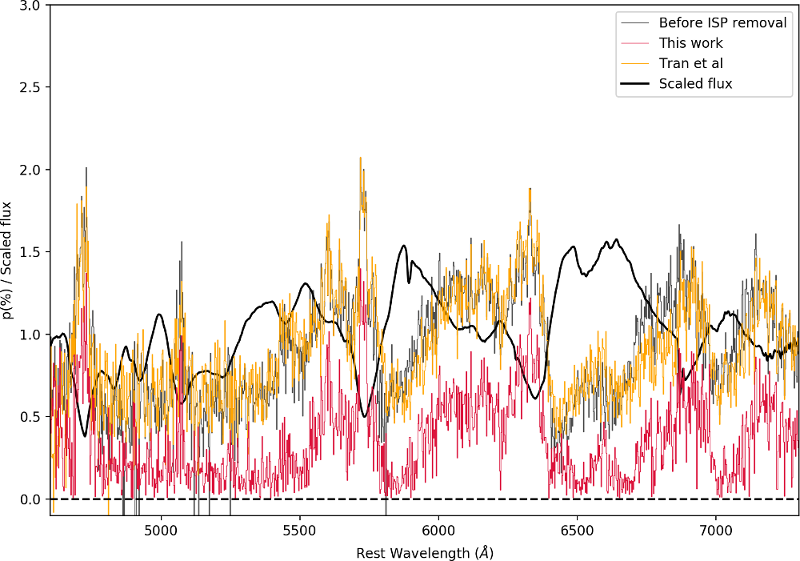}
    \caption[SN 1993J comparison]{\label{93Jfig:isp_comp} Comparison of the intrinsic degree of polarisation for SN 1993J on April 30 as obtained by \citeauthor{tran97} (orange) and in this work (red). The polarisation before ISP removal is also shown in grey, as well as the scaled flux spectrum in black.}
\end{figure}

A comparison of the intrinsic polarisation of SN 1993J on April 30 as obtained in this work and as estimated by \cite{tran97} is given in Figure \ref{93Jfig:isp_comp}.
On the whole, our ISP correction lowers the overall $p$ levels of SN 1993J. 
In particular, the near zero degree of polarisation associated with the blue parts of the spectrum (where line blanketing occurs) and in the spectral regions associated with strong emission lines is consistent with the expectation that these spectral regions should show little polarisation. 

It is also in line with the remark of \cite{chornock11}:
when comparing their data of SN 2008ax to other IIb SNe (including SN 1993J), they observed that the level of polarisation on April 30 near 6600\r{A} in the  $\mathrm{H\alpha}$ peak (\about0.7 percent) reported by \cite{tran97}  was high.
They suggested that a different ISP estimate for SN 1993J could lower the polarisation levels of SN 1993J in this spectral region, and this is seen to happen with our ISP estimate (see Figure \ref{93Jfig:isp_comp}).

Overall, we are confident that our ISP removal method is an improvement on previous methods and we use our estimates to correct the data we have for SN 1993J.

\newpage
\section{Intrinsic Polarisation}
\label{93Jpol}

\subsection{Degree of polarisation}
\label{93Jsec:p}
\begin{figure}
\center
	\includegraphics[width=15cm]{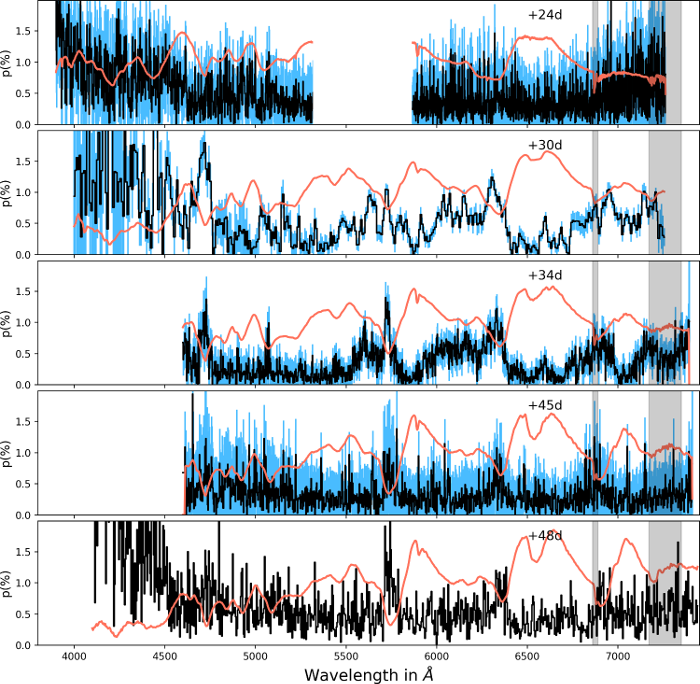}
    \caption[SN 1993J intrinsic polarisation]{\label{93Jfig:93j_pol} ISP corrected degree of polarisation of SN 1993J (black) and scaled flux spectrum (red). The errors on the polarisation are shown in blue, although note that we do not have errorbar for the last epoch. The grey shaded areas show the telluric line regions.}
\end{figure}

In Figure \ref{93Jfig:93j_pol} the ISP corrected degree of polarisation of SN 1993J is presented alongside the flux spectrum from +9 days to +48 days (with respect to explosion date). 
At our first epoch the data are completely dominated by noise, but the data at +30 and +34 days show very clear line polarisation features. 

\begin{table}
\centering
\caption[SN 1993J polarisation peaks]{\label{93Jtab:pol}Summary of the main polarisation peaks in the data of SN 1993J. The bin size can be found in Table \ref{93Jtab:obs}.}
\begin{tabular}{c c c c}
\hline
Element & Wavelength / Velocity & $p$  & P.A \\
 & (\r{A}/\kms) & (percent) & (\degree)\\
\hline\\
\multicolumn{4}{c}{April 26: +30 days} \\
\\
$\mathrm{H\beta}$ & 4720 / $-$8,800\kms & 1.8$\pm$0.3 &  18$\pm$6\\
HV He\,{\sc i} $\lambda5876$ & 5632 / $-$12,500\kms & 0.93$\pm$0.11 & 43.0$\pm$0.3 \\
He\,{\sc i} $\lambda5876$ & 5728 / $-$7,550\kms & 1.02$\pm$0.08 &  46.3$\pm$0.2\\
HV $\mathrm{H\alpha}$ & 5912 / $-$30,000\kms & 0.60$\pm$0.12 &  23$\pm$5\\
HV $\mathrm{H\alpha}$ & 6040 / $-$24,000\kms & 0.91$\pm$0.12 &  32$\pm$ 2\\
HV $\mathrm{H\alpha}$ & 6072 / $-$22,500\kms & 0.94$\pm$0.11 & 37$\pm$1 \\
HV $\mathrm{H\alpha}$ & 6152 / $-$18,800\kms & 0.90$\pm$0.11 & 43.4$\pm$0.2 \\
$\mathrm{H\alpha}$ & 6304/ $-$11,800\kms & 1.20$\pm$0.13 & 45.9$\pm$0.1\\
HV He\,{\sc i} $\lambda6678$  & 6456/ $-$10,000\kms & 0.40$\pm$0.13 & 115$\pm$7 \\
He\,{\sc i} $\lambda6678$  & 6600/ $-$3,500\kms & 0.63$\pm$0.07 & 88$\pm$74\\
He\,{\sc i} $\lambda7065$  & 6952/ $-$4,800\kms & 0.98$\pm$0.11 & 47$\pm$0.2\\

\\
\multicolumn{4}{c}{April 30: +34 days} \\
\\
$\mathrm{H\beta}$ & 4695-4740 / $-10,500$ to $-7,500$\kms & 0.90$\pm0.08$ &  24\degree$\pm2$\\
HV He\,{\sc i} $\lambda5876$ & 5590-5620 / $-14,600$ to $-13,200$\kms & 0.78$\pm0.05$ & 34\degree$\pm1$ \\
He\,{\sc i} $\lambda5876$ & 5715-5740 / $-8,200$ to $-7,000$\kms & 1.20$\pm0.08$ & 32\degree$\pm1$ \\
$\mathrm{H\alpha}$ & 6332 / $-10,550$\kms & 1.17$\pm$0.23 & 36\degree$\pm2$ \\
\hline

\end{tabular}
\end{table}

\subsubsection{+30 days}
At +30 days, helium and hydrogen polarisation features are strong and complex, showing multiple peaks. 
The wavelength, degree of polarisation $p$ and P.A. of the strongest features are summarised in Table \ref{93Jtab:pol}.

The strongest peak at that epoch is that of $\mathrm{H\beta}$; it is very broad, spanning \about 100\r{A} and reaching a polarisation of level of 1.8 percent. 
$\mathrm{H\alpha}$ also exhibits a very broad feature, extending from \about 5960 to 6415 \r{A}.
The absorption minimum of the P Cygni profile of $\mathrm{H\alpha}$ is correlated with a strong peak reaching a maximum polarisation of $p=1.20\pm0.13$ percent at 6304\r{A}.
Three other distinct peaks are superposed onto the broad hydrogen feature at 6040, 6072, and 6152\r{A}, all with $p$\about0.9 percent. 
They are coincident with dips in the blue shoulder of the $\mathrm{H\alpha}$ absorption line profile, and I interpret these as high velocity (HV) components of hydrogen overlapping the main P Cygni profile and its associated inverse P Cygni in the spectropolarimetry. 
Additionally, a very sharp peak can be observed at 5912\r{A}, offset in the blue compared to the main $\mathrm{H\alpha}$ features. 
This could be the result of an even higher velocity component of hydrogen (\about 30,000\kms). 

He\,{\sc i} $\lambda5876$ also shows a complex multi-peaked feature at +30 days. 
The two main components are found at 5632 and 5728\r{A} reaching a similar degree of polarisation, $p$ \about 1 percent (see Table \ref{93Jtab:pol} for more detail).
The 5632\r{A} peak coincides with the bottom of the absorption component of He\,{\sc i} $\lambda5876$, whereas the other peak is associated with a blue shoulder in the profile of the absorption component of the P Cygni profile. 
As a result we refer to this second peak as a HV component of the He\,{\sc i} $\lambda5876$ polarisation.
Weaker features ($p$\about 0.5 percent) are found both on the blue and red side of those peaks, suggesting asymmetries at a wide range of depth in the ejecta. 

Additionally, the spectral region associated with the $\mathrm{H\alpha}$ emission, although it is mostly depolarised, shows two peaks that are likely associated with He\,{\sc i} $\lambda6678$ (Table \ref{93Jtab:pol}), which is seen to start developing in the flux spectrum at this epoch.  
Moreover, a polarisation peak is associated with the onset of the He\,{\sc i} $\lambda7065$ in the flux spectrum (see Table \ref{93Jtab:pol} and Figure \ref{93Jfig:93j_pol}).

On the whole, the polarisation at +30 days is dominated by complex hydrogen and helium features.

\subsubsection{+34 days}

The data at +34 days suffer from more noise than those at +30 days. 
This is due to a combination of effects: the supernova was fading, the telescope was smaller, as are the bins. 
Nonetheless, very clear peaks similar to that observed at +30 days can be seen in the spectropolarimetric data at this epoch. 
For all features, excluding $\mathrm{H\alpha}$, the degree of polarisation and P.A. are calculated from the weighted average of the normalised Stokes parameters in the corresponding range indicated in Table \ref{93Jtab:pol}.
The errors are propagated from the errors on the weighted mean of the Stokes parameters.

At this epoch, hydrogen still exhibits a strong $\mathrm{H\beta}$ peak and a broad $\mathrm{H\alpha}$ feature topped by a peak associated with the main absorption line. 
The $\mathrm{H\alpha}$ feature actually seems to extend further in the blue than it did at +30 days: down to \about5910 \r{A}. 

The He\,{\sc i} $\lambda5876$ still shows two main peaks with degree of polarisation similar to that observed at +30 days (see Table \ref{93Jtab:pol}): one associated with the absorption minimum in the flux spectrum, and the other associated with a blue shoulder in the profile of the absorption line, which is interpreted as a HV component.  
Once again the depolarised region corresponding to the $\mathrm{H\alpha}$ emission line is topped by a peak of He\,{\sc i} $\lambda6678$, \about 6600\r{A}.
Additionally, the broad polarisation feature extending from 6715\r{A} to 7000\r{A} could be a signature of He\,{\sc i} $\lambda7065$.

All together, the polarisation data at +34 days still show very strong hydrogen and helium features.

\subsubsection{Late times}
At +45 days and +48 days, the main He\,{\sc i} $\lambda5876$ and $\mathrm{H\alpha}$ peak remain visible but the noise and errorbars are such that we will refrain from making quantitative estimates of the level of polarisation for these lines. 
The decrease in signal to noise ratio is caused by the fading of the supernova, as well as the smaller telescope aperture in the case of the +48 days data.
Overall, the polarised signal at these epochs does seem lower, which would most likely due to the decrease in electron density in the ejecta.

\subsection{Stokes $q-u$ plane}
\label{93Jsec:qu}
The behaviour of spectropolarimetric data on the $q-u$ plots can reveal both large-scale and small-scale asymmetries (see \citealt{WW08} for a review). 
When plotting the whole data, an alignment on the $q-u$ plane ordered with wavelength (dominant axis) can be interpreted as the of result bi-axial geometry, from an oblate photosphere or from an off-centre energy source in a spherical envelope -- see case (i) and case (iii) in Section \ref{introsec:pol_origin}.
Additionally, changes in polarisation angle across the wavelength range of specific spectral lines (loops), is a sign of departure from axial symmetry. 

\begin{landscape}
\begin{figure}
	\includegraphics[width=18cm]{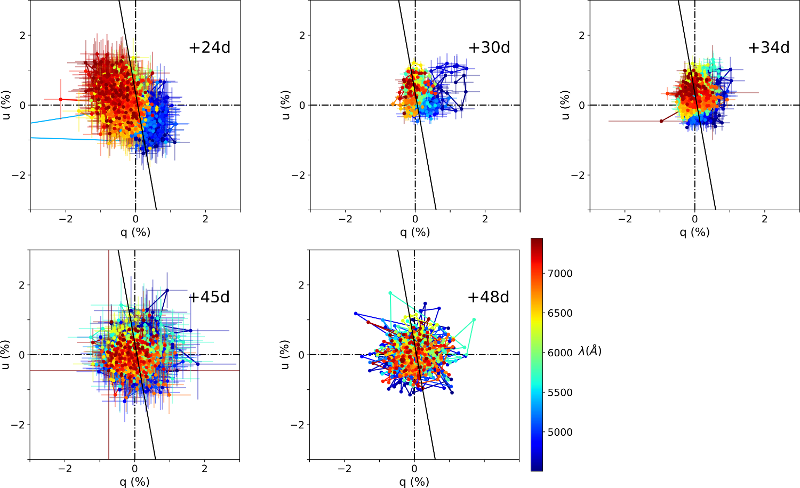}
    \caption[SN 1993J $q-u$ plane -- whole wavelength range]{\label{93Jfig:qu_whole} ISP corrected polarisation of SN 1993J on the $q-u$ plane for April 20 to May 14 from a starting wavelength of 4500\r{A}. The colorbar shows wavelength.  The black line shows the dominant axis obtained when fitting the data at +30 days and excluding the $\mathrm{H\beta}$ feature.}
\end{figure} 
\end{landscape}

In Figure \ref{93Jfig:qu_whole} we present the ISP corrected data of SN 1993J from +24 days to +48 days.
At the first epoch, we can see some alignment in the data, but due to the level of noise at this time caution is required. 
At +30 days and +34 days, some alignment may also be present.
Note that the blue points deviating from the main locus of the data at +30 days belong to the $\mathrm{H\beta}$ line, which is discussed below. 
At +45 days and +48 days the data seem to form a circular cloud centred on the zero point, indicating no bi-axial geometry and no significant continuum polarisation.

To check for the presence of a dominant axis we perform ODR fits on all of our data sets. 
In the present data, ODR was not able to find suitable solutions for most epochs, apart from the polarisation at +30 days, with the best fit found at angle of 101\degree$\pm2$\degree. 
Note that this fit did not include the bluer data points corresponding to $\mathrm{H\beta}$.
Although the regression did not converge at + 34 days, the dominant axis found at +30 days seems to be a good fit to +34 days. 
It is possible that the size of the errors caused the fitting algorithm not to converge, and that binning the data at +34 days could have helped identify a dominant axis. 
Unfortunately we are unable to rebin the data we were provided, since we do not have the original flux data.
Binning should be applied on the flux before we apply the transformations described in Section \ref{datredsec:linear_pol} to find the Stokes parameters. 

\begin{figure}
	\includegraphics[width=15cm]{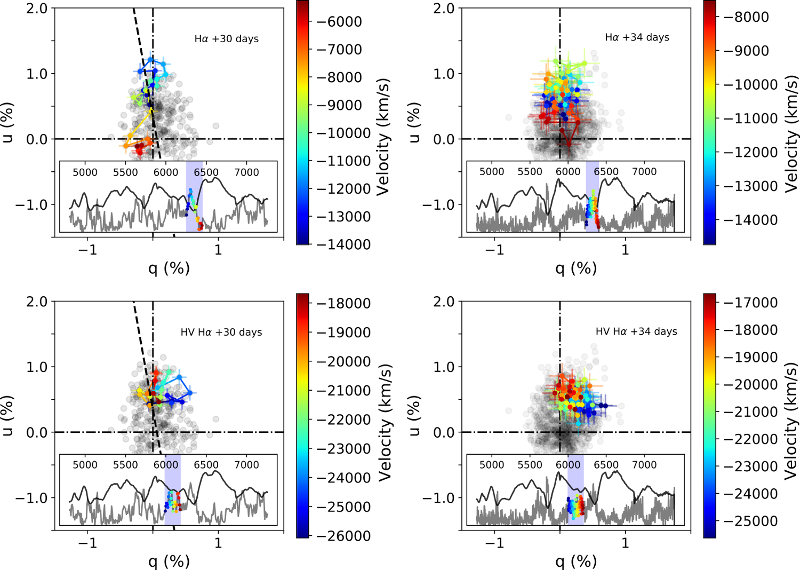}
    \caption[SN 1993J $q-u$ planes -- $\mathrm{H\alpha}$]{\label{93Jfig:quha} ISP corrected $\mathrm{H\alpha}$ polarisation of SN 1993J on April 26 (+30 days) and April 30 (+34 days). The colorbar shows velocity. The inset plots show the flux spectrum in dark grey and the polarisation in light grey except for the data points shown on the $q-u$ plots, which are shown in with the corresponding colormap. }
\end{figure}

\begin{figure}
	\includegraphics[width=\columnwidth]{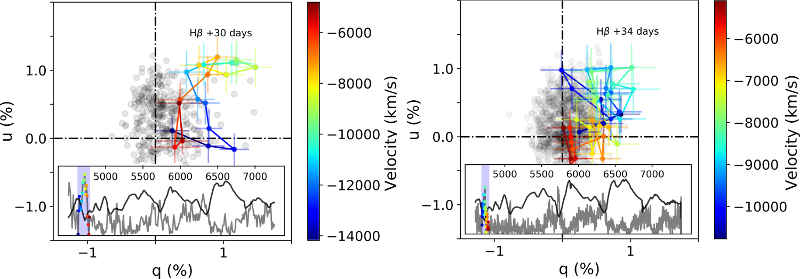}
    \caption[SN 1993J $q-u$ planes -- $\mathrm{H\beta}$]{\label{93Jfig:quhb} ISP corrected He\,{\sc i} $\mathrm{H\beta}$ polarisation of SN 1993J on April 26 (+9 days) and April 30 (+13 days). The colorbar shows velocity. The inset plots show the flux spectrum in dark grey and the polarisation in light grey except for the data points shown on the $q-u$ plots, which are shown in with the corresponding colormap.}
\end{figure}

\begin{figure}
	\includegraphics[width=15cm]{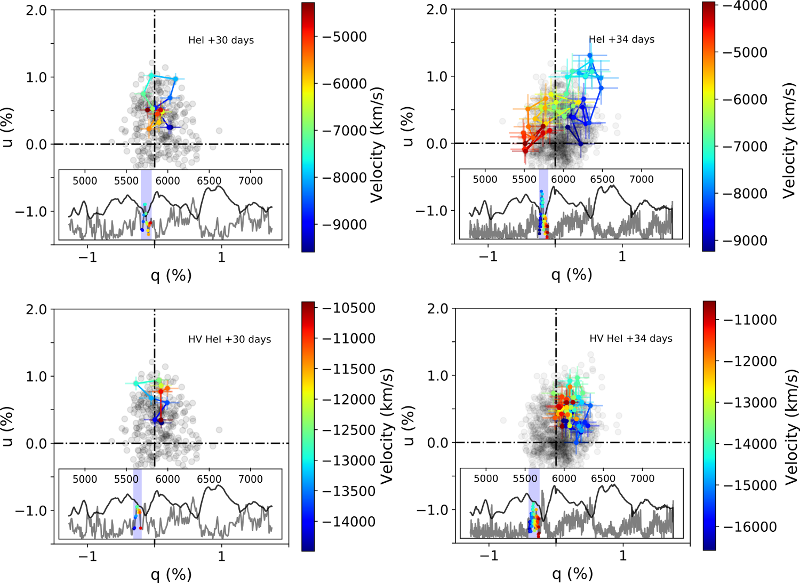}
    \caption[SN 1993J $q-u$ planes -- He\,{\sc i}]{\label{93Jfig:quhe} ISP corrected He\,{\sc i} $\lambda5876$ polarisation of SN 1993J on April 26 (+9 days) and April 30 (+13 days). The colorbar shows velocity. The inset plots show the flux spectrum in dark grey and the polarisation in light grey except for the data points shown on the $q-u$ plots, which are shown in with the corresponding colormap. }
\end{figure}

In Figures \ref{93Jfig:quha}, \ref{93Jfig:quhb} and \ref{93Jfig:quhe} we present the $\mathrm{H\alpha}$, $\mathrm{H\beta}$ and He\,{\sc i} $\lambda5876$ data on $q-u$ plots. 
We only show the +30 days and + 34 days data since no statistically significant line features were observed at other epochs (see Figure \ref{93Jfig:93j_pol}).

The data corresponding to the main $\mathrm{H\alpha}$ peak show both at +30 and +34 days a linear configuration on the $q-u$ plane.
This is indicative that this line forming region probes a bi-axial ejecta configuration. 
The putative HV features of $\mathrm{H\alpha}$, on the other hand, show clear rotations in P.A. at +30 and +34 days. 
Whether these wavelengths do correspond to HV hydrogen or not, they probe significant deviations from bi-axial symmetry. 

In Section \ref{93Jsec:p}, we saw that the strongest polarisation features at +30 days was that of $\mathrm{H\beta}$.
The $\mathrm{H\beta}$ data at this epoch also shows one of the most complex loop in this thesis, exhibiting a mixture of successive anti-clockwise and clockwise rotation. 
Significant departures from a bi-axial geometry are therefore evidenced by this loop.
At +34 days the error bars are larger, as is the noise.
There may be some rotation of the P.A. and departures from bi-axial symmetry, but the case is not as clear as for that of $\mathrm{H\beta}$ at +30 days.

The He\,{\sc i} $\lambda5876$ line at +30 and +34 days was associated with two main polarisation peaks, the first corresponding to the absorption flux minimum and the second interpreted as a HV component, associated with a blue shoulder in the helium line profile (see Section \ref{93Jsec:p}).
Both at +30 and +34 days, the main helium features show clear anti-clockwise loops on the $q-u$ plane. 
The loop at +34 days is more developed than 4 days prior, showing 180\degree rotation across the feature and a greater amplitude. 
Consequently, the main He\,{\sc i} $\lambda5876$ peak probes departures from bi-axial geometry. 
The high velocity components, on the other hand a both epochs show very little P.A. rotation if any.
Departures from axial symmetry are therefore less prominent, if at all present, than in the deeper layers of this line forming region.

Possible implications for these observation are discussed in Section \ref{93Jsec:disc_comp}.


\section{Discussion}
\label{93Jdisc}
\subsection{Assumptions in ISP estimates and alternative methods}
\label{93Jsec:disc_isp}

In Section \ref{93Jisp} we presented a new method to estimate the ISP in the data of SN 1993J.
It differs from previous methods employed by \cite{trammell93} and \cite{tran97} in a number of ways. 
Firstly, the historical analyses used the assumption that $\mathrm{H\alpha}$ emission was completely depolarised. 
This idea stems from the fact that the emission component of P Cygni profiles are the result of light being scattered into the observers line of sight by resonant scattering, which is a depolarising process. 
However, the superposition of an He\,{\sc i} $\lambda6678$ absorption component and its associated polarisation can render the assumption of complete depolarisation in this region of the spectrum invalid. 
To derive their ISP, \cite{trammell93} and \cite{tran97} used the data on April 20 (+26 days) and justified their assumption by saying that no helium component could be seen. 
However, the flat top profile of $\mathrm{H\alpha}$ on April 20 (see Figure \ref{93Jfig:93j_pol}) suggests that some He\,{\sc i} $\lambda6678$ is present, putting the assumption of complete depolarisation into question.

In order to compute our ISP, we assumed that the data at our latest epoch was representative of the ISP.
This is valid if the electron density in the ejecta has decreased sufficiently such that intrinsic supernova polarisation is negligible.
This assumption, however, is not completely exact, as line polarisation features remain somewhat visible at +48 days.
Consequently, some latent  SN  continuum polarisation may have biased our ISP estimate. 

Another difference between our ISP and previous estimates is the wavelength dependence. 
In \cite{trammell93}, a Serkowski type law was used, whereas in \cite{tran97} negligible wavelength dependence was assumed.
In the former case, this poses the question of the validity of the Serkowski law  \citep{serkowski73} to account for the dust properties in other galaxies and in the vicinity of SNe. 
We now know of a number of cases where the form of the ISP in the line of sight to the SN have been shown to be significantly different from that of the Milky Way (e.g.\citealt{patat15}, Chapter \ref{chpt:11hs}), and therefore this assumption should be used with care. 
Using the approximation of a constant ISP can be sufficient if the wavelength dependency is very shallow and consequently negligible given the noise levels. 
   
The ISP Stokes parameters we derive in Section \ref{93Jisp} do not make use of the Serkowski assumption, and are expressed as linear functions of wavelength (see Eqs \ref{93Jeq:qisp} and \ref{93Jeq:uisp}).
This also results in a wavelength dependence in the P.A. of the ISP.
This can be caused by the superposition in the line of sight of dust clouds with different particle sizes and orientations, or by the superposition of ISP and some latent intrinsic SN polarisation, as mentioned above \citep{coyne66}.

Despite the caveats associated with our assumptions, the ISP estimates we derived result in ISP removed data (see Figure \ref{93Jfig:93j_pol}) which show a behaviour that is consistent with expectations.
As mentioned in Section \ref{93Jisp}, the blue region of the spectrum between 4600 and \about 5500\r{A} is mostly depolarised, apart from some peaks of line polarisation.
This is consistent with the expected depolarisation in that part of the spectrum caused by line blanketing \citep{howell01}. 
Moreover, $\mathrm{H\alpha}$ emission is associated with polarisation that is close to zero, apart from peaks associated with the absorption component of He\,{\sc i} $\lambda6678$. 
Finally, our new ISP solves the issue of the high level of polarisation (\about0.7 percent) observed near 6600\r{A} in the \cite{tran97} data on April 30, which was highlighted by \cite{chornock11} and which they suggested could be solved if different ISP values were to be used. 

As a result we are confident that the method described in Section \ref{93Jisp} was successful in estimating the ISP present in the SN 1993J data. 
Nevertheless, this method should be applied with care and with full consideration of the caveats mentioned above, particularly when fitting data that show residual intrinsic SN polarisation.

An alternative way to estimate the ISP is to make use of the assumption that the blue part of the spectrum is completely depolarised due to line blanketing \citep{howell01}, and simply average the Stokes parameters in this range. 
To test this method we applied it to April 20, 26, 30 and May 11. 
A weighted average of the Stokes parameters was performed; the errors on the weighted mean were also calculated, and propagated through to the degree of polarisation and polarisation angle. 
The degree of polarisation was also debiased. 
The wavelength ranges used and resulting ISP values are given in Table \ref{93Jtab:pol_comp} alongside the literature ISP values and equivalent 5200\r{A} ISP values obtained from Eqs \ref{93Jeq:qisp} and \ref{93Jeq:uisp}, for comparison purposes.
Note that the range used for the average on April 20 is different from the other epochs because no data is available between 5300\r{A} and 5500\r{A}. 

On the whole, the estimates in Stokes parameters, degree of polarisation and polarisation angle using this method vary from epoch to epoch. 
This is inconsistent with the behaviour of the ISP, which is constant on short timescales, and shows that the results were biased by intrinsic polarisation. 
It is interesting to note that the values obtained from May 11 (+48 days) are very close to the ISP values we derived at 5200\r{A}. 
This can be understood as the result of a decreasing amount of intrinsic polarisation at later dates due to the decreasing electron density, causing the observed polarisation to be dominated by ISP. 

The main issue with this method is that, although depolarisation of the continuum through line blanketing is a real effect, this region of the spectrum does sometime show significant line polarisation, as can be clearly seen in SN 1993J at +30 days, or in SN 2008aq at +27 days (see Chapter \ref{chpt:08aq}).
Consequently, although keeping in mind the expected effects of line blanketing makes for a good sanity check, it is difficult in practice to use it to find reliable ISP values. 

\begin{table}
\centering
\caption{ \label{93Jtab:pol_comp} Comparison of ISP values obtained using different methods.}
\begin{tabular}{c c c c c c}

\hline
Date & Wavelength & $q$ & $u$ & $p$ & P.A.\\
\hline
\multicolumn{6}{c}{Assuming depolarisation in the blue} \\
April 20 & 4500-5300\r{A} & 0.58$\pm0.03$ & 0.19$\pm0.03$ & 0.61$\pm0.03$ & 9\degree$\pm$5\degree \\
April 26 & 4800-5500\r{A} & 0.61$\pm0.05$ & 0.65$\pm0.04$ & 0.89$\pm0.04$ & 23\degree$\pm$1\degree \\
April 30 & 4800-5500\r{A} & 0.42$\pm0.03$ & 0.53$\pm0.03$ & 0.68$\pm0.02$ & 26\degree$\pm$1\degree \\
May 11 & 4800-5500\r{A} & 0.28$\pm0.04$ & 0.45$\pm0.04$ & 0.53$\pm0.04$ & 29\degree$\pm$1\degree \\
\multicolumn{6}{c}{sigma clipping method} \\
May 14 & 5200\r{A} & 0.30$\pm0.15$ & 0.60$\pm0.15$& 0.63$\pm0.15$ & 32\degree$\pm$3\degree\\
\multicolumn{6}{c}{Tran et al.} \\
April 20 & 6390-6890 & 0.60 & -0.19 & 0.63 & 171\degree\\
\multicolumn{6}{c}{Trammell et al.} \\
April 20 & $\mathrm{H\alpha}$ emission & 0.55 & -0.95 & 1.1$\pm$0.1 & 150\degree$\pm$0.1\degree\\

\hline

\end{tabular}
\end{table}

\subsection{Interpretation of the data and comparison to previous studies}
\label{93Jsec:disc_comp}

Previous studies of the spectropolarimetric properties of SN 1993J were performed by \cite{trammell93} and \cite{tran97}.
These analyses were done at a time where intrinsic polarisation had only been observed in one supernova -- SN 1987A -- and therefore mostly focus on whether intrinsic polarisation was present in the supernova and on quantifying the degree of continuum polarisation.
In order to do this, \cite{trammell93} and \cite{tran97} use the same method of averaging the data across a large range of wavelengths. 
In the former case, they use the range 4900-6800\r{A} and find $p$ = 1.6 percent (ISP removed) on April 20, whereas in the latter they average their data over 4800-6800\r{A} and find $p$\about1 percent on April 26.

Since this wavelength range encompasses both the He\,{\sc i} $\lambda5876$ and $\mathrm{H\alpha}$ lines, this method will capture the presence of intrinsic polarisation, however it is not suitable to isolate the continuum polarisation from the line polarisation. 
In order to estimate the continuum independently, it is necessary to identify a region of the spectrum that is devoid of strong lines.
A good spectral range for this purpose would be between the 7630\r{A} telluric lines and the onset of the calcium infra red triplet absorption component (e.g. Chapter \ref{chpt:08aq}).
However, this region of the spectrum is not accessible in these data sets (see Figure \ref{93Jfig:93j_pol} in Section \ref{93Jsec:p}), and we therefore did not attempt to calculate a continuum polarisation.

Information about the geometry of the ejecta can be inferred from the $q-u$ plots presented in Section \ref{93Jsec:qu}.
The data as a whole showed marginal elongation on the $q-u$ planes (see Figure \ref{93Jfig:qu_whole}) at the first 3 epochs, indicating some degree of axial symmetry in the ejecta.
The $\mathrm{H\alpha}$ data at +30 and +34 days between about $-$14,000\kms and $-$6,000\kms showed linear configurations consistent with bi-axial geometries. 
If the features interpret as HV $\mathrm{H\alpha}$ are indeed hydrogen features, then the P.A. variation they show on the $q-u$ plane $>-$18,000\kms indicate a departure from axial symmetry at higher depths. 
This could either be a line specific effect or the result of the global geometry. 
Since no other lines probe such high velocities, it is not possible to distinguish between these two possibilities at the present time. 

Although the main peak of $\mathrm{H\alpha}$ at +30 days behaves as expected in the case of a bi-axial geometry, $\mathrm{H\beta}$ shows a very complex loop --and so a complex geometry that significantly departs from axial symmetry-- at similar velocities, see Figure \ref{93Jfig:quhb}.
Therefore, the main feature of $\mathrm{H\alpha}$ probes a different geometry from that of $\mathrm{H\beta}$ at the same depth. 
This means that global geometry effects are not the only source of polarisation, and some line specific geometries must come into play. 

In the case of He\,{\sc i} $\lambda5876$, loops are seen at a depth below  $-$9,000\kms at +30 and +34 days, which contrasts with the linear configuration followed by the $\mathrm{H\alpha}$ data at similar depths. 
In addition, the helium loop strengthens between +30 and +34 days, as the helium lines in the spectrum become more prominent. 
Consequently, these polarisation features seem to probe line specific geometries and departures from the global geometry.

On the whole, SN 1993J does appear to have some degree of global bi-axial geometry, and considerable line specific departures from this axial symmetry. 
The bi-axial geometry could be the result of an off-centre energy source as suggested in \cite{hoflich95}. 
Inhomogeneities in the distribution of hydrogen and helium could result in uneven obscuration of the underlying photosphere, causing the observed departures from axial symmetry. 

In order to better understand the geometry of the ejecta of SN 1993J, careful modelling will be required, but this is beyond the scope of this Thesis.

\section{Conclusion}
\label{93Jconc}

In this Chapter we presented a preliminary re-analysis of spectropolarimetric data of SN 1993J from +24 to +48 days with respect to explosion date obtained from private correspondence with H Tran (through J. R. Maund). 
We estimated the ISP using a method independent of those used by \cite{trammell93} and \cite{tran97} and found our values to be inconsistent with the literature.
Because these studies used assumptions that we now know are not necessarily true, and since we find our ISP answering some issues raised by \cite{chornock11} concerning the intrinsic polarisation of SN 1993J, we are confident that our estimates are improvements on literature values. 

In the ISP removed data at +30 and +34 days we found significant line polarisation features of hydrogen and helium with degree of polarisation \about 1 percent in $\mathrm{H\alpha}$ and He\,{\sc i} $\lambda5876$.
At +30 days,  $\mathrm{H\beta}$ showed a degree of polarisation as high as 1.8 percent. 
Possible HV features of $\mathrm{H\alpha}$ and He\,{\sc i} $\lambda5876$ were also identified although spectral synthesis has not yet been performed to confirm or disprove this. 

The $q-u$ plots of SN 1993J revealed marginal bi-axial geometry at +30 and +34 days.
Loops were seen in $\mathrm{H\alpha}$, $\mathrm{H\beta}$ and  He\,{\sc i} $\lambda5876$ at different depths.
The non-homogeneous behaviour of the loops from line to line suggests that line-specific geometries are the cause for departures from bi-axial geometry. 
More analysis and modelling are needed to fully explore the new intrinsic polarisation of SN 1993J presented here. 

In the following chapter we will move away from the Type IIb SNe, and present the best spectropolarimetric data set obtained for a Type Ic-bl (to this day).

\chapter{A choked jet in Type Ic-bl SN~2014ad?}

\label{chpt:14ad} 

\lhead{\emph{SN 2014ad}} 

\section{Introduction}

As mentioned in Section \ref{introsec:classification}, broad-lined type Ic SNe (SNe Ic-bl)  show extremely broad and blended spectral features compared to ``normal" type Ic SNe (e.g. SN~1997X, see \citealt{munari98}; SN~1997ef,  see \citealt{iwamoto00}; or SN~2002ap, see \citealt{mazzali02}), as a result of significantly greater ejecta velocities. 
A study of 17 type Ic SNe and 21 SNe Ic-bl conducted by \cite{modjaz16} has shown that the mean peak expansion velocity is \about 10,000~\kms faster in Ic-bl than in type Ic SNe over all epochs.

SNe Ic-bl have been of particular interest over the past two decades, owing to their relation to long Gamma-Ray Bursts (GRBs) and X-Ray Flashes (XRFs) -- see Section \ref{introsec:classification}.
Not all SNe Ic-bl are part of a SN/GRB pair, however, which, as stated by \cite{soderberg06}, cannot be explained solely through viewing angle effects.
This implies that the progenitors and/or explosion mechanisms giving rise to GRB/SNe and GRB-less SNe Ic-bl are distinct in some way. 

The main model proposed to produce GRBs is the Collapsar model, whereby the rapidly rotation core of the massive star collapses to a black hole and forms an accretion disk which collimates jets.
It has also been suggested that type Ic-bl SNe without GRBs could also be powered by jets that have not broken out of the envelope, potentially due to shorter central engine lifetimes (see Section \ref{introsec:explosion}). 

SN~2014ad is a type Ic-bl SN without a detected GRB. 
In this Chapter we use spectropolarimetry to infer geometrical properties of its ejecta and relate them to potential explosion mechanisms.
We present the best spectropolarimetric data set obtained yet for a SN Ic-bl, extending over 7 epochs, and ranging from $-$2 days to +66 days with respect to V-band maximum.
We also report  8 epochs of spectroscopy ranging from $-$2 days to +107 days. 
Note that this Chapter is an update on our publication of the SN 2014ad spectropolarimetric data in \cite{stevance17}.
The main difference is in the size of the error bars, as we realised after publication that the error treatment in the IRAF utilities we used was incorrect, and caused a significant over-estimate of the uncertainties.
We are pleased to report that the conclusions drawn in \cite{stevance17} are completely unaffected.  

In Section \ref{14adsec:obs} we give an account of our observations and data reduction; in Section \ref{14adsec:res} we present our spectroscopic and spectropolarimetric results and in Section \ref{14adsec:analysis} we discuss interstellar polarisation and analyse the intrinsic polarisation of SN~2014ad using $q-u$ plots and synthetic V-band polarimetry. 
Our results and analysis are then discussed in Section \ref{14adsec:disc} and conclusions are given in Section \ref{14adsec:conc}. 

\section{Discovery and Observations}
\label{14adsec:obs}

\begin{landscape}
\begin{figure}
\centering
\includegraphics[width=15cm]{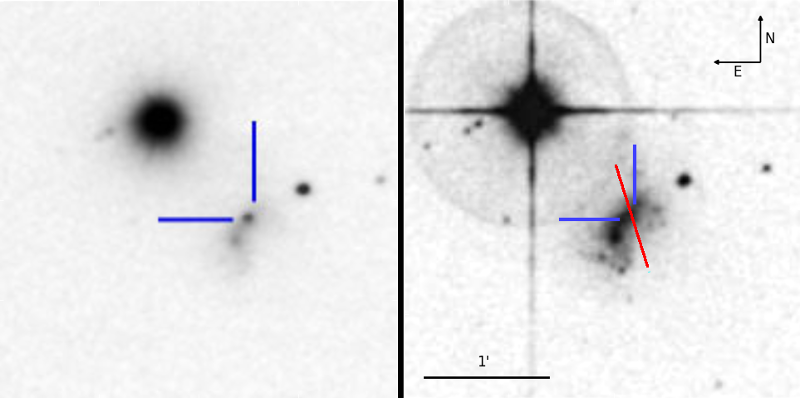}
\caption[SN 2014ad in MRK 1309 and P.A.]{Images of MRK 1309 with (left panel) and without (right panel) SN~2014ad. The image on the right hand side is a SERC image from 03 August 1989  retrieved via Aladin. The image of SN~2014ad in MRK 1309 on the left-hand side was obtained by Stan Howerton from eleven unfiltered 20 second exposures acquired on 3.1 May 2014 using a Celestron 11-inch CST and a Orion StarShoot Deep Space Monochrome Imager III. The red line superposed onto the SERC image is the direction of the interstellar polarisation (ISP) determined for MRK 1309 (see section \ref{14adsec:isp}).  }
\label{14adfig:14ad}
\end{figure}
\end{landscape}

SN~2014ad was discovered by \citet{14ad} on 12.4 March 2014 in public images from the Catalina Sky Survey (CSS, \citealt{CRTS}) at RA = 11:57:44.44 and $\delta$ = \mbox{-10}:10:15.7. 
It is located in MRK 1309 (see Figure \ref{14adfig:14ad}), with a recessional velocity of 1716~\kms \citep{HIPASS06}. 
Spectropolarimetric observations of SN~2014ad were conducted  with FORS2 (see Section \ref{datredsec:FORS}) under the program ID: 093.D-0820 (PI: J. Maund). 
The order sorting filter GG435 was used, which made all wavelengths below 4450\,\r{AA} inaccessible to us. 
Linear spectropolarimetric data of SN~2014ad were obtained at 7 epochs between 22 March 2014 and 29 May 2014. 
Circular spectropolarimetric data were also acquired on 29 March 2014 and 11 April 2014. 
Additionally, spectroscopic data were taken on July 2014; a summary of observations is given in Table \ref{14adtab:obs}. 

\begin{table}
\centering
\caption[Observations of SN 2014ad]{\label{14adtab:obs} VLT Observations of SN~2014ad.  The epochs are given relative to the estimated V-band maximum. }
\begin{tabular}{c c c c c}
\hline\hline
Object & Date & Exp. Time & Epoch & Airmass \\
 & (UT) & (s) & (days) & (Avg.)\\
\hline
\hline
\multicolumn{5}{c}{Linear Spectropolarimetry} \\
\hline
SN~2014ad & 2014 March 22 & 16 $\times$ 310 & $-$2 & 1.069 \\
CD-32d9927\footnotemark[1]  & 2014 March 22 & 2$\times$10 & - & 1.394\\
 \\ 
SN~2014ad & 2014 March 29 & 16 $\times$ 345 & +5 & 1.057 \\
LTT4816\footnotemark[1] & 2014 March 29 & 95 & - & 1.15 \\
 \\
SN~2014ad & 2014 April 11 & 12 $\times$ 345 & +18 & 1.037 \\
LTT4816\footnotemark[1] & 2014 April 11 & 95 & - & 1.117 \\
 \\
SN~2014ad & 2014 April 26 & 8 $\times$ 855 & +33 & 1.072 \\
LTT4816\footnotemark[1]  & 2014 April 26 & 95 & - & 1.109 \\
 \\
SN~2014ad & 2014 May 06 & 8 $\times$ 855 & +43 & 1.061 \\
LTT4816\footnotemark[1]  & 2014 May 06 & 95 & - & 1.41 \\
  \\
SN~2014ad & 2014 May 21 & 8 $\times$ 855 & +58 & 1.080 \\
LTT4816\footnotemark[1]  & 2014 May 21 & 95 & - & 1.159 \\
 \\
SN~2014ad & 2014 May 29 & 8 $\times$ 855 & +66 & 1.092 \\
LTT4816\footnotemark[1]  & 2014 May 29 & 95 & - & 1.172 \\
\hline
\multicolumn{5}{c}{Circular Spectropolarimetry}  \\
\hline
SN~2014ad & 2014 March 29 & 2 $\times$ 345 & +5 & 1.057 \\
LTT4816\footnotemark[1] & 2014 March 29 & 95 & - & 1.15 \\
 \\
SN~2014ad & 2014 April 11 & 4$\times$ 345 & +18 & 1.037 \\
LTT4816\footnotemark[1] & 2014 April 11 & 95 & - & 1.117 \\
\hline
\multicolumn{5}{c}{Spectroscopy} \\
\hline
SN~2014ad & 2014 July 09 & 3 $\times$ 1180 & +107 & 1.38 \\
LTT4816\footnotemark[1] & 2014 July 10 & 2 $\times$ 40 & - & 1.38\\
\hline\hline
\end{tabular}
\end{table}
\footnotetext[1]{Flux Standard}

All the observations were taken using the $300V$ grism, providing a spectral resolution of 12~\r{A} (as determined from arc lamp calibration frames). 
The spectropolarimetric data were reduced according to the prescriptions described in Chapter \ref{chpt:datred}.
In order to increase the level of signal-to-noise the spectropolarimetric data were re-binned to 45\,\r{A}, which did not affect the resolution considering the breadth of the spectral features. 

\section{Results}
\label{14adsec:res} 
\subsection{Lightcurve}

\begin{figure}
\centering
\includegraphics[width=9cm]{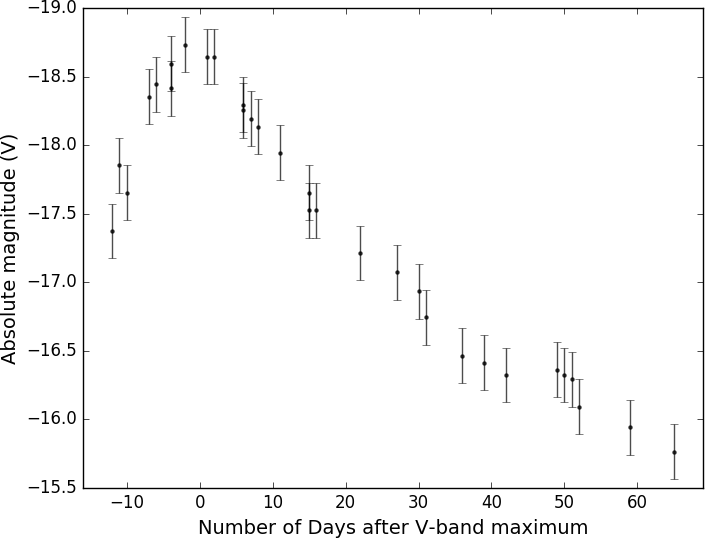}
\caption[Sn 2014ad Light Curve]{\label{14adfig:LC} De-reddened lightcurve of SN~2014ad from 12 March 2014 to 28 May 2014, using $A_v = 0.51$ mag and a distance modulus = 31.96 mag. Details of photometric acquisition can be found in \cite{howerton17}.}
\end{figure}

V-band observations were acquired between 12 March 2014 and 28 May 2014 (for details see \citealt{howerton17}). 
A luminosity distance of 24.7 Mpc was calculated from the redshift of MRK1309 (z=0.005723, \citealt{HIPASS06}), and we estimated that SN~2014ad reached maximum with an absolute V-band magnitude of $-$18.8 $\pm$ 0.2 mag on 24 March 2014 (considering an extinction $A_V = 0.51$ based on the redenning quoted in Section \ref{14adsec:isp} and a value of R = 3.1).
All epochs subsequently mentioned in this paper are quoted with respect to this date. 
The absolute V-band magnitude was corrected for extinction using the Milky Way and host galaxy reddening we calculated (see section \ref{14adsec:isp}).
An exhaustive analysis of the lightcurve is not the main focus of this work, and we refer the reader to \cite{sahu18}, who have now published exhaustive optical photometry and spectroscopy of SN 2014ad.
It is worth noting that they found an absolute V-band magnitude and maximum date that were consistent with what is presented here and in \cite{stevance17}. 

\subsection{Flux Spectroscopy}
\label{14adsec:syn++}
\begin{figure}
\centering
\includegraphics[width=9cm]{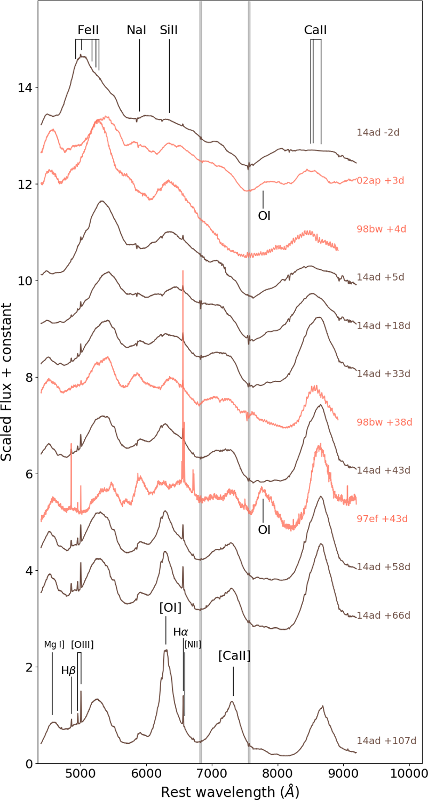}
\caption[Comparison of spectra of SN 2014ad to other Ic-bl SNe]{\label{14adfig:spctr}Flux spectrum of SN~2014ad at 8 epochs (brown) corrected for recessional velocity. We compare with the flux spectra of the three SNe Ic-bl which most closely matched SN~2014ad according to Gelato\footnotemark[4]: SN~1997ef (43 days after V-band maximum), SN~1998bw (at +4 days and +38 days), as well as SN~2002ap (3 days post V-band maximum).  The grey bands highlight telluric features. SN~1998bw is known for its connection to GRB 980425 \citep{woosley99}, SN~1997ef was thought to be the optical counterpart of GRB 971115, but the significance of their correlation is much lower than for SN~1998bw/GRB 980425 \citep{iwamoto00}. The narrow lines in the spectra of SN~2014ad are associated with the interstellar medium of the host galaxy.}
\end{figure}
\footnotetext[4]{https://gelato.tng.iac.es/}

The spectral features of SN~2014ad are very broad, particularly at early times, which is characteristic of SNe Ic-bl (see Figure \ref{14adfig:spctr}). 
At 5 days after V-band maximum, SN~2014ad shows broader features than SN~2002ap at +3 days, that are nearly as broad as SN~1998bw at +4 days. 
Although the spectral lines become less broad as time passes, SN~2014ad still shows features that are much broader than that of SN~1997ef 43 days after V-band maximum. Additionally, at +33d and +43d SN~2014ad is very similar to SN~1998bw at +38d.

\begin{figure}
\centering
\includegraphics[width=8.5cm]{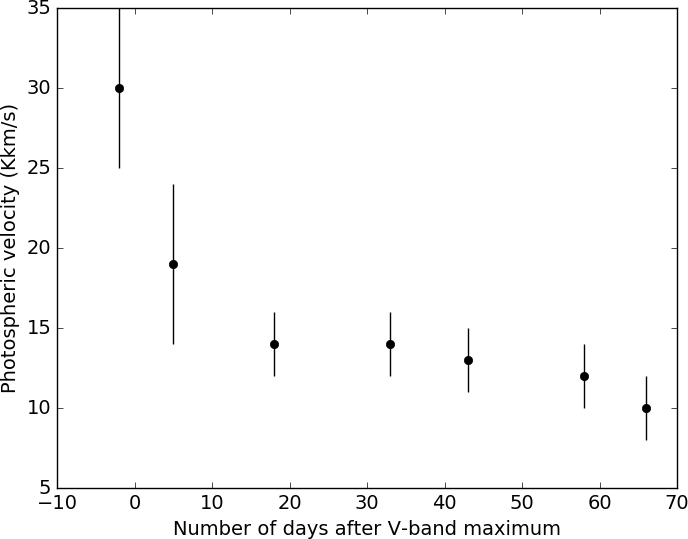}
\caption[SN 2014ad photospheric velocity]{\label{14adfig:phot_vel} Time evolution of the photospheric velocity of SN~2014ad, as measured by fitting the spectra using SYN++}
\end{figure}

In CCSNe, the line centre velocity of the Fe\,{\sc ii} $\lambda$5169 absorption component is commonly used as a proxy for the photospheric velocity (e.g. \citealt{modjaz16}). 
The extreme line blending occurring in the spectra of SN~2014ad, however, made using this feature unreliable \citep{liu16}. 
Consequently, we used \textsc{syn++} \citep{syn++} to fit the Fe\,{\sc ii} blend in our spectra of SN~2014ad and determine the photospheric velocity, as well as confirm line identification. 
\textsc{syn++}  is a radiative transfer code that assumes spherical symmetry and no electron scattering. 
It allows the user to choose the ions to be added to the synthetic spectra, and change parameters such as the photospheric velocity, the opacity, the temperature and the velocity of each ion, in order to construct the best fit to the data.
We found a photospheric velocity as high as 30,000 $\pm$ 5,000~\kms at $-$2 days, decreasing to 10,000 $\pm$ 2,000~\kms by +66 days (see Figure \ref{14adfig:phot_vel}).
The photospheric velocities later reported by \cite{sahu18} were consistent with our estimates, see their figure 10.

\begin{figure}
\centering
\includegraphics[width=8.5cm]{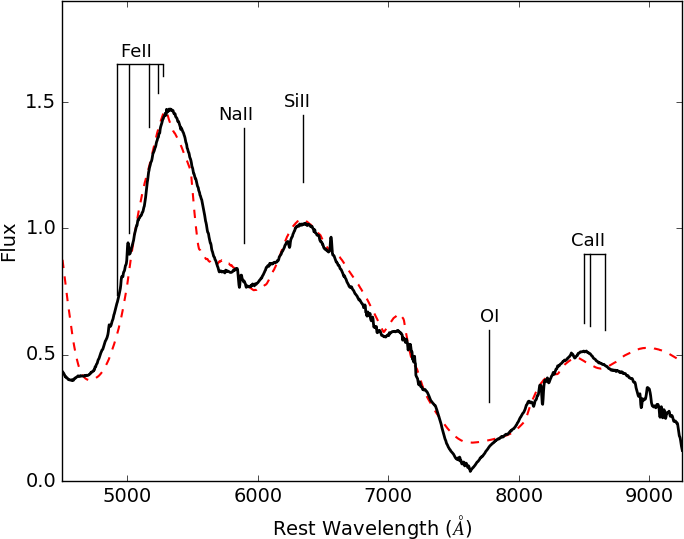}
\caption[SN 2014ad SYN++ fit]{\label{14adfig:syn++} Spectrum of SN~2014ad at 5 days after V-band maximum (solid black) and model obtained using \textsc{syn++} (dashed red). The model includes Fe\,{\sc ii}, Na\,{\sc i}, Si\,{\sc ii}, O\,{\sc i} and Ca\,{\sc ii}.}
\end{figure}

The extreme line blending caused by the high ejecta velocities also made line identification challenging. 
At all epochs, best \textsc{syn++}  fits were obtained for synthetic spectra containing Fe\,{\sc ii}, Na\,{\sc i}, Si\,{\sc ii}, O\,{\sc i} and Ca\,{\sc ii}. 
It was not possible to obtain uniformly good fits, especially at later epochs due to the appearance of forbidden lines of Mg\,{\sc i}], [O\,{\sc i}] and [Ca\,{\sc ii}]. 
We were however able to find good fits of the Fe\,{\sc ii} complex at all epochs; Figure \ref{14adfig:syn++} shows our fit to the spectrum of SN~2014ad at +5 days. 

At the earlier epochs, the spectrum is dominated by a blend of iron lines in the region 4500-5500~\ang, and a deep absorption component with a minimum at \about 7500~\ang\, corresponding to a blend of O\,{\sc i} $\lambda$7774 and Ca\,{\sc ii} infrared (IR) triplet. 
In the first epoch the emission component of the Ca\,{\sc ii}\,+\,O\,{\sc i} blend is flat topped, but it becomes prominent by +5 days. 
At this epoch, a strong Si\,{\sc ii} $\lambda$6355 emission has also emerged. 
Over time, line blending diminishes as opacity decreases and we are probing the deeper, slower ejecta, however the O\,{\sc i} $\lambda$7774 never completely separates from the Ca II IR triplet as it does in SN~1997ef. 
At later epochs, the strength of the iron blend progressively decreases, and the Ca\,{\sc ii} IR triplet emission strengthens. 
By 43 days after maximum, we can see the spectrum of SN~2014ad starting to transition from the photospheric phase towards the nebular phase: The [O\,{\sc i}] emission at \about 6300~\ang\,  dominates over the Si\,{\sc ii} $\lambda$6355 emission observed at previous epochs, and the absorption of the  Ca\,{\sc ii} IR triplet becomes increasingly flat-bottomed as the P-Cygni profile fades and is replaced by just an emission feature. 
At +107 days, SN~2014ad has a typical nebular spectrum, with (semi-)forbidden lines of Mg\,{\sc I}] $\lambda$4571, [O\,{\sc I}] $\lambda$6300 and [Ca\,{\sc II}] $\lambda\lambda$7291,7824.

\subsection{Linear Spectropolarimetry}
\label{14adsec:lin.specpol}

\begin{figure}
\centering
\includegraphics[width=10cm]{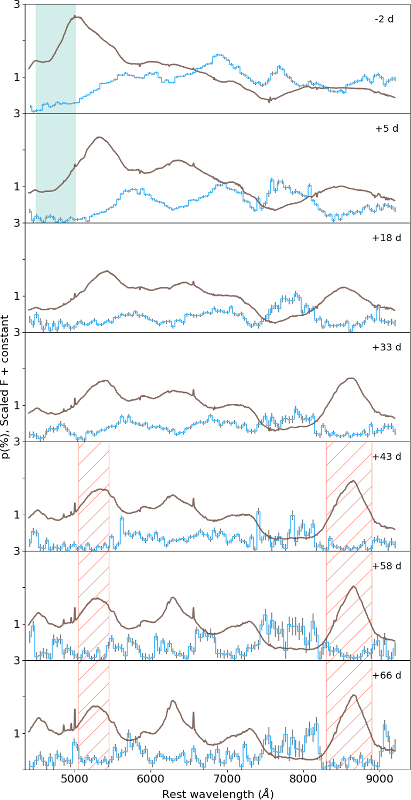}
\caption[Flux and polarisation of SN 2014ad]{\label{14adfig:flu_n_pol} Flux spectra (brown) and the degree of polarisation (blue) of SN~2014ad from $-$2 days to 66 days after V-band maximum. The polarisation spectra were binned to 45~\ang\, in order to increase the signal-to-noise ratio. The data presented here are not corrected for ISP. The ranges used to determine the ISP are indicated by the shaded green region -- method (i) -- and hashed orange region-- method (ii). For more detail see Section \ref{14adsec:isp}. For line IDs see Figure \ref{14adfig:spctr}.}
\end{figure}

The degree of polarisation $p$ and flux for each of our 7 epochs of linear spectropolarimetry are shown in Figure \ref{14adfig:flu_n_pol}. 
At \mbox{$-$2} days, 5 peaks can be seen in the polarisation spectra at \about5672\r{A} ($p=1.10 \pm 0.03$ percent), 6030\r{A} ($p=1.10 \pm 0.02$ percent -- averaged between 5985\r{A} and 6075\r{A}), 6880\r{A} ($p= 1.60 \pm 0.03$ percent),  7685\r{A} ($p= 1.30 \pm 0.05$ percent), and 8130\r{A} ($p= 0.83 \pm 0.03$ percent  -- averaged between 8093\r{A} and 8181\r{A}). 

The 6880\r{A} peak shows the highest level of polarisation recorded in our data set. This feature is likely the result of O\,{\sc i} $\lambda$~7774 at $-$34,500~\kms. 
As for the other features, the 5672\r{A} peak in polarisation is associated with Si\,{\sc ii} $\lambda$6355 with velocity $-$32,200~\kms, the 7685~\ang\, feature arises from Ca\,{\sc ii} at $-$30,900~\kms, and the peak seen at 8130\r{A} could be O\,{\sc i} $\lambda$9264 at $-$36,600~\kms. 
The origin of the 6030\r{A} feature, however, remains unclear. 
The photospheric velocity at $-$2 days found using \textsc{syn++} was 30,000 $\pm$ 5,000~\kms (see Section \ref{14adfig:phot_vel}), therefore the line-forming region yielding the Ca\,{\sc ii} IR, Si\,{\sc ii} and O\,{\sc ii} $\lambda$7774 polarisation features must be close to the photosphere at the first epoch.

By 5 days post-maximum, the 5672 and 6030~\ang\, features have merged, yielding a broader peak centred on 5780~\ang\, with polarisation $p=0.88\pm0.04 $  percent (calculated as the average polarisation between  5670 and 5900 \ang). 
Similarly, the 7685~\ang\, (Ca\,{\sc ii}) and 8130~\ang\, (O\,{\sc i}) peaks seen in the first epoch are replaced by a broader feature extending between \about 7590~\ang\, and \about 8050~\ang, with $p=0.90\pm0.15 $  percent (the average between 7580 and 8130\ang). 
The O\,{\sc i} $\lambda$7774 peak is still present, but has moved \about50~\ang\, to the red (now at 6920~\ang) and its polarisation has decreased to $p=1.05\pm0.102$  percent (averaged between 6870\r{A} and 6965\r{A}).
The first peak at \about 5780~\ang\, is consistent with Si\,{\sc ii} $\lambda$6355 at $-$27,000~\kms, and the O\,{\sc i} $\lambda$7774 velocity is now $-$32,900 \kms, therefore both elements show a decrease in velocity of \about 2,000~\kms. 
The photospheric  velocity at this epoch was found to be 19,000 $\pm$ 5000~\kms (see Figure \ref{14adfig:phot_vel}), suggesting that the line-forming region is more detached from the photosphere by the second epoch than in the first epoch.

The amplitude of the Si\,{\sc ii} $\lambda$6355 and O\,{\sc i} $\lambda$7774 polarisation peaks decreases drastically by 18 days after V-band maximum, with $p=0.48\pm0.04 $  percent and $p=0.60\pm0.05 $  percent respectively. 
The Ca\,{\sc ii} IR/O\,{\sc i}~$\lambda9264$ peak, however, remains constant with $p=0.92\pm0.07 $  percent. 
Its degree of polarisation then decreases to \about0.6\% by +33 days, and is increasingly noisy, whereas the first 2 peaks remain about constant. 

\subsection{Circular Spectropolarimetry}
\label{14adsec:circ_pol}

\cite{wolstencroft72} suggested that if the core of a progenitor collapses to a neutron star it may have sufficiently large magnetic fields to induce detectable circular polarisation from bremsstrahlung or synchrotron emission. 
They cautioned that the degree of circular polarisation would be significantly influenced by the precise conditions of the explosion, and no strong evidence supporting the presence of circular polarisation in CCSNe has been detected so far. 

\begin{figure}
\centering
\includegraphics[width=13cm]{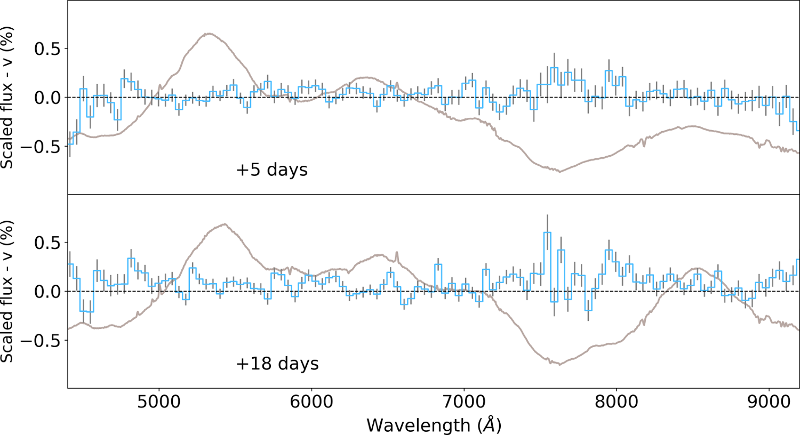}
\caption[Circular polarisation of SN 2014ad]{\label{14adfig:circ_pol}Circular polarisation (blue) binned to 45~\ang\, of SN~2014ad on 2014 March 29 (+5 days) and 2014 April 11 (+18 days). The spectrum at the corresponding epochs are under-plotted in brown.}
\end{figure}

We investigated the potential presence of circular polarisation in SN~2014ad. 
Circular spectropolarimetry was acquired on 29 March 2014 and 11 April 2014 and is presented in Figure \ref{14adfig:circ_pol}. 
At both epochs we found that the average circular polarisation (0.025 and 0.069 at epoch 2 and 3, respectively) was lower than the standard deviation (0.104 and 0.128 at epoch 2 and 3, respectively) over the range 4500-9300~\ang. 
We therefore conclude there is no detectable circular polarisation in SN 2014ad.

\newpage
\section{Analysis}
\label{14adsec:analysis}

\subsection{Spectropolarimetry}


\subsubsection{Interstellar Polarisation}
\label{14adsec:isp}
As we have done in in previous chapter, we must quantify the ISP in order to isolate the polarisation that is intrinsic to the SN.
An upper limit on the ISP can be estimated if we assume a standard Serkowski-Galactic type ISP \citep{serkowski73}. The degree of polarisation due to the ISP ($p_{\text{ISP}}$) is then related to the total reddening $E(B-V)_{\text{total}}$ = $E(B-V)_{\text{MW}}$ + $E(B-V)_{\text{host}} $  by Eq. \ref{eq:serk_plim}.

The reddening can be estimated from the Na\,{\sc i} D lines in the spectra following \cite{poznanski12}. 
In our spectra of SN~2014ad, only the Na\,{\sc i} D line arising from the Milky Way contribution is visible. 
In order to put an upper limit on the reddening occurring in MRK 1309, we calculated a detection limit at  3$\sigma$ using our spectral resolution of FWHM = 12~\ang\, for Na\,{\sc i} D at the recessional velocity of the host galaxy. 
Our estimates for the reddening are $E(B-V)_{\text{MW}}$ = 0.14 mag, and $E(B-V)_{\text{Host}} \le$ 0.023 mag, yielding an upper limit for the ISP $p_{\text{ISP}} \le$ 1.56 \%. By applying the Serkowski law to the host galaxy we implicitly assumed that the size and composition  of the dust in the host galaxy are the same as that of the Milky Way, which may not be true.

The Milky Way component of the ISP can also be evaluated by finding Milky Way stars near the line-of-sight towards SN~2014ad. 
If we assume that their intrinsic polarisation is null, then the polarisation measured for these stars is purely due to the Galactic ISP. 
Within 2 degrees of SN~2014ad, two stars were found in the \cite{Heiles} catalogue: HD 104304 with $p =$ 0.08 ($\pm$ 0.035) \% and HD 104382 with $p=$ 0.11 ($\pm$ 0.066)~\%. 
Using the parallaxes found in the Gaia data release DR2 \citep{gaia_parallaxes} we calculate distances of $12.69 \pm 0.02$ pc and $366 \pm 10$ pc for HD 104304 and  HD 104382.
With a Galactic latitude of 50.5\degree, however, SN~2014ad is located far from the Galactic plane, and we do not expect significant dust alignment along the line-of-sight. 
Consequently we concluded that a lower limit for the Milky Way component of the ISP is $\sim$ 0.1 \%. 

In order to directly quantify the ISP from the polarisation data obtained for SN~2014ad, assumptions must be made. We use two different assumptions to calculate two independent estimates of the ISP: complete depolarisation in line blanketed regions and complete depolarisation at late times.
\begin{enumerate}
\item[(i)]Strong depolarisation is observed at short wavelengths (see Figure \ref{14adfig:flu_n_pol}) due to line blanketing by Fe\,{\sc ii} and Sc\,{\sc ii} lines \citep{wang01, howell01}, and if we assume complete depolarisation of the SN~light, the resulting observed degree of polarisation is $p_{\text{ISP}}$. 
We calculated the polarisation at $-$2 days and +5 days over the wavelength range 4500-5000\,\r{A} (the green shaded regions in Figure \ref{14adfig:flu_n_pol}), yielding $q_{\text{ISP}}$ = 0.19 ($\pm$0.06)\%, $u_{\text{ISP}}$ = 0.12 ($\pm$0.08)\%  and $p_{\text{ISP}}$ = 0.21 ($\pm$0.07)\% at $-$2 days and $q_{\text{ISP}}$ = 0.04 ($\pm$0.09)\%, $u_{\text{ISP}}$ = 0.04 ($\pm$0.07)\% and $p_{\text{ISP}}$ = 0.06 ($\pm$0.08)\% at +5 days. 
Because ISP should be constant with time, we average the two sets of Stokes parameters and obtain $q_{\text{ISP}}$ = 0.14 ($\pm$0.05)\% and $u_{\text{ISP}}$ = 0.075 ($\pm$0.05)\%, which correspond to $p_{\text{ISP}}$ = 0.15 ($\pm$0.05)\% and a polarisation angle $\theta_{\text{ISP}}$ = 14\degree ($\pm$18\degree), which is within the limits previously established. 
We considered the line blanketing regions at the first two epochs only, because the higher ejecta velocities result in greater line blending and their spectropolarimetric data have better levels of signal-to-noise as SN~2014ad was at its brightest.\\

\item[(ii)]At late times, when the ejecta start transitioning to the optically-thin nebular phase (in the case of SN 2014ad, from +43 days), electron scattering does not dominate anymore and the intrinsic polarisation of the supernovae is expected to tend to zero. 
Consequently, if we assume complete depolarisation in the SN ejecta at late times, any level of polarisation observed must be due to ISP. 
We focus our analysis on the polarisation that correlates with strong emission lines because emission lines are often associated with depolarisation even when electron scattering is significant.
With this in mind, we pick 3 wavelength ranges (5050-5450~\r{A} and 8300-8900~\r{A};  see hashed orange region in Figure \ref{14adfig:flu_n_pol}) corresponding to the \mbox{(semi-)}forbidden lines of magnesium and calcium, and we average the Stokes parameters across these ranges. 
Doing this for epochs 5, 6 and 7 yielded 2 sets of $q_{\text{ISP}}$ and $u_{\text{ISP}}$ values which were averaged, yielding $q_{\text{ISP}}$ = 0.04 ($\pm$0.05)\%, $u_{\text{ISP}}$ = 0.12 ($\pm$0.04)\%, $p_{\text{ISP}}$ = 0.11 ($\pm$0.04)\% and $\theta_{\text{ISP}}$ = 36\degree ($\pm$4\degree). 
These values are consistent with the ones found using the line blanketing assumption.
\end{enumerate}

The two methods described above therefore resulted in ISP estimates that were small and consistent with each other. We decided to adopt the Stokes parameters values calculated with the first method for our subsequent ISP corrections since they were estimated using polarisation spectra at early days with better signal-to-noise ratio.


\subsubsection{Polarisation in the Stokes q-u plane}
\label{14adsec:q-u}
\begin{landscape}
\begin{figure}
\centering
\includegraphics[width=22cm]{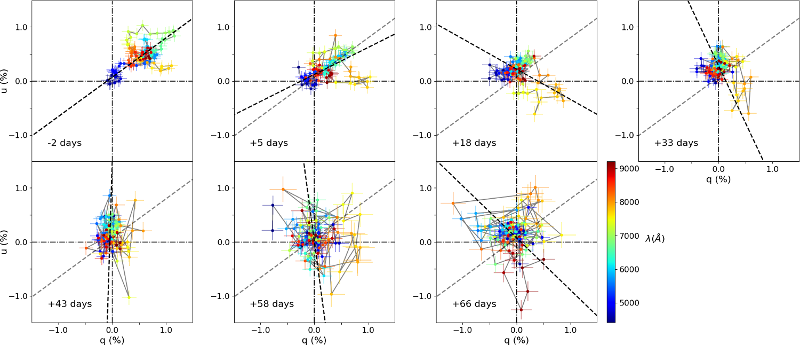}
\caption[$q-u$ plots (whole) of SN 2014ad]{\label{14adfig:qu}Stokes $q-u$ planes of SN~2014ad in 7 epochs. The polarisation data were binned to 45 \ang and the colour scheme represents wavelength. The dominant axis was calculated at each epoch using the Orthogonal Distance Regression (ODR) method, and is shown by the overlaid black dashed line. The grey dashed line is the dominant axis found at $-$2 days superposed on the data of subsequent epochs for comparison purposes. The data presented here were corrected for ISP using the values derived in Section \ref{14adsec:isp}.}
\end{figure}
\end{landscape}

The ISP corrected data were plotted on the Stokes $q-u$ plane for each epoch as shown in Figure \ref{14adfig:qu}, where the colour scale represents wavelength. 

At 2 days before V-band maximum the data are aligned along a very clear dominant axis with P.A. $= 16.6^{\circ} \pm 2.2 ^{\circ}$ (corresponding to an angle of 33\degree on the $q-u$ plane) as shown by the over-plotted ODR fit (see Figure \ref{14adfig:qu}). 
All ODR fits were performed using the entire wavelength range. 
A loop at \about7000~\ang\, can be seen following the dominant axis orientation. From our identification in Section \ref{14adsec:lin.specpol}, this feature is associated with O\,{\sc i} $\lambda$7774. 

Our data at +5 days show that the P.A. of the dominant axis (P.A $= 10.3^{\circ} \pm 2.5^{\circ}$) is similar to the first epoch, and both dominant axes are consistent with the data. The loop at \about 7000~\ang\, has become less prominent, but is still distinguishable. 
Another loop arises at \about 7700~\ang\, corresponding to the Ca\,{\sc ii} IR triplet, which runs almost perpendicular to the dominant axis. This indicates a significant departure from axial symmetry in the distribution of calcium in the form of large clumps.

By +18 days the presence of a dominant axis is less pronounced as the overall degree of polarisation decreases, but a Pearson test performed on the data -- excluding the Ca II loop -- yielded a coefficient of 0.50, supporting the presence of a linear correlation. 
The ODR fit includes the Ca II IR component and seems to follow the direction of the strong calcium loop, but it is not an accurate representation of the data as a whole. With the exception of the Ca\,{\sc ii} IR triplet, the data follow a direction similar to that of the dominant axis found in the first epoch. 
The O\,{\sc i} $\lambda$7774 loop has completely disappeared, but the calcium feature is still very strong. 
At later epochs, the calcium loop weakens and the data as a whole cluster towards the origin as the overall level of polarisation decreases. 
The ODR fits of epochs 4 to 7 were plotted on their respective $q-u$ plane for completeness, but the Pearson correlation calculated for the data at these dates ($-0.04$, 0.08, $-0.03$ and $-0.08$, respectively) revealed that the presence of a linear correlation is very unlikely.

\subsubsection{Rotation of the Stokes parameters}
\label{14adsec:rot_stokes}
\begin{figure}
\centering
\includegraphics[width=15cm]{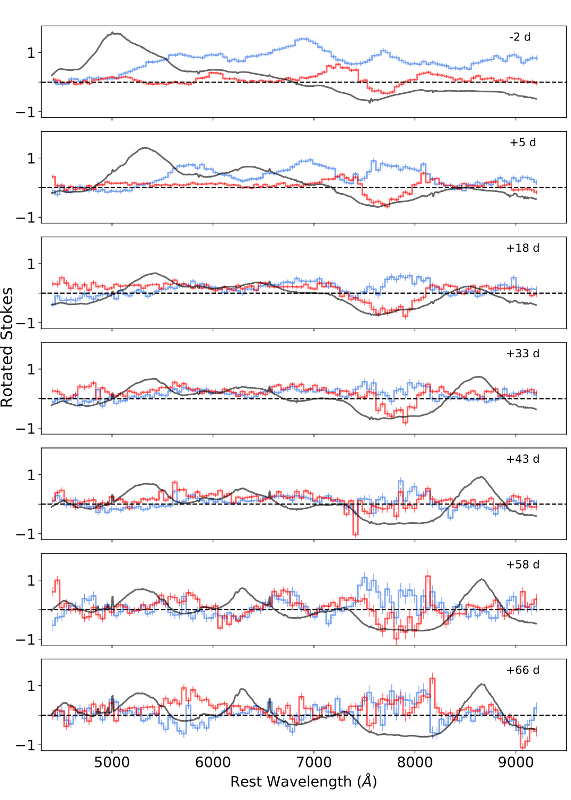}
\caption[Rotated Stokes Parameters of SN 2014ad]{$P_d$ (blue) and $P_o$ (red), as defined by Equation \ref{14adequ:rot}, corresponding to our 7 epochs of spectropolarimetry. The spectropolarimetric data were binned to 45 \ang~and corrected for the ISP before being rotated. The flux spectrum (unbinned) is plotted (grey) at each epoch. For line IDs see Figure \ref{14adfig:spctr}. }
\label{14adfig:Pd_Po}
\end{figure}

The data presented on the $q-u$ plane can be rotated  in order to obtain two new components which correspond to the Stokes parameters projected onto the dominant axis ($P_d$) and the axis orthogonal to the dominant axis ($P_o$). If we consider the rotation matrix of the form:
\begin{equation}
\mathbf{R(\theta_{\text{rot}})}=
\begin{pmatrix}
\cos\theta_{\text{rot}} & -\sin\theta_{\text{rot}} \\
\sin \theta_{\text{rot}} & \cos\theta_{\text{rot}}\\
\end{pmatrix}
\end{equation}
then the rotated Stokes parameters are given by:
\begin{equation}\label{14adequ:rot}
\begin{pmatrix}
P_d \\
P_o\\
\end{pmatrix}
=\mathbf{R(\theta_{\text{rot}})} \cdot
\begin{pmatrix}
q\\
u\\
\end{pmatrix}
\end{equation}
where $\theta_{\text{rot}} = -2 \times \theta_{\text{dom}}$. 

The direction of the dominant axis found using ODR at the first epoch is consistent with the data at epochs 2 and 3, and is therefore chosen to perform the rotation, corresponding to \mbox{$\theta_{\text{rot}}$ = 33\degree}. 
The resulting rotated Stokes parameters $P_d$ and $P_o$ were then plotted against wavelength, as shown in Figure \ref{14adfig:Pd_Po} (the colour scale matches that used in Figure \ref{14adfig:qu}).

Overall, at each epoch the orthogonal parameter $P_o$ is within 1$\sigma$ of zero along most of the spectrum, except within the wavelength ranges corresponding to loops seen in the $q-u$ plane (see Figure \ref{14adfig:qu}). 
This shows that our choice of dominant axis is appropriate and confirms our interpretation that the ODR fit of the data at +18 days was dominated by a prominent loop rather than the direction of the dominant axis consistent with the majority of the data. 
In Figure \ref{14adfig:qu}, one can see that the data at +18 days are sightly offset with respect to the chosen dominant axis (indicated by the grey dotted line -- identical to the dominant axis found using ODR in the first epoch) but seem parallel. This behaviour is also seen in Figure \ref{14adfig:Pd_Po} at +18 days, where in $P_o$ most of the data (apart from the Ca\,{\sc ii} loop between \about 7200-8200~\ang) run parallel to the dominant axis ($P_o$=0\%) but are shifted up by about 0.2\%.

The significant departures from null polarisation in $P_o$ correspond to loops in the $q-u$ plane. 
At $-$2 days, the main feature is found between \about 7000~\ang\, and \about 8300~\ang. 
It deviates from the dominant axis by \about 0.5\% on either side, and is most likely caused by a mixture of O\,{\sc i}~$\lambda$7774 and Ca\,{\sc ii}~$\lambda$8567. 
At +5 days the deviation from the dominant axis associated with O\,{\sc i}+Ca\,{\sc ii} becomes more prominent between 7500~\ang\, and 8000~\ang, but still extends from \about 7000~\ang\, and up to \about 8300~\ang. 
By +18 days the loop only departs on one side of the dominant axis and now starts at \about 7200~\ang, which may indicate that the oxygen component has reduced significantly and the feature is now dominated by the Ca\,{\sc ii} component. 
Two weeks later, the amplitude of the loop has started to decrease and by 43 days after V-band maximum it is not distinguishable anymore.

\subsubsection{Loops in the $q-u$ plane}
\label{14adsec:loops}
\begin{figure}
\centering
\includegraphics[width=15cm]{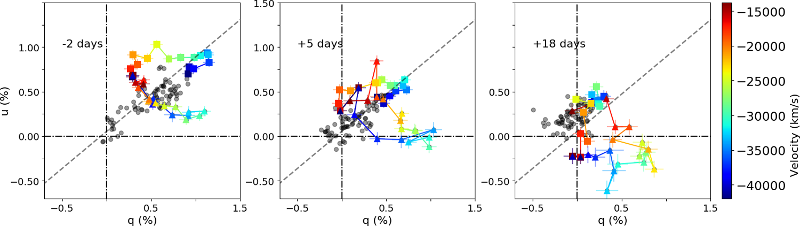}
\caption[$q-u$ plots of calcium and oxygen in SN 2014ad]{Stokes $q-u$ planes of SN~2014ad in the first three epochs (see also Figure \ref{14adfig:qu}); the spectropolarimetric data were binned to 45 \ang. The colour scheme represents ejecta velocity. The dominant axis (dashed grey line) is the dominant axis as calculated by ODR at the first epoch. The data marked by rectangles correspond to O\,{\sc i} $\lambda$7774 and the data marked by triangles is Ca\,{\sc ii} IR. For comparison the data over the range 4500-9300\ang\, were plotted (grey points). For clarity error bars were only plotted for the O\,{\sc i} $\lambda$7774 and the Ca\,{\sc ii} IR data. The data were corrected for ISP (see Section \ref{14adsec:isp}). }
\label{14adfig:loops}
\end{figure}

In Figure \ref{14adfig:loops} we show the O\,{\sc i} $\lambda$7774 and Ca\,{\sc ii} IR loops on the $q-u$ planes at $-$2, +5 and +18 days, where the colour scale now represents ejecta velocity. 
At $-$2 days the O\,{\sc i} feature dominates and the Ca\,{\sc ii} IR feature forms a line oriented away from the dominant axis. 
By +5 days the Ca\,{\sc ii} IR feature strengthens and becomes a loop oriented in the same direction as the line observed at $-$2 days, while the oxygen loop becomes less prominent. 
At +18 days the O\,{\sc i} data is in the same locus as the majority of the data, and the Ca\,{\sc ii} IR loop has strengthened again. At $-$2, +5 and +18 days, the oxygen and calcium loops are consistently formed anti-clockwise with increasing wavelength on the $q-u$ plane.

The fact that the calcium and oxygen loops evolve separately and in different directions indicates that they break axial symmetry in different ways.
A possible explanation is partial occultation of a bi-axial photosphere (supported by the presence of a dominant axis) by separate line forming regions.
This corresponds to a mixture of case (i) and case (ii) described in Section \ref{introsec:pol_origin}. 
Furthermore, the oxygen and calcium features never coincide simultaneously in their Stokes parameters and in velocity space, which could indicate overlap of the line-forming regions. This suggests that the line-forming regions of O\,{\sc i} and Ca\,{\sc ii} are very distinct from each other both in radial velocity (= radius) and on the plane of the sky.

\subsection{V-band Polarisation}
\label{14adsec:vpol}
Because the spectral features of SN 2014ad were so broad, we could not isolate a line-free region of the spectrum to estimate the continuum polarisation. 
We therefore followed \cite{leonard06} who use use V-band polarimetry as a proxy for the overall evolution of the polarisation. 
The V-band polarisation of SN~2014ad was calculated from the spectropolarimetric data by weighting all our spectra by the transmission function of the Johnson V filter before performing the polarimetry calculations. 
\textsc{pysynphot}\footnote[6]{https://pysynphot.readthedocs.io/en/latest/} \citep{pysynphot} was used to calculate the weighted spectrum for each ordinary and extra-ordinary ray, which were subsequently processed with FUSS to calculate the resulting degree of polarisation and normalised Stokes parameters.

\begin{figure}
\centering
\includegraphics[width=15cm]{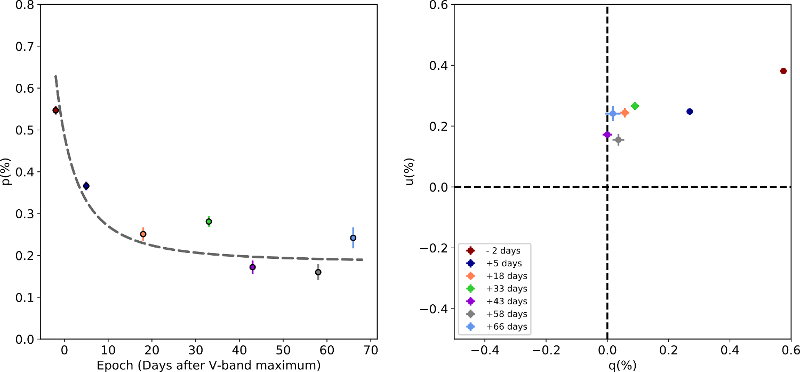}
\caption[V band polarisation of SN 2014ad]{\label{14adfig:pv} \textbf{Left panel:} Degree of V-band polarisation against time (i.e. epoch in days after V-band maximum). The blue line is a fit of the form $p$ = 40.3 t$^{-2}$ + 0.15, yielding a reduced $\chi^2$ = 4.5. The colours represent different epochs (see legend on right panel). \textbf{Right panel:} V-band polarisation of SN~2014ad in the $q-u$ plane. The colours represent different epochs (see legend).}
\end{figure}

The resulting V-band polarisation for all 7 epochs is plotted against time and on the $q-u$ plane, see Figure \ref{14adfig:pv}. 
Were the decrease in the degree of polarisation caused solely by dilution due to the expansion of the ejecta, one would expect the polarisation to follow $p = \alpha \times t^{-2} + \beta$, where $t$ is time and $\alpha$ and $\beta$ are constants.

We tentatively fitted this relationship to our V-band polarisation using Monte Carlo methods to find the best values of $\alpha$ and $\beta$. 
For the best fit the reduced $\chi^2$ is 4.5, and the constants are found to be $\alpha$ = 40.3 $\pm$ 0.3 and $\beta$ = 0.15 $\pm$ 0.004. 
The fit is plotted as a blue line in Figure \ref{14adfig:pv}. The V-band polarisation of SN~2014ad seems to exhibit a behaviour that is not completely inversely proportional to time squared, with significant departures form our best fit.
This suggests that the decrease in polarisation is not just a manifestation of ejecta expansion \citep{leonard06}. 

As seen in the right panel of Figure \ref{14adfig:pv} the polarisation angle remains nearly constant between the first two epochs (19.0 $\pm$ 0.6\degree and 29.1 $\pm$ 0.1\degree), then increases at epoch 3 (56.07 $\pm$ 0.05\degree), and subsequently stays constant. 
This behaviour could reflect a difference in geometry between the outermost layers, which are probed at early times ($-$2 and 5 days), and deeper layers of the ejecta, which are observed at later times (from +18 days through to +66 days) as the photosphere recedes through the envelope. 
This interpretation should however be considered with caution as the broad band-pass used here encompasses both continuum and line polarisation. 

\section{Discussion}
\label{14adsec:disc}

\subsection{Evolution of the shape of SN 2014ad}
\label{14adsec:shape}

Quantifying the level of continuum polarisation of SN~2014ad was made difficult by the extreme line blending, but with a maximum polarisation $p_{\text{max}}$ = 1.60 $\pm$0.03 \% the axis ratio of the photosphere could not be smaller than \about 0.72 \citep{hoflich91}. 
A clear dominant axis is visible in the $q-u$ plots at $-$2 and +5 days, implying that the ejecta possess a strong axis of symmetry at early days. By +18 days, the data has receded towards the origin of the $q-u$ plots as the polarisation of SN~2014ad has decreased (see Figure \ref{14adfig:flu_n_pol}), but a dominant axis is still present (section 4.1.2 and 4.1.3). 
The overall decrease in $p$ and the disappearance of the dominant axis seem to indicate that the deeper ejecta are more spherically symmetric than the outer envelope, with the data at +33 days indicating a maximum axis ratio of \about0.89. 
The $q-u$ plots of the last 3 epochs do not offer more insight as the data are dominated by noise. 

Significant departures from the identified dominant axis can be seen in the form of loops in the $q-u$ plots at epoch 1 through 4, indicating that the line-forming regions are made of large clumps that are asymmetrically distributed.  
At $-$2 days features of O\,{\sc i} $\lambda$7774 and Ca\,{\sc ii} IR are observed (see Figures \ref{14adfig:qu} and \ref{14adfig:Pd_Po}), with the O\,{\sc i} feature being slightly more prominent and pointing to a direction opposite to that of Ca\,{\sc ii}. 
This suggests that the two line-forming regions are inhomogeneously mixed with the rest of the ejecta and spatially distinct from each other. 
At +5 days the O\,{\sc i} loop has weakened --but is still present-- and the Ca\,{\sc ii} loop has strengthened. 
By +18 days the O\,{\sc i} loop has completely disappeared and the Ca\,{\sc ii} feature has strengthened even further. 
This trend could be explained by the calcium layer being distributed deeper into the ejecta than the oxygen layer, which is consistent with the fact that we found Ca\,{\sc ii} to have a lower velocity than O\,{\sc i} (see Section \ref{14adsec:lin.specpol}). 
By +33 days the Ca\,{\sc ii} loop has also started weakening but is still visible, and from +43 days the data are dominated by noise (see Figures \ref{14adfig:flu_n_pol}, \ref{14adfig:qu} and \ref{14adfig:Pd_Po}). 

\begin{figure}
\centering
\includegraphics[width=9cm]{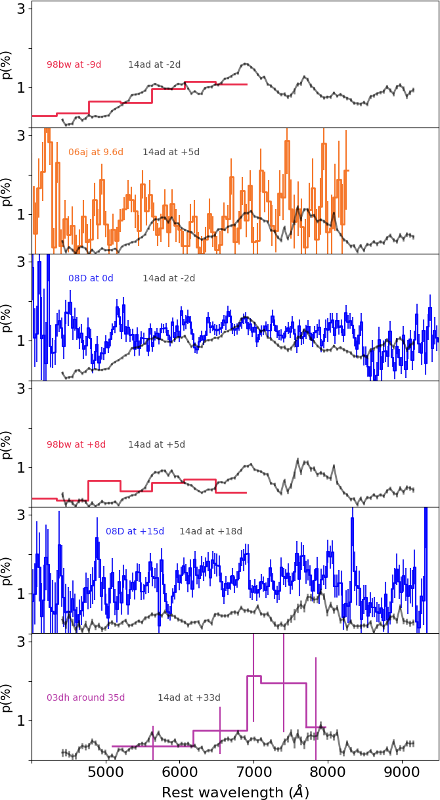}
\caption[Comparison of polarisation of SN 2014ad to other SNe]{\label{14adfig:comp_pol} Polarisation data (uncorrected for ISP) of SN~1998bw (GRB/Ic-bl), SN~2003dh (GRB/Ic-bl), SN~2006aj (XRF/Ic) and SN~2008D (Ib) compared to the data of SN~2014ad binned to 45 \ang at similar epochs. All epochs are given with respect to V-band maximum, except SN~2003dh which is quoted with respect to the GRB detection date. The data of SN 1998bw were binned to have errors $<$~0.1\% \citep{patat01}.}
\end{figure}

Over the past 20 years spectropolarimetric measurements have been obtained for a number of Ib/c, and SNe Ic-bl with and without GRBs. 
The maximum level of linear polarisation detected for SN~2014ad is not unusually high compared to previous examples and is similar to SN~1998bw (see Table \ref{14adtab:comp_table}); it is also the only SN~for which constraints on the circular polarisation have been established. 
The spectropolarimetric data of SN~2014ad is, to this writing, the best data set obtained for a SN~of this type, both in its low level of noise and its large time coverage. 
For comparison, the data of SN~1998bw (GRB/Ic-bl), SN~2003dh (GRB/Ic-bl), SN~2006aj (XRF/Ic) and SN~2008D (Ib) were plotted along our data at similar epochs, see Figure \ref{14adfig:comp_pol}. 
Comparison of the polarisation data of SN~2014ad to that of other SNe can be made difficult by the high levels of noise (e.g. SN~2006aj), broad bins (e.g. SN~1998bw and SN~2003dh) or large error bars (e.g. SN~2003dh) present in their data.

\begin{landscape}
\begin{table}
\centering
\caption[Polarisation comparison of SN 2014ad to other SNe]{\label{14adtab:comp_table} Maximum degree of polarisation observed in SN~2014ad compared to other broad-lined type Ic and normal Ib/c SNe.$^*$Unless otherwise stated the dates are given with respect to V-band maximum.$^{\dagger}$ Uncorrected for ISP.}
\begin{tabular}{l l c c c c c}
\hline\hline
Type & Name & GRB & Date$^*$ & p$_{\text{max}}$(\%) & $\lambda_{\text{max}}^{\dagger}$ (\ang) & Reference\\
\hline
Ic-bl & SN~2014ad & \xmark &$-$2 days & 1.60 $\pm0.03$ & 6880 & Section \ref{14adsec:lin.specpol}\\
Ic-bl & SN~2003dh & \cmark &+34 d after GRB & 2 $\pm0.5$ & 6500-7500 & \cite{kawabata03}\\
Ic-bl & SN~2006aj & \xmark (XMF) &+9.6 days & 3.8 $\pm0.5$ & \about 4200 & \cite{maund06aj}\\
Ic-bl & SN~1998bw & \cmark & $-$9 days & 1.1 & 6000-6200 & \cite{patat01}\\
Ic & SN~1997X  & \xmark &+ 11 days & 7 & 5000-6000 & \cite{wang01}\\
Ib/c & SN~2005bf& \xmark & 16 days after first V-band maximum & 4.5 & \about 3800 & \cite{maund05bf}\\
Ic & SN~2008D & \xmark &+ 15 days & 3 $\pm0.5$ & \about 8500 & \cite{maund08D}\\

\hline\hline
\end{tabular}
\end{table}
\end{landscape}

As seen in Figure \ref{14adfig:comp_pol}, the polarisation of SN~1998bw before maximum follows a trend similar to that seen in our data for SN~2014ad, showing a global increase in polarisation from 4000~\ang\, to 6000~\ang. 
On the other hand, the data sets of SN~1998bw at +8 days and SN~2014 at +5 days show little resemblance, apart from a similar level of median polarisation over the range 4000~\ang\, to 7000~\ang. 
In particular, there is a strong discrepancy around 5000~\ang\, where SN~1998bw shows a significant (\about 0.75\%) level of polarisation, whereas SN~2014ad exhibits polarisation that is close to zero. The degree of polarisation observed in the data of SN~1998bw at these wavelengths goes against the assumption that line blanketing due to iron lines in this region causes near complete depolarisation. 
In the case of SN 2014ad, however, we are confident that this assumption was justified, since the ISP values derived from it were consistent with ISP estimates calculated under the assumption of complete depolarisation at late times (see Section \ref{14adsec:isp}).

Another comparison of interest is that of the temporal evolution of the degree of polarisation of SN~2014ad and the normal type Ib SN~2008D (see \citealt{maund08D} and references therein). 
In Figure \ref{14adfig:comp_pol}, we see that around V-band maximum both SNe show similar levels of polarisation. 
By +15 days, the level of polarisation of SN~2008D has remained approximately constant whereas that of SN~2014ad has significantly decreased. 
This indicates that the overall geometry of the photosphere of SN~2008D remains the same closer to the core, whereas SN~2014ad shows evidence of more spherical ejecta in its interior.

When plotted on $q-u$ planes, the spectropolarimetric data of SN~2005bf (Ib/c), SN~2006aj (Ic-bl/XRF), SN~2008D (Ib) and SN~2002ap (Ic-bl) show more scatter and less well defined dominant axes than seen in SN~2014ad at $-$2 days and +5 days (\citealt{wang03,maund05bf,maund06aj,maund08D}). 
Additionally, an O\,{\sc i} $\lambda$7774 loop is present in the data of SN~2002ap. 
Its deviation from the rest of the data is very strong at $-$6 days and $-$2 days but subsequently weakens as Ca\,{\sc ii} IR features emerge.
By 3 days after V-band maximum the O\,{\sc i} loop in SN~2002ap has nearly disappeared. 
This evolution is akin to that of the O\,{\sc i} $\lambda$7774 feature in SN~2014ad and our interpretation that the calcium layer is deeper than the oxygen layer is consistent with \cite{wang03}. 
This suggests that the calcium seen in the data is not primordial but a product of stellar nucleosynthesis and that the ``onion" structure of the stellar interior is partially maintained. 
On the other hand we also see significant disruption of this structure since the oxygen and calcium line-forming regions were found to be spatially distinct from each other.

As a whole, the spectropolarimetric data reveal ejecta with significant axial symmetry that remain stable for more than 2 weeks after maximum light, and a more spherical interior uncovered by +43 days, whereas one might expect stronger axial symmetry towards the core in the case of a jet driven explosion. 
Compositional asymmetries are also present, resulting from the partial disruption of the interior structure of the progenitor. 
Whether these characteristics are consistent with a GRB driven explosion would require modelling, which is beyond the scope of this paper.

\subsection{Spectral modelling and photospheric velocity}
\label{14adsec:phot_vel}
In order to find values of the photospheric velocity and check our line identification, we created synthetic spectra with \textsc{syn++} (see Section \ref{14adsec:syn++}). 
Best results were obtained when using Fe\,{\sc ii}, Na\,{\sc i}, Si\,{\sc ii}, O\,{\sc i} and Ca\,{\sc ii}, but we also tried to fit our data using additional ions. 
One of our attempts included magnesium, and although it helped suppress the \about 7100~\ang\, peak, it also weakened the Ca\,{\sc ii} emission making the fit to SN~2014ad less accurate. 
We also tentatively added helium to the synthetic spectrum, in the hope that it may help fit the double dip between the iron and silicon emission. 
This approach was unsuccessful, and we believe that sodium is the most likely cause of that double dip, although in our fit (shown in Figure \ref{14adfig:syn++}) the Na\,{\sc i} peak is slightly blue-shifted compared to our data. This may be due to the limitations of a one-dimensional code, which does not account for asphericities in the ejecta.  

The photospheric velocities of SN~2014ad were obtained from model fitting and ranged from $-$30,000 $\pm$ 5,000~\kms at $-$2 days to $-$10,000 $\pm$ 2,000~\kms at 66 days (see Figure \ref{14adfig:phot_vel}). 
\cite{modjaz16} studied the spectral properties of 17 type Ic SN, 10 Ic-bl without observed GRBs and 11 Ic-bl with GRBs, and reported their Fe\,{\sc ii} $\lambda$5169 absorption velocities (used as proxy for the photospheric velocity), see their Figure 5. 
They also showed that Ic-bl in a GRB/SN~pair tend to have ejecta velocities \about~6000~\kms greater than for SNe Ic-bl without GRB counterparts. 
Our values of the photospheric velocity for SN~2014ad fit within the regime of the Ic-bl associated with GRBs. With a velocity reaching $-$30,000 $\pm$ 5,000~\kms at $-$2 days it even surpasses the average photospheric velocity of SNe Ic-bl with GRBs at the same epoch by approximately $-$10,000~\kms.
Additionally, it is very similar to the expansion velocity observed in SN~1998bw at maximum light \citep{iwamoto98}. 
The ejecta velocities observed in SN~2014ad are therefore consistent with SNe that have been associated with GRBs.

Irrespective of the way one measures velocities from the flux spectrum (e.g. line fitting, absorption line minimum), the calculated values correspond to the projected velocities, which may differ from velocities depending on viewing angle. 
In order to put constraints on the estimates made with the 1D code we used to calculate the photospheric velocity, we take as a limiting case that of a spheroid with axis ratio (0.72) given by the highest polarisation recorded for SN 2014ad in Section \ref{14adsec:lin.specpol} (1.60 \%; \citealt{hoflich91}), and calculate the velocity for a region of the photosphere at the pole of the oblate spheroid. 
The maximum photospheric velocity obtained at $-$2 days was $-$21 750 $\pm$ 3600~\kms, and decreased to $-$13 775 $\pm$ 3600~\kms by +5 days, which are still within the SNe Ic-bl with GRB regime. 

\textsc{syn++} assumes an infinitely narrow photosphere which does not overlap with the line forming regions, and therefore does not account for electron scattering, as mentioned in section \ref{14adsec:syn++}. 
High levels of polarisation are however detected at early days, which indicate the significant role played by electron scattering in the formation of the spectrum. Consequently, our interpretation of just the flux spectrum using  \textsc{syn++} may be incomplete. 

\subsection{Metallicity}
\label{14adsec:met}

\cite{modjaz08}, \cite{levesquekewleyberger10} and \cite{graham15} -- among others -- have compared the metallicity of the host galaxies of Ic-bl with and without GRBs. It has been shown that SNe Ic-bl with GRBs tend to arise in galaxies with lower metallicity than Ic-bl without GRBs, but that the two populations somewhat overlap.

\begin{figure}
\centering
\includegraphics[width=8.5cm]{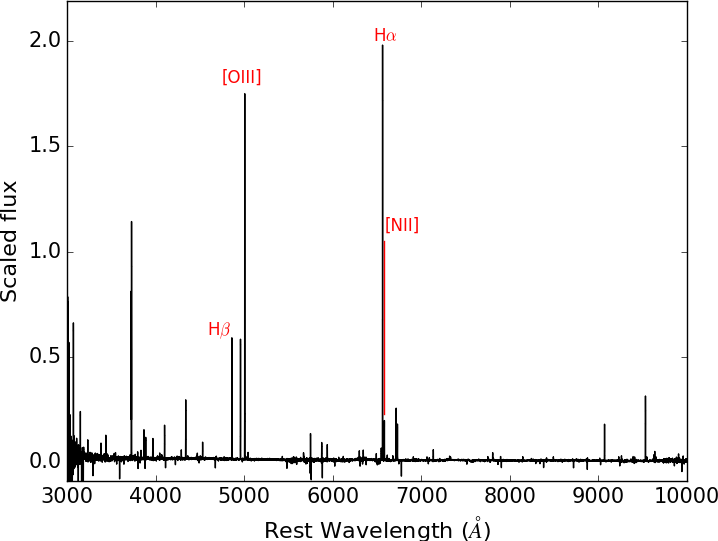}
\caption[X shooter spectrum of SN 2014ad]{\label{14adfig:xshoot} X-shooter spectrum of the host environment of SN~2014ad obtained on 22 May 2015, 424 days after V-band maximum. The features identified correspond to the lines used for metallicity calculations. }
\end{figure}

We compared the oxygen abundance of the host galaxy of SN~2014ad to a sample of 32 SNe and GRBs listed in Table \ref{14adtab:met}. 
They include three dark GRBs and 6 bursts for which the presence of a SN~in the afterglow could not be ruled out. 
We used the $O3N2$ diagnostic described by \cite{pp04} as:
\begin{equation}\label{eq:pp04}
12 + \log (\text{O/H}) = 8.73 - 0.32 \times O3N2,
\end{equation}
where O3N2 $\equiv$ log [([O\,{\sc iii}] $\lambda$ 5007 / \hbeta) / ([N\,{\sc ii}] $\lambda$ 6583 / \halpha)].
We derived the required line fluxes from public spectroscopic X-shooter observations of SN~2014ad obtained on 22 May 2015 \footnote[7]{Under programme 095.D-0608(A). PI: J.Sollerman}, or 424 days after V-band maximum, in which no SN~features are visible. 
The exposure time was 900 seconds for the blue arm (2970~\ang\, to 5528~\ang) and 960 seconds for the visible arm (5306~\ang\, to 10140~\ang). 
The data were reduced using the X-shooter pipeline by Dr. Marvin Rose, and the combined spectrum can be seen in Figure \ref{14adfig:xshoot}. 
The derived line fluxes are reported in Table \ref{14adtab:met}. 
We calculated the  metallicity of the host of SN~2014ad using our own routine and recomputed oxygen abundances of the comparison objects from values of the line fluxes reported in previous works for the \hbeta, [O\,{\sc iii}] $\lambda$ 5007, \halpha and [N\,{\sc ii}]~$\lambda$~6584 lines (see \citealt{modjaz08,graham15}; and references therein). 

The line fluxes, calculated metallicities and absolute B-band magnitudes for the host galaxies of our comparison objects and SN~2014ad are given in Table \ref{14adtab:met}. 
It is crucial to compare metallicities computed with the same diagnostic as different metallicity indicators yield slightly different values.
We therefore re-computed the oxygen abundance for all but two (SN~2002bl and SN~2007qw) of our comparison SNe as we could not find values for the line fluxes of \hbeta, [O\,{\sc iii}] $\lambda$5007, \halpha and [N\,{\sc ii}] $\lambda$6584. We used our own routines and line fluxes taken from the literature. Unfortunately a lot of the flux values reported did not have errors. 
We assumed uncertainties of the order 10\% of the nominal value of the flux, which is consistent with \cite{modjaz08}. 
It should be noted, however, that the main source of error when calculating the $O3N2$ metallicity are the systematic errors (0.14 dex) of this particular diagnostic \citep{pp04}. 
 
\begin{figure}
\centering
\includegraphics[width=8.5cm]{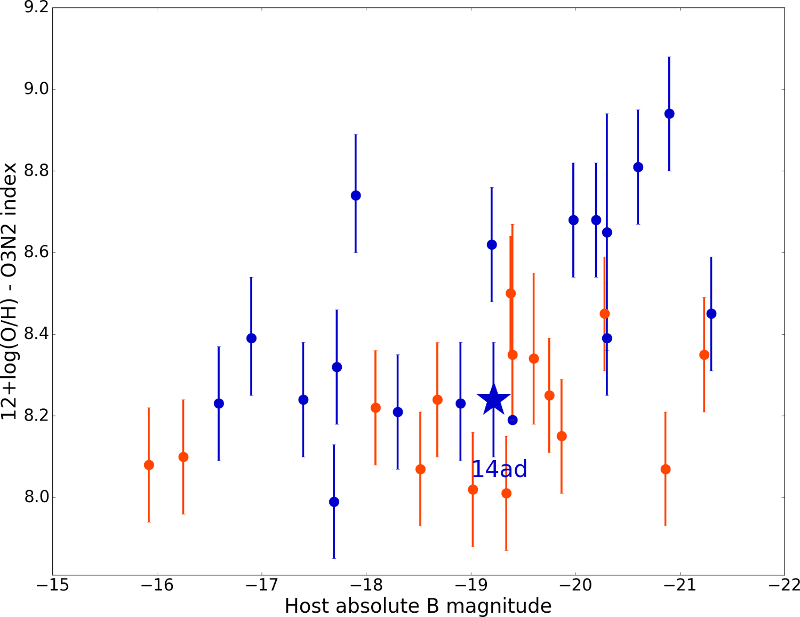}
\caption[Metallicity of SN 2014ad and other SNe and GRBs]{\label{14adfig:met} Comparison of oxygen abundance with host B-band absolute magnitude for 18 SNe Ic-bl (blue) and 15 GRBs (orange). SN 2014ad is represented by a star marker. }
\end{figure}

The oxygen abundance of the host environment of SN~2014ad was found to be \metal = 8.24 $\pm0.14$ (\about a third Solar and intermediate between the LMC and SMC), which is comparable to both populations of Ic-bl with and without GRBs (see Table \ref{14adtab:met} and Figure \ref{14adfig:met}).

\begin{table}
\tiny
\centering
\begin{tabular}{l c c c c c c c }
\hline \hline 
Ic-bl and GRB Hosts & \hbeta & [OIII] $\lambda$5007 & \halpha & [NII] $\lambda$6584 & \metal & M$_B$ & Spectrum
\\\hline 
SN1997ef$^*$ & 478 $\pm$ 51 & 223 $\pm$ 30 & 1730 $\pm$ 170 & 581 $\pm$ 62 & 8.68$\pm$ 0.14 & -20.2$^1$ & Flux$^1$ \\
SN2003jd$^*$ & 229 $\pm$ 24 & 342 $\pm$ 36 & 752 $\pm$ 76 & 98 $\pm$ 11 & 8.39 $\pm$ 0.14 & -20.3$^1$ & Flux$^1$ \\
SN2005kz$^*$ & 656 $\pm$ 12 & 115 $\pm$ 17 & 1660 $\pm$ 170 & 1300 $\pm$ 130 & 8.94 $\pm$ 0.14 &-20.9$^1$ & Flux$^1$ \\
SN2005nb$^*$ & 376 $\pm$ 45 & 455 $\pm$ 52 & 1510 $\pm$ 150 & 299 $\pm$ 33 & 8.45$\pm$ 0.14 & -21.3$^1$ & Flux$^1$ \\
SN2005kr & 8.20 $\pm$ 0.83 & 24.3 $\pm$ 2.4 & 28.4 $\pm$ 2.8 & 2.41 $\pm$ 0.26 & 8.24 $\pm$ 0.14 & -17.4$^1$ & Flux$^1$ \\
SN2005ks &  69.9 $\pm$ 7.3 & 51.2 $\pm$ 5.5 & 272 $\pm$ 27 & 92.3 $\pm$ 9.3 & 8.62 $\pm$ 0.14 & -19.2$^1$ & Flux$^1$ \\
SN2006nx & 6.9 $\pm$ 1.1 & 20.9 $\pm$ 2.2 & 33.1 $\pm$ 3.4 & 2.73 $\pm$ 0.51 & 8.23 $\substack{+0.15 \\ -0.14}$ & -18.9$^1$ & Flux$^1$ \\
SN2006qk & 43.7 $\pm$ 5.0 & 15.3 $\pm$ 2.6 & 230 $\pm$ 23 & 86.2 $\pm$ 8.8 & 8.74 $\substack{+0.15 \\ -0.14}$ & -17.9$^1$ & Flux$^1$ \\
SN2007I & 28.7 $\pm$ 2.9 & 57.2 $\pm$ 7.1 & 119 $\pm$ 13 & 20.0 $\pm$ 3.2 & 8.39 $\substack{+0.15 \\ -0.14}$ & -16.9$^1$ & Flux$^1$ \\
SN2002ap & 1.285 & 0.212 & 6.007 & 1.812 & 8.81 $\pm$ 0.14 & -20.6$^1$ & Hbeta$^2$\\
SN2007ce & 108.4 & 594.8 & 285.1 & 7.813 & 7.99 $\pm$ 0.14 & -17.69$^3$ & Hbeta$^4$\\
SN2008iu & 117.1 & 745.2 & 255.1 & 43.25 & 8.23 $\pm$ 0.14 & -16.59$^3$ & Hbeta$^4$\\
SN2009bb & 37.36 & 17.56 & 155.3 & 51.59 & 8.68 $\pm$ 0.14 & -19.98$^5$ & Flux$^5$\\
SN2010ah & 104.0 & 186.9 & 359.3 & 33.88 & 8.32 $\pm$ 0.14 & -17.22$^3$ & Hbeta$^4$\\
SN2010ay & 271.6 & 905.0 & 837.2 & 67.20 & 8.21 $\pm$ 0.14 &  -18.30$^6$ & Flux$^3$\\
\textbf{SN2014ad} & \textbf{9.35 $\pm$ 0.33 }& \textbf{28.53 $\pm$ 0.38 }& \textbf{33.25 $\pm$ 0.14} & \textbf{3.09 $\pm$ 0.11} & \textbf{8.24 $\pm0.14$} & \textbf{-19.22} & \textbf{Flux}\\
GRB980425/1998bw & 219.69 & 851.4 & 713.7 & 68.7 & 8.22 $\pm$ 0.14 & -18.09$^1$ & Hbeta$^1$\\
GRB991208/... & 3.493 & 5.848 & 17.04 & 0.852 & 8.24 $\pm$ 0.14 & -18.68$^1$ & Flux$^7$\\
GRB010921/... & 9.518 & 29.65 & 40.11 & 1.858 & 8.15 $\pm$ 0.14 & -19.87$^1$ & Flux$^{7,8}$\\
GRB011121/... & 17.13 & 16.64 & 65.61 & 2.0 & 8.25 $\pm$ 0.14 & -19.75$^1$ & Flux$^9$\\
GRB020819B/Dark & 3.0 $\pm$ 0.9 & 9.7 $\pm$0.7 & 14.5 $\pm$ 1.2 & 2.7$\pm$1.7 &  8.34 $\substack{+0.21\\ -0.16}$& -19.6$^1$ & Flux$^{10}$\\
GRB020903/unnamed & 44.0 & 335.0 & 168.0 & 7.2 & 8.01 $\pm$ 0.14 & -19.34$^1$ & Flux$^{11}$\\
GRB030329/2003dh & 1 & 3.40 & 2.74 & 0.1 & 8.10 $\pm$ 0.14 & -16.52$^1$ & Hbeta$^{7,12}$\\
GRB031203/2003lw & 1 & 6.36 & 2.82 & 0.15 & 8.07 $\pm$ 0.14 & -18.52$^1$ & Hbeta$^{12}$\\
GRB050824/... & 2.529 & 15.45 & 7.6 & 0.2797 & 8.02 $\pm$ 0.14 & -19.02$^1$ & Flux$^{13}$\\
GRB050826/... & 28.65 & 36.22 & 85.22 & 14.41 & 8.45 $\pm$ 0.14 & -20.28$^1$ & Flux$^7$\\
GRB051022/Dark & 25.29 & 59.57 & 104.99 & 15.97 & 8.35 $\pm$ 0.14 & -21.23$^1$ & EQW$^{3}$\\
GRB060218/2006aj & 68.83 & 229.9 & 170.5 & 5.122 & 8.08 $\pm$ 0.14 & -15.92$^1$ & Flux$^{14}$\\
GRB060505/Dark & 4.553 & 5.500 & 22.85 & 5.185 & 8.50 $\pm$ 0.14 & -19.38$^{3,15}$ & Flux$^{15}$\\
GRB070612A/... & 36.60 & 41.01 & 152.0 & 1.520 & 8.07 $\pm$ 0.14 & -20.86$^3$ & Flux$^7$\\
GRB120422A/SN2012bz$^*$ & 0.5 $\pm$ 0.4 & 1.9 $\pm$ 0.2 & 2.4 $\pm$ 0.1 & 0.6 $\pm$ 0.2 & 8.35$\substack{+0.32 \\ -0.16}$ & - 19.4$^{16}$ & Flux$^{16}$ \\ \hline 
SN2002bl$^*$ & ... & ... & ... & ... & 8.65 $\pm$ 0.3$^{17}$ & -20.3$^{17}$ & ...\\
SN2007qw$^*$ & ... & ... & ... & ...  & 8.19 $\pm$ 0.01$^{17}$& -19.4$^{17}$ & ...\\
\hline \\

\end{tabular}
\caption[Metallicity of type Ic-bl and GRB SNe]{\label{14adtab:met} Line fluxes, O3N2 metallicity \citep{pp04} and B-band absolute magnitude of the host environments of type Ic-bl SNe and GRBs. The SNe associated with the GRBs are given when known, "Dark" indicates the absence of an optical counterpart, and an ellipse "..." is shown when a SN~cannot be ruled out because the searches conducted were not deep enough. The emission line measurements correspond to the values for the centre of the host galaxy, unless the object is marked with an asterisk ($^*$), in which case they coincide with the host environment of the object within the host galaxy. The line fluxes and equivalent widths are all in units of 10$^{-17}$ erg s$^{-1}$ cm$^{-2}$. When the values found of the literature did not have associated errors, we assumed an uncertainty \about 10\%. The last column indicates whether the values for a given object correspond to the flux values, the equivalent width, or the flux value normalised with respect to \hbeta (Note that the \hbeta column of an \hbeta normalised flux might not be 1 or 100 if the galactic extinction correction was applied to normalised values in the literature). The metallicity of SN~2002bl and SN~2007qw are taken directly from (17). \textbf{References:} (1) \cite{modjaz08}; (2) \cite{ferguson98}; (3) \cite{graham13}; (4) \cite{sanders12}; (5) \cite{levesquesoderberg10}; (6) SDSS-mpg; (7) \cite{levesquekewleyberger10}; (8) \cite{price12}; (9) \cite{garnavich03}; (10) \cite{perley17}; (11) \cite{hammer06}; (12) \cite{sollerman05}; (13) \cite{mcglynn07}; (14) \cite{levesquebergerkewley10}; (15) \cite{thone08}; (16) \cite{schulze14}; (17) \cite{modjaz11}. }
\end{table}

\subsection{Late time [O\,{\sc i}] line profile}

\begin{figure}
\centering
\includegraphics[width=8.5cm]{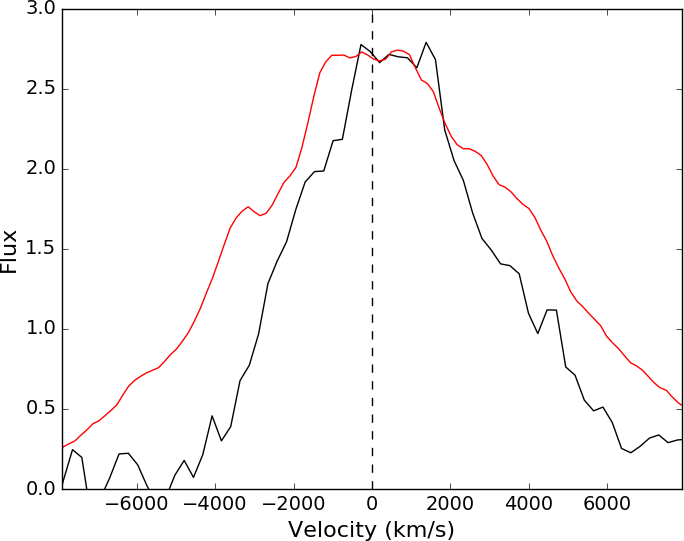}
\caption[Forbidden oxygen line of SN 2014ad]{\label{14adfig:OI} Comparison of the [O\,{\sc i}] $\lambda\lambda$6300,6364 line profile of SN~2014ad (red) at 107 days post V-band maximum to SN~1997dq (black) +262 days after estimated outburst \citep{mazzali04}. }
\end{figure}

The shape of the [O\,{\sc i}] line varies significantly from one SN~to another, ranging from single peaks to symmetric double peaks and asymmetric double peaks \citep{maeda08,milisavljevic10}. 
In the case of SN~2014ad, the [O\,{\sc i}] emission at +107 days mostly resembles a flat-topped (from $-$1000~\kms to +1000~\kms)  single peak profile, as seen in SN~1997dq (see Figure \ref{14adfig:OI}; \citealt{mazzali04}). Additionally, a secondary peak is noticeable at $-$3200~\kms.

The flat top is characteristic of a spherical shell type structure (e.g. \citealt{ignace00, ignace06, maeda08}), however the presence of a blue-shifted feature also indicates contribution from asymmetrically distributed oxygen \citep{milisavljevic10}. 
Given the low velocity and mostly spherical characteristic of this oxygen line-forming region, it must be separate from the line-forming region responsible for the high velocity asymmetric O\,{\sc i} $\lambda$7774 seen at early days.

Additionally we used the [O\,{\sc i}] line profile as a template to fit the Mg\,{\sc i}] $\lambda$4571 and the [Ca\,{\sc ii}] $\lambda\lambda$7291,7824 lines also present in the spectrum at +107 days (see Figure \ref{14adfig:spctr}). 
We found that the FWHM of the feature centred on the Mg\,{\sc i}] $\lambda$4571 line is \about 13,000 km/s, a factor of 2 broader than the feature we identify as [O\,{\sc i}]. 
Either this is due to blending of lines or a significantly different distribution within the ejecta of the supernova. 
In other supernovae (e.g. \citealt{spyromilio94}) these features have had very similar profiles. The red side of the calcium forbidden line is well fitted by the [O\,{\sc i}] profile, however the blue side is broader, also suggesting blending with other spectral features. 

\subsection{Schematic of SN 2014ad}

\begin{figure}
\centering
\includegraphics[width=13cm]{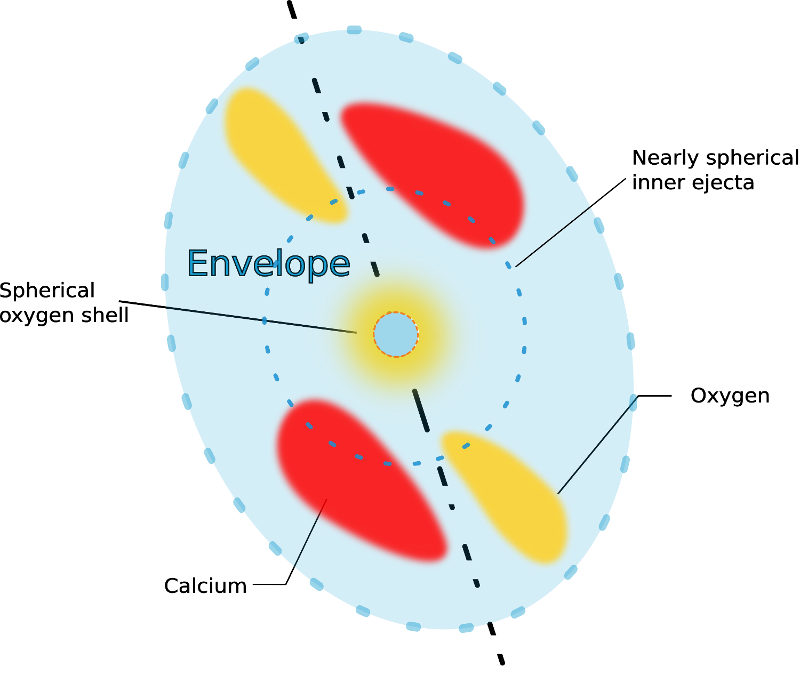}
\caption[Cartoon of SN 2014ad]{\label{14adfig:toymodel} Toy model of the ejecta of SN~2014ad. The dashed black line represents the dominant axis. Note that the range of the P.A. (0\degree to 180\degree) results in an artificial degree of symmetry.}
\end{figure}

From our interpretation of the spectropolarimetric and spectroscopic data of SN~2014ad we produced a qualitative picture of the structure of the ejecta (see Figure \ref{14adfig:toymodel}). 
The dominant axis present in the Stokes parameters at early days suggests a significant departure from spherical symmetry, but with axial symmetry in the geometry of the outer ejecta. 
This could translate as an ellipsoidal envelope, the minimum axis ratio of which (0.72) can be constrained by the maximum degree of polarisation we recorded at the first epoch (1.60\%). 
Departures from axial symmetry were observed in the behaviour of oxygen and calcium suggesting that these line-forming regions are made of large clumps whose angular distribution changes with radius.
Additionally, the behaviour of the oxygen and calcium features in the  $q-u$ planes reveals that the line-forming regions for the two species must be distinct, and their temporal evolution indicates that the oxygen clumps have higher velocities and dissipate before those of calcium. 
The decrease in the overall degree of polarisation over time and the disappearance of polarisation features at later epochs suggest that the asymmetries of the outer envelope are not replicated deeper in the ejecta. 
It is important to note that since the Stokes parameters $q$ and $u$ are quasi-vectors (i.e. the P.A. ranges from 0\degree to 180\degree) the sketch exhibits an artificial degree of symmetry that may not be representative of the actual structure of the ejecta.

Jet induced models have shown that the distribution of oxygen and calcium tends to be limited to an equatorial torus \citep{khokhlov99}, however in SN~2014ad the line-forming regions are distinct, which is inconsistent with these models. 
Additionally the Fe blends in the early spectra SN~2014ad correspond to near zero levels of polarisation; \cite{maund05bf} associated the lack of iron polarisation with the absence of jets, as they argued that one would expect to see such signatures if core material was brought up by a jet. 
Note that this was a qualitative argument and not a theoretical constraint.
Furthermore, \cite{maund05bf} used helium as a tracer for the asymmetric nickel distributions that excite it (as helium is excited non-thermally in supernvoa ejecta), which could be caused by a stalled jet in the envelope; unfortunately the lack of helium in SN~2014ad denies us any information about the nickel. 
The decreasing level of polarisation over time is also difficult to reconcile with the ``choked" jet model proposed for SN~2005bf and SN~2008D (\citealt{maund05bf,maund08D}). 
Those conclusions, however, were drawn on limited observations of those SNe and serve to highlight the importance of multi-epoch spectropolarimetry.

\section{Conclusion}
\label{14adsec:conc}

We report 7 epochs of spectropolarimetry for the Ic-bl SN~2014ad ranging from $-$2 days to +66 days, as well as 8 epochs of spectroscopy from $-$2 days to +107 days. 
This is to our knowledge the best spectropolarimetric data set available for a Ic-bl SN. A maximum degree of polarisation of 1.60 $\pm$0.03 \% was detected at $-$2 days, and the level of ISP was found to be 0.15 $\pm$0.06\%. 
A clear dominant axis in the $q-u$ plots corresponding to the first 3 epochs was found with a P.A. \about 20\degree, indicating significant axial symmetry. 
The decrease in the overall level of polarisation over time and the disappearance of the dominant axis at +33 days suggests that the inner ejecta are more spherical than the outer layers. 

Polarisation features associated with O\,{\sc i} and Ca\,{\sc ii} show departures from the dominant axis in the form of loops suggesting line-forming regions in the form of large clumps that are asymmetrically distributed. 
Since the oxygen and calcium loops are oriented in different directions the two line-forming regions must be distinct from each other, but both are moving very rapidly. 
Additionally, due to the separate temporal evolution of their polarisation signatures (oxygen appears and disappears earlier than calcium) and different velocities (oxygen is faster than calcium)  we conclude that the calcium layer is deeper into the ejecta than this oxygen layer. 
This means that the calcium we observe is not primordial but a product of stellar nucleosynthesis and that the ``onion" structure of the progenitor is at least partially preserved. 

At +107 days the [O\,{\sc i}] line profile suggests the presence of a deeper oxygen line-forming region in a near spherical shell, unlike the oxygen observed at earlier dates. 
This is consistent with the decreasing levels of polarisation detected deeper in the ejecta. 

Using X-shooter data at +424 days we calculated the metallicity of the host galaxy of SN~2014ad and found it was rather low and consistent with the population of Ic-bl with GRB counterparts. 
It should be noted that it is also close to the metallicity of the host galaxy of two other Ic-bl without GRBs with similar B-band magnitude: SN~2006nx and SN~2007qw. 
Because of the overlap in the two populations, the low metallicity of the host environment of SN~2014ad is inconclusive. 

It is not clear from the spectropolarimetric data of SN~2014ad whether the explosion was driven by jets: the axial symmetry at early epochs is consistent with such scenario, but the more spherical inner ejecta is more difficult to reconcile with the presence of a jet. 
The high velocity oxygen and calcium line-forming regions follow directions that are similar to that of the dominant axis, which would correlate with the direction of jets, should they be present.
Hydrodynamical simulations of asymmetrical explosions produced by \cite{maeda02} to model SN 1998bw showed low-velocity oxygen being ejected in the equatorial plane, which is inconsistent with our picture of SN 2014ad. 
On the other hand, fully jet-driven explosion models by \cite{couch11} revealed intermediate mass elements being entrained by the outflow and as a result roughly following the direction of the jets, which is consistent with the behaviour of the high velocity calcium and oxygen in SN 2014ad. 
The potential outflow may not have fully broken out of the envelope and produced a GRB, but could have been sufficiently powerful to accelerate the ejecta as fast as 30,000 $\pm$ 5,000~\kms and create the strong alignment observed in the $q-u$ planes until 18 days after V-band maximum. 
This scenario is consistent with the conclusions of \cite{lazzati12}, who showed that for an engine of short enough life time the jets may dissipate through the envelope before reaching the surface, resulting in a relativistic SN~explosion but no GRB. 
It is, however, difficult to reconcile the case of ``choked" jets with nearly spherical ejecta, as observed in SN~2014ad. 
If SN~2014ad did have a GRB counterpart, it may have been collimated away from our line of sight, and without radio observations we cannot exclude or support this scenario with complete certainty.

In the following Chapter we will move away from SN spectropolarimetry and look at potential progenitors of type Ic-bl.

\chapter{Spectropolarimetry of Galactic WO stars}

\label{chpt:wos} 

\lhead{\emph{Spectropolarimetry of Galactic WO stars}} 

\section{Introduction}
Oxygen sequence WR stars, or WO stars, are a subset of helium deficient WR stars which are expected to explode as Type Ic SNe (see Section \ref{introsec:ccsne}). 
Additionally, they are very close to core helium exhaustion (e.g \citealt{langer12}), and are therefore the final evolutionary phase before core collapse, with a timescale of only a few thousand years \citep{tramper15}. 

Some of these stars are expected to result in Ic-bl SNe, and potentially produce GRBs via the collapsar model (see Section \ref{introsec:classification}).
One of the conditions for the successful creation of a collapsar is sufficient angular momentum ($j\geq 3\times10^{16}$ cm$^2$/s -- \citealt{macfadyen99}), and therefore searching for rapid rotation on WO stars could provide information on their fate. 

As described in Section \ref{introsec:wrpol}, the presence of a line effect in WR stars can be used to probe an asymmetric wind geometry. 
Evidence for such an effect has previously been reported in Galactic, Small Magellanic Cloud (SMC) and Large Magellanic Cloud (LMC) WN and WC stars, with amplitudes ranging from \about 0.3 to \about 1 percent, and with an incidence of \about 20 percent in the Milky Way and \about 10 percent in the LMC \citep{harries98, vink07, vink17}.
The study of \cite{vink17} included spectropolarimetric observations for a binary WO star in the SMC but detected no line effect in this object. 
Because WO stars are believed to be only a few thousand years away from core collapse \citep{tramper15}, it is crucial to determine their rotational properties in order to constrain explosion models of Type Ic-bl and GRB-SNe. 

This chapter investigates WR93b (WO3, \citealt{drew04}) and WR102 (WO2, \citealt{tramper15}), whose time to explosion are estimated to be \about 9000 and 1500 years, respectively \citep{tramper15}.
They were selected on the basis that they are single Galactic WO stars accessible from the Very Large Telescope (VLT) in Chile.
This work was published in \cite{stevance18}.

\section{Observations and Data Reduction}
Spectropolarimetric observations of WR93b and WR102 (see Table \ref{wotab:obs}) were conducted under the ESO programme ID 079.D-0094(A) (P.I: P. Crowther).
The data were collected on 2007 May 02 with the VLT of the European Southern Observatory (ESO) using the Focal Reducer and low-dispersion Spectrograph (FORS1) in the dual-beam spectropolarimeter ``PMOS" mode \citep{appenzeller98}.
The 300V grism was used in combination with a 1" slit, providing a spectral range 2748--9336\r{A}, and a resolution of 12\r{A} (as measured from the CuAr arc lamp calibration). 
No order sorting filter was used.
Additionally, the observations were taken under median seeing conditions of FWHM $=0.7$". 
Linear spectropolarimetric data of the targets were obtained at 4 half-wave retarder plate angles: 0\degree, 22.5\degree, 45\degree and 67\degree.
The data were reduced in the standard manner using IRAF and {\sc FUSS} following the description of Chapter \ref{chpt:datred}.   
To improve the signal to noise ratio the data were binned to 15\r{A} which, given the breadth of the emission lines (FWHM \about 200 \r{A}), would not prevent us from resolving any potential line effect.
Intensity spectra of WR93b and WR102 were retrieved by adding the flux spectra of each ordinary and extra-ordinary ray. 
Since no spectrophotometric standard was observed, however,  the flux spectra could not be calibrated and Stokes I (in counts) are used throughout this chapter. 
Note that that Stokes I for both targets is presented unbinned. 
Calibrated, de-reddened Xshooter spectra of WR93b and WR102 ranging from 3000 to 25000\r{A} can be found in \cite{tramper15}.

\begin{table}[h!]
\centering
\caption{\label{wotab:obs} VLT FORS1 Observations of WR93b and WR102.}
\begin{tabular}{c c c c}
\hline\hline
Object & Date & Exp. Time & Airmass \\
 & (UT) & (s)  & (mean)\\
\hline
WR93b & 2007 May 02 & 48 $\times$ 150 & 1.12\\
WR102 & 2007 May 02 & 48 $\times$ 70 & 1.06\\
BD -12\degree 5133 & 2007 May 02 & 8 $\times$ 12 & 1.14\\
\hline\hline
\end{tabular}
\end{table}

\begin{landscape}
\begin{figure*}
	\includegraphics[width=17cm]{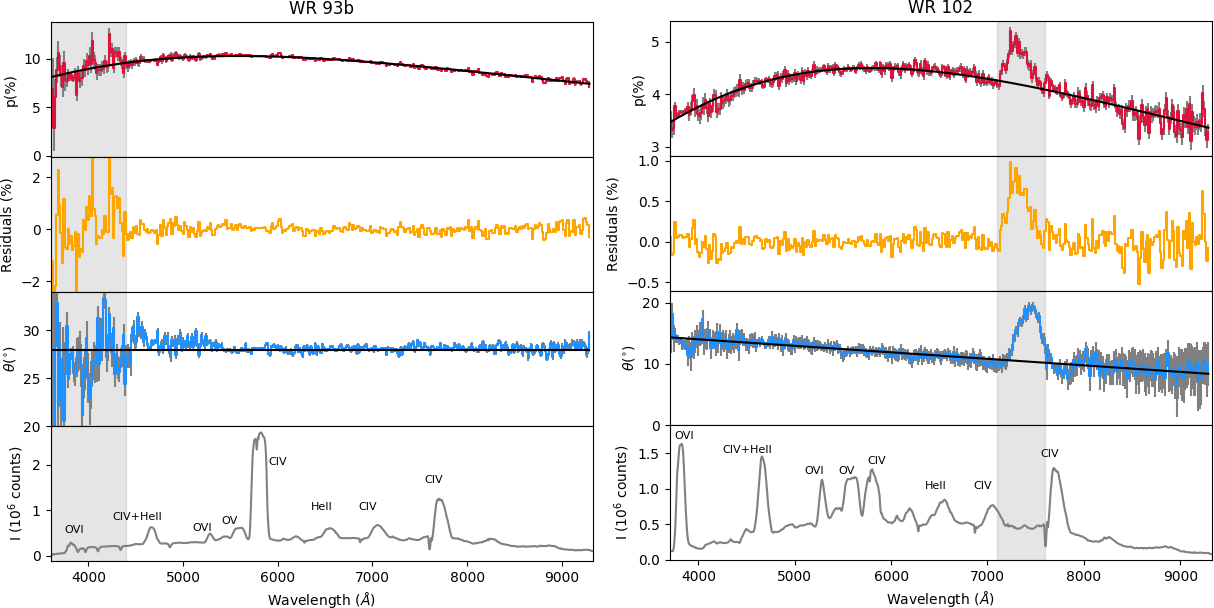}
    \caption[WO star polarisation]{\label{wofig:pol} Spectropolarimetric data of WR93b (left) and WR102 (right). The panels, from top to bottom, respectively contain: The degree of polarisation (red) and best Serkowski law fit (black); the residuals after Serkowski fit removal (orange); the polarisation angle (blue); Stokes I (grey) and line identification. The grey areas represent the discrepant regions of the spectrum described in Section \ref{wosec:obs_prop}.}
\end{figure*}
\end{landscape}

\section{Polarisation of WR93b and WR102}
\label{wosec:pol}
\subsection{Observational properties}
\label{wosec:obs_prop}

The reduced spectropolarimetric data of WR93b and WR102 are presented in Figure \ref{wofig:pol}.
Before commenting on the characteristics of the polarisation of the targets, it is worth discussing specific spectral regions with noisy or spurious features:
in WR93b, the degree of polarisation ($p$) and polarisation angle ($\theta$) show greater levels of noise in the blue parts of the spectrum, particularly below 4400\r{A};
in the data of WR102, a broad peak in the degree of polarisation (\about 1 percent deviation)  and polarisation angle (\about 8\degree deviation) is seen between 7100 and 7600\r{A}. 
The latter feature cannot be a line effect as it is located in a region of the spectrum that is devoid of strong lines, and although its origin remains unknown, the investigations detailed in Section \ref{wosec:disc_102} clearly highlight its spurious nature. 
Given these considerations, the analysis presented in the rest of this chapter was performed without the spectral region below 4400\r{A} in WR93b, and the range 7100-7600\r{A} in WR102.

Excluding these wavelength ranges, the polarisation measured for the data of WR93b and WR102 slowly rises up to an amplitude of \about 4.5 percent and \about 10 percent (respectively) between 5000 and 6000\r{A}, and then decreases redward of the peak.
Their polarisation angles have distinct behaviours: WR93b exhibits a constant polarisation angle of $\theta = 28.2 \pm 0.5$\degree, whereas the polarisation angle of WR102 shows a downward trend which can be fit with a line of the form $\theta = -1.08\times10^{-3}\lambda + 18.4$ degrees. 
A non zero $\Delta \theta/\Delta \lambda$ can be explained either by the superposition of intrinsic and interstellar polarisation \citep{coyne74} or by the presence in the line of sight of dust clouds with different particle sizes and alignments \citep{coyne66}.
The merit of these interpretations is discussed in Section \ref{wosec:disc_serk}.

\subsection{7300\r{A} feature in WR102}
\label{wosec:disc_102}
\begin{figure}
	\includegraphics[width=\columnwidth]{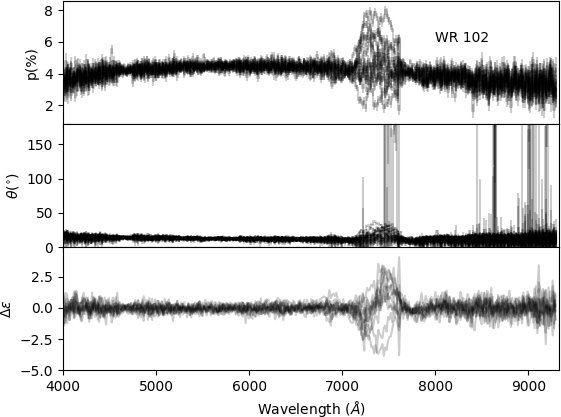}
    \caption[WR 102 discrepancy]{\label{wofig:102eps} Superposed plots of the degree of polarisation (p), polarisation angle ($\theta$) and instrumental signature correction difference ($\Delta \epsilon$) as calculated for each of the 12 sets of 4 half-wave retarder plate angles of WR102.}
\end{figure}

In Section \ref{wosec:obs_prop} a prominent feature in WR102 spanning the wavelength range 7100-7600\r{A} was highlighted. 
This peak was labelled as spurious and disregarded in the analysis. 
Here, the steps taken in investigating the nature of this feature are detailed. 

The first step was to explore whether the spike in polarisation is intrinsic to the data. 
The fact that the feature is not associated with any strong line (as would be expected for a line effect present in the WR star) is however inconsistent with this idea. 
Additionally, plotting the polarisation calculated for each of the 12 sets of observations at 4 half-wave retarder plate angles (see Figure \ref{wofig:102eps}), reveals that the shape and orientation (peak or trough) is highly variable from set to set, and no one set is consistent with another. 
The difference in instrumental polarisation corrections ($\Delta \epsilon$, see Section \ref{datred:deps}) can be used to formally investigate the presence of spurious data in this region of the spectrum. 
As previously mentioned, this property is expected to be 0 for polarisation signals $\le 20$ percent, and deviations indicate that the polarisation does not reflect a real signal. 
In Figure \ref{wofig:102eps} it is clearly visible that $\Delta \epsilon$ is consistent with 0 across the spectrum but deviates significantly in the wavelength range that coincides with the discrepancy. 

As the possibility that this feature was intrinsic to WR102 was ruled out, further investigation was required to understand its nature. 
The presence of a potential detector artefact was explored by visually examining the 2D images, but none were found. 
Additionally, the data reduction of the polarisation standard BD -12\degree 5133 (observed immediately after WR102) showed no inconsistencies from the expected signal, and none of the WR93b data sets (observed before WR102) exhibited features such as that seen in WR102. 

Finally, the calculation of the Stokes parameters from the ordinary and extra-ordinary fluxes was performed as follows.
\begin{itemize}
\item[(i)] Calculating $q$ and $u$ for each set of 4 half-wave retarder plate angles and computing the average.\footnote{Method normally used by FUSS}
\item[(ii)] Averaging the normalised flux differences for each half-wave retarder plate angle and calculating $q$ and $u$.
\item[(iii)] Averaging the ordinary and extraordinary rays for each half-wave retarder plate angle to then calculate the normalised flux differences and Stokes parameters.
\end{itemize}
As expected, all these methods yielded consistent results including a visible discrepant feature around $7100-7600$ \r{A}.

Ultimately, the origin of this feature was not unveiled, as it did not seem to be either intrinsic to WR102 nor to be an issue with FORS1. 
However, since the rest of the data were unaffected the analysis was carried out disregarding the affected region of the spectrum.

\subsection{Interstellar Polarisation and limits on the line effect}
\label{wosec:serkfits}

The dust present in our Milky Way has a tendency to align along the magnetic field of the Galaxy which causes the light that passes through these dusty regions to become polarised (see Section \ref{introsec:intro_ISP}). 
Both WR93b and WR102 are close to the Galactic plane and are obscured by a large amount of dust, as is evidenced by the high reddening values reported by \cite{tramper15}: $E(B-V) = 1.26$ and $E(B-V) = 1.94$ mags for WR93b and WR102, respectively.
This results in high levels of ISP.

In order to retrieve the signal intrinsic to the targets it would be ideal to quantify this ISP and remove it.
To tackle this problem, one approach is to look at nearby standard stars that are intrinsically unpolarized, as any polarisation detected for these objects can then be attributed to ISP.
Unfortunately, the closest standard polarisation stars in the \cite{Heiles} catalogue were located over a degree away from our targets, making them inadequate probes of the interstellar medium between us and WR93b and WR102.

In this context a precise estimate on the ISP in the direction of the targets is not possible, however for the task at hand an absolute measure is not actually necessary.
Indeed, as we are looking for the presence of a line effect, such as that observed in \cite{harries98}, all that is needed is to investigate whether the polarisation associated with emission lines shows any deviation from the \emph{observed} continuum, which could be a blend of ISP and intrinsic continuum polarisation.
We can fit this underlying signal and remove it, to then investigate whether the residual data are consistent with noise or if a line effect can be detected.

To this end we can use our knowledge of the shape of the ISP in the Milky Way, which is described by Serkowski's law (\citealt{serkowski73} -- see Eq. \ref{introeq:serk}).
We performed fits of the degree of polarisation of WR93b and WR102 with $p_{\text{max}}$, $\lambda_{\text{max}}$ and K as free parameters using a non-linear least-squares optimiser (SciPy - \citealt{scipy}). 
The best fitting parameters are summarised in Table \ref{wotab:fits} and the corresponding fits are shown in Figure \ref{wofig:pol}.
Should there be any intrinsic continuum polarisation, the shape of $p(\lambda)$ will remain unchanged since polarisation arising from Thomson scattering (such as can be the case in WR winds) has no wavelength dependence (see Section \ref{introsec:pol_origin}).
Consequently a blend of ISP and intrinsic continuum polarisation will still be described by Eq. \ref{introeq:serk}, although the values of $p_{\text{max}}$ would differ from the ISP-only case. 
These fits can be subtracted to the polarisation data observed for WR93b and WR102 to yield the residuals seen in Figure \ref{wofig:pol}.
No evidence of a line effect is visible in either WR93b or WR102.

\begin{table}[h!]
\centering
\caption{\label{wotab:fits} Best fitting parameters of the Serkowski law fits to WR93b and WR102}
\begin{tabular}{c c c c }
\hline\hline
Object & $p_{\text{max}}$  & $\lambda_{\text{max}}$  & K\\
 & (percent) & (\r{A}) & -- \\
\hline
WR93b & 10.31 $\pm$ 0.01  &  5577 $\pm$ 18 & 1.25 $\pm$ 0.02\\
WR102 & 4.50 $\pm$ 0.01 & 5795 $\pm$ 15 & 1.30 $\pm$ 0.03\\
\hline\hline
\end{tabular}

\end{table}

In order to place limits on the maximum amplitude a line effect could have while still being undetected, we measure the standard deviation of the residual polarisation in the wavelength ranges corresponding to strong line regions, see Table \ref{wotab:limits}.

\begin{landscape}

\begin{table}
\centering
\caption[WO star polarisation limits]{\label{wotab:limits} 1$\sigma$ limits on the polarisation that could remain undetected in the signal associated with the strong emission lines of WR93b and WR102. The intensity at peak relative to continuum $I_{\ell}$ as well as the derived upper limit on the intrinsic continuum polarisation (see Section \ref{wosec:lim_pol}) are also given. }
\begin{tabular}{c c c c c }
\hline\hline
Line & Wavelength Range & 1$\sigma$ (percent)  & $I_{\ell}$ &  $P_{\text{cont}}$ (percent) \\
\hline
\multicolumn{5}{c}{WR93b} \\
\hline
C\,{\sc iv} $\lambda 5808$ &  ($5705-5920$)   &  0.066 & 6.20 & $<$0.077\\
C\,{\sc iv} $\lambda 7724$  & ($7640-7840$) & 0.118 & 2.13 & $<$0.173\\
\hline
\multicolumn{5}{c}{WR102} \\
\hline
C\,{\sc iv} $\lambda 4659$  + C\,{\sc iv} $\lambda 4686$  + He\,{\sc ii} $\lambda 4686$ &  ($4600-4775$) & 0.05 & 2.74 &$<$0.068\\
O\,{\sc vi} $\lambda 5290$ &  ($5225-5360$) & 0.03 & 1.13 &  $<$0.057\\
O\,{\sc v}  $\lambda 5590$  & ($5510-5690$) & 0.035 &  0.79 & $<$0.079\\
C\,{\sc iv} $\lambda 5808$   & ($5705-5930$)  & 0.05 & 1.15 & $<$0.093\\
He\,{\sc ii} $\lambda 6560$ + C\,{\sc iv} $\lambda 6560$ &  ($6380-6670$)  & 0.06 & 0.69 & $<$0.147\\
C\,{\sc iv}  $\lambda 7724$  & ($7640-7870$)  & 0.11 & 1.95 & $<$0.166\\

\hline\hline
\end{tabular}
\end{table}

\end{landscape}

\subsection{Upper limit on the continuum polarisation}
\label{wosec:lim_pol}
The line effect being searched for is caused by a dilution of intrinsic continuum polarisation $P_{\text{cont}}$ by the unpolarized line flux. 
Under the assumption that the flux of the emission lines is completely unpolarized, the continuum polarisation is related to the amplitude of the line effect ($\Delta P$) and the peak intensity relative to the continuum ($I_{\ell}$) by the relationship given in \cite{harries98}:
\begin{equation}\label{woeq:pcont}
P_{cont} = \Delta P \frac{I_{\ell}+1}{I_{\ell}}
\end{equation} 

As already stated in Section \ref{wosec:serkfits}, there is no indication of a line effect in the polarisation data of WR93b and WR102, however it is possible to make use of the limits placed on the amplitude of a potentially undetected line effect (see Table \ref{wotab:limits}) and Eq. \ref{woeq:pcont} to put constrains on the continuum polarisation in our targets.
The measurements of $I_{\ell}$ for the strong emission lines of WR93b and WR102 are given in Table \ref{wotab:limits}.
An upper limit on the continuum that could remain undetected for each of these lines can then be estimated by substituting into Eq. \ref{woeq:pcont} the measured $I_{\ell}$ and the standard deviation value (in the corresponding wavelength range) in place of $\Delta P$. 
The resulting values of $P_{\text{cont}}$ are given in Table \ref{wotab:limits}, and represent the maximum level of polarisation that could be left unseen for each  \textit{individual line}. 
These upper limits on $P_{\text{cont}}$ vary greatly from line to line, which is not unexpected as the proxy for $\Delta P$ is the standard deviation measured for a residual signal that is dominated by noise.

In order to select a final upper limit from the values calculated it should be noted that, as previously mentioned, the continuum polarisation is not expected to be wavelength dependent.
This means that any amount of polarisation that goes undetected in one strong line will also be present in every other line and potentially be detectable. 
As a results, the effective 1$\sigma$ limits on the values of $P_{\text{cont}}$ for WR93b and WR102 will be the lowest values obtained from Eq. \ref{woeq:pcont} and reported in Table \ref{wotab:limits}, regardless of which line they were derived for.
Consequently, the upper limits on the continuum polarisation $P_{\text{cont}}$ of WR93b and WR102 are 0.077 percent and 0.057 percent, respectively. 


\subsection{ISP and Serkowski fits}
\label{wosec:disc_serk}
It is now clear that the  continuum polarisation of WR102 and WR93b, should there be any, is low ( $<0.077$ and $<0.057$ percent, respectively). 
This implies that the polarisation signal detected is overwhelmingly dominated by ISP, and the fitting parameters (summarised in Table \ref{wotab:fits}) are representative of the ISP. 
Also, it is now reasonable to conclude that the wavelength dependency of the polarisation angle of WR102 mentioned in Section \ref{wosec:obs_prop} is likely caused by two dust clouds overlapping in our line of sight with different particle sizes and grain orientation, rather than a superposition of ISP and continuum polarisation. 

In the case of both WR93b and WR102, $\lambda_{\text{max}}$ is found to be very close to the median value of 0.545 $\mu$m observed by \cite{serkowski75} in the Milky Way.
Regarding $K$, however, quantitatively comparing the present estimates to the value assumed by \citeauthor{serkowski75} is delicate since they provide no errors, and it is not clear which other values were tested and ruled out.
On the whole our values of $K$ are close to the estimate of \citeauthor{serkowski75}, and they are not statistically inconsistent with each other. 
Performing a three parameter fit of the Serkowski law rather than assume $K = 1.15$ is an approach that has been employed by multiple studies (e.g. \citealt{martin99, patat15})
and is motivated by the fact that the best value of $K$ can be highly variable from target to target, even within the Milky Way (e.g. \citealt{martin92}). 

Lastly, the \citeauthor{serkowski75} study also reported the relation R \about $5.5 \times \lambda_{\text{max}}$ ($\mu$m), which can be used to estimate the total to selective extinction from our estimates of $\lambda_{\text{max}}$. 
Values of R = 3.1 and R = 3.2 are found for WR93b and WR102, respectively. 
These values are consistent with the estimates of \cite{tramper15} and standard interstellar medium values.

\section{Discussion}
\label{wosec:disc}
\subsection{Rate of line effect in WR stars}

In Section \ref{wosec:serkfits} upper limits on the amplitude of the line effect that could go undetected in the polarisation data of WR93b and WR102 were derived (see Table \ref{wotab:limits}).
All of these limits have an amplitude significantly smaller than the amplitudes of the line effect observed in previous studies ($<$0.3 percent, e.g. see \citealt{harries98, vink07}), consequently it is safe to conclude that there is no line effect in either WR93b or WR102. 
Including the studies of \cite{harries98}, \cite{vink07} and \cite{vink17}, a sample of 82 WR stars have been observed with spectropolarimetry: 54 WN stars, 25 WC, stars and 3 WO stars.
This sample includes Galactic, LMC and SMC WR stars. 
Of these, 11 stars showed a line effect: 9 WN stars, 2 WC star and no WO stars.
The incidence of the line effect on WR star populations in the Milky Way, SMC and LMC is summarised in Table \ref{wotab:line_effect}; the errors quoted on the percentages are the 68 percentile of the binomial confidence interval.
Across all types, 13.4$\pm2.6$ percent of WR stars exhibit a line effect, however when separated into sub-types we see that 16.6$\pm3.4$ percent of WNs showed a line effect, whereas only 8$\pm3.7$ percent of WCs and none of the WO stars did.
It is important to note, however, that these values were obtained from data sets that are not uniform in quality, and should therefore be considered with caution.

\begin{table*}[h!]
\centering
\caption[Rate of line effect in WR stars]{\label{wotab:line_effect} Incidence of WR stars found to have a line effect as a percentage of number of WR star observed. The error are 68 percentile binomial confidence intervals. For each category N indicates the total number of stars in the sample. \textbf{References:} 1: \protect\cite{coyne88}, 2:\protect\cite{whitney89}, 3:\protect\cite{schulte94}, 4:\protect\cite{harries98}, 5: This work, 6: \protect\cite{vink07},  7: \protect\cite{vink17}  }
\begin{tabular}{l c c c c c c c c c c}
\hline\hline
Location & \multicolumn{2}{c}{All types} & \multicolumn{2}{c}{WN}  &  \multicolumn{2}{c}{WC} & \multicolumn{2}{c}{WO} & References\\
 & (\%) & N & (\%) & N & (\%) & N & (\%) & N  & \\
\hline
Milky Way & 19.4$\pm 4.8\%$  & 31 &  23.8$\pm  6.3\%$ & 21 &  12.5$\pm 8.0\%$ & 8 & 0$\pm - \%$ & 2 & 1,2,3,4,5\\
LMC & 10.3$\pm 3.3\%$ & 39 & 13.6$\pm  5.0 \%$ & 22 & 5.9$\pm  3.9 \%$ & 17 & -- & 0 & 6, 7\\
SMC & 8.3$\pm 5.4\%$ & 12 & 9.0$\pm 5.9\%$  & 11 & -- & 0 & 0$\pm - \%$ & 1 & 7 \\
\hline
Total & 13.4$\pm 2.6\%$ & 82 & 16.6$\pm 3.4\%$ & 54 & 8$\pm 3.7\%$ & 25 & 0$\pm -\%$ & 3 &\\

\end{tabular}

\end{table*}

On the whole, it seems clear that younger WR stars (i.e. nitrogen sequence) shows a higher rate of line effect, although the apparent statistical significance is subject to the caveat mentioned above.
This is in agreement with the study of \cite{vink11}, who pointed out the strong correlation between line-effect WR stars and the presence of ejecta nebulae.
The latter are the aftermath of a recent strong mass-loss episode undergone by the stars during a red super giant or LBV phase, and are therefore associated with WR sub-types in which hydrogen is present (i.e. late WN stars). 
This is consistent with the idea that heavy mass loss also causes angular momentum loss, meaning that rapid rotation can be dissipated as a WR star further evolves. 

\subsection{Rotational velocities of WR93b and WR102}

Stellar rotation, if rapid enough, can have dramatic effects on surrounding winds, resulting in flow trajectories that decrease the polar wind densities and increase wind densities at the equator: this is known as the Wind Compressed Disk model \citep{bjorkman93, owocki94}.

At smaller rotation speeds, mild departures from a spherical flow are expected to cause Wind Compression Zones (WCZ -- \citealt{ignace96}). 
This model can be used to investigate the dependence of polarisation on stellar rotation speed, and therefore deduce rotational velocity limits from the polarisation limits calculated. 

The analytical derivation and development of this model was done by Professor Richard Ignace, and are therefore omitted here. 
More details can be found in our publication \cite{stevance18}. 

\begin{figure}
	\includegraphics[width=\columnwidth]{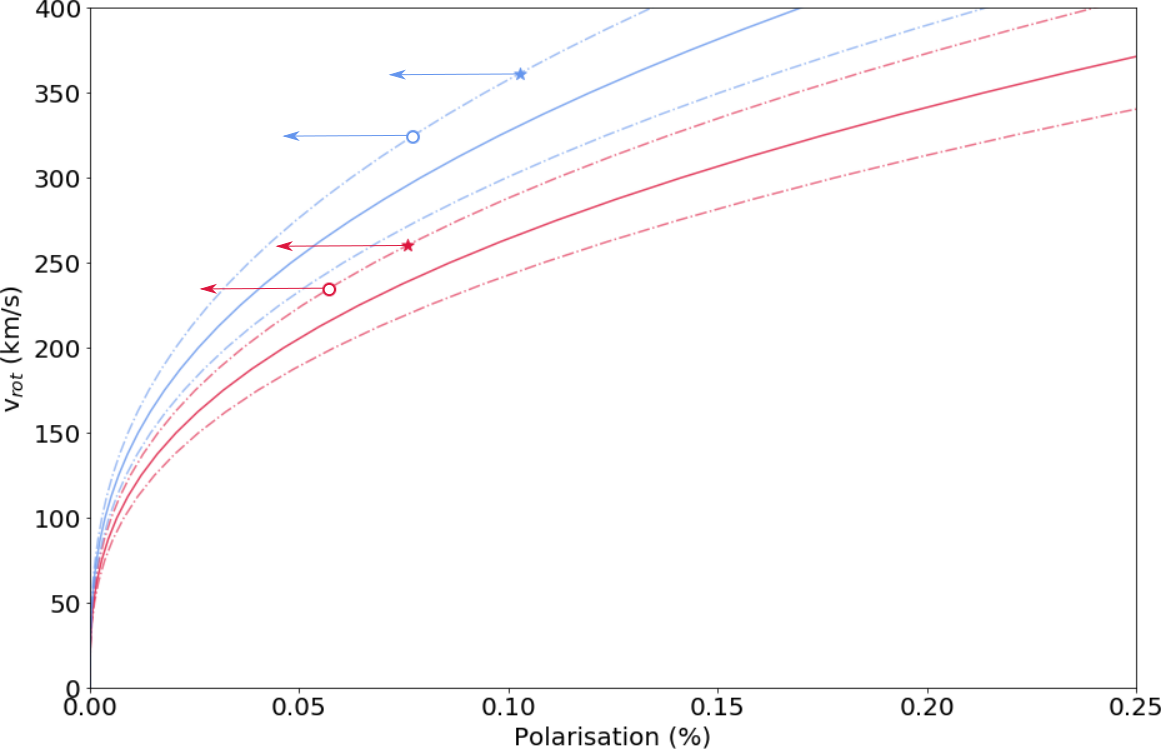}
    \caption[Rotational velocity limits]{\label{wofig:vrotpol} Relation between polarisation and rotational velocity for WR93b (blue), and WR102 (red). The dashed lines represents 1 $\sigma$ limits. The polarisation limits derived in Section \ref{wosec:lim_pol} are marked by the open circles while the polarisation limits scaled by $<\sin i>^2$ are represented by the star markers.  }
\end{figure}

The WCZ models provide a prediction of the net linear polarisation as a function of $w_{\rm rot} = v_{\rm rot}/v_\infty$.  
Terminal wind speeds of 5000 km\,s$^{-1}$ are reported for both WR93b and WR102 \citep{tramper15}, and were used to obtain the solid lines shown on Figure \ref{wofig:vrotpol}.
Note that the curve is for a model that assumes the axisymmetric wind is viewed edge-on. 
This configuration maximises the observed polarisation and therefore minimises the limit placed on the rotation velocity. 
Although it is unlikely for either WR93b or WR102 to be edge-on systems, it is also unlikely that either (or that both) would be observed near pole-on.
Therefore, in addition to considering the calculated polarisation limits (which represent the edge on case), the polarisation limits found in Section \ref{wosec:lim_pol} were scaled for $<\sin^2 i> = 0.75$ to represent the expected value of a star with random inclination.  
The intersection of the polarisation limits and the upper limits on the curves of WR93b and WR102 are indicated by the markers on Figure \ref{wofig:vrotpol}.  
The corresponding rotational velocities are summarised in Table \ref{wotab:vrot}.

\begin{table}[h!]
\centering
\caption[Rotational velocity limits]{\label{wotab:vrot} Summary of the upper limits on the rotational velocities of WR93b and WR102 obtained for a 90\degree inclination and a random inclination, as well as the resulting limit on $v_{\text{rot}}/v_{\text{crit}}$ and specific angular momentum $j=v_{\rm rot}R_{*}$. }
\begin{tabular}{c c c c c}
\hline\hline
 WR & Scaled p  & $v_{\text{rot}}$ & $v_{\text{rot}}/v_{\text{crit}}$ & $\log(j)$ \\
 &(\%) & (km\,s$^{-1}$) & (\%) & (cm$^2$/s) \\
\hline
\multicolumn{5}{c}{$i=90$\degree}\\
\hline
93b & $<$0.077 & $<$324 & $<$19 & $<$18.0 \\
102 & $<$0.057 & $<$234 & $<$10 & $<$17.6\\
\hline
\multicolumn{5}{c}{$<\sin i>$}\\
\hline
93b & $<$0.190 & $<$457 & $<$26 & $<$18.1 \\
102 & $<$0.141 & $<$327 & $<$14 & $<$17.7\\
\hline
\end{tabular}
\end{table}

It should be noted that the wind compression effect as considered here does not include the role of additional rotational effects such as gravity darkening or oblateness. 
For the case of fast rotating B stars, \cite{cranmer95} found that the inclusion of these non-radial force components would weaken the wind compression effect, which would lead to rotational velocity underestimates of 5-10 percent.
Unfortunately no similar study exists for the specific case of WR stars and such investigation is beyond the scope of this chapter, however considering the B star case our upper limits on rotational velocities may be even higher if other rotational effects could be taken into account.

Lastly, it is worth pointing out that our upper limits on the rotational velocity of WR102 are much lower than the rotational velocity of \about 1000 km\,s$^{-1}$ inferred by \cite{sander12} from studying the shape of the emission lines in the spectrum.
The model used by \cite{sander12} used a number of assumptions and physical stellar parameters that are not consistent with the present study, which could explain the difference in values.
Most notably they assumed spherical symmetry and used physical parameters significantly different from those adopted here.
Additionally, assuming a rotational velocity of 1000\kms for WR102 would result in $v_{\rm rot}/v_{\text{crit}} = 44^{+10}_{-8}$ percent (see Section \ref{wosec:vcrit})  which is much greater than observed in Galactic WR stars that did show a line effect \citep{harries98}.
These factors suggest that $v_{\rm rot}=$ 1000\kms may be an overestimate for WR102, warranting a re-evaluation of the \citeauthor{sander12} approach using  parameters based on Gaia DR2 distance.

\subsection{$v_{\text{rot}}/v_{\text{crit}}$ and specific angular momentum $j$}
\label{wosec:vcrit}
In order to fairly compare the rotation of the targets to other WR stars it is useful to consider the ratio $v_{\rm rot}/v_{\text{crit}}$. 
The values of $v_{\text{crit}}$ can be calculated using 
\begin{equation}\label{woeq:vcrit}
v_{\text{crit}} = \sqrt{\frac{\text{GM}_{\star}}{\text{R}_{\star}} \times (1-\Gamma_e)},
\end{equation}
where M$_{\star}$ is the stellar mass, R$_{\star}$ is the stellar radius, and $\Gamma_e$ is the Eddington factor \citep{langer98}.
Both the Eddington factor and the stellar mass are functions of the stellar luminosity \citep{schaerer92, vink15}, thus $v_{\text{crit}}$ can be calculated from the stellar luminosity (L$_{\star}$) and the stellar radius R$_{\star}$. 
We find $\Gamma_e$ = 0.12 and 0.11 for WR93b and WR102, respectively.

The recent DR2 data release of Gaia parallaxes \citep{gaia_parallaxes} allowed new distances to WR93b and WR102 to be calculated:  2.3 $\pm 0.3$ kpc and 2.6 $\pm 0.2$ kpc, respectively (Rate et al. in prep). 
These estimates can be used to calculate updated values of stellar parameters from the \citeauthor{tramper15} values (see their table 4) which are summarised in Table \ref{wotab:phys_param}.
The corresponding values of $v_{\text{crit}}$ are found to be: $v_{\text{crit}}=1734^{+545}_{-415}$ \kms and $v_{\text{crit}}=2286^{+528}_{-429}$ \kms for WR93b and WR102, respectively. 
The resulting limits on the values of $v_{\text{rot}}/v_{\text{crit}}$ for a 90\degree and a random inclination are also summarised in Table \ref{wotab:vrot}.
%

\begin{table}
\centering
\caption[Physical parameters WR 93b and WR 102]{\label{wotab:phys_param} Summary of the physical parameters of WR93b and WR102 and their calculated critical velocities.}
\begin{tabular}{c c c c c c }
\hline\hline
 WR & M$_{\star}$ & log(L$_{\star}$) &  R$_{\star}$ & $v_{\text{crit}}$ & d\\
 &\msol & \lsol  &\rsol & km\,s$^{-1}$ & kpc\\
\hline

93b & $7.1^{+2.4}_{-1.8}$ & $4.96 \pm 0.22 $ & $0.39^{+0.11}_{-0.09}$ & $1734^{+545}_{-415}$ & $2.3 \pm 0.3$\\
102 & $7.0^{+1.8}_{-1.4}$ & $4.95 \pm 0.17 $ & $0.23^{+0.05}_{-0.04}$ & $2286^{+528}_{-429}$ & $2.6 \pm 0.2$\\
\hline

\end{tabular}
\end{table}

Additionally, based on the calculated upper limits of $v_{\rm rot}$ for WR93b and WR102, upper limits for their specific angular momentum ($j$) can be calculated and compared to the threshold of the collapsar scenario  ($j\geq 3\times10^{16}$ cm$^2$/s -- \citealt{macfadyen99}) as well as other Galactic WR stars \citep{grafener12}.
The values of $j$ calculated for both targets for an inclination of 90\degree and a random inclination are summarised in Table \ref{wotab:vrot}. 

It is found that both limits on $v_{\text{rot}}/v_{\text{crit}}$ and $j$ are very similar to the values calculated for Galactic WR stars showing a line effect \citep{harries98,grafener12}. 
Note that the rotational velocities inferred for the WR stars in \cite{harries98} and \cite{grafener12} were calculated using spectroscopic and photometric variability, which rely on fewer assumptions than the method employed here and are therefore more robust. 
We can see that the limits placed on $j$ for our WO stars exceed the threshold for the collapsar model, and therefore cannot exclude WR93b and WR102 being LGRB progenitors.
However LGRBs are seen to prefer low metallicity environments \citep{graham13}, making this scenario highly unlikely.

Finally, these results indicate that the absence of a line effect is not necessarily synonymous of insignificant rotation. 
When investigating the presence of rotation of WR stars using spectropolarimetry, caution is therefore required when interpreting non detections as the absence of a line effect is not a direct measure of the absence of rapid rotation.


\section{Conclusions}
\label{sec:conclusion}
This chapter presented FORS1 spectropolarimetric data of WR93b and WR102, which is the first spectropolarimetric data set obtained for Galactic WO stars.
The main results of this work are:
\begin{enumerate}
\item[1)] No line effect is found for either WR93b and WR102.
\item[2)] Upper limits for continuum polarisation of $P_{\text{cont}} < 0.077$ percent and $P_{\text{cont}} < 0.057$ percent were deduced for WR93b and WR102, respectively.
\item[3)] The corresponding upper limits on the rotational velocity for an edge-on case and a velocity law $\beta=1$ are $v_{\rm rot}<324$~\kms and $v_{\rm rot}<234$~\kms, for WR93b and WR102, respectively.
\item[4)] Upper limits on $v_{\text{rot}}/v_{\text{crit}}$ were found: $<$19 percent and $<$10 percent for WR93b and WR102, respectively. 
\item[5)] Lastly we calculated upper bounds for the specific angular momentum of WR93b and WR102: log($j$)$<$18.0 and  log($j$)$<$17.6 (cm$^2$/s), respectively. 
These values do not exclude the collapsar model and therefore it is not possible to constrain the fate of WR93b and WR102, although the preference of LGRBs for low-metallicity environment makes this outcome highly unlikely.
\end{enumerate}

The upper limits on $v_{\text{rot}}/v_{\text{crit}}$ and log($j$) were found to be similar to values found for Galactic WR stars showing a prominent line effect. 
Consequently this shows that the absence of a line effect is not necessarily synonymous with the absence of rapid rotation.

\chapter{Conclusions}

\label{chpt:conclusions} 

\lhead{\emph{Conclusions}} 

In this last chapter we summarise the work presented in this thesis, offer some general remarks on our findings and briefly address future prospects.

\section{Summary of the new supernova spectropolarimetric data sets}
For over 30 years now, spectropolarimetry has been a fantastic technique to probe to the geometry of CCSNe.
Spectropolarimetric observations, however, are challenging.
Consequently the number of multi-epoch spectropolarimetric data sets for CCSNe remains low: only a dozen in the past three decades (see Table \ref{introtab:sn_specpol}).
The work presented in this thesis includes multi-epoch data for three new SNe, therefore increasing this sample by 20 percent. 
The type IIb SN 2008aq was observed at two epochs (+16 days and +27 days after explosion), while we obtained 7 epochs of spectropolarimetry for both the type IIb SN 2011hs (from $-$3 days to +40 days with respect to V band maximum) and the type Ic-bl SN 2014ad (from $-$2 days to +66 days with respect to V band maximum). 
Furthermore, the latter is to the best of our knowledge the most exhaustive data set obtained for a SN of this type up to this point.

\subsection{Type IIb SN 2008aq and SN 2011hs} 

In Chapter \ref{chpt:08aq}, we found that SN 2008aq exhibited a low degree of continuum polarisation at the first epoch ($p$\about0.2 percent).
By the second epoch, the helium spectral features had significantly strengthened, and the continuum polarisation had risen to ($p$\about1.3 percent). 
Additionally, the $q-u$ plots of SN 2008aq showed very little elongation at the first epoch, but a clear dominant axis by  +27 days after explosion. 
These results show a departure from sphericity that increased significantly as the photosphere receded to the deeper part of the ejecta into the helium layers. 
This is similar to the behaviour observed in SN 2001ig \citep{maund01ig}.

On the $q-u$ plane, the spectropolarimetric line features of $\mathrm{H\alpha}$ and He\,{\sc i}$\lambda5876$ develop clear loops at +27 days. 
This demonstrates a departure from axial symmetry.
Additionally, while the loop of helium evolves clockwise on the $q-u$ plot, the hydrogen loop shows an anti-clockwise rotation.
Consequently, the geometrical effects causing these loops are distinct and reveal line-specific asymmetries. 
These could be caused by partial obscuration of an ellipsoidal photosphere by clumpy line forming regions.

The case of SN 2011hs (see Chapter \ref{chpt:11hs}) was substantially different from that of SN 2008aq. 
Significant continuum polarisation was detected at the first two epochs ($p=0.55\pm0.12$ percent and $p=0.75\pm0.11$ percent at $-3$ and +2 days, respectively).
As opposed to SN 2008aq, whose polarisation increased to $>$ 1 percent as helium lines strengthened, the continuum polarisation of SN 2011hs decreased by \about 0.25 percent +10 days after V-band maximum, and further declined by +18 days. 
This is reminiscent of SN 2011dh, whose continuum polarisation was \about 0.45 percent at 11 and 6 days before V-band maximum, but then decreased down to $p <$ 0.2 percent by +8 days.

The continuum polarisation of SN 2011hs also showed a progressive rotation from epoch 2 to 4 (after which the continuum P.A. could not be evaluated precisely).
This was interpreted as the presence of a third axis coming into play causing a break from bi-axial geometry. 
The rotation of the continuum P.A. was followed by that of helium, hydrogen and calcium line features, indicating that global geometry significantly impacted the line polarisation.
One possible interpretation is the presence of an off-axis energy source present deep within the ejecta at early days, below an ellipsoidal photosphere (which is bi-axial). 
As the photosphere recedes, the off-axis energy source would become exposed over time, causing the continuum and line polarisation to rotate.  
A correlation between helium, hydrogen and continuum P.A. was also observed for SN 2011dh, and the preferred interpretation of \cite{mauerhan15} was that of nickel plumes coming from the centre of the ejecta.
Asymmetric nickel distributions could also be the off-axis energy source we refer to in the case of SN 2011hs.

Furthermore, the distinct evolution of the hydrogen, helium and calcium loops over time in SN 2011hs indicates that line specific geometries are also a contributor to the line polarisation, in addition to the effects of the global geometry. 
Anisotropic distributions of the line forming regions causing partial occultation of the photosphere could therefore also be present in SN 2011hs. 

On the whole, these two data sets (i.e. SN 2008aq and SN 2011hs) exhibit very different behaviours and reveal drastically different geometries. 
This brings further evidence that the spectropolarimetry of type IIb SNe, and their geometry, can vary significantly on a case by case basis.

\subsection{Type Ic-bl SN 2014ad}

In Chapter \ref{chpt:14ad} we presented the spectropolarimetric data for the type Ic-bl SN 2014ad.
The spectropolarimetric line features were extremely broadened by the very high ejecta velocities (\about 30,000\kms at $-2$ days), making line identification difficult.
We were also not able to isolate a region of the spectrum devoid of strong lines to measure the continuum polarisation. 

Overall, the degree of polarisation is seen to go down over time, and the clear dominant axis visible on the $q-u$ plots at the first epoch becomes significantly less prominent epoch after epoch.
Therefore, we can still conclude that the outer ejecta has significant global elongation and becomes more spherical towards the centre. 
Additionally, the [O\,{\sc i}] line at 107 days after V-band maximum shows a flat topped profile, centred on the rest wavelength of the line, which is characteristic of the presence of a spherical oxygen shell in the deep ejecta. 

The evolution of the polarisation associated with oxygen and calcium on the $q-u$ plots between $-2$ days and $+18$ days shows that they arise from very distinct geometries. 
This suggests that the line forming regions are clumpy and anisotropically distributed at different depths. 
Oxygen is seen to evolve earlier than calcium, when the photosphere probes parts of the ejecta further out. 
Furthermore, the calculated oxygen velocities are higher than that of calcium.
We concluded that the 'onion' layer of the progenitor is partially conserved in SN 2014ad, and that the calcium likely originated from nucleosynthesis. 

Our spectropolarimetric analysis did not allow us to conclude whether SN 2014ad did or did not have jets, although the distributions of the oxygen and calcium layers was consistent with jet-driven explosion models of \cite{couch11}.

\section{The challenge of determining the ISP}

Quantifying the ISP is a crucial step to retrieving the intrinsic polarisation of the object of interest, and was very important to the analysis of the above studies. 
Over the past three decades, and in this thesis, a number of methods have been used, most of them involving the assumption that some part of the spectrum is completely depolarised. 

Using the polarisation associated with the $\mathrm{H\alpha}$ emission component as a proxy for ISP is a technique that was used in the case of SN 1993J by \cite{trammell93} and \cite{tran97}, and that we employed in Chapter \ref{chpt:08aq} for SN 2008aq.
Unfortunately, the assumption of complete depolarisation in the $\mathrm{H\alpha}$ emission has previous been shown to not necessarily be accurate \citep{tanaka09}.
In the case of SN 2008aq, the ISP corrected data exhibited very unexpected behaviours that were most likely the result of an inaccurate ISP estimate.
For this particular SN, the ISP was very low, and therefore it was possible to work with ISP uncorrected data. 

This is not always the case, however.
In SN 2011hs, the SN polarisation features were only revealed once the ISP had been removed. 
In such cases, being able to reliably determine the interstellar component is not only important, it is essential to being able to perform a subsequent analysis. 
A number of methods were attempted to estimate the ISP in the data of SN 2011hs.
We tried to decompose the polarisation of strong emission lines to separate the continuum polarisation from the line polarisation. 
Under the assumption that intrinsic line polarisation is zero, the values found for its component would actually reveal the ISP. 
Even if this assumption was completely valid, the method requires an estimate of the continuum flux across a line, and this is extremely difficult to do accurately. 
In the end, we were not able to obtain results from this technique. 

Another approach is to consider the strong emission lines at late time. 
The decrease in intrinsic continuum flux and polarisation at these dates rendered the assumption that line polarisation equated to ISP more robust. 
This was successfully used in SN 2014ad by averaging the stokes parameters in the strong emission line regions.
In SN 2011hs, we fitted a line through the data points associated with the strong emission lines to obtain an ISP estimate. 
This was also satisfactory, although the errors on the fit were quite large and propagated to the ISP corrected data. 

The final method used to remove the ISP of SN 2011hs was to fit with a straight line the spectropolarimetric data at the last epoch across the whole spectrum, using sigma clipping to remove outliers. 
This is independent of any assumption on the emission lines, but only works if the fitted data is taken at an epoch were the intrinsic SN polarisation is negligible compared to that induced by the ISM.

One last approach mentioned in this thesis was the use of the spectral region affected by line blanketing (blue parts of the spectrum \about5000\r{A}).
This was used in SN 2014ad successfully, but one major caveat to this method is that it cannot be used reliably if line polarisation arises in this region of the spectrum (e.g. see SN 2008aq at +27 days and Table \ref{93Jtab:pol_comp}).

On the whole, estimating the ISP is difficult. 
All methods rely on assumptions, and they cannot be universally used without some degree of healthy scepticism. 
In some cases were the ISP is low, like SN 2008aq  or SN 2011dh,  it is preferable not to attempt to remove the ISP, since a wrong estimate can severely impact the data (see Chapter \ref{chpt:08aq}, \citealt{mauerhan15}).

\section{To use or not to use a toy model: that is the question}

In order to further interpret the observed spectropolarimetry of SN and provide a more detailed picture of their ejecta, models are required. 
Fully hydrostatic models with radiative transfer are not yet available.
Monte Carlo toy models have been used in the past by \cite{maund05hk}, and \cite{reilly16, reilly17} as a way to provide insight on the geometries resulting in the observed polarisation. 
Unfortunately, these models were used without performing a full exploration of parameter space, and one of our main goals was to do this methodically to understand the range of possible solutions to a given observation. 

These toy models are two dimensional and work under a number of assumptions. 
An ellipsoidal photosphere (surface of last scattering) is created by randomly generating photon packets and attributing each of them a luminosity and a polarisation according to their position. 
Some fine tuning is required for the polarisation of a photosphere with a given axis ratio to reproduce the results of more exhaustive models such as those of \cite{hoflich91}.
The full set of formulae used can be found in Chapter \ref{chpt:tm}. 

Line polarisation in these models is only considered to arise from partial obscuration, and therefore do not take into account the potential effects of global geometry on line polarisation (see Section \ref{introsec:pol_origin}).
Line forming regions are placed in front of the photosphere, and are given a very high optical depth such that all the underlying photon packets are removed from subsequent calculations. 
In our models we used one circular line forming region, described by 3 parameters: its radius ($r$) and the Cartesian coordinates of its centre ($x$, $y$).
After applying our line forming region to our ellipsoidal photosphere we averaged the Stokes parameters $q$ and $u$, and calculated the ratio of the flux after line absorption to the total (i.e. continuum) flux. 
This could then be compared to the values observed in the data.   
Note that the models of \cite{reilly16,reilly17} included more complex geometries. 
As a first approach, however, we focused on the most simple case (a single circular absorption region), as one of our goals was to explore parameter space methodically.

In order to do this, we used a grid-search approach and created look-up tables to summarise the Stokes parameters and flux ratio simulated for each set of three parameters. 
Observations could then be compared to these values.
Because it would have been overly time consuming to create a fine enough grid to provide 1$\sigma$ solutions to all our observations, we first isolated the region of parameter space, including 3$\sigma$ solutions, and then ran more models with a finer grid focused on that particular region of parameter space. 

We used our models on the helium, hydrogen and calcium data of SN 2011hs from $-$3 to +10 days (see Chapter \ref{chpt:tm}).
Although solutions were found for all the observables we attempted to model, we also found a great number of degeneracies.
Even more degeneracies would occur if more parameters were considered.
This is a concern for previous models that did not fully explore parameter space whilst using more complex geometries. 
Additionally, we also found good solutions at +10 days, although our analysis of SN 2011hs in Chapter \ref{chpt:11hs} revealed that the global geometry was tri-axial rather than bi-axial, invalidating the assumption that the photosphere can be represented as a simple projected ellipse. 
Consequently, this type of toy model can be misleading and should be used with great care, if at all. 

\section{Spectropolarimetry of WO stars and progenitor link}

Spectropolarimetry has been used extensively to study the asymmetries of WR star winds.
The presence of a line effect (peaks or troughs associated with strong emission lines) has been used as a proxy for rapid rotation, as a wind distorted by rapid rotation would yield net continuum polarisation. 

In Chapter \ref{chpt:wos}, we studied for the first time the spectropolarimetry of two Galactic WO stars: WR93b and WR102.
WO stars in particular are candidate progenitors for type Ic-bl SNe, and the collapsar model by which they are thought to create these SNe requires enough angular momentum to power the central engine \citep{macfadyen99}·

We found no line effect in either WR93b nor WR102.
We used a Serkowski law fit to the data to remove the contribution of the ISP and continuum, to help us place upper limits on the detectability of the line effect.
This in turn allowed us to derive upper limits for the continuum polarisation of $P_{\text{cont}} < 0.077$ percent and $P_{\text{cont}} < 0.057$ for WR93b and WR102, respectively. 

Using an analytical model by Richard Ignace (private communications), we then constrained the rotational velocity to $v_{\rm rot}<324$~\kms and $v_{\rm rot}<234$~\kms, for WR93b and WR102, respectively.
Note that this is for an edge-on case  and a velocity law $\beta=1$.
These values correspond to upper limits on $v_{\text{rot}}/v_{\text{crit}}$ of $<$19 percent and $<$10 percent for WR93b and WR102, respectively. 
Additionally, from the above rotational velocities we can derive limits on the specific angular momentum of log($j$)$<$18.0 and log($j$)$<$17.6 (cm$^2$/s), for WR93b and WR102 respectively. 
These are not sufficient to exclude the possibility of a collapsar event, and we could not further constrain the fate of WR93b and WR102.

Lastly, the upper limits on $v_{\text{rot}}/v_{\text{crit}}$ and log($j$) were found to be similar to values found for Galactic WR stars showing a prominent line effect. 
Consequently this shows that the absence of a line effect is not necessarily synonymous with the absence of rapid rotation.

\section{Future work}

In addition to the work summarised in the previous sections, we also provided a new estimate of the ISP of SN 1993J and preliminary re-analysis of its spectropolarimetric data (see Chapter \ref{chpt:93J}).
Further analysis is required and discussions are on-going with my collaborators. 

As mentioned previously, one of the main road-blocks to exhaustively interpreting SN spectropolarimetry is the need for better modelling.
3D hydrostatic models with radiative transfer that can reproduce both the polarisation and the flux spectra will be the key to extracting more information from the spectropolarimetric data sets already in the literature, and the ones to come in the future.

On the whole, the number of multi-epoch data sets available for CCSNe still remains limited, now counting 15 objects. 
It would be beneficial to obtain time coverage at a higher cadence as it would offer better depth resolution.
Additionally, being able to follow up SNe very soon after explosion will be made easier by initiatives such as the Zviky Transient Factory and the Large Synoptic Survey Telescope. 
Currently, SN 2008ax is our earliest sample, observed only 6 days after explosion. 

The fact that we are now finding SNe at a greater rate than ever before, however, will not necessarily result in a dramatic increase in the number of high-quality multi-epoch spectropolarimetric data sets. 
Indeed, these observations remain very challenging, requiring long exposure times, even on 8 meter class telescopes. 
Additionally, spectropolarimetric analysis is difficult, particularly the ISP removal process. 
This will require a lot of human input and is not easily automated. 
On the whole, although spectropolarimetry is a very useful technique, the diversity seen in the currently available data shows that a much larger sample will be required to find global trends within and across SN sub-types. 
This will be challenging, and will require an observing and analysis strategy that is scalable to the new era of Big Data Astronomy.

\end{onehalfspace}

\begin{singlespace}

\bibliographystyle{mn2e} 

\end{singlespace}  

\onehalfspacing
\appendix

\clearpage



\chapter{Investigating the original reduction of SN 2008aq}
\label{app:08aq}

\lhead{\emph{Investigating the original reduction of SN 2008aq}}

Unlike the data presented in Chapter \ref{chpt:08aq}, originally published data in \cite{stevance16} were not reduced by myself using {\sc FUSS}.
Instead, the analysis was performed on polarisation that had been pre-reduced. 

Ulteriorly, the raw data of SN 2008aq were reduced using {\sc FUSS}, revealing a number of discrepancies which brought me to investigate the original data further (details below).
The issues identified being minor, they do not affect the main conclusions of our paper, however during the writing of this thesis I chose to move away from the original reduction and to instead use my own; I motivate this choice in this appendix. 

\section{Discrepant values in +16 days data}
The most visible difference between the original reduction and my own was the presence of a peak in polarisation at +16 days associated with the Ca\,{\sc ii} infrared triplet in the original data reduction that could not be reproduced.
In order to investigate the nature of this peak, one can look at the behaviour of the difference in instrumental signature correction (\deps), which is expected to be \about 0  percent for small values of $p$  ($<20$ percent -- see Section \ref{datred:deps}). 
Values of \deps that strongly deviate from zero indicate that the polarisation data associated with these bins are not representative of a real signal. 
Consequently, comparing the degree of polarisation from original and the new data reduction to their respective \deps is a good way to check for discrepancies in both reductions. 

\begin{figure}[h!]
\centering
\includegraphics[width=15cm]{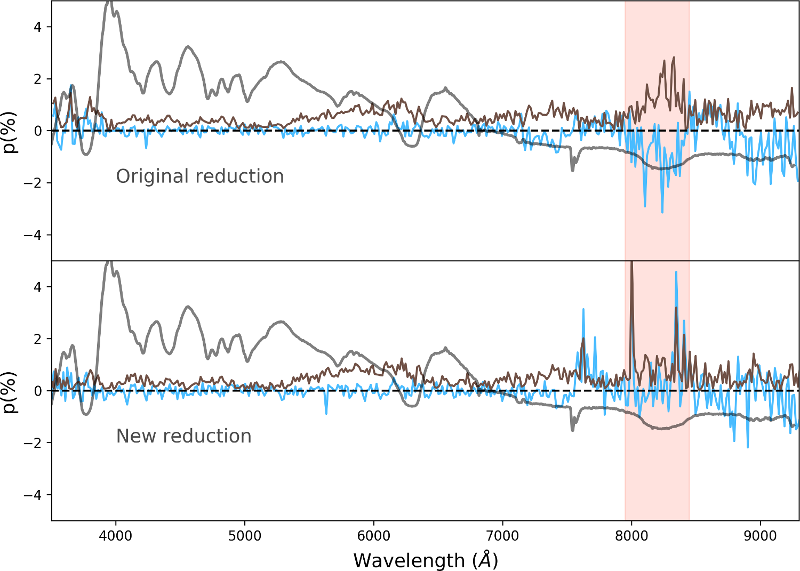}
\caption[Comparison of the original and revised polarisation and $\Delta \epsilon$]{Comparison of the degree of polarisation (brown) and \deps (blue) of the original and new data reduction of SN 2008aq. The flux spectrum is shown in grey for reference. A non-zero \deps indicates that the polarisation data is not representative of a real signal. The shaded region highlights the absorption feature of the calcium infrared triplet.}
\label{app08aqfig:08aq_deps}
\end{figure}

Figure \ref{app08aqfig:08aq_deps} shows $p$ (brown), \deps (blue) and the flux spectrum (grey) of SN 2008aq at +16 days obtained from the original data reduction and the new one. 
Both data sets show increased noise in the red part of the spectrum which is expected. 
The new reduction exhibits a few strong outliers in $p$ at 7630, 8005, 8350, and 8410 \r{A}, which are matched by spikes in \deps. 
Apart from these points, the degree of polarisation is quite flat (albeit noisy) across the calcium triplet feature. 
However, the original reduction does seem to show a polarisation feature (brown) across the absorption feature of Ca\,{\sc ii}. 
Unfortunately, this feature is matched closely by a departure from zero in \deps (blue). 

Therefore we must conclude that the calcium feature initially reported in \cite{stevance16} with an amplitude of 1.5$\pm0.6$ percent is not real, or at least not as pronounced as previously thought.

\section{Errors on the degree of polarisation in the original reduction}
\label{app08aqsec:deps}
Additionally, in the original reduction the errors on the degree of polarisation  were inconsistent with the errors on the Stokes parameters (i.e parsing $\Delta q$ and $\Delta u$ in Equation Equation \ref{introeq:pol_err} did not reproduce the reported $\Delta p$).

To investigate this discrepancy, I recalculated $p$ and the associated errors from the Stokes parameters in the original reduction (using Equation \ref{introeq:pol_err}) and compared it to the values of $p$ and $\Delta p$ reported in the original polarisation data.
The errors on $p$ as calculated directly from the Stokes parameters are much grater (\about 15 times) than the original errors on $p$. 

To understand where such a difference could originate from, I attempted to reproduce the reported errors on $p$ by modifying Equation \ref{introeq:pol_err}.
Being told that the issue arose from a missing factor of $p$ in the denominator of Equation \ref{introeq:pol_err} (Maund, private correspondence), I tested two modifications of Equation \ref{introeq:pol_err}: 
\begin{equation}\label{app08aq:pol_err1}
\Delta p = \sqrt{(q \Delta q)^2 + (u \Delta u)^2}
\end{equation}
\begin{equation}\label{app08aq:pol_err2}
\Delta p = \sqrt{(1/p)\times(q \Delta q)^2 + (u \Delta u)^2}.
\end{equation}

\begin{figure}[h!]
\centering
\includegraphics[width=13cm]{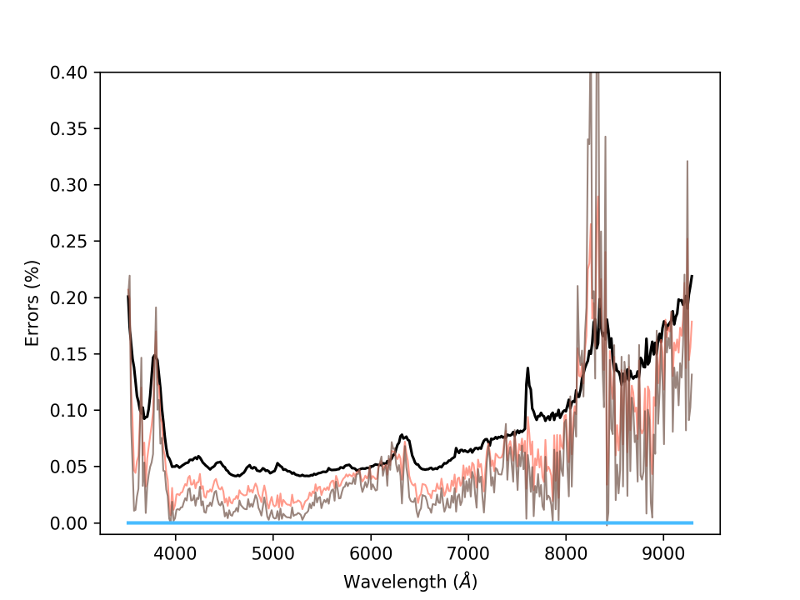}
\caption[Residuals of error reproduction]{Residuals when comparing recalculated errors on $p$ and originally reported values of $\Delta p$.
The black line shows the residuals for the recalculated $\Delta p$ using Equation \ref{introeq:pol_err}, the red line shows the residuals for errors calculated using Equation \ref{app08aq:pol_err2}, the brown line shows the residuals for errors on $p$ calculated using Equation \ref{app08aq:pol_err1} and, lastly, the blue line shows the residuals when comparing the originally reported errors to a reproduction using Equation \ref{app08aq:pol_err2} and a missing factor of 10.}
\label{app08aqfig:08aq_errors}
\end{figure}

The calculations were performed with the original values of $q$ and $u$, as well as their errors $\Delta q$ and $\Delta u$, as they would have been used to calculated the original values of $p$ and $\Delta p$.
Figure \ref{app08aqfig:08aq_errors}, shows the residuals obtained from subtracting the original values of  $\Delta p$ at epoch 1 to the attempted reproductions. 

It is clear that the residuals when comparing the original errors on $p$ to the ones calculated from Equation \ref{app08aq:pol_err2} divided by 10 (blue) are null. 
Qualitatively, the average of the residuals is found to be 2.19$\times10^{-15}$ percent which is consistent with python rounding errors.
A similar result is found when performing the same procedure on the data at the second epoch. 
Consequently, it must be concluded that the original errors on $p$ were underestimated by a factor of $10\sqrt{p}$


\clearpage


\chapter{List of Publications}
\label{app:papers}

\lhead{\emph{List of Publications}}

This is a list of publications resulting from the work I did during my PhD or to which I consider I significantly contributed to.

\begin{itemize}

\item 	\textbf{Stevance, H. F.}; Maund, J. R.; Baade, D.; Bruten, J.; Cikota, A.; Höflich, P.; Wang, L.; Wheeler, J. C.; Clocchiatti, A.; Spyromilio, J.; Patat, F.; Yang, Y.; Crowther, P., 2019, MNRAS, 485, 102, "The 3D shape of SN 2011hs" \footnote{Presented in Chapter \ref{chpt:11hs}}.

\item 	\textbf{Stevance, H. F.}; Ignace, R.; Crowther, P. A.; Maund, J. R.; Davies, B.; Rate, G., 2018, MNRAS, 479, 4535, "Probing the rotational velocity of Galactic WO stars with spectropolarimetry"\footnote{Presented in Chapter \ref{chpt:wos}}.

\item \textbf{Stevance, H. F.}; Maund, J. R.; Baade, D.; Höflich, P.; Howerton, S.; Patat, F.; Rose, M.; Spyromilio, J.; Wheeler, J. C.; Wang, L., 2017, MNRAS, 469, 1897, "The evolution of the 3D shape of the broad-lined Type Ic SN 2014ad"\footnote{Updated in Chapter \ref{chpt:14ad}}.

\item 	\textbf{Stevance, H. F.}; Maund, J. R.; Baade, D.; Höflich, P.; Patat, F.; Spyromilio, J.; Wheeler, J. C.; Clocchiatti, A.; Wang, L.; Yang, Y.; Zelaya, P., 2016, MNRAS, 461, 2019, "Spectropolarimetry of the Type IIb SN 2008aq"\footnote{Updated in Chapter \ref{chpt:08aq}}.

\item 	Higgins, A. B.; Wiersema, K.; Covino, S.; Starling, R. L. C.; \textbf{Stevance, H. F}.; Wyrzykowski, Ł.; Hodgkin, S. T.; Maund, J. R.; O'Brien, P. T.; Tanvir, N. R., 2019, MNRAS, 482, 5023, "SPLOT: a snapshot survey for polarized light in optical transients".

\end{itemize}\clearpage
\includepdf[pages={1}]{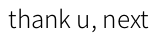}

\end{document}